Technical Report OBU-ECM-AFM-2018-02

# Datamorphic Testing: A Methodology for Testing AI Applications

**Hong Zhu[(*)], Dongmei Liu[(+)], Ian Bayley[(*)], Rachel Harrison[(*)], and Fabio Cuzzolin[(*)]**

[(*)] School of Engineering, Computing and Mathematics,
Oxford Brookes University, Oxford, UK
Emails: {hzhu, ibayley, rachel.harrison, fabio.cuzzolin}@brookes.ac.uk

[(+)] School of Computer Science and Engineering,
Nanjing University of Science and Technology, Nanjing 210094, China
E-mail: dmliukz@njust.edu.cn

27 December 2018

## Abstract

With the rapid growth of the applications of machine learning (ML) and other artificial intelligence (AI) techniques, adequate testing has become a necessity to ensure their quality. This paper identifies the characteristics of AI applications that distinguish them from traditional software, and analyses the main difficulties in applying existing testing methods. Based on this analysis, we propose a new method called *datamorphic testing* and illustrate the method with an example of testing face recognition applications. We also report an experiment with four real industrial application systems of face recognition to validate the proposed approach.

# Table of Contents

# Datamorphic Testing: A Methodology for Testing AI Applications

**Hong Zhu[(*)], Dongmei Liu[(+)], Ian Bayley[(*)], Rachel Harrison[(*)], and Fabio Cuzzolin[(*)]**
[(*)] School of Engineering, Computing and Mathematics, Oxford Brookes University, Oxford, UK
Emails: {hzhu, ibayley, rachel.harrison, fabio.cuzzolin}@brookes.ac.uk
[(+)] School of Computer Science and Engineering, Nanjing University of Science and Technology, Nanjing 210094, China
E-mail: dmliukz@njust.edu.cn.

***Abstract*.** With the rapid growth of the applications of machine learning (ML) and other artificial intelligence (AI) techniques, adequate testing has become a necessity to ensure their quality. This paper identifies the characteristics of AI applications that distinguish them from traditional software, and analyses the main difficulties in applying existing testing methods. Based on this analysis, we propose a new method called *datamorphic testing* and illustrate the method with an example of testing face recognition applications. We also report an experiment with four real industrial application systems of face recognition to validate the proposed approach.

## I. MOTIVATION

We have seen a rapid growth in the application of machine learning (ML), data mining and other artificial intelligence (AI) techniques to software systems in recent years [1]. Typical examples of such applications include driverless vehicles, face recognition and fingerprint recognition in security control, workload pattern learning in computer cluster operation, personalization of social media networking for business intelligence, situation recognition and action rule learning in healthcare, smart homes and smart cities, etc. Many such applications are also closely integrated with robotics, the Internet-of-Things, Big Data, and Edge, Fog and Cloud computing, which all require automated and optimized data collection and processing. AI techniques, especially ML, have been widely regarded as a promising solution to the underlying hard computational problems. All these applications must be thoroughly tested to ensure their quality [2].

For traditional software applications, testing is efficient and effective. Automated software testing tools like the XUnit automated test framework [3, 4] such as JUnit and GUI-based automated testing tools such as Selenium[(1)] are widely used across the industry. Software testing methods and techniques are employed systematically in the best practice. Research has demonstrated that a large proportion of faults can be detected before product release. Software products that pass testing can be reliable and trustworthy if testing is performed thoroughly. Testing plays a central role in software engineering, for example, in test-driven development, which is widely used in industry [5].

However, the current practice of testing AI applications lags far behind the maturity of testing traditional software applications. Testing has become a grave challenge because the distinctive features of AI applications disqualify existing software testing methods, techniques and tools.

This paper examines the difficulties when testing AI applications using existing software testing methods and techniques, and proposes a novel approach called *datamorphic testing*.

The paper is organized as follows. Section II discusses the challenges of AI applications currently confronting software test engineers. Section III introduces the basic concepts of *datamorphic testing* method. Section IV discusses the process of datamorphic testing and various testing strategies of datamorphic testing. Section V reports an experiment with four real industry ML applications of face recognition. Section VI concludes the paper with a discussion of related work and future work.

## II. THE CHALLENGES

In this section, we first identify the distinctive features of AI and ML applications. Then, we discuss the applicability of existing software testing methods and techniques.

### *A. Distinctive Features of AI Applications*

Testing AI applications is different from testing traditional software applications for the following reasons.

#### *(1) Uncertainty*

Like all software systems, the development of an AI application is subject to influences from uncertainties, which are factors that may cause random errors in the software. Testing as a quality assurance measure prevents the uncertainties from having a negative impact on software quality, detects the defects in the intermediate and final products due to uncertainties, and eliminates the defects from the product.

Three types of uncertainties have been recognized for the development of traditional software systems [6]. They are:

a) *Godel type uncertainties* due to incompleteness of the specifications of software system;
b) *Heisenberg type uncertainties* due to users' uncertainty about their requirements before they actual see and use the product;

(1) https://www.seleniumhq.org

c) *Pragmatic uncertainties* due to the software engineers' random errors during engineering process.

Existing software testing, verification and validation techniques and other software engineering techniques and methods have been developed to deal with these known types of uncertainty. No doubt, these types of uncertainty are also associated with AI applications, but they exhibit themselves in new forms. For example, the sources of pragmatic uncertainty in developing ML applications include the choices of training data, the selection of models and the fine-tuning of the model parameters. These affect the performances of ML applications significantly, and it is an art rather than a science. This gives a large new scope for pragmatic uncertainty.

Moreover, AI applications are subject to a new type of uncertainty:

d) *Algorithmic uncertainty* due to the randomness in the algorithms used in the development and operation of the software.

For example, ML applications are subject to the randomness associated with training algorithms. For instance, in the training of neural networks, the weights associated with the links between neurons may be initialized at random. Research has shown that the initial random state of a neural network may have significant impact on the final result. Many other AI algorithms, such as in genetic algorithms, multi-agent systems and swam intelligence, etc., also employ randomness. Due to the fundamental differences of the uncertainties associated to AI applications to those associated to traditional software, the testing methods and techniques developed so far are neither efficient nor effective for dealing with them. How to control such uncertainties and how to assess the quality of a product that is subject to algorithmic uncertainty in the development process are still open problems in software engineering.

*(2) Adaptive Behaviour*

AI techniques are often employed in situations that the application is required to adapt its behavior in a dynamic environment like the Internet and cloud. A typical example is online learning, such as chat-bots, which learn from conversations with online users and apply what is learnt to future conversations. For such online learning applications, it is difficult (if not impossible) to know the quality of training data in advance. Another example of online learning is cluster optimization in cloud computing, which can collect data about a system's workload, learn a workload distribution model, and predict workloads to optimize cloud resource usages by, for example, switching on servers before demand peaks. Such applications demonstrate a high level of intelligence by behaving adaptively according to the runtime situation. Mathematically speaking, such software can be regarded as a higher-order function, which takes a set of data as input and generates a function. It is particularly difficult to design test data for such systems using existing testing techniques, which are developed to deal with first-order functions.

On the other hand, testing may play a crucial role in not only the development but also the operation of such systems, for example, by performing online testing and controlling the direction of adaptation. This requires testing to be performed automatically.

*(3) Incomprehensibility*

A neural network is hard, if not impossible, to interpret logically, although much research is directed towards this problem in order to extract rules from neural networks, for example. Existing program-based testing, validation and verification techniques cannot be applied to design test cases according to the details of a neural network. How to judge or measure test adequacy according to test's coverage of the neural network structure is also an open problem.

Some machine learning algorithms do produce results that are interpretable and comprehensible. For example, decision tree learning produces a decision tree as a predictive model for classification. Decision trees are comprehensible and logically explainable. In fact, decision trees and decision tables have been used in software development for a long time. There are a few well-defined testing techniques targeting such software artifacts. Such techniques can also be applied or adapted for testing ML applications by generating test cases and measuring test adequacy.

*(4) Lack of testable specification*

ML applications are often difficult (if not impossible) to define with testable and verifiable specifications. By this, we mean an intrinsic specification (either formal or informal) based on which test cases can be generated and test results can be checked for correctness objectively. The following are typical examples of requirements statements of ML applications.

- *Face recognition*, i.e. software that detects and recognises human beings in picture and identifies who the person is by referring to a database of facial images;
- *Fingerprint recognition*, which compares a fingerprint against a set of known fingerprints stored in a database to identify a person;
- *Object recognition*, which identifies the objects in an image or video;
- *Driverless vehicles*, which reads signals from sensors and video cameras mounted on cars or other types of vehicles, recognises road conditions and the traffic, learns and predicts the pedestrian and other vehicles' moving trajectories, and controls a vehicle's motion without input from a human driver.

Statements like the above are neither testable nor verifiable because they cannot be refined to the necessary level of clarity and detail needed by engineers to design test cases and to check test results. Significant creative efforts are required to bridge the gap between such statements and testing activities.

Given these distinctive features of AI applications, we review existing software testing methods and techniques and analyse their applicability to AI applications next.

*B. Applicability of Existing Software Testing Methods*

Existing software testing techniques can be roughly classified into the following four types, but in practice they are often used in combination.

*(1) Specification Based Approach*

In this approach, test cases are designed according to the specification of the software under test, test results are checked against the specification, and test adequacy are measured according to how well the specified functions of the software are exercised. Specifications can be represented in (a) mathematical and formal notations, (b) semi-formal graphical notations, or (c) natural language for defining the functional and non-functional requirements. The last of these, using structured natural language, is the most widely used. For example, in test-driven development methodology, functional requirements are specified by user stories from which test cases and/or executable test code are derived manually from narrative descriptions of software functions. The format of such user stories, e.g. JBehave[(2)], focuses on user-computer interactions, where each step is a fairly simple computation. In contrast, the computations realized by ML are normally much more complicated. Consequently, user story descriptions in a format like JBehave are insufficient for deriving test cases and for checking test results for ML applications.

Formal and semi-formal specifications make it possible to test and check test results automatically. However, existing formal methods (like process algebras, Petri-nets, state-charts, Z and B etc.) rely on state-based model of computation, which mismatch the computation models of AI algorithms like connectionist, swam intelligence, game theory, rule-based systems, heuristic rules, multi-agent systems based on modal logic of belief, desire and intention, etc. So, we need a new approach. One possibility is algebraic specification and its generalization, *metamorphic relations*, which is independent of computation models.

*(2) Program Based Approach*

In this approach, test cases are derived from the program code and test adequacy is measured according to the coverage of the program code with respect to various adequacy metrics. Examples include (a) *control flow testing* to achieve statement coverage, branch coverage and various path coverage criteria, (b) *data flow testing* to satisfy def-use path coverage criteria, etc., or (c) *predicate testing* to exercise the predicates in the code (such as the Boolean conditions in if-then-else statements) to achieve Modified Condition/Decision Coverage (MC/DC) predicate coverage criterion, etc., (d) *mutation testing* is another typical program-based testing method in which bugs are injected into the program code to generate mutants of the original program and test adequacy is measured by the percentage of such bugs that can be detected by the test data.

Control flow and data flow testing methods are only applicable to programs that can be modeled in flow graphs. Thus, they are not directly applicable to neural networks and many other AI applications. Predicate testing can only be applied if the code contains predicates. Therefore, it may be adapted for testing decision tree learning models, but not neural networks. Mutation testing techniques rely on the analysis of faults in code to develop a set of mutation operators that inserts faults into code to generate mutants. The principle of mutation testing is applicable to all AI techniques, including neural networks. However, research is required to develop a set of meaningful mutation operators for various AI models. This may lead to useful metrics for measuring test adequacy for neural networks.

*(3) Usage Based Approach*

In this approach, the usage of the system is analysed to identify the input and output space, or more generally, the human computer interactions. Test cases are designed to explore the input/output space, for example, according to the risks associated with various subdomains of the space or certain types of points in the space. Probabilistic models are often used to model the users' behaviour. For example, profiles of users' inputs may be used to model the probability distribution of the input data. Markov chains have been used to model events in GUI operations. Typical examples of usage based testing techniques include combinatorial testing, GUI exploratory testing, random testing, fuzz testing and data mutation testing.

Random and fuzz testing generate test input data randomly. They are efficient and effective techniques for data spaces that are structurally simple, but less suitable for input spaces that are structurally highly complicated like human face images, diagrams, natural language sentences, etc. This includes most AI applications. Even if the input data are simple in structure, checking the correctness of the test result on random input could be a nontrivial problem when the computation is complicated. This is also the case for most AI applications.

Combinatorial testing is concerned with the situation

(2) https://jbehave.org

where the software has a large number of input variables while each variable can have a relatively small number of different values. This leads to a combinational explosion of test cases. Combinatorial testing techniques reduce the cost of testing while maintain test effectiveness by designing test suite that cover a subset of combinations, such as 2-way combinations. It might be possible to adapt this technique to test certain AI applications that have high dimensional input spaces.

Data mutation testing was developed to deal with structurally complicated input spaces [7]. The basic idea is to develop a set of operators on software input data so that when they are applied to an existing test case, a set of new test cases can be generated as the mutants of the original test cases. This is another testing method that inspired the approach proposed in this paper.

*(4) Error Based Approach*

In this approach, testing is based on a good understanding of the errors commonly made by software engineers. Test cases are designed to check if any of these errors are residual in the product. A typical example is category partitioning testing. A common programming error is to shift or rotate the boundary condition by a small amount. For example, searching an element in an array often misses the last one or the first one. Therefore, a testing method is to select test data on each boundary line of the subdomains as well as some points near the boundary. Such testing techniques target the pragmatic uncertainty based on the knowledge about common errors made by software engineers.

Since ML applications are developed by training a model rather than designing and coding in the traditional way, error-based testing techniques like category partitioning testing make little sense. Although the principle of error-based testing is applicable, many research questions remain: what are the common errors in training ML models? How will such errors exhibit themselves in the resulting product? And, how can such errors be detected by test cases?

In summary, existing software testing techniques and methods cannot be applied to AI applications due to their new features. It is difficult and expensive to generate test cases, to check the correctness of test results, and to measure test adequacy. Existing testing techniques and methods cannot effectively detect errors in ML models and thus are unable to ensure the quality of software products. New testing methods and techniques are required to take the specific features of ML/AI applications and their development processes into account.

In the next section, we will outline the proposed method datamorphic testing for ML and AI applications, addressing the problem of how to generate test data and how to check the correctness of test results.

## III. Basic Concepts of Datamorphic Testing

We first define the key concepts underlying the proposed testing method and illustrate them using face recognition as an example. Here, facial recognition is the problem of determining whether an image portrays somebody whose facial image is stored in a database of known persons.

### *A. Datamorphism*

A *datamorphism* is a transformation that derives new test data, called *mutants*, from existing test data.

Let $D$ and $C$ be the input and output domains of a program $P$ under test, respectively, $V(x_1, x_2, \dots, x_k, l)$ be a predicate on the set $D^k \times L$, where $L$ is a given set of parameters, and $k \geq 0$ is a given natural number.

**Definition 1. (Datamorphism)**

A *k-ary datamorphism* $\varphi$ is a mapping from $D^k \times L$ to $D$ such that for all $\vec{x} = (x_1, x_2, \dots, x_k) \in D^k$, $l \in L$, if $V(\vec{x}, l) = true$, we have that $\varphi(\vec{x}, l) \in D$. The elements $l$ in set $L$ are called the *parameters* of the datamorphism. $V$ is called *applicability condition* of the datamorphism. □

For example, for the face recognition application, the input domain of the application contains images of human faces. The following examples of datamorphisms are applicable to human facial images:

1) Add a pair of glasses;
2) Add makeup;
3) Change hairstyle;
4) Change hair colour.

Figure 1 shows the results of applying these datamorphisms to photos; (a) is the original photo[(3)], (b) adds a pair of glasses to (a), (c) adds a pair of sunglasses, (e) – (g) add makeup, (h) changes the hairstyle and colour. Images (i) and (j) are obtained by transforming (a) into black-and-white and into watercolours, respectively.

Note that some transformations are not meaningful in all contexts. For example, automated passport control requires people to remove items that obscure the face so adding glass is inapplicable there.

Note that these datamorphisms could be implemented as program components. For example, images (b)-(h) in Figure 1 were obtained by using a mobile phone app called *YouCam* and images (i) and (j) were obtained using another mobile phone app called *Prima*. In the experiment reported in Section V we also use the facial attribute inverting operators provided by the AttGAN system [8].

### *B. Metamorphism*

Not only do datamorphisms provide a means of test case generation, they are also a powerful means to specify the required functions of the application in terms of relationships between test cases and the expected outputs. Such a relationship is called a *metamorphic relation* [9, 10].

**Definition 2. (Metamorphic Relation)**

[3] From the public dataset at URL http://vis-www.cs.umass.edu/lfw/

Let $k \geq 1$ be a natural number. A *k-ary metamorphic relation M for program P* is a relation on $D^k \times C^k$ such that program $P$ is correct on input $\vec{x} = (x_1, .., x_K) \in D^k$ implies that the relation $M(\vec{x}, P(x_1), \ldots, P(x_k))$ holds, where $P(x)$ is program $P$'s output on input $x$. □

The following is a typical example of metamorphic relation for the *Sin*($x$) function, where $x$ and $y$ can be any real number.

$$(x + y = \pi) \Rightarrow Sin(x) = Sin(y).$$

The above equation defines a binary metamorphic relation (i.e. $k = 2$) for the $Sin$ function. A typical form of metamorphic relations is

$$V(x_1, \ldots, x_n) \Longrightarrow R(x_1, \ldots, x_n, P(x_1), \ldots, P(x_n)),$$

where $V(x_1, \ldots, x_n)$ is a relation on the input data $x_1, \ldots, x_n$, $P(x_1), \ldots, P(x_n)$ are the corresponding outputs, and $R(x_1, \ldots, x_n, y_1, \ldots, y_n)$ is a relation on them both. It asserts that for all input data $x_1, \ldots, x_n$ satisfying condition $V$, called the *applicability condition*, then the outputs from program $P$ must satisfy condition $R$, called the *correctness condition*. The condition $V$ is called the *applicability condition*. Condition $R$ is called the *correctness condition*.

A metamorphic relation is an assertion about software correctness in terms of an expected relationship between inputs and outputs. It is a very flexible and expressive means of specifying software functions.

To apply a metamorphic relation in the above form in software testing, one must generate test cases $a_1, \ldots, a_n$ that satisfy condition $V(a_1, \ldots, a_n)$, for example, by searching on the input space or by solving constraints. Then, the program $P$ under test is executed on test cases $a_1, \ldots, a_n$ to obtain outputs $P(a_1), \ldots, P(a_n)$. Finally, the condition $R$ is checked on the outputs to determine whether the program is correct on these test cases.

The two main difficulties of using metamorphic relations in software testing are finding a suitably set of metamorphic relations that are effective for detecting faults in the software under test and finding test cases that satisfy the applicability conditions of the metamorphic relations. These difficulties can be eased if metamorphic relations are combined with datamorphisms as shown in [11].

**Definition 3. (Metamorphism)**

Let $\psi \neq \emptyset$ be a given set of datamorphisms on the input domain of program $P$. A metamorphic relation $M$ is called a *metamorphism*, if it can be presented in the following form.

$$R(P(x_1), \ldots, P(x_k), P(x'_1), .., P(x'_m))$$

where $x'_i = \varphi_i(\vec{z_i}, l_i)$, $\vec{z_i}$ is a subset of $\{x_1, \ldots, x_k\}$, $\varphi_i \in \psi$ for all $i = 1, \ldots, m$. We say that the metamorphism is defined on $\varphi_1, \varphi_2, \ldots, \varphi_m$. □

A metamorphism asserts that for all input data $x_1$, … , $x_k$, the correctness condition $R$ holds on $P(x_1)$, … , $P(x_k), P(x'_1), .., P(x'_m)$. For the $Sin$ function example, the following is a metamorphism for the datamorphism $\varphi(x) = \pi - x$.

$$Sin(x) = Sin(\pi - x).$$

With metamorphisms, testing can be performed automatically by, first, applying the datamorphism on existing test cases to obtain mutant test cases. Then, the program is executed on the seed and mutant test cases. Finally, the results of the test executions are checked against the correctness condition to determine the correctness of the program. This avoids searching for test cases or constraint solving.

In many cases, metamorphism can be easily derived from the meaning of the datamorphism. For example, let $AddGlasses(x)$ denote the datamorphism of adding a pair of glasses on a facial image. A metamorphism for the face recognition application $FaceOf(x)$ can be formally defined on $AddGlasses(x)$ as follows.

$$FaceOf(x) = FaceOf(AddGlasses(x)).$$

This metamorphism states that if the face recognition application recognises a person in an image, then, after adding a pair of glasses by editing the image, the software should still recognize the person. Therefore, applying this metamorphism to image (a) and (b) in Figure 1, we expect a face recognition application will identify that the images are of the same person.

Metamorphisms like the above are actually software formal specifications. Testing an AI application can be automated if datamorphisms and metamorphisms can be implemented in software. For example, as stated earlier, it is possible to edit an image and generate a mutant of an image as shown in Figure 1. Then, by feeding both the original and the mutant images to a facial recognition application, the correctness of the application can be checked by comparing the outputs of these two test cases. If the outputs are identical, the application passes the test;

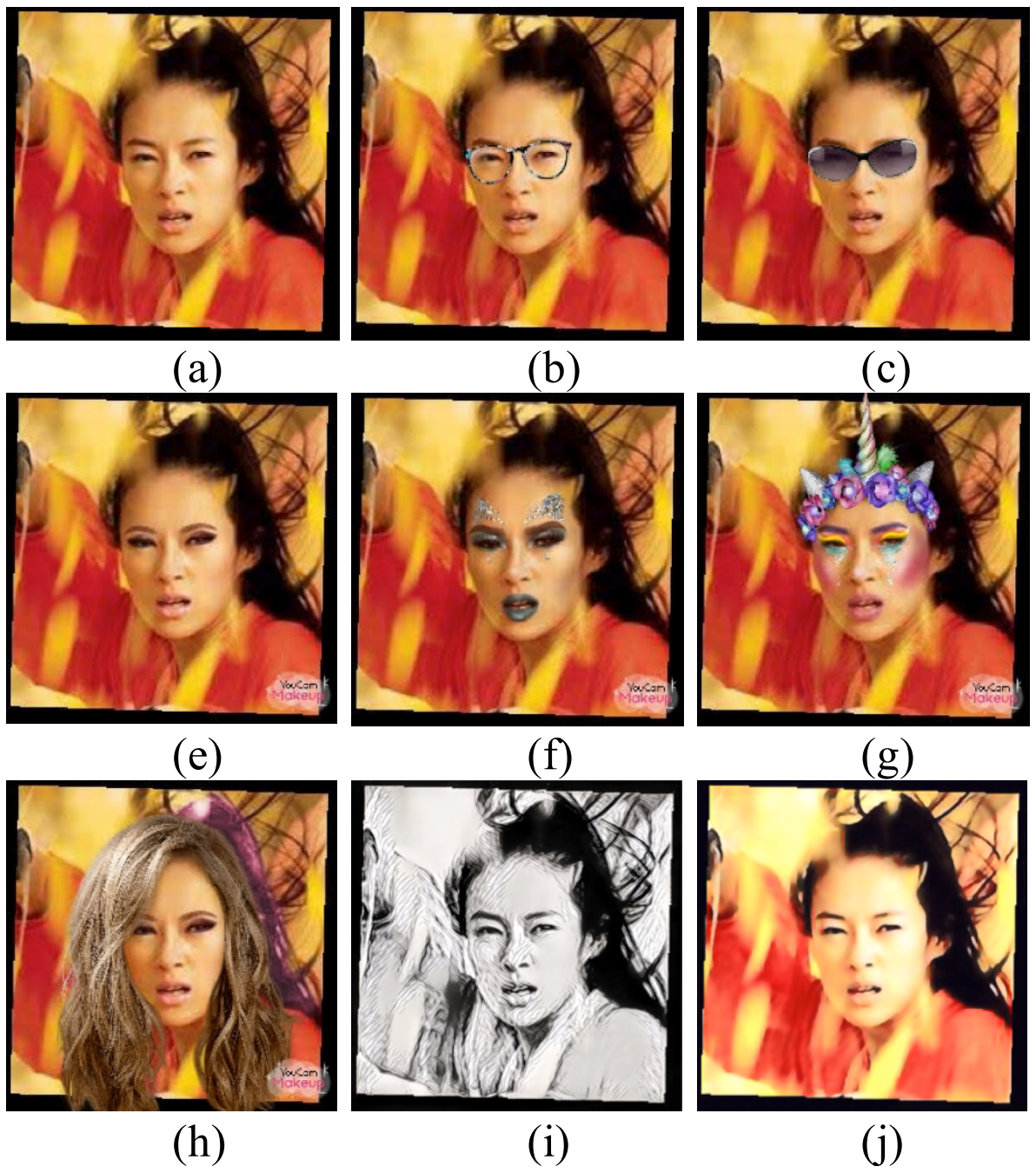

**Figure 1. Examples of Datamorphisms on Images**

otherwise, an error is detected.

### *C.* Seed *Test* Cases

For a datamorphism to be useful, we must have a set of known test cases, called *seed* test cases, or simply *seeds*.

In the face recognition example, a *seed* could be an image of a person's face. Such a set of seeds is normally available for testers of many AI applications, for example, as training data for an ML application. The seeds could be a subset of such training data selected at random or according to certain criteria. However, seeds alone are inadequate. Our method uses the seeds to generate more test cases to make an adequate test of the application. Seeds are not necessarily labeled with the expected outputs unless the datamorphisms require such labels.

A datamorphism may well be applicable to mutants, especially when the mutants are generated by a different datamorphism. For example, in Figure 1, (h) is obtained by applying a datamorphism on mutant (e).

In summary, our testing framework consists of three elements: a set of seed test cases, a set of datamorphisms and a set of metamorphisms.

**Definition 4. (Datamorphic Test Framework)**

Let $D$ be the input domain of a program $P$ under test. A datamorphic test framework $\mathcal{F}$ is an ordered triple $\langle \mathcal{S}, \mathcal{D}, \mathcal{M} \rangle$, where $\mathcal{S} \subseteq D$ is a finite subset of $D$. The elements of $\mathcal{S}$ are called the *seed test cases*, or simply *seeds*. $\mathcal{D}$ is a finite set of datamorphisms, and $\mathcal{M}$ is a finite set of metamorphisms. □

The next section discusses how to construct a datamorphic test framework and how such a test framework can be used with different strategies.

## IV. Testing Process and Strategies

### *A. Process of Datamorphic Testing*

As illustrated in Figure 2, the datamorphic testing process consists of three stages.

The first stage is *analysis* of the testing problem in order to design a datamorphic test framework. In this stage, seed test cases, datamorphisms and metamorphisms are identified. These three elements are closely related to each other, thus should be engineered systematically.

Analysis starts by identifying the operating conditions of the application. For face recognition application at an international airport's border control, for example, the input to the software is a photo from a camera fitted on an automatic passport checking machine and the photo of the passport holder retrieved from the information contained in a smart passport. The person may be of any ethnic background, age or gender. Normally, the person should be facing the camera directly without glasses and without heavy makeup, etc. However, in reality, people may wear glasses (even sunglasses), and have makeup. The photo could be many years old and taken from an unusual angle and the camera may have dust on its lens, etc. These form a variety of operating conditions of the application. Real-world use of the software could be a combination of these aspects. Adequate testing of the application must cover all such combinations. Directly collecting test data to achieve adequate testing could be a challenge.

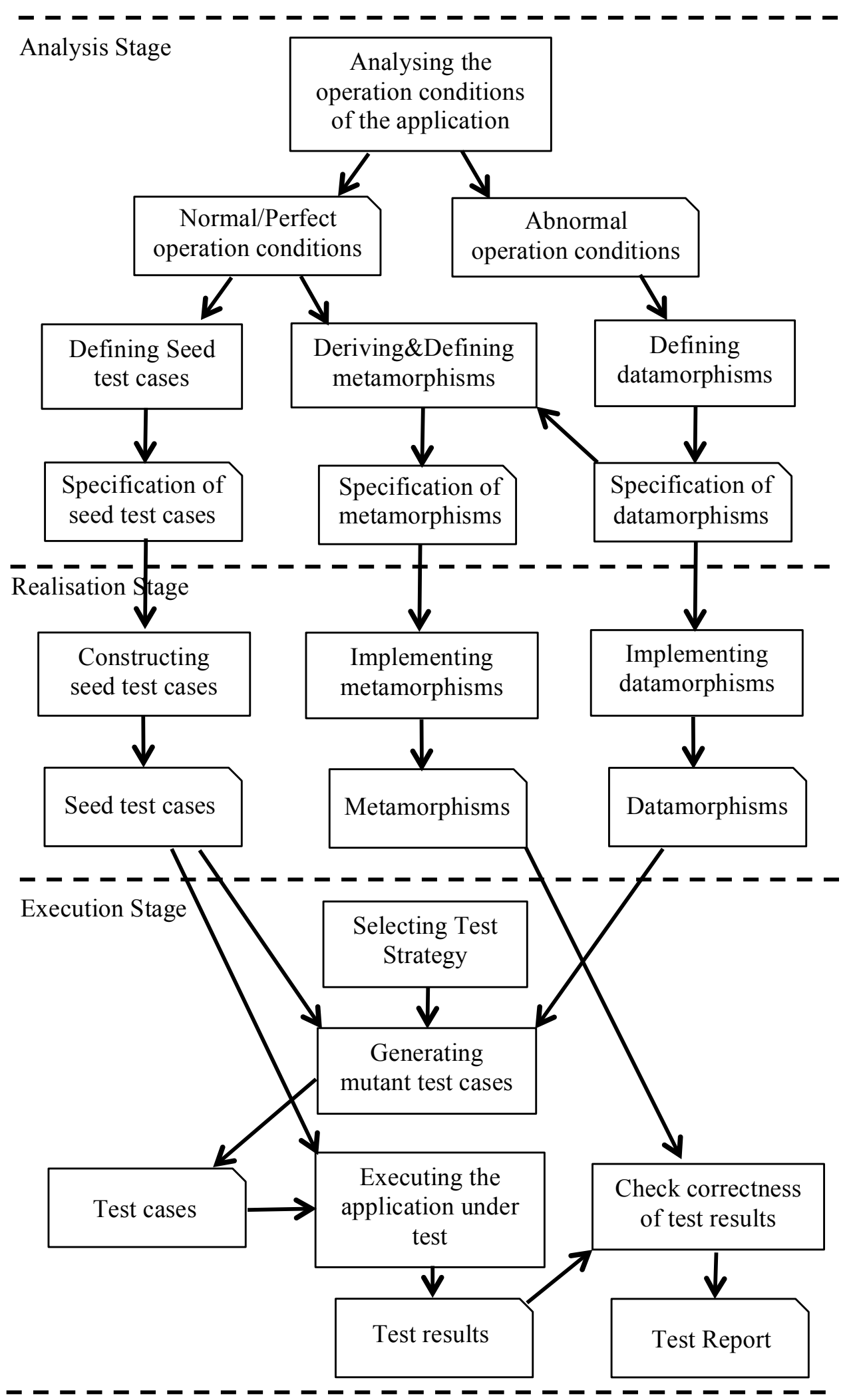

**Figure 2. Datamorphic Testing Process**

Our approach to solve this problem is to take a subset of the operating conditions as the normal conditions, and the others as abnormal conditions. The seed test cases are constructed from various combinations of the normal operating conditions. For example, we consider the person in front of the automatic passport checking machine is in normal operating condition if he/she does not wear glasses, has no makeup and the passport photo is taken recently. He/she could be of different ethnic background, in a different age group, and of different gender. Photos with combinations of these factors are sought as the seed tests.

The other operating conditions are regarded as "abnormal". For example, for the face recognition application to passport control, abnormal operating conditions include the situations that the passenger wears glasses, makeup, changes hair colour, the passport photo is out of date, and the camera lens is dusty, or the camera points to the person at a different angle, etc. Test cases representing abnormal operation conditions are obtained by transforming seed test cases into mutant test cases. Each of these abnormal conditions in the operation of the software is therefore a candidate for datamorphism.

For an abnormal operating condition to be a datamorphism, it must also be feasible to implement a transformation on the input data. Otherwise, the operating condition must be added to the set of "normal" operating condition and the corresponding test cases must be obtained directly as seeds.

Once the normal and abnormal operating conditions are identified, the corresponding changes of a datamorphism on the output should be identified, thus metamorphisms can be derived. For example, adding a pair of glasses should still recognize the person.

The first stage should finish with a set of specifications of the seed test cases, the datamorphisms and the metamorphisms. These specifications can be in natural language, but should be detailed enough for performing the nest steps of the process.

The second stage of datamorphic testing is realization. In this stage, the actual test data of the seeds are constructed, and datamorphisms and metamorphisms are realized.

A datamorphism can be realised by developing software that takes test data as input and generates new test data. Sometimes, applications already available can be used as datamorphisms as shown by the face recognition example. A datamorphism can also be realized by manually editing test cases. If datamorphisms cannot be realized, seed test cases must be obtained directly to represent the corresponding operating conditions.

A metamorphism can typically be implemented as code that invokes the application with seed and mutant test cases, and then stores and/or compares the result according to the metamorphism.

The final stage of datamorphic testing is execution, in which the test is executed according to a test strategy, which is discussed in detail in the next subsection.

### B. *Datamorphic Testing Strategies*

There are many strategies of generating test cases using datamorphisms.

- *Exhaustive*

An exhaustive strategy is to generate all possible mutant test cases by repeatedly applying the datamorphisms on the seeds until no more new mutants can be generated. Formally, let $\mathcal{T}$ be the set of test cases. $\mathcal{T}$ is initialized to the set of seed test cases. For each datamorphism, it is applied on every seed with all possible parameters to generate a set of mutants. These mutants are added to the set $\mathcal{T}$ of test cases while duplicated test cases are removed. This mutant generation process is repeated until no new test cases can be added.

Exhaustive testing may generate a huge number of test cases from a small set of seeds. In some cases, there may be an infinite number of test cases that can be generated from a finite number of seeds. Therefore, it is desirable to select a subset of such exhaustive test set. The following are some examples.

- *Combinatorial*

Assume that there is a set $\psi$ of $n>0$ unary datamorphisms. A mutant test case $m$ obtained by a sequence of applications of datamorphisms $\varphi_1, \varphi_2, \dots, \varphi_l \in \psi$ on a seed test case $s \in \mathcal{S}$ is represented as $\varphi_1 \circ \varphi_2 \circ \dots \circ \varphi_l(s)$. A set $\mathcal{T}$ of test cases is said to be 1-way combinatorial complete, if for every seed test case $s$ and every datamorphism $\varphi$, there is a test case $t \in \mathcal{T}$ such that $t = \cdots \circ \varphi \circ \dots (s)$. A set $\mathcal{T}$ of test cases is said to be 2-way combinatorial complete, if for any ordered pair of datamorphisms $\varphi_1, \varphi_2 \in \psi$, for every seed test case, there is a test case $t \in \mathcal{T}$ such that $t = \cdots \cdot \varphi_1 \circ \dots \circ \varphi_2 \circ \dots (s)$. Similarly, we can define $k$-way combinatorial completeness for every $k > 2$. Moreover, we say a set of test cases is 0-way combinatorial complete, if it contains all seed test cases. A set of test cases satisfies the $k$-way combinatorial coverage criterion, if it is $n$-way combinatorial complete for all $n = 0, \dots, k$. Similarly, we can define $k$-way combinatorial completeness and $k$-way combinatorial coverage criteria for a set of non-unary datamorphisms.

As in traditional combinatorial testing, the number of test cases in a test that satisfies the $k$-way combinatorial coverage criterion can be significantly smaller than number of all combinations of $k$ datamorphisms on all seed test cases. For many ML applications, the datamorphisms, like those for transformation of facial images, are often commutative, associative, and idempotent. Thus, the number of test cases to satisfy a combinatorial coverage criterion can be much smaller.

- *Optimal*

This strategy is inspired by genetic algorithms. Consider the set of seed test cases as the initial population and the datamorphisms as mutation operators. At each step in test case generation process, select a subset of the current population and a subset of datamorphisms to generate new mutants and add them into the population. The selection can be guided by a fitness function to either achieve maximal fitness in a fixed number of cycles, or until the fitness level peaks, or until the population reaches a certain number. Depending on the definition of the fitness function and the choice of termination condition, various kinds of optimization of the test set can be obtained.

- *Random*

A basic strategy to is select a seed test case and a datamorphism at random to generate one mutant a time until a total number of test case are generated or the testing is adequate according to some adequacy criterion.

- *Exploratory*

For classification and clustering problems, a practical goal of testing is often to find out the boundary between two classes. This can be done by defining binary datamorphisms to seek the boundary points and thereby find the Pareto front. For example, assume that $P(x)$ is a program that classifies an input real number $x$ into two classes $A$ and $B$. If there are two test cases $a$ and $b$ such that $P(a) \neq P(b)$, a datamorphism $Mid(x, y) = \frac{x+y}{2}$ can be applied to generate a test case $c = Mid(a, b)$. If $P(a) \neq P(c)$, then another test case $d = Mid(a, c)$ will be generated and the program tested on; otherwise, test case $d = Mid(b, c)$ is generated, and tested on. This process repeats iteratively until the distance between two test cases is small enough. In general, a test strategy may use the program output on test cases to determine which datamorphism to apply and on which test case. The principles of *search-based testing* apply [13].

## V. Experiment

In this section, we report an experiment to demonstrate the validity of datamorphic testing method.

### A. *Goal of the experiment*

The goal of the experiment is to investigate the validity of test case generation by applying datamorphisms on images for testing face recognition. The research questions to be answered are:

*RQ1*: Is an image generated by applying datamorphisms on an existing image a valid test case for face recognition applications?

*RQ2*: Are the test results on a ML application valid when using mutant test data obtained by applying datamorphisms in seed test cases?

### B. *Design of the experiment*

The experiment consists of three key elements: (a) the AI application to be tested; (b) the dataset used to select seed test cases and the real test data to be used in comparison with mutant test cases; (c) the transformations on test cases to be used as the datamorphisms.

(1) Applications under test

We have selected four real ML applications of the same kind from industry, i.e. four face recognition applications. They are:

a) Tencent Face Recognition(4)

b) Baidu Face Recognition(5)

c) Face++ online face recognition(6)

d) SeetaFace face recognition.

The first three are online services invoked through APIs written in Java. SeetaFace is an open source project based on openCV. The project is cloned from GitHub(7) and installed on our local computer system. It is written in C++ and our invocation code is also in C++.

These applications are used to evaluate the validity of test cases generated by datamorphisms on facial images as well as to demonstrate the applicability and cost efficiency of the testing method. The employment of multiple ML applications of the same kind but developed by independent vendors enables us to demonstrate the reusability and generalizability of the datamorphisms and test strategies.

(2) Datasets

Two public datasets are used in our experiment.

- CelebA(8), which contains ten thousand identities, each of which has twenty images. There are two hundred thousand images in total. We selected randomly 200 (i.e. 1%) images of different identities from the dataset as the seed test cases to generate mutant test cases.
- PubFig(9), which contains 58,797 images of about 200 people also collected from the internet. For each individual, the dataset contains multiple images (about 300 on average) taken in completely uncontrolled situations as non-cooperative subjects with large variation in pose, lighting, expression, scene, camera, imaging conditions and parameters, etc. This unique feature of the dataset makes it ideal to compare the images generated by applying datamorphisms.

(3) Datamorphisms and Metamorphisms

Instead of manually operating the YouCam APP as in Section III, we used the open source GitHub project AttGAN(10), which implements a set of 13 facial attribute editing operators [8]. Each operator takes a facial image as input and generates a new image that changes a facial attribute. They are listed in Table 1 and illustrated in Figure 3. These operators are used as the datamorphisms.

The metamorphisms used in the experiments are

$$FaceSimile(x, T(x)) \geq 80\%$$

where $T(x)$ is any of the datamorphisms given in Table 1, $FaceSimile$ is any of the four face recognition applications. For each of them, $FaceSimile(x, y)$ returns a number in the interval [0,100] as the similarity score between two facial images $x$ and $y$.

4 URL: https://ai.qq.com/product/face.shtml#detect

5 URL: https://aip.baidubce.com/rest/2.0/face/v3/match

6 URL: https://api-cn.faceplusplus.com/facepp/v3/compare

7 URL: https://github.com/seetaface/SeetaFaceEngine

8 URL: https://github.com/LynnHo/AttGAN-Tensorflow

9 URL: http://www.cs.columbia.edu/CAVE/databases/pubfig/

10 URL: https://github.com/LynnHo/AttGAN-Tensorflow

**Table 1. AttGAN's Face Attribute Editing Operators**

| Operation | Meaning |
|---|---|
| Bald | Change the facial image into bald |
| Bangs | Add bangs to the facial image |
| Black Hair | Change the hair colour into black |
| Blond Hair | Change the hair colour into blond |
| Brown Hair | Change the hair colour into brown |
| Bushy Eyebrows | Change the eyebrows to be bushy |
| Eyeglasses | Add eyeglasses to the image |
| Male | Change the image from female to male |
| Mouth Open | Change the mouth to be slightly open |
| Mustache | Add or remove mustache to the facial image |
| Beard | Add or remove beard |
| Pale Skin | Make the skin tone to be pale |
| Young | Change the image to look younger |

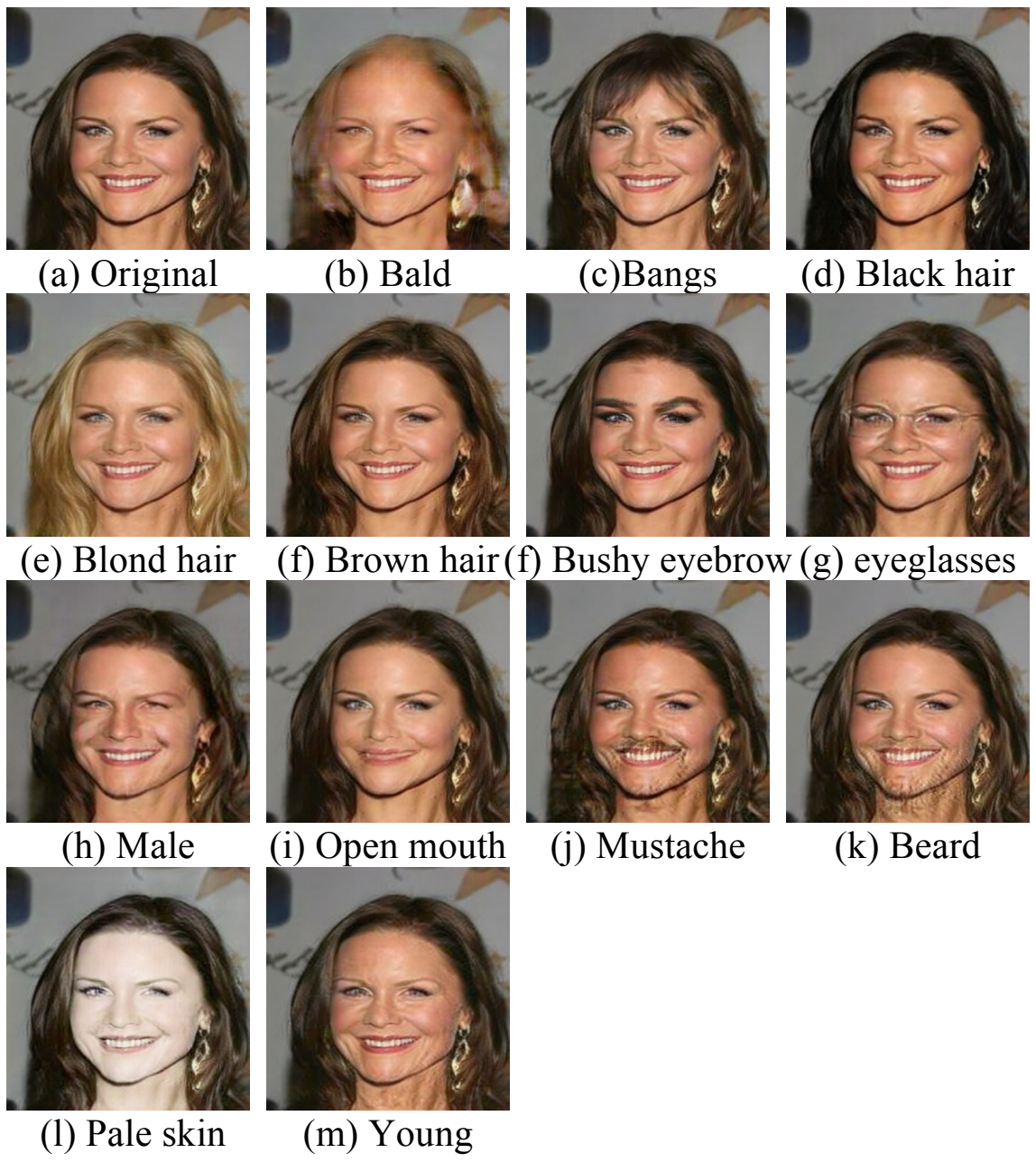

**Figure 3. Illustration of The Image Transformations**

*C. Execution of The Experiment*

The experiment consists of the following 3 steps.

1) Generation of mutant test cases

The mutant test cases are generated by using the AttGAN software on 200 images selected at random from the CelebA dataset.

The validity of the AttGAN algorithm for inverting facial images on various attributes has been intensively studied via cross-validation [8] on two large-scale labeled datasets, which clearly demonstrated that the resulting facial images achieved their purposes from the machine learning and image processing points of view. Our purpose differs from their experimental study. It is to validate the use of modified images as test cases for face recognition applications. Therefore, each facial image in the selection from the CelebA dataset is used as the seed, and 13 mutants generated by applying the transformations listed in Table 1 are used as mutant test cases. A total of 2,600 mutant test cases were generated.

2) Testing on generated test cases

The face recognition applications are tested on the mutant test cases against the seed test cases.

These mutants were input to four face recognition ML applications to obtain a measure of the similarity between the seed and the mutant, which is a numerical score in the range between 0 to 100. The raw data of the test results can be found in Appendix A.

3) Testing on real images

To validate the result of the testing on these mutant test cases, we selected at random 13 images for each individual from the PubFig dataset. We use these real images to test the face recognition applications and obtained their recognition accuracies. A total of 2600 real images were used as the real test cases. See Appendix B for details of the test results.

*D. Analysis of The Results*

The data collected from the experiments are analysed to answer the research questions RQ1 and RQ2.

- *Validity of Using Generated Images as Test Data*

To answer research question RQ1, we analyse how close the generated test cases are with respect to the original image. For each type of mutant, the average of similarity scores indicates how well the generated test case is close to the original image in the eyes of the ML application. The distribution of average similarity scores and their standard deviations over different image operators are shown in Figure 4 (a) and (b). Details of the data can be found in Appendix C.

The results show that the overall average similarity scores are between 80.32 and 99.70 for different face recognition applications. The smallest standard deviation is 1.51 while the largest standard deviation is 7.07. Therefore, we can conclude that facial images generated by applying such image processing algorithms to change various attributes of the image are very close to the real images, thus valid as test cases.

There are a small number of cases where recognition fails, as shown in Figure 4(c). This is either because the application does not recognize any face in the image or because there is a timeout in the transmission of image data to the servers on the Cloud. Both of these cases are ignored when calculating recognition accuracy.

- *Test Effectiveness*

To answer research question RQ2, we compare the test results obtained by using generated mutant test cases and the results obtained by using real images.

Our experiments demonstrated that using mutant test cases to test facial recognition applications can

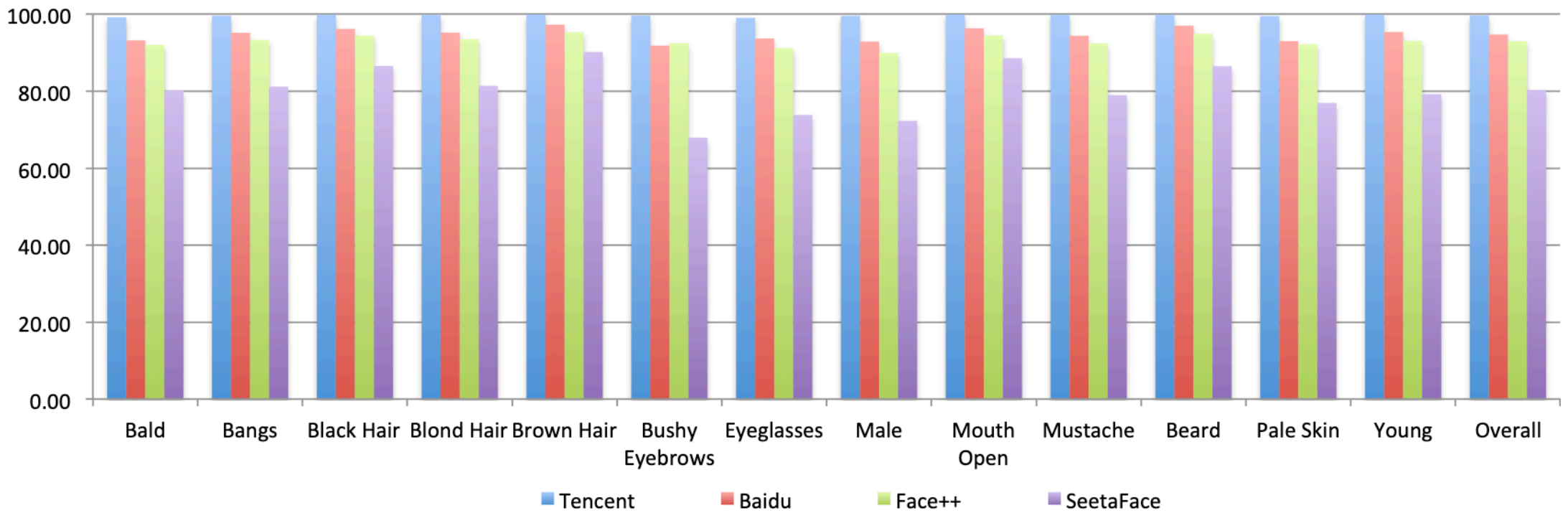

(a) Average Similarity Scores between Seeds and Mutants

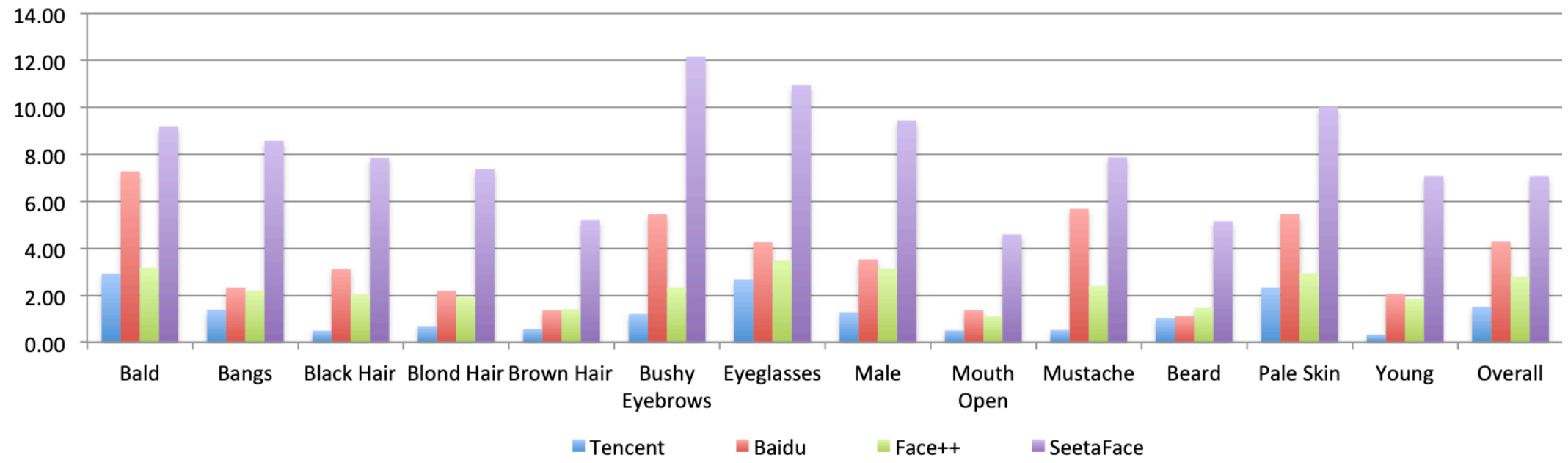

(b) Standard Deviations of The Similarity Scores

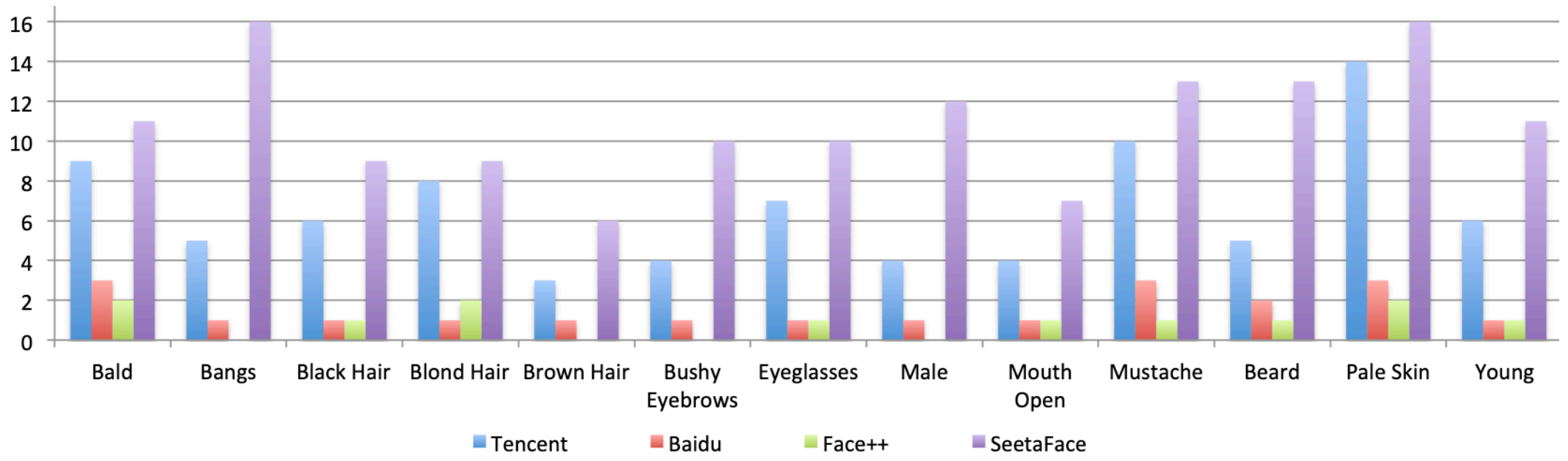

(c) Number of Failures

**Figure 4. Similarity Between Seeds and the Mutants**

differentiate their recognition capability; see Figure 5.

The results show that the Tencent Face Recognition gives the highest overall average (99.70) of similarity scores between the seeds and mutants, while SeetaFace has the lowest overall average (80.29) of similarity scores. Face++ and Baidu Face Recognition are very close on overall average scores, 93.09 and 94.75, respectively. These results are highly correlated to the overall average scores of the testing with real images. The Pearson's correlation coefficient between them is 0.99. The standard deviations are also highly correlated with a Pearson's correlation coefficient of 0.82. Therefore, we can conclude that the test results obtained by using mutant test cases is valid.

## VI. Conclusion

In this paper, we proposed a new software testing method called datamorphic testing and explored its applicability to testing ML applications. An experiment has shown it is a valid approach.

### *A. Related work*

The proposed approach is an improvement, generalization and integration of many data centric testing methods.

Data mutation testing was proposed by Shan and Zhu to test software whose input is structurally complex [7]. A test framework in data mutation testing consists of a set of know test cases called seeds and a set of data mutation operators applicable on the seed test cases. This inspired the seed test cases and datamorphisms of our approach. An empirical study of data mutation testing method was conducted by testing a modeling tool. It demonstrated high fault detection ability. Datamorphic testing improves upon data mutation testing by integrating it with metamorphic testing. It advocates that data mutation operators should be engineered together with metamorphic relations, so that the correctness of the software on test cases can be automatically checked.

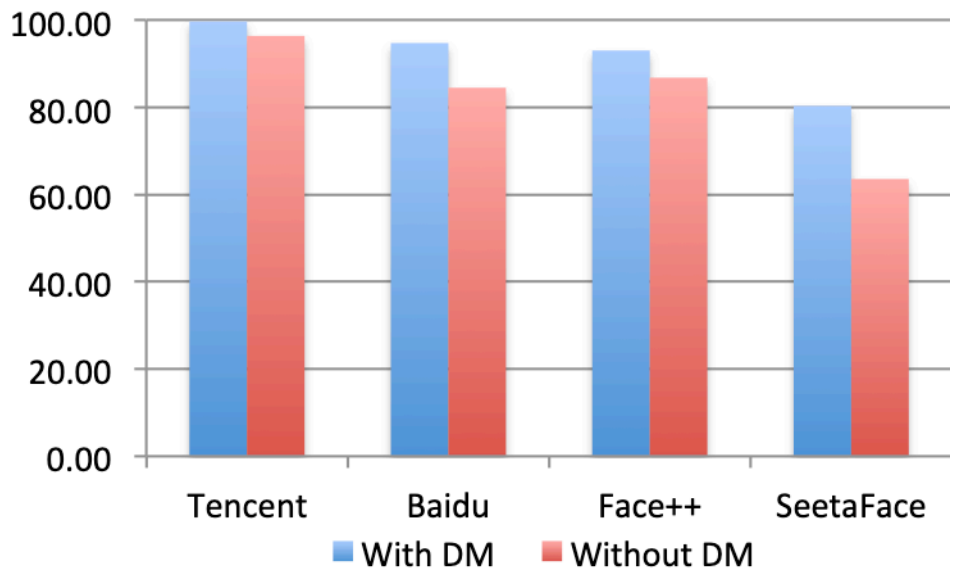

(a) Overall Average Scores

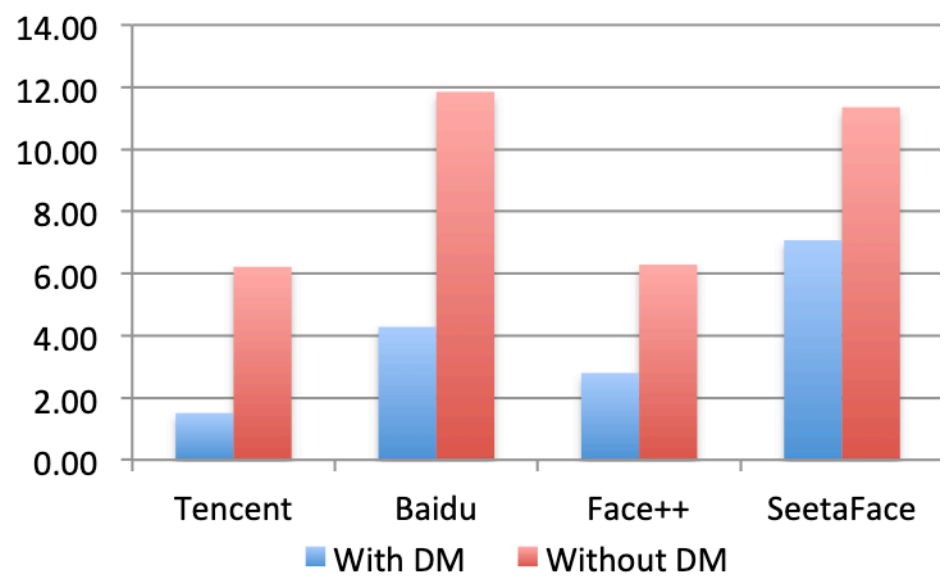

(b) Overall Standard Deviations

**Figure 5. Testing on Mutants vs on Real Test Cases**

Metamorphic testing was proposed by Chen, *et al.* [9], who introduced the notion of metamorphic relations and explored its use in software testing. It has been an active research topic since then, with researchers empirically studying the effectiveness of the testing method. Recent surveys of the work on this topic can also be found [10].

Zhu has integrated data mutation testing and metamorphic test methods and showed that using data mutation operators, metamorphic relations can be easily derived [11], overcoming a long-standing barrier to practical use. Zhu also reported on an automated testing tool called JFuzz to support the use of integrated data mutation and metamorphic testing methods for Java unit testing. Datamorphic testing improves the applicability of metamorphic testing by allowing systematic derivation of datamorphisms and metamorphisms, as illustrated by the example above and the experiment. Moreover, JFuzz implements a simple fixed strategy of using data mutation operators and metamorphisms in testing. In contrast, the datamorphic testing method proposed in this paper regards testing strategy as a variable element.

Fuzz testing is a type of random testing [12], in which a test case is randomly modified and the correctness criterion is that the software does not crash. It is a special trivial case of datamorphic testing. Thus, datamorphic testing can significantly increase testing effectiveness by specifying more meaningful modifications of the test cases with datamorphisms and more accurate correctness conditions in metamorphisms.

Search-based testing regards testing as an optimization problem, to maximize the test effectiveness or test coverage by searching on the space of test cases [13]. Genetic algorithms are often employed to realize such optimization. Datamorphic testing can be combined with search-based software testing to search for optimised test sets, for example, by using datamorphisms as means of generating a population of test cases, and coverage of metamorphisms as the optimisation target. As discussed in Section IV.B, the principle of search-based testing can be applied to form specific testing strategies of datamorphism testing.

Testing ML applications is rarely reported in the research literature. An exception is DeepTest [142], which is a software tool for testing deep neuron network (DNN) driven autonomous cars. It automatically generates test cases leveraging real-world changes in driving conditions like rain, fog, lighting conditions, etc. via image transformations. Metamorphic relations are defined based on such image transformations and used to detect erroneous behaviours. DeepTest can be understood as a datamorphic testing tools for a special type of ML applications. It demonstrates that such synthetic test cases can be realistic and capable of finding a large number of erroneous behaviors under different realistic driving conditions, many of which led to potentially fatal crashes in three top performing DNNs in the Udacity self-driving car challenge. Most existing testing techniques for DNN-driven vehicles are heavily dependent on the manual collection of test data under different driving conditions. This is prohibitively expensive as the number of test conditions increases. DeepTest shows that the approach to generate a large number of realistic test cases can be cost efficient. However, it is unclear how the seed test cases reported in [14] are selected, how transformations on test cases are identified, and what test strategy were implemented in DeepTest to generate mutant test cases. These are some of the issues that this paper attempts to address.

The main contribution of this paper is a theoretical framework of a testing method for AI applications, which unifies the existing data centric testing methods and techniques. It lays a foundation for an engineering methodology and automated testing tool for AI applications.

### B. Further work

We are developing an automated testing tool to support datamorphic testing method for AI applications. It aims to show the feasibility of datamorphic testing with significant improvement in test effectiveness and efficiency.

We are further investigating practical techniques in the framework of datamorphic testing. These techniques include (a) adequacy criteria for testing AI applications, (b) various test strategies and heuristics rules for controlling the testing activities, and (c) test process models that integrate testing with the development process of AI applications.

We are conducting further experiments with different testing strategies and empirical studies with different real AI applications in order to develop practical guidelines on the uses of the testing techniques and automated tools based on empirical evidence.

## Acknowledgement

The work reported in this paper is partially supported by the National Science Foundation of China (Grant No. 61502233).

Appendix A. Results of Testing on Mutant Test Case with Datamorphisms

## 1. Results of Testing Tencent Face Recognition with Datamorphisms

| Image ID | Bald | Bangs | Black_Hair | Blond_Hair | Brown_Hair | Bushy_Eyebr | Eyeglasses | Male | Mouth_Sligh | Mustache | No_Beard | Pale_Skin | Young | Average | StDev |
|---|---|---|---|---|---|---|---|---|---|---|---|---|---|---|---|
| 1 | 100 | 100 | 100 | 100 | 100 | 100 | 100 | 100 | 100 | 100 | 100 | 100 | 100 | 100.0000 | 0 |
| 2 | 98 | 100 | 100 | 100 | 100 | 100 | 100 | 100 | 100 | 100 | 100 | 94 | 99 | 99.3077 | 1.7022 |
| 3 | 100 | 100 | 97 | 100 | 100 | 94 | 100 | 100 | 100 | 100 | 100 | 100 | 100 | 99.3077 | 1.7974 |
| 4 | 100 | 100 | 100 | 100 | 100 | 100 | 100 | 100 | 100 | 100 | 100 | 100 | 100 | 100.0000 | 0.0000 |
| 5 | 100 | 100 | 100 | 100 | 100 | 98 | 97 | 98 | 100 | 98 | 100 | 100 | 100 | 99.3077 | 1.1094 |
| 6 | 100 | 100 | 100 | 100 | 100 | 100 | 96 | 98 | 100 | 100 | 100 | 100 | 100 | 99.5385 | 1.1983 |
| 7 | 100 | 100 | 100 | 100 | 100 | 100 | 100 | 97 | 100 | 100 | 100 | 100 | 100 | 99.7692 | 0.8321 |
| 8 | 100 | 100 | 100 | 100 | 100 | 97 | 96 | 100 | 100 | 100 | 100 | 100 | 100 | 99.4615 | 1.3301 |
| 9 | 98 | 100 | 100 | 100 | 100 | 100 | 100 | 99 | 100 | 100 | 100 | 100 | 100 | 99.7692 | 0.5991 |
| 10 | 100 | 90 | 99 | 94 | 100 | 100 | 100 | 100 | 100 | 100 | 100 | 100 | 100 | 98.6923 | 3.0926 |
| 11 | | | | | | | | | | | | | | #DIV/0! | #DIV/0! |
| 12 | 100 | 100 | 100 | 100 | 100 | 100 | 95 | 100 | 100 | 100 | 100 | 100 | 100 | 99.6154 | 1.3868 |
| 13 | 100 | 100 | 100 | 100 | 100 | 100 | 100 | 100 | 100 | 100 | 100 | 100 | 100 | 100.0000 | 0.0000 |
| 14 | 100 | 100 | 100 | 100 | 100 | 100 | 99 | 99 | 100 | 100 | 100 | 100 | 100 | 99.8462 | 0.3755 |
| 15 | 100 | 100 | 100 | 100 | 100 | 100 | 100 | 100 | 100 | 100 | 100 | 100 | 100 | 100.0000 | 0.0000 |
| 16 | 100 | 100 | 100 | 100 | 100 | 100 | 100 | 100 | 100 | 100 | 100 | 100 | 100 | 100.0000 | 0.0000 |
| 17 | 100 | 100 | 100 | 100 | 100 | 100 | 100 | 100 | 100 | 100 | 100 | | 100 | 100.0000 | 0.0000 |
| 18 | 100 | 100 | 100 | 100 | 100 | 100 | 100 | 100 | 100 | 100 | 100 | 100 | 100 | 100.0000 | 0.0000 |
| 19 | 99 | 100 | 100 | 100 | 100 | 100 | 100 | 100 | 100 | 100 | 100 | 100 | 100 | 99.9231 | 0.2774 |
| 20 | 100 | 100 | 100 | 100 | 100 | 100 | 100 | 100 | 100 | 100 | 100 | 100 | 100 | 100.0000 | 0.0000 |
| 21 | 100 | 100 | 100 | 100 | 100 | 100 | 99 | 99 | 100 | 100 | 100 | 100 | 100 | 99.8462 | 0.3755 |
| 22 | 100 | 100 | 100 | 100 | 100 | 100 | 100 | 100 | 100 | 100 | 100 | 100 | 100 | 100.0000 | 0.0000 |
| 23 | 100 | 100 | 100 | 100 | 100 | 100 | 100 | 100 | 100 | 100 | 100 | 100 | 100 | 100.0000 | 0.0000 |
| 24 | 100 | 100 | 100 | 100 | 100 | 97 | 87 | 100 | 100 | 100 | 100 | 100 | 100 | 98.7692 | 3.6321 |
| 25 | 97 | 100 | 100 | 100 | 100 | 100 | 100 | 100 | 100 | 100 | 100 | 100 | 100 | 99.7692 | 0.8321 |
| 26 | 99 | 100 | | 100 | 100 | 100 | 100 | 100 | 100 | 100 | 100 | 100 | 100 | 99.9167 | 0.2887 |
| 27 | 97 | 100 | 100 | 100 | 100 | 100 | 100 | 100 | 100 | 100 | 100 | 100 | 100 | 99.7692 | 0.8321 |
| 28 | 100 | 100 | 100 | | 100 | 100 | 100 | 100 | 100 | 100 | 100 | 100 | 100 | 100.0000 | 0.0000 |
| 29 | 100 | 100 | 100 | 100 | 100 | 100 | 100 | 100 | 100 | 100 | 100 | 100 | 100 | 100.0000 | 0.0000 |
| 30 | 100 | 100 | 100 | 100 | 100 | 100 | 100 | 100 | 100 | 100 | 100 | 100 | 100 | 100.0000 | 0.0000 |
| 31 | 100 | 100 | 100 | 100 | 100 | 100 | 100 | 100 | 100 | 100 | 100 | 100 | 100 | 100.0000 | 0.0000 |
| 32 | 100 | 100 | 100 | 100 | 100 | 100 | 100 | 100 | 100 | 100 | 100 | 100 | 100 | 100.0000 | 0.0000 |
| 33 | 100 | 100 | 100 | 100 | 100 | 100 | 100 | 98 | 100 | 100 | 100 | 100 | 100 | 99.8462 | 0.5547 |
| 34 | 100 | 100 | 100 | 100 | 100 | 100 | 100 | 100 | 100 | 100 | 100 | 100 | 100 | 100.0000 | 0.0000 |
| 35 | 100 | 94 | 100 | 100 | 100 | 100 | 100 | 100 | 100 | 100 | 100 | 100 | 100 | 99.5385 | 1.6641 |
| 36 | 87 | 100 | | | 92 | 100 | 100 | | | | 100 | | | 96.5000 | 5.6480 |
| 37 | 100 | 100 | 100 | 100 | 100 | 100 | 100 | 98 | 100 | 100 | 100 | 100 | 100 | 99.8462 | 0.5547 |
| 38 | 94 | 100 | 100 | 96 | 100 | 100 | 94 | 94 | 100 | 100 | 100 | 100 | 100 | 98.3077 | 2.6890 |
| 39 | 100 | 100 | 100 | 100 | 100 | 100 | 100 | 100 | 100 | 100 | 100 | 100 | 100 | 100.0000 | 0.0000 |
| 40 | 100 | 100 | 100 | 100 | 100 | 100 | 100 | 100 | 100 | 97 | 100 | 100 | 100 | 99.7692 | 0.8321 |
| 41 | | 100 | 100 | 100 | 100 | 100 | 100 | 100 | 100 | 100 | 100 | 97 | 100 | 99.7500 | 0.8660 |
| 42 | 96 | 100 | 100 | 100 | 100 | 100 | 100 | 100 | 100 | 100 | 100 | 100 | 100 | 99.6923 | 1.1094 |
| 43 | 94 | 99 | 100 | 99 | 100 | 100 | 94 | 94 | 100 | 100 | 100 | | | 98.1818 | 2.7136 |
| 44 | 100 | 100 | 100 | 100 | 100 | 100 | 100 | 100 | 100 | 100 | 100 | 100 | 100 | 100.0000 | 0.0000 |
| 45 | 100 | 100 | 100 | 100 | 100 | | 100 | 94 | 100 | 100 | 100 | 100 | 100 | 99.5000 | 1.7321 |
| 46 | 100 | 100 | 100 | 100 | 100 | 100 | | 100 | 100 | 100 | 100 | 100 | 100 | 100.0000 | 0.0000 |
| 47 | 100 | 100 | 100 | 100 | 100 | 100 | 100 | 100 | 100 | 100 | 100 | 100 | 100 | 100.0000 | 0.0000 |
| 48 | 100 | 100 | 100 | 100 | 100 | 100 | 100 | 100 | 100 | 100 | 100 | 100 | 100 | 100.0000 | 0.0000 |
| 49 | 100 | 100 | 100 | 100 | 100 | 100 | 100 | 100 | 100 | 100 | 100 | 100 | 100 | 100.0000 | 0.0000 |
| 50 | 100 | 100 | 100 | 100 | 100 | 100 | 100 | 100 | 100 | 100 | 100 | 100 | 100 | 100.0000 | 0.0000 |
| 51 | 74 | 100 | 100 | 100 | 100 | 100 | 100 | 100 | 100 | 100 | 100 | 100 | 100 | 98.0000 | 7.2111 |
| 52 | 100 | 98 | 100 | | 100 | 100 | 100 | 100 | 100 | 100 | 100 | 96 | 100 | 99.5000 | 1.2432 |
| 53 | 100 | 100 | 100 | 100 | 100 | 100 | 100 | 100 | 100 | 100 | 100 | 100 | 100 | 100.0000 | 0.0000 |
| 54 | 100 | 100 | 99 | 100 | 100 | 99 | 100 | 100 | 100 | 100 | 100 | 100 | 100 | 99.8462 | 0.3755 |
| 55 | 100 | 100 | 100 | 100 | 100 | 97 | 94 | 100 | 100 | 100 | 100 | 100 | 100 | 99.3077 | 1.7974 |
| 56 | 100 | | 100 | | 100 | 98 | 96 | 100 | 100 | | 100 | 100 | 100 | 99.4000 | 1.3499 |
| 57 | 100 | 100 | 100 | 100 | 100 | 100 | 100 | 100 | 100 | 100 | 100 | 100 | 100 | 100.0000 | 0.0000 |
| 58 | 100 | 100 | 100 | 100 | 100 | 100 | 100 | 100 | 100 | 100 | 100 | 100 | 100 | 100.0000 | 0.0000 |
| 59 | 96 | 100 | 100 | 100 | 100 | 100 | 100 | 100 | 100 | 100 | 100 | 100 | 100 | 99.6923 | 1.1094 |
| 60 | 100 | 100 | 100 | 100 | 100 | 100 | 100 | 100 | 100 | 100 | 100 | 100 | 100 | 100.0000 | 0.0000 |
| 61 | | 94 | 100 | 97 | 100 | 100 | 94 | 94 | 100 | 100 | 100 | 95 | 96 | 97.5000 | 2.7469 |
| 62 | | | | 100 | 100 | | | | | | | 100 | | 100.0000 | 0.0000 |
| 63 | 100 | 100 | 100 | 100 | 100 | 99 | 94 | 100 | 100 | 97 | 100 | 100 | 100 | 99.2308 | 1.7867 |
| 64 | 100 | 100 | 100 | 100 | 100 | 100 | 100 | 100 | 100 | 100 | 100 | 100 | 100 | 100.0000 | 0.0000 |
| 65 | 100 | 100 | 100 | 100 | 100 | 100 | 100 | 100 | 100 | 100 | 100 | 100 | 100 | 100.0000 | 0.0000 |
| 66 | 100 | 100 | 100 | 100 | 100 | 100 | 100 | 100 | 100 | 100 | 100 | 100 | 100 | 100.0000 | 0.0000 |
| 67 | 100 | 100 | 100 | 100 | 100 | 100 | 100 | 100 | 100 | 100 | 100 | 100 | 100 | 100.0000 | 0.0000 |
| 68 | 100 | 100 | 100 | 100 | 100 | 100 | 100 | 100 | 100 | 100 | 100 | 100 | 100 | 100.0000 | 0.0000 |
| 69 | 100 | 94 | 100 | 100 | 100 | 100 | 96 | 100 | 100 | 100 | 100 | 100 | 100 | 99.2308 | 1.9215 |
| 70 | 100 | 100 | 100 | 100 | 100 | 100 | 100 | 100 | 100 | 100 | 100 | 100 | 100 | 100.0000 | 0.0000 |
| 71 | 94 | 100 | 100 | 100 | 100 | 100 | 100 | 100 | 100 | 100 | 100 | | 100 | 99.5000 | 1.7321 |
| 72 | 100 | 100 | 100 | | 100 | 100 | 100 | 98 | 100 | 100 | 100 | 100 | 100 | 99.8333 | 0.5774 |
| 73 | 100 | 100 | 100 | 100 | 100 | 100 | 100 | 100 | 100 | 100 | 100 | 100 | 100 | 100.0000 | 0.0000 |
| 74 | 100 | 100 | 100 | 100 | 100 | 100 | 100 | 100 | 100 | 100 | 100 | 100 | 100 | 100.0000 | 0.0000 |
| 75 | 100 | 100 | 100 | 100 | 100 | 100 | 100 | 100 | 100 | 100 | 100 | 100 | 100 | 100.0000 | 0.0000 |
| 76 | 100 | 100 | 100 | 100 | 100 | 100 | 100 | 100 | 100 | 100 | 100 | 100 | 100 | 100.0000 | 0.0000 |
| 77 | 100 | 100 | 100 | 100 | 100 | 100 | 100 | 100 | 100 | 100 | 100 | 100 | 100 | 100.0000 | 0.0000 |
| 78 | 100 | 100 | 100 | 100 | 100 | 100 | 100 | 100 | 100 | 100 | 100 | 100 | 100 | 100.0000 | 0.0000 |
| 79 | 100 | 100 | 100 | 100 | 100 | 100 | 100 | 100 | 100 | 100 | 100 | 100 | 100 | 100.0000 | 0.0000 |
| 80 | 99 | 98 | 100 | 100 | 100 | 100 | 98 | 98 | 100 | 100 | 100 | 100 | 100 | 99.4615 | 0.8771 |
| 81 | 100 | 100 | 100 | 100 | 100 | 100 | 100 | 100 | 100 | 100 | 100 | 100 | 100 | 100.0000 | 0.0000 |
| 82 | 100 | 100 | 100 | 100 | 100 | 100 | 100 | 100 | 100 | 100 | 100 | 100 | 100 | 100.0000 | 0.0000 |
| 83 | 100 | 100 | 100 | | 100 | 100 | 100 | 100 | 100 | 97 | 100 | | 100 | 99.7273 | 0.9045 |
| 84 | 100 | 100 | 100 | 100 | 100 | 100 | 100 | 100 | 100 | 100 | 100 | 100 | 100 | 100.0000 | 0.0000 |
| 85 | 100 | 100 | 100 | 100 | 100 | 100 | 100 | 100 | 100 | 100 | 100 | 100 | 100 | 100.0000 | 0.0000 |
| 86 | 100 | 100 | 100 | 100 | 100 | 100 | 100 | 100 | 100 | 100 | 100 | 100 | 100 | 100.0000 | 0.0000 |
| 87 | 100 | 100 | 100 | 100 | 100 | 100 | 100 | 100 | 100 | 100 | 100 | 100 | 100 | 100.0000 | 0.0000 |
| 88 | 100 | 100 | 100 | 100 | 100 | 98 | 100 | 100 | 100 | 100 | 100 | 100 | 100 | 99.8462 | 0.5547 |
| 89 | 100 | 100 | 100 | 100 | 100 | 100 | 100 | 100 | 100 | 100 | 100 | 100 | 100 | 100.0000 | 0.0000 |
| 90 | 100 | 100 | 100 | 100 | 100 | 100 | 100 | 100 | 100 | 100 | 100 | 100 | 100 | 100.0000 | 0.0000 |
| 91 | 100 | 100 | 100 | 100 | 100 | 100 | 100 | 100 | 100 | 100 | 100 | 100 | 100 | 100.0000 | 0.0000 |
| 92 | 98 | 99 | 100 | 100 | 100 | 100 | 100 | 100 | 100 | 100 | 100 | 99 | 100 | 99.6923 | 0.6304 |
| 93 | 100 | 100 | 100 | 100 | 100 | 100 | 100 | 100 | 100 | 100 | 100 | 100 | 100 | 100.0000 | 0.0000 |
| 94 | 90 | 100 | 100 | 100 | 100 | 100 | 100 | 99 | 100 | 100 | 100 | 100 | 100 | 99.1538 | 2.7642 |
| 95 | | 100 | 100 | 100 | 100 | 100 | | 100 | 100 | 100 | 100 | | | 100.0000 | 0.0000 |
| 96 | 100 | 98 | 100 | 100 | 100 | 100 | 99 | 100 | 100 | 100 | 100 | 100 | 100 | 99.7692 | 0.5991 |
| 97 | 100 | 100 | 100 | 100 | 100 | 100 | 100 | 100 | 100 | 100 | 100 | 100 | 100 | 100.0000 | 0.0000 |
| 98 | 100 | 100 | 100 | 100 | 100 | 100 | 100 | 100 | 100 | 100 | 100 | 100 | 100 | 100.0000 | 0.0000 |
| 99 | 100 | 100 | 100 | 100 | 100 | 100 | 100 | 100 | 100 | 100 | 100 | 100 | 100 | 100.0000 | 0.0000 |
| 100 | 100 | 100 | 100 | 100 | 100 | 100 | 100 | 100 | 100 | 100 | 100 | 100 | 100 | 100.0000 | 0.0000 |
| 101 | 100 | 100 | 100 | 100 | 100 | 100 | 100 | 100 | 100 | 100 | 100 | 100 | 100 | 100.0000 | 0.0000 |
| 102 | 87 | 100 | 100 | 100 | 100 | 100 | 97 | 100 | 100 | 99 | 100 | 100 | 100 | 98.6923 | 3.6144 |
| 103 | | | | 100 | | 100 | 100 | 100 | 100 | | 99 | | 100 | 99.8571 | 0.3780 |
| 104 | 100 | 100 | 100 | 100 | 100 | 100 | 100 | 100 | 100 | 100 | 100 | 100 | 100 | 100.0000 | 0.0000 |
| 105 | 98 | 100 | 100 | 100 | 100 | 100 | 97 | 100 | 100 | 100 | 100 | 100 | 100 | 99.6154 | 0.9608 |
| 106 | 100 | 100 | 100 | 100 | 100 | 98 | 100 | 100 | 100 | 100 | 100 | 100 | 100 | 99.8462 | 0.5547 |
| 107 | 100 | 100 | 100 | 100 | 100 | 100 | 100 | 100 | 100 | 100 | 100 | 100 | 100 | 100.0000 | 0.0000 |
| 108 | 100 | 100 | 100 | 100 | 100 | 100 | 98 | 94 | 100 | 100 | 100 | 100 | 100 | 99.3846 | 1.7097 |
| 109 | 100 | 100 | 100 | 100 | 100 | 100 | 100 | 100 | 100 | 100 | 100 | 100 | 100 | 100.0000 | 0.0000 |
| 110 | 100 | 100 | 100 | 100 | 100 | 100 | 100 | 100 | 100 | 100 | 100 | 97 | 100 | 99.7692 | 0.8321 |
| 111 | 100 | 100 | 100 | 100 | 100 | 100 | 97 | 100 | 100 | 100 | 100 | 100 | 100 | 99.7692 | 0.8321 |
| 112 | 100 | 100 | 100 | 100 | 100 | 100 | 100 | 100 | 100 | 100 | 100 | 100 | 100 | 100.0000 | 0.0000 |
| 113 | 100 | 100 | 100 | 100 | 100 | 100 | 100 | 100 | 100 | 100 | 100 | 100 | 100 | 100.0000 | 0.0000 |
| 114 | 100 | 100 | 100 | 100 | 100 | 100 | 95 | 100 | 100 | 100 | 100 | 100 | 100 | 99.6154 | 1.3868 |
| 115 | 100 | 100 | 100 | 100 | 100 | 100 | 100 | 100 | 100 | 100 | 100 | | 100 | 100.0000 | 0.0000 |
| 116 | 100 | 100 | 100 | 100 | 100 | 100 | 97 | 100 | 100 | 100 | 100 | 100 | 100 | 99.7692 | 0.8321 |
| 117 | 100 | 100 | 100 | 100 | 100 | 100 | 100 | 100 | 100 | 100 | 100 | 100 | 100 | 100.0000 | 0.0000 |
| 118 | 100 | 100 | 100 | 100 | 100 | 100 | 100 | 100 | 100 | 100 | 100 | 100 | 100 | 100.0000 | 0.0000 |
| 119 | 100 | 100 | 100 | 100 | 100 | 100 | 100 | 100 | 100 | 100 | 100 | 100 | 100 | 100.0000 | 0.0000 |
| 120 | 100 | 100 | 100 | 100 | 100 | 100 | 100 | 100 | 100 | 100 | 100 | 100 | 100 | 100.0000 | 0.0000 |
| 121 | 100 | 100 | 100 | 100 | 100 | 100 | 100 | 100 | 100 | 100 | 100 | 100 | 100 | 100.0000 | 0.0000 |
| 122 | | 97 | 95 | 100 | 100 | 98 | | 100 | 100 | | | | 100 | 98.7500 | 1.9086 |
| 123 | 100 | 100 | 100 | 100 | 100 | 100 | 100 | 100 | 100 | 100 | 100 | 100 | 100 | 100.0000 | 0.0000 |
| 124 | 100 | 100 | 100 | 100 | 100 | 100 | 100 | 100 | 100 | 100 | 100 | 100 | 100 | 100.0000 | 0.0000 |
| 125 | 100 | 100 | 98 | 100 | 100 | 100 | 100 | 100 | 100 | 100 | 100 | 100 | 100 | 99.8462 | 0.5547 |
| 126 | 100 | 100 | 100 | 100 | 100 | 100 | 100 | 100 | 100 | 100 | 100 | 100 | 100 | 100.0000 | 0.0000 |
| 127 | 100 | 100 | 100 | 100 | 100 | 100 | 99 | 100 | 100 | 100 | 100 | 100 | 100 | 99.9231 | 0.2774 |
| 128 | 100 | 100 | 100 | 100 | 100 | 98 | 100 | 100 | 100 | 100 | 100 | 100 | 100 | 99.8462 | 0.5547 |
| 129 | 100 | 100 | 100 | 100 | 100 | 100 | 97 | 100 | 100 | 100 | 100 | | 100 | 99.7500 | 0.8660 |
| 130 | 100 | 100 | 100 | 100 | 100 | 100 | 100 | 100 | 100 | 100 | 100 | 83 | 100 | 98.6923 | 4.7150 |
| 131 | 100 | 100 | 100 | 100 | 100 | 100 | 100 | 100 | 100 | 100 | 100 | 100 | 100 | 100.0000 | 0.0000 |
| 132 | 98 | 99 | 100 | 100 | 100 | 96 | 100 | 99 | 100 | 100 | 100 | 100 | 100 | 99.3846 | 1.1929 |
| 133 | | | | | | | | | | | | | | #DIV/0! | #DIV/0! |
| 134 | 100 | 100 | 100 | 100 | 100 | 100 | 100 | 100 | 100 | 100 | 100 | 100 | 100 | 100.0000 | 0.0000 |
| 135 | 98 | 100 | 100 | 100 | 100 | 100 | 100 | 100 | 100 | 100 | 100 | 99 | 100 | 99.7692 | 0.5991 |
| 136 | 98 | 100 | 100 | 100 | 100 | 99 | 83 | 100 | 100 | 100 | 100 | 100 | 100 | 98.4615 | 4.6836 |
| 137 | | 100 | 98 | 96 | 100 | 100 | 100 | 100 | 94 | 96 | 91 | 94 | 100 | 97.4167 | 3.1467 |
| 138 | 100 | 100 | 100 | 100 | 100 | 94 | 83 | 96 | 100 | 100 | 100 | 100 | 100 | 97.9231 | 4.8727 |
| 139 | 100 | 100 | 100 | 100 | 100 | 100 | 100 | 100 | 100 | 100 | 100 | 100 | 100 | 100.0000 | 0.0000 |
| 140 | 100 | 100 | 100 | 100 | 100 | 100 | 100 | 100 | 100 | 100 | 100 | 100 | 100 | 100.0000 | 0.0000 |
| 141 | 100 | 100 | 100 | 100 | 100 | 100 | 100 | 100 | 100 | 100 | 100 | 100 | 100 | 100.0000 | 0.0000 |
| 142 | 100 | 100 | 100 | 100 | 100 | 100 | 100 | 95 | 100 | 98 | 100 | 100 | 100 | 99.4615 | 1.4500 |
| 143 | 100 | 100 | 100 | 100 | 100 | 100 | 100 | 100 | 100 | 100 | 100 | 100 | 100 | 100.0000 | 0.0000 |
| 144 | 100 | 100 | 100 | 100 | 100 | 100 | 100 | 100 | 100 | 100 | 100 | 100 | 100 | 100.0000 | 0.0000 |
| 145 | 100 | 100 | 100 | 100 | 100 | 100 | 100 | 100 | 100 | 100 | 100 | 100 | 100 | 100.0000 | 0.0000 |
| 146 | 100 | 100 | 100 | 100 | 100 | 100 | 100 | 100 | 100 | | 100 | 100 | 100 | 100.0000 | 0.0000 |
| 147 | 100 | 100 | 100 | 100 | 100 | 100 | 100 | 100 | 100 | 100 | 100 | 100 | 100 | 100.0000 | 0.0000 |
| 148 | 100 | 100 | 100 | 100 | 100 | 100 | 100 | 100 | 100 | 100 | 100 | 100 | 100 | 100.0000 | 0.0000 |
| 149 | 100 | 100 | 100 | 100 | 100 | 100 | 100 | 100 | 100 | 100 | 100 | 100 | 100 | 100.0000 | 0.0000 |
| 150 | 100 | 100 | 100 | 100 | 100 | 100 | 100 | 99 | 100 | 100 | 100 | 100 | 100 | 99.9231 | 0.2774 |
| 151 | 100 | 100 | 100 | 100 | 100 | 100 | 100 | 97 | 100 | 100 | 100 | 100 | 100 | 99.7692 | 0.8321 |
| 152 | 94 | 97 | 100 | 98 | 100 | 100 | 100 | 100 | 100 | 99 | 100 | 86 | 100 | 98.0000 | 4.0208 |
| 153 | 95 | 100 | 100 | 100 | 100 | 100 | 100 | 100 | 100 | 100 | 100 | 100 | 100 | 99.6154 | 1.3868 |
| 154 | 100 | 94 | 100 | 100 | 100 | 100 | 97 | 100 | 100 | 100 | 100 | | 100 | 99.2500 | 1.8647 |
| 155 | 100 | 100 | 100 | 100 | 100 | 97 | 99 | 100 | 100 | 100 | 100 | 100 | 100 | 99.6923 | 0.8549 |
| 156 | 98 | 100 | 100 | 100 | 100 | 100 | 98 | 100 | 100 | 100 | 100 | 100 | 100 | 99.6923 | 0.7511 |
| 157 | 100 | 100 | 100 | 100 | 100 | 100 | 100 | 100 | 100 | 100 | 100 | 100 | 100 | 100.0000 | 0.0000 |
| 158 | 100 | 100 | 100 | 100 | 100 | 100 | 100 | 100 | 100 | 100 | 100 | 100 | 100 | 100.0000 | 0.0000 |
| 159 | 100 | 99 | 100 | 100 | 100 | 94 | 97 | 100 | 100 | 100 | 100 | 100 | 100 | 99.2308 | 1.7867 |
| 160 | 82 | 99 | 99 | 97 | 100 | 91 | 94 | 97 | 96 | 99 | 89 | 80 | 98 | 93.9231 | 6.6013 |
| 161 | 94 | 100 | 100 | 100 | 100 | 100 | 85 | 100 | 100 | 100 | 100 | 100 | 100 | 98.3846 | 4.3501 |
| 162 | 100 | 100 | 100 | 100 | 100 | 100 | | 100 | 100 | | | 98 | 100 | 99.8000 | 0.6325 |
| 163 | 100 | 100 | 100 | 100 | 100 | 100 | 100 | 100 | 100 | 100 | 100 | 100 | 100 | 100.0000 | 0.0000 |
| 164 | 100 | 100 | 100 | 100 | 100 | 100 | 100 | 100 | 100 | 100 | 100 | 100 | 100 | 100.0000 | 0.0000 |
| 165 | 100 | 100 | 100 | 100 | 100 | 100 | 100 | 100 | 100 | 100 | 100 | 100 | 100 | 100.0000 | 0.0000 |
| 166 | 100 | 100 | 100 | 100 | 100 | 100 | 100 | 100 | 100 | 100 | 100 | 100 | 100 | 100.0000 | 0.0000 |
| 167 | 100 | 100 | 100 | 100 | 100 | 100 | 100 | 99 | 100 | 100 | 100 | 100 | 100 | 99.9231 | 0.2774 |
| 168 | 100 | 100 | 100 | 100 | 100 | 100 | 100 | 100 | 100 | 100 | 100 | 100 | 100 | 100.0000 | 0.0000 |
| 169 | 99 | 94 | 100 | 100 | 100 | 100 | 88 | 100 | 100 | 100 | 100 | | 100 | 98.4167 | 3.7040 |
| 170 | 100 | 100 | 100 | 100 | 100 | 100 | 99 | 100 | 100 | 100 | 100 | 100 | 100 | 99.9231 | 0.2774 |
| 171 | 100 | 100 | 100 | 100 | 100 | 100 | 100 | 100 | 100 | 100 | 100 | 100 | 100 | 100.0000 | 0.0000 |
| 172 | 100 | 100 | 98 | 100 | 100 | 98 | 100 | 100 | 100 | 100 | 100 | 100 | 100 | 99.6923 | 0.7511 |
| 173 | 100 | 100 | 100 | 100 | 100 | 100 | 100 | 100 | 100 | 100 | 100 | 100 | 100 | 100.0000 | 0.0000 |
| 174 | 100 | 94 | 100 | 100 | 100 | 99 | 100 | 98 | 100 | 100 | 100 | 100 | 100 | 99.3077 | 1.7022 |
| 175 | 100 | 100 | 100 | 100 | 100 | 100 | 100 | 100 | 100 | 100 | 100 | 100 | 100 | 100.0000 | 0.0000 |
| 176 | 100 | 100 | 100 | 100 | 100 | 100 | 100 | 100 | 100 | 100 | 100 | 100 | 100 | 100.0000 | 0.0000 |
| 177 | 100 | 100 | 100 | 100 | 100 | 100 | 100 | 100 | 100 | 100 | 100 | 100 | 100 | 100.0000 | 0.0000 |
| 178 | 100 | 100 | 100 | 100 | 100 | 100 | 100 | 100 | 100 | 100 | 100 | 100 | 100 | 100.0000 | 0.0000 |
| 179 | 100 | 100 | 100 | 100 | 100 | 100 | 100 | 100 | 100 | 100 | 100 | 100 | 100 | 100.0000 | 0.0000 |
| 180 | 100 | 100 | 100 | 100 | 100 | 100 | 98 | 100 | 100 | 100 | 100 | 100 | 100 | 99.8462 | 0.5547 |
| 181 | 100 | 100 | 100 | 100 | 100 | 100 | 100 | 100 | 100 | 100 | 100 | 100 | 100 | 100.0000 | 0.0000 |
| 182 | 100 | 100 | 100 | 100 | 100 | 100 | 100 | 100 | 100 | 100 | 100 | 100 | 100 | 100.0000 | 0.0000 |
| 183 | 100 | 100 | 100 | 100 | 100 | 100 | 100 | 100 | 100 | | 100 | 100 | 100 | 100.0000 | 0.0000 |
| 184 | 100 | 96 | 100 | 98 | 100 | 100 | 96 | 100 | 100 | 100 | 100 | 94 | 100 | 98.7692 | 2.0878 |
| 185 | 100 | 100 | 100 | 100 | 100 | 100 | 100 | 100 | 100 | 99 | 100 | 100 | 100 | 99.9231 | 0.2774 |
| 186 | 100 | 100 | 100 | 100 | 100 | 100 | 100 | 100 | 100 | 100 | 100 | 100 | 100 | 100.0000 | 0.0000 |
| 187 | 100 | 100 | 100 | 100 | 100 | 100 | 100 | 100 | 100 | 100 | 100 | 100 | 100 | 100.0000 | 0.0000 |
| 188 | 100 | 100 | 100 | 100 | 100 | 100 | 100 | 100 | 100 | 100 | 100 | 100 | 100 | 100.0000 | 0.0000 |
| 189 | 100 | 100 | 100 | 100 | 100 | 100 | 100 | 100 | 100 | 100 | 100 | 100 | 100 | 100.0000 | 0.0000 |
| 190 | 100 | 100 | 100 | 100 | 100 | 100 | 100 | 94 | 100 | 100 | 100 | 100 | 100 | 99.5385 | 1.6641 |
| 191 | 100 | 100 | 100 | 100 | 100 | 100 | 100 | 100 | 100 | 100 | 100 | 100 | 100 | 100.0000 | 0.0000 |
| 192 | 100 | 100 | 100 | 100 | 100 | 100 | 100 | 100 | 100 | 100 | 100 | 99 | 100 | 99.9231 | 0.2774 |
| 193 | 100 | 100 | 100 | 100 | 100 | 100 | 100 | 100 | 100 | 100 | 100 | 100 | 100 | 100.0000 | 0.0000 |
| 194 | 100 | 94 | 100 | 100 | 100 | 94 | 100 | 100 | 100 | 100 | 100 | 100 | 100 | 99.0769 | 2.2532 |
| 195 | 100 | 100 | 100 | 100 | 100 | 100 | 98 | 100 | 100 | 100 | 100 | 100 | 100 | 99.8462 | 0.5547 |
| 196 | 100 | 100 | 100 | 100 | 100 | 100 | 100 | 100 | 100 | 100 | 100 | 100 | 100 | 100.0000 | 0.0000 |
| 197 | 100 | 100 | 100 | 100 | 100 | 100 | 100 | 94 | 100 | 100 | 100 | 100 | 100 | 99.5385 | 1.6641 |
| 198 | 100 | 100 | 100 | 100 | 100 | 100 | 100 | 100 | 100 | 100 | 100 | 100 | 100 | 100.0000 | 0.0000 |
| 199 | 100 | 100 | 100 | 100 | 100 | 100 | 100 | 100 | 100 | 100 | 100 | 100 | 100 | 100.0000 | 0.0000 |
| 200 | 100 | 100 | 100 | 100 | 100 | 100 | 100 | 100 | 100 | 100 | 100 | 100 | 100 | 100.0000 | 0.0000 |
| 201 | 96 | 100 | 100 | 100 | 100 | 100 | 100 | 100 | 100 | 100 | 100 | 100 | 100 | 99.6923 | 1.1094 |
| Average | 99.203125 | 99.627551 | 99.9128205 | 99.8704663 | 99.959596 | 99.6598985 | 99.0721649 | 99.5888325 | 99.9492386 | 99.8900524 | 99.8928571 | 99.5240642 | 99.9641026 | 99.7025 | |
| StDev | 2.91509712 | 1.38824963 | 0.49491199 | 0.69132142 | 0.56853524 | 1.20841691 | 2.68655228 | 1.28113357 | 0.51255854 | 0.52661331 | 1.01463648 | 2.33790839 | 0.3270354 | 1.5080 | |
| Count | 192 | 196 | 195 | 193 | 198 | 197 | 194 | 197 | 197 | 191 | 196 | 187 | 195 | | |
| Not Recognised | 9 | 5 | 6 | 8 | 3 | 4 | 7 | 4 | 4 | 10 | 5 | 14 | 6 | 85 | |

*Note:* *The original data are in the format of integers in the range between 0 and 100.*
*The data shows the similairity between the mutant images and the original images.The higher the score is, the more similar of the image to the original. The cell is left empty if failed to recognise a face.*
*The data in the red cells are the overall average and standard deviation of similarity scores, respectively.*

**2. Results of Testing Baidu Face Recognition with Datamorphisms**

| ID | Bald | Bangs | Black_Hair | Blond_Hair | Brown_Hair | Bushy_Eyebr | Eyeglasses | Male | Mouth_Sligh | Mustache | No_Beard | Pale_Skin | Young | Average | StDev | Not Recognised |
|---|---|---|---|---|---|---|---|---|---|---|---|---|---|---|---|---|
| 1 | 93.0723648 | 96.1601486 | 96.4501648 | 97.7339172 | 98.4625397 | 92.9880066 | 96.4094391 | 96.1647568 | 97.2691193 | 97.4248352 | 98.1586533 | 96.6934357 | 98.3145065 | 96.5616837 | 1.76006272 | 0 |
| 2 | 91.6354752 | 97.0456085 | 97.0442352 | 95.2690125 | 96.3038788 | 94.1517105 | 95.8644943 | 93.5421448 | 94.85672 | 95.2379532 | 97.2010727 | 70.4935532 | 91.7775497 | 93.109493 | 7.03833532 | 0 |
| 3 | 94.5047302 | 95.5327988 | 95.7782898 | 97.0555115 | 98.3067322 | 91.8055039 | 95.3230896 | 94.1678391 | 97.3341446 | 95.913559 | 98.4977951 | 96.9787598 | 97.3116837 | 96.0392644 | 1.84914481 | 0 |
| 4 | 95.9726563 | 95.9020233 | 98.1240616 | 96.5533218 | 98.7613754 | 82.6860962 | 96.560051 | 93.8718567 | 97.488472 | 96.2151794 | 97.9195023 | 95.5554733 | 97.1303482 | 95.5954167 | 4.08094884 | 0 |
| 5 | 93.1503983 | 93.3662491 | 94.4981613 | 90.6517716 | 91.9542007 | 89.5172501 | 76.0771866 | 89.2585754 | 93.0293198 | 81.8344421 | 96.2122498 | 80.4754944 | 92.3819733 | 89.415944 | 6.10165197 | 0 |
| 6 | 95.6603394 | 94.9706802 | 97.2061386 | 91.4824219 | 97.0156479 | 93.368248 | 92.9856033 | 93.109726 | 94.4100189 | 96.5347366 | 97.2284012 | 96.0430985 | 93.6661606 | 94.8985555 | 1.88052203 | 0 |
| 7 | 92.9900971 | 93.5469132 | 97.9602203 | 92.2581406 | 98.1401367 | 91.7457275 | 92.9236527 | 77.7750473 | 97.5745316 | 96.1190949 | 96.9414673 | 93.8052979 | 95.3195877 | 93.6230704 | 5.26327689 | 0 |
| 8 | 91.6528015 | 96.4232712 | 93.7717514 | 94.9591293 | 97.2629471 | 82.3256149 | 80.8067932 | 90.3063355 | 97.5064926 | 93.7327423 | 97.0079041 | 95.5628204 | 96.1151352 | 92.8795184 | 5.46965069 | 0 |
| 9 | 94.9060898 | 98.000473 | 98.3521271 | 95.2277908 | 98.3730087 | 96.1902084 | 96.9955521 | 90.442337 | 96.4658661 | 92.2446976 | 96.9344635 | 95.2964935 | 96.835228 | 95.8664874 | 2.32785265 | 0 |
| 10 | 95.4530487 | 90.4710541 | 98.223114 | 94.1851196 | 96.6316834 | 91.9185028 | 92.0929031 | 91.4099579 | 96.450676 | 96.8280106 | 94.6704636 | 92.5937958 | 96.3107834 | 94.4030087 | 2.47817465 | 0 |
| 11 | 93.780983 | 96.1578293 | 97.5466538 | 95.9163818 | 98.0663528 | 94.7996368 | 95.3074265 | 95.3307343 | 96.7800903 | 94.4623947 | 97.5838165 | 92.9155579 | 96.7149658 | 95.7971403 | 1.55342539 | 0 |
| 12 | 95.1820831 | 94.5591965 | 97.0997391 | 95.2913742 | 96.1291275 | 93.5933609 | 80.1766663 | 95.572937 | 95.2142334 | 97.342514 | 95.659317 | 93.9903259 | 95.4338303 | 94.2495927 | 4.35784236 | 0 |
| 13 | 95.9514237 | 96.9283066 | 97.3093262 | 95.5873489 | 98.5549545 | 94.3253403 | 92.5302963 | 94.4932098 | 95.0392838 | 96.1028519 | 98.1887283 | 95.4664764 | 94.6339417 | 95.7778068 | 1.67390033 | 0 |
| 14 | 90.9821014 | 96.2140274 | 97.5524368 | 95.1008453 | 98.7279129 | 94.6370697 | 92.359848 | 90.9961014 | 93.6714401 | 94.194931 | 97.3536453 | 93.7903137 | 92.4748612 | 94.4658103 | 2.47076226 | 0 |
| 15 | 92.6949921 | 95.0473023 | 95.79142 | 96.9118729 | 97.6390076 | 95.9229126 | 92.0566483 | 93.8877029 | 96.3787766 | 95.3260727 | 97.5732651 | 94.548996 | 96.0316696 | 95.3700491 | 1.71828818 | 0 |
| 16 | 92.5055161 | 95.5948715 | 93.3901062 | 95.0574646 | 96.8528442 | 93.4379883 | 93.4866104 | 90.7149277 | 97.9754639 | 95.4086609 | 98.4472961 | 94.4641571 | 96.3255463 | 94.8970349 | 2.20807428 | 0 |
| 17 | 95.3372803 | 96.7721405 | 91.6278839 | 95.9220429 | 96.214592 | 78.1570129 | 96.7852783 | 93.2923203 | 95.2413406 | 97.2378464 | 98.2205811 | 95.2199783 | 97.3583679 | 94.4143589 | 5.19077994 | 0 |
| 18 | 91.8662338 | 95.0997009 | 96.3258667 | 96.1093292 | 98.0309448 | 93.945816 | 96.3763351 | 95.6577301 | 97.8896332 | 94.2418823 | 96.6002502 | 91.8964615 | 96.3379974 | 95.4137063 | 1.96058083 | 0 |
| 19 | 91.2189484 | 95.694458 | 97.7918091 | 94.1373901 | 98.0799637 | 94.4788208 | 95.8486633 | 91.7346268 | 95.6063232 | 94.5035477 | 98.0885468 | 91.3039246 | 96.4100952 | 94.992086 | 2.41969961 | 0 |
| 20 | 95.6532059 | 95.791893 | 92.2677002 | 97.2494049 | 96.1040268 | 92.4335022 | 95.01828 | 94.1616516 | 96.2212448 | 94.6412277 | 97.1610947 | 97.9919815 | 95.3324432 | 95.3867428 | 1.71989713 | 0 |
| 21 | 92.4382248 | 94.1398926 | 96.1738892 | 94.2156906 | 97.2587585 | 93.3543472 | 93.286911 | 91.7549973 | 96.1456223 | 95.9002152 | 97.7116242 | 90.7515335 | 95.2750855 | 94.4928301 | 2.13751257 | 0 |
| 22 | 95.9650269 | 96.9640961 | 97.3117371 | 96.2857895 | 98.380333 | 95.5708618 | 95.1933518 | 90.6342621 | 97.286293 | 94.1205139 | 97.464119 | 93.4876251 | 95.6707764 | 95.7180604 | 2.05634612 | 0 |
| 23 | 95.08358 | 95.1192474 | 97.9548721 | 95.708992 | 98.8624878 | 93.5706863 | 93.0375671 | 92.7281189 | 94.4043808 | 95.1362 | 96.7899475 | 94.8523789 | 94.352356 | 95.2000627 | 1.79646908 | 0 |
| 24 | 91.829628 | 94.8010254 | 95.1448364 | 93.8829498 | 94.0098419 | 91.3385391 | 71.7374573 | 94.4149399 | 95.5200195 | 94.8686142 | 96.271286 | 94.5006714 | 94.3435898 | 92.5125691 | 6.38607548 | 0 |
| 25 | 95.2629471 | 94.4130478 | 97.6974487 | 95.0402527 | 96.152092 | 94.3705978 | 94.0194168 | 96.521553 | 97.8738022 | 95.5770569 | 97.2490692 | 89.1621704 | 94.172966 | 95.1932631 | 2.247598 | 0 |
| 26 | 92.2734451 | 97.0857086 | 97.857872 | 96.2118988 | 98.7267685 | 94.8210144 | 94.3089447 | 91.2852097 | 96.1168137 | 94.8071289 | 97.3003006 | 95.4216766 | 94.4510956 | 95.4359906 | 2.12529014 | 0 |
| 27 | 89.5600281 | 97.1910172 | 96.565445 | 97.196907 | 98.4111862 | 97.3964386 | 93.2222443 | 95.5018845 | 97.0685272 | 94.4313278 | 97.8900452 | 95.9996796 | 96.4579163 | 95.914819 | 2.37455427 | 0 |
| 28 | 97.2849731 | 96.5270081 | 98.3390503 | 97.7125168 | 98.7322693 | 93.5980682 | 97.4800491 | 97.4520645 | 98.7808151 | 95.8555145 | 98.5539093 | 96.5060272 | 96.5416489 | 97.1818396 | 1.42987174 | 0 |
| 29 | 95.0684738 | 93.9677048 | 92.8939743 | 95.7899551 | 96.1792679 | 91.7775879 | 95.9322586 | 94.7923203 | 96.7347488 | 96.0499191 | 97.0249405 | 93.7128449 | 94.7307663 | 94.9734433 | 1.54781296 | 0 |
| 30 | 92.0416565 | 93.9739838 | 95.1695938 | 95.2501526 | 97.7674332 | 76.8468933 | 95.519043 | 91.0271759 | 97.4704132 | 95.4813919 | 98.1560211 | 95.0863037 | 95.7902832 | 93.8138727 | 5.49074996 | 0 |
| 31 | 90.5874023 | 92.2332993 | 70.2584839 | 94.9592438 | 91.9522705 | 77.9148483 | 92.499527 | 91.7767563 | 96.2387314 | 93.3840256 | 95.4249191 | 94.7427826 | 95.7828522 | 90.5965494 | 7.69216679 | 0 |
| 32 | 94.2871933 | 94.2450409 | 96.6030655 | 95.6421738 | 97.1824875 | 91.6892624 | 95.4281235 | 96.4463196 | 95.1285629 | 94.0666122 | 96.8990707 | 93.9375839 | 95.8122101 | 95.1821313 | 1.52509613 | 0 |
| 33 | 96.5516663 | 96.5901108 | 98.8258896 | 96.7596664 | 98.3302841 | 93.9594955 | 94.3975525 | 92.1034927 | 97.5411377 | 96.1873627 | 98.3332901 | 93.9286499 | 96.5968781 | 96.1619597 | 2.01494429 | 0 |
| 34 | 95.3388367 | 95.4298859 | 97.4652329 | 94.2810593 | 97.1685867 | 93.4265442 | 97.1254273 | 93.1348724 | 96.7004623 | 94.2515411 | 96.9505081 | 95.7335281 | 95.6371613 | 95.5879728 | 1.46556477 | 0 |
| 35 | 98.0434494 | 94.2305374 | 97.8048096 | 96.3945084 | 98.4772415 | 90.8396606 | 92.4482575 | 95.9735718 | 98.5495529 | 99.0222702 | 98.884964 | 96.4889755 | 95.8206253 | 96.3829557 | 2.55641628 | 0 |
| 36 | | | | | | | | | | | | | | #DIV/0! | #DIV/0! | 13 |
| 37 | 95.9578781 | 95.9886322 | 98.298378 | 94.8024521 | 98.2608185 | 96.7656021 | 93.378891 | 93.2156601 | 96.5529709 | 95.4672165 | 97.3907852 | 95.8392944 | 96.8581009 | 96.0597446 | 1.58845977 | 0 |
| 38 | 92.7824936 | 93.7535629 | 97.2904434 | 92.7577744 | 96.3076325 | 95.1428909 | 77.4441147 | 87.4028626 | 96.6692581 | 92.2506485 | 98.3034363 | 93.9280548 | 93.9531784 | 92.922027 | 5.41425503 | 0 |
| 39 | 94.33638 | 96.9607391 | 97.9958344 | 96.7498627 | 97.7535172 | 94.2649612 | 90.024025 | 96.4722138 | 97.5869827 | 97.4743424 | 97.4891281 | 93.3294373 | 97.3940964 | 95.98704 | 2.34946224 | 0 |
| 40 | 94.3224716 | 97.3297424 | 96.9480972 | 96.7576065 | 98.2981033 | 94.8849182 | 96.476593 | 95.2171555 | 94.9306488 | 93.3167725 | 96.4522324 | 95.6836472 | 94.2533035 | 95.7593302 | 1.42494513 | 0 |
| 41 | | 94.7207718 | 96.6697388 | 95.1296692 | 96.2186585 | 94.2233505 | 93.35289 | 95.0828552 | 94.2075806 | 94.2490845 | 96.5132752 | 94.9313736 | 92.5799255 | 94.8232644 | 1.23064684 | 1 |
| 42 | 85.9081726 | 96.1647034 | 95.837738 | 94.1493378 | 96.9812775 | 79.7439728 | 94.6827393 | 92.0969086 | 95.886467 | 93.8064957 | 96.1741486 | 94.9449921 | 95.1788178 | 93.1965978 | 4.93278352 | 0 |
| 43 | 65.3233719 | 91.9051514 | 95.497612 | 91.0356445 | 92.9744415 | 90.7717056 | 69.0494385 | 88.7159271 | 95.4628601 | 94.474411 | 97.2307205 | 92.1803131 | 78.9836426 | 87.9696338 | 10.2816542 | 0 |
| 44 | 94.097435 | 96.8552551 | 97.2193375 | 95.8505096 | 98.2084961 | 93.1324387 | 95.7073288 | 92.0186386 | 97.1865234 | 94.2917633 | 95.8744202 | 97.0078278 | 95.0798492 | 95.5792172 | 1.7981454 | 0 |
| 45 | 95.5047913 | 96.3153 | 96.8359604 | 92.5833511 | 96.9890442 | 93.9653702 | 93.8194199 | 90.009964 | 95.9094772 | 93.5066071 | 96.3414841 | 93.7173843 | 94.9432373 | 94.6493378 | 1.98795328 | 0 |
| 46 | 96.8258286 | 97.9231567 | 94.1697464 | 97.4294052 | 98.4400101 | 91.5084 | 96.9485931 | 95.4187851 | 98.1589432 | 97.8213654 | 98.5101318 | 94.7607346 | 97.5592041 | 96.5749465 | 2.0633317 | 0 |
| 47 | 95.2139511 | 95.8525391 | 94.6388855 | 96.4147797 | 97.4364853 | 90.3511581 | 95.4975205 | 92.4887238 | 97.1454544 | 95.1758957 | 97.4666214 | 94.7355881 | 95.500412 | 95.2244627 | 1.98725743 | 0 |
| 48 | 94.5053177 | 95.1719132 | 96.4065628 | 92.9859543 | 95.3577042 | 95.6300812 | 94.1185532 | 91.5143814 | 96.4163666 | 93.5966568 | 96.3405457 | 90.2766571 | 92.1272202 | 94.1883011 | 1.99019992 | 0 |
| 49 | 93.5231171 | 96.0236816 | 98.5196228 | 93.8345566 | 97.8268356 | 95.6923294 | 94.9424591 | 91.6825714 | 96.6615677 | 95.5989304 | 97.7854919 | 95.6495132 | 95.5709 | 95.6393521 | 1.88781915 | 0 |
| 50 | 96.2618256 | 96.8117447 | 98.1220627 | 96.9494247 | 97.3872452 | 96.1222305 | 96.951004 | 95.1652603 | 96.7029648 | 96.4776077 | 97.3094788 | 92.2821579 | 97.4667206 | 96.4622867 | 1.45215452 | 0 |
| 51 | 65.4039612 | 94.5200119 | 93.2115402 | 93.3943176 | 96.1489258 | 95.1828384 | 91.619812 | 71.9858322 | 93.9769058 | 93.0833435 | 97.2016525 | 92.8346558 | 94.8968353 | 90.2662025 | 9.77801862 | 0 |
| 52 | 95.2960205 | 91.1978226 | 97.3260803 | 96.9540558 | 97.6616745 | 94.4846649 | 97.748291 | 97.2343063 | 98.5703964 | 96.6325531 | 96.8466034 | 71.93927 | 94.9562607 | 94.372923 | 7.00701213 | 0 |
| 53 | 96.4123077 | 96.4437027 | 93.2414703 | 95.6652985 | 97.3400955 | 91.0242004 | 93.3530655 | 93.4614868 | 97.524025 | 95.4219666 | 98.5758667 | 95.4834366 | 95.6621246 | 95.3545421 | 2.08865954 | 0 |
| 54 | 95.608757 | 97.005188 | 91.2219315 | 97.0380173 | 97.0225906 | 77.9322586 | 95.482338 | 91.0352325 | 98.0871201 | 95.5439224 | 97.5440903 | 96.2673111 | 96.8606567 | 94.3576472 | 5.40194438 | 0 |
| 55 | 94.2354202 | 96.2806473 | 97.7537308 | 96.8028488 | 98.4191513 | 91.3036499 | 91.942894 | 79.3823547 | 96.2536011 | 93.9528656 | 96.6794434 | 95.0213089 | 94.722168 | 94.0576988 | 4.87930448 | 0 |
| 56 | 93.1574783 | 91.878006 | 96.3934708 | 92.8746185 | 94.8257294 | 87.8881226 | 95.2609634 | 94.1657105 | 96.3757706 | 92.157814 | 96.2989349 | 90.4055786 | 94.5839768 | 93.5589365 | 2.5317517 | 0 |
| 57 | 94.3810501 | 96.2648697 | 96.7231979 | 92.9454575 | 96.5707398 | 94.3314896 | 91.7088547 | 90.835434 | 94.2547226 | 93.9656143 | 97.5072632 | 93.5342178 | 95.9130096 | 94.5335324 | 2.00682369 | 0 |
| 58 | 97.8080368 | 97.4304886 | 95.1007996 | 97.7544403 | 97.5414734 | 88.2865067 | 96.5502014 | 94.7529755 | 98.9398041 | 97.4112625 | 98.1665344 | 95.2650147 | 97.6173172 | 96.3557581 | 2.73514775 | 0 |
| 59 | 83.3308258 | 95.225502 | 93.978859 | 95.4646759 | 96.651207 | 93.8435898 | 92.3560867 | 91.7153702 | 95.7100525 | 92.4886627 | 95.8266449 | 79.769928 | 95.8081131 | 92.4745783 | 5.13577021 | 0 |
| 60 | 94.812294 | 95.304245 | 97.416153 | 94.6833496 | 97.3618774 | 92.5160065 | 90.6985703 | 91.3663788 | 94.7952042 | 93.2037582 | 96.8428116 | 92.7690048 | 92.8670883 | 94.2028263 | 2.18803119 | 0 |
| 61 | | 90.873848 | 95.9354248 | 91.136795 | 94.5330734 | 87.1174316 | 86.8468018 | 75.8849106 | 95.4116058 | 94.6844406 | 95.3932343 | 92.5700684 | 83.1967621 | 90.2986997 | 6.08617376 | 1 |
| 62 | 93.7132187 | 95.46595 | 96.598877 | 95.7804642 | 97.5252914 | 93.216713 | 94.6682587 | 93.5583344 | 95.0840149 | 94.8348618 | 96.7997742 | 96.6082459 | 94.5869522 | 95.2646889 | 1.35104606 | 0 |
| 63 | 94.8032761 | 96.1090241 | 97.7599182 | 95.6810837 | 97.97789 | 91.1033859 | 94.0677033 | 93.9992142 | 97.1367798 | 94.655159 | 97.5438156 | 94.6400223 | 96.8317261 | 95.5622306 | 1.95369395 | 0 |
| 64 | 15.2467728 | 95.0198441 | 95.8922958 | 96.422142 | 97.4107513 | 90.844635 | 92.3285065 | 90.6355209 | 94.1614456 | 90.1230621 | 96.5747299 | 91.6629181 | 94.3960419 | 87.7475897 | 21.9228992 | 0 |
| 65 | 96.0715027 | 96.2281418 | 96.1722641 | 96.838562 | 98.8619919 | 92.0752182 | 92.4972611 | 93.5718155 | 97.338501 | 94.039917 | 97.584465 | 95.5812912 | 96.3146439 | 95.6288904 | 2.0243962 | 0 |
| 66 | 94.2082748 | 95.5347977 | 98.1974335 | 93.5892334 | 97.9792938 | 95.428566 | 96.4389877 | 94.0154953 | 96.7235642 | 93.4385529 | 97.2914124 | 93.3622742 | 97.0290909 | 95.6336136 | 1.76696851 | 0 |
| 67 | 96.9570999 | 95.8382721 | 98.2755814 | 96.8210526 | 97.8821793 | 91.4748535 | 97.8544922 | 97.3590927 | 97.6463242 | 97.192131 | 98.6743393 | 93.1053925 | 94.1753311 | 96.4043186 | 2.17882005 | 0 |
| 68 | 95.8783875 | 97.9033432 | 93.9455566 | 97.2072449 | 98.0579605 | 77.1345902 | 97.0680237 | 94.8775635 | 96.8622971 | 96.639595 | 96.8603363 | 94.7405472 | 95.5771713 | 94.8271244 | 5.45916393 | 0 |
| 69 | 92.4609222 | 92.2993088 | 96.8748016 | 92.7339554 | 96.6608963 | 92.699791 | 92.589241 | 92.0607758 | 95.6210327 | 94.0650101 | 96.6342545 | 90.7267151 | 94.6898346 | 93.8551184 | 2.04407407 | 0 |
| 70 | 90.7687836 | 96.5593567 | 91.5919952 | 96.3471756 | 96.0364838 | 93.466568 | 96.263092 | 94.5174484 | 96.9703827 | 96.1253662 | 97.8100128 | 96.4418564 | 96.9644547 | 95.3740751 | 2.16375594 | 0 |
| 71 | 67.7022095 | 91.9101257 | 93.9598694 | 92.9995041 | 94.2791443 | 92.6471329 | 92.4955216 | 94.0263825 | 93.4661789 | 94.4932251 | 94.8814545 | | 91.7954102 | 91.2213465 | 7.47499298 | 1 |
| 72 | 95.0994568 | 95.5039673 | 96.0121841 | 95.4561234 | 97.4649506 | 91.4058304 | 95.1674347 | 90.4016266 | 96.8451538 | 97.2653961 | 97.2149277 | 96.0441895 | 95.5828247 | 95.3433897 | 2.14093375 | 0 |
| 73 | 96.4516068 | 96.9097061 | 97.9366913 | 95.517952 | 97.7718048 | 96.5172577 | 95.7657929 | 92.8634949 | 97.2989121 | 92.96315 | 96.9832382 | 94.1892319 | 94.7531281 | 95.8401513 | 1.70028944 | 0 |
| 74 | 95.9374237 | 96.0220795 | 96.827179 | 94.6161652 | 98.0216904 | 95.2744217 | 95.990448 | 92.6884613 | 95.0468826 | 92.5113602 | 96.9953461 | 94.5221634 | 95.0030212 | 95.3428186 | 1.57630605 | 0 |
| 75 | 93.2048264 | 96.8257904 | 97.9099808 | 96.7982636 | 99.0729981 | 95.6607513 | 97.2738419 | 94.1657562 | 98.4268189 | 96.2722931 | 97.7391815 | 96.6067123 | 98.0031509 | 96.7661819 | 1.65990245 | 0 |
| 76 | 94.2534561 | 97.2571106 | 96.4062653 | 96.7181702 | 98.9856644 | 94.874321 | 96.9799347 | 92.8989105 | 94.1737747 | 96.528656 | 97.1579666 | 96.1027985 | 96.8135376 | 96.0885051 | 1.62265159 | 0 |
| 77 | 97.0588913 | 97.2756882 | 96.3942261 | 97.6193237 | 97.8063583 | 94.1053543 | 97.3143005 | 95.9934921 | 97.3935623 | 96.8650894 | 97.3923874 | 95.8093109 | 95.5107346 | 96.6568246 | 1.05683239 | 0 |
| 78 | 96.9847412 | 93.7929688 | 97.1506043 | 96.6953354 | 98.4192352 | 93.4171372 | 95.4176712 | 92.4543076 | 96.3725662 | 93.8765259 | 97.1883545 | 92.9009018 | 95.6918488 | 95.4124768 | 1.92256599 | 0 |
| 79 | 95.8282242 | 96.9435196 | 98.3506928 | 97.94944 | 98.9551239 | 96.0128174 | 96.6628494 | 94.9520416 | 98.3867493 | 97.8357163 | 98.852623 | 96.0045319 | 96.9364929 | 97.2054479 | 1.28149916 | 0 |
| 80 | 92.7724381 | 91.3867188 | 96.8234024 | 92.3743973 | 94.0947113 | 77.5707398 | 78.3260498 | 91.784195 | 94.5714188 | 91.8323364 | 96.4850769 | 94.5334015 | 94.1205902 | 91.2827289 | 6.15748898 | 0 |
| 81 | 95.5575256 | 96.6740875 | 92.8353119 | 95.74189 | 97.0341339 | 84.206131 | 95.4018326 | 92.8507156 | 96.7453232 | 95.3980331 | 97.4663315 | 94.6065064 | 96.8309479 | 94.7191362 | 3.48263507 | 0 |
| 82 | 95.4991226 | 96.9996567 | 98.3117752 | 94.6379852 | 97.6072159 | 96.2162857 | 95.0736084 | 90.6994629 | 96.4639969 | 91.3695526 | 97.0720902 | 96.6479721 | 95.8003464 | 95.5691593 | 2.25251216 | 0 |
| 83 | 96.1552963 | 93.8312531 | 97.5961914 | 94.4505081 | 96.8864441 | 94.5734024 | 91.9589996 | 94.3777161 | 96.168869 | | 97.1058655 | | 94.3750763 | 95.2254202 | 1.69560917 | 2 |
| 84 | 95.2706909 | 97.4654312 | 98.4497223 | 96.2926407 | 98.6715927 | 98.023468 | 97.8728714 | 97.0808716 | 96.0308456 | 94.4038162 | 97.7490921 | 95.8173599 | 97.1968384 | 96.9480955 | 1.29300267 | 0 |
| 85 | 96.4171219 | 98.1930161 | 94.320282 | 97.7987976 | 97.7862473 | 90.8630295 | 96.4825974 | 95.7027435 | 95.7610703 | 96.1922607 | 98.5357056 | 93.3974762 | 97.7687531 | 96.093777 | 2.18426171 | 0 |
| 86 | 95.548111 | 95.0834503 | 96.9739838 | 95.436348 | 97.3384552 | 93.9209137 | 93.3457413 | 93.7795181 | 97.3631287 | 94.2570801 | 95.8116226 | 94.6761017 | 96.2789002 | 95.370258 | 1.35421256 | 0 |
| 87 | 97.0841217 | 95.4258575 | 97.2008209 | 97.8613586 | 98.9729843 | 92.0763321 | 97.4778214 | 95.0932465 | 97.4672775 | 96.7711182 | 98.0867462 | 97.0286179 | 97.9651337 | 96.808572 | 1.76006954 | 0 |
| 88 | 94.6973953 | 93.8209381 | 94.5316391 | 93.9942322 | 95.0284729 | 85.9593506 | 92.8664703 | 93.5926209 | 95.9939804 | 93.6721726 | 96.984642 | 93.8765106 | 95.4032135 | 93.8785876 | 2.62592659 | 0 |
| 89 | 97.1609345 | 96.5307999 | 97.1860275 | 96.0234757 | 95.7758789 | 95.1944428 | 95.7975617 | 96.2968979 | 97.1662827 | 93.1750946 | 96.8376007 | 95.4567261 | 95.7550659 | 96.0274453 | 1.08981038 | 0 |
| 90 | 96.3733292 | 96.8666306 | 96.2418671 | 98.3330841 | 98.1712418 | 92.0250702 | 96.8675537 | 95.5209503 | 97.7094116 | 90.6023026 | 95.7950287 | 95.7194672 | 96.5275269 | 95.9041126 | 2.24026403 | 0 |
| 91 | 91.6803131 | 95.5642166 | 96.795166 | 93.5561142 | 97.7067337 | 95.2033997 | 93.4950867 | 94.3381043 | 94.9808044 | 97.7686462 | 97.1758499 | 95.6177063 | 93.9623795 | 95.2188093 | 1.82466072 | 0 |
| 92 | 90.9763413 | 82.9396896 | 96.5057221 | 75.918602 | 95.1986694 | 91.3219528 | 91.9575958 | 86.89534 | 94.3349686 | 95.3140259 | 92.951683 | 82.2656555 | 94.6387482 | 90.0937688 | 6.26247591 | 0 |
| 93 | 95.4716492 | 97.2160492 | 91.2381439 | 96.6036453 | 95.7793503 | 92.8896637 | 96.4789658 | 95.4995422 | 97.1746216 | 95.8247147 | 96.2457275 | 97.0065918 | 96.8327866 | 95.7124194 | 1.75838005 | 0 |
| 94 | 90.8841095 | 96.9378662 | 97.0607224 | 95.7730866 | 98.0465698 | 95.6057282 | 93.6796646 | 90.5587463 | 96.4508362 | 92.0796814 | 97.2833939 | 96.6001282 | 95.3866577 | 95.1036301 | 2.50123323 | 0 |
| 95 | 92.9489975 | 95.9031906 | 96.9369888 | 94.9852524 | 98.1361694 | 96.0687485 | 95.3999481 | 95.3262482 | 97.0494843 | 94.9021454 | 97.5278778 | 79.7124252 | 95.6368332 | 94.6564853 | 4.68330619 | 0 |
| 96 | 94.3756867 | 94.0946884 | 98.1186447 | 96.1253204 | 97.6423035 | 93.1419373 | 90.2438736 | 94.9290772 | 96.9580994 | 96.2737656 | 97.6955872 | 94.1694641 | 96.6815033 | 95.419227 | 2.22142744 | 0 |
| 97 | 95.4064407 | 96.6696167 | 98.7077179 | 94.5897141 | 98.5396957 | 96.3999329 | 94.6885681 | 94.9210663 | 97.4389877 | 95.4796372 | 98.2935181 | 96.3989868 | 95.6402969 | 96.3980138 | 1.45619838 | 0 |
| 98 | 97.0278778 | 95.685997 | 97.8025894 | 96.7837448 | 98.3235855 | 93.3290558 | 96.8034363 | 95.547348 | 97.5487671 | 95.154007 | 97.8498383 | 95.7865982 | 96.3953552 | 96.464477 | 1.36429585 | 0 |
| 99 | 95.2005539 | 95.8521042 | 98.5009689 | 97.3176498 | 98.7182159 | 93.9475784 | 95.1017609 | 90.6944809 | 97.6948471 | 95.2012024 | 98.1728058 | 94.6256104 | 95.7639389 | 95.9070552 | 2.22247179 | 0 |
| 100 | 95.7385635 | 97.4257813 | 98.8295212 | 94.0194244 | 98.5586777 | 96.6442413 | 97.1498489 | 95.0024414 | 96.8749542 | 95.6594925 | 98.2391205 | 94.6463928 | 96.6710053 | 96.573805 | 1.50572789 | 0 |
| 101 | 96.085144 | 95.6497192 | 98.5868683 | 96.2316742 | 98.5886078 | 93.9579773 | 93.75737 | 91.5908966 | 97.9179688 | 96.4328995 | 98.1298599 | 91.5247192 | 96.0418015 | 95.7304236 | 2.41010548 | 0 |
| 102 | 76.8312607 | 95.6882324 | 94.3954697 | 95.1384125 | 95.875885 | 95.1399841 | 90.1836472 | 96.4745636 | 95.1225815 | 91.7456055 | 95.9425964 | 93.7301025 | 93.7881012 | 93.0812648 | 5.19483384 | 0 |
| 103 | 93.4089661 | 96.9244232 | 94.1626434 | 96.384285 | 97.2859802 | 94.3321838 | 96.9576645 | 96.6296845 | 96.2602387 | | | 93.1919022 | 96.4328308 | 95.6337093 | 1.53313964 | 2 |
| 104 | 95.9923172 | 96.594101 | 98.3734818 | 97.4818115 | 98.8634415 | 94.7846985 | 95.1317368 | 93.6569748 | 97.1363831 | 95.7138596 | 97.7410355 | 95.8775253 | 96.6777725 | 96.4634722 | 1.4772846 | 0 |
| 105 | 94.5274506 | 94.1994019 | 95.7734146 | 92.3625717 | 95.5847244 | 94.012497 | 85.5182037 | 95.8760605 | 94.3261643 | 96.2799149 | 92.9020996 | 93.4404602 | 96.0635147 | 93.912806 | 2.81511154 | 0 |
| 106 | 91.8480301 | 95.9373932 | 97.5329361 | 96.1356278 | 97.3687134 | 87.7774353 | 95.6876984 | 93.2872391 | 95.8113098 | 94.7781372 | 96.78125 | 96.664093 | 94.7125244 | 94.947876 | 2.6832596 | 0 |
| 107 | 93.8296127 | 93.6370239 | 97.250618 | 96.2626419 | 96.3052445 | 92.6976013 | 92.6692658 | 92.366478 | 94.5756683 | 94.4109192 | 97.3987961 | 92.6824265 | 95.5124512 | 94.584519 | 1.80116747 | 0 |
| 108 | 95.1528397 | 95.5596237 | 96.7105331 | 92.5124741 | 97.1702423 | 96.4636231 | 90.1956329 | 86.9634628 | 95.8544388 | 93.7095718 | 96.4978256 | 93.8046875 | 95.9982224 | 94.3533214 | 2.96644743 | 0 |
| 109 | 92.2654877 | 93.8589478 | 95.9719467 | 94.9847031 | 97.2121735 | 91.5681229 | 91.3295975 | 93.4000244 | 95.7795715 | 96.1124115 | 95.1729507 | 94.1901016 | 92.0140381 | 94.1430828 | 1.92199455 | 0 |
| 110 | 93.8972702 | 93.6831589 | 96.0277176 | 94.3051605 | 97.6891937 | 94.8017044 | 93.0715942 | 93.6518402 | 97.0832596 | 95.695137 | 97.2963715 | 93.7442169 | 94.0695496 | 95.0012442 | 1.5768025 | 0 |
| 111 | 92.7516556 | 96.8337555 | 97.4652939 | 97.2934418 | 98.5971298 | 93.2453156 | 93.190033 | 95.3948517 | 96.8059311 | 94.7773819 | 97.5218964 | 97.2831116 | 96.1946716 | 95.9503438 | 1.90995555 | 0 |
| 112 | 96.2099075 | 96.5570908 | 97.6850967 | 95.679451 | 98.1044464 | 95.5265198 | 95.5449371 | 92.7216873 | 96.6761398 | 95.5124283 | 98.1392136 | 95.8463669 | 95.1202393 | 96.1018096 | 1.43869064 | 0 |
| 113 | 97.4446182 | 97.4330978 | 96.3049622 | 97.2755737 | 98.0094833 | 91.9288178 | 95.8572006 | 94.5889511 | 98.1639938 | 95.8036957 | 97.8236618 | 94.8765945 | 96.4316483 | 96.3032538 | 1.75082767 | 0 |
| 114 | 95.6204376 | 93.7114945 | 97.7878494 | 94.8266449 | 98.0343475 | 90.9494247 | 93.9783554 | 95.4467926 | 94.867157 | 96.1594849 | 97.5502701 | 92.9924927 | 95.9544678 | 95.2214784 | 2.01813634 | 0 |
| 115 | 92.1307449 | 96.1915436 | 92.3736267 | 96.9826126 | 98.2347183 | 95.7325516 | 95.2454681 | 94.8323288 | 96.830658 | 95.6523438 | 96.4334106 | 94.1036911 | 96.3527069 | 95.4689542 | 1.760094 | 0 |
| 116 | 91.9606171 | 94.8319092 | 97.0213776 | 94.1148224 | 96.8373261 | 94.9294891 | 91.9517517 | 94.4870911 | 96.6046982 | 96.1064987 | 96.9654465 | 94.1242218 | 95.3132248 | 95.0191134 | 1.72225543 | 0 |
| 117 | 93.1820831 | 96.3984299 | 95.5888825 | 97.2037506 | 98.8876114 | 93.439682 | 97.5733643 | 94.4335403 | 95.2548523 | 95.6780014 | 98.0158005 | 91.5654526 | 96.8363571 | 95.6967545 | 2.11013162 | 0 |
| 118 | 94.7554016 | 97.5973892 | 97.0856857 | 96.6442184 | 97.5169678 | 94.3061523 | 94.9854202 | 92.9185791 | 94.0927353 | 93.6496353 | 96.0570984 | 96.5131531 | 95.1562958 | 95.4829794 | 1.52901135 | 0 |
| 119 | 95.6083527 | 94.9122162 | 97.4228287 | 95.7922592 | 98.5077896 | 76.8293152 | 96.4081879 | 92.5104981 | 97.1799698 | 93.9068451 | 96.8483582 | 95.4638367 | 96.7839661 | 94.4749556 | 5.52805114 | 0 |
| 120 | 94.9879532 | 95.6933289 | 98.1944275 | 96.4112091 | 98.6191025 | 91.7384644 | 96.4438934 | 93.5537262 | 97.2977676 | 97.0197144 | 98.0583038 | 93.2429657 | 96.6076355 | 95.989884 | 2.08635114 | 0 |
| 121 | 94.4492416 | 95.2134705 | 98.0562439 | 94.6343613 | 98.149559 | 91.4054947 | 93.4992981 | 91.6470413 | 96.6532517 | 94.1508637 | 96.9195938 | 95.0989838 | 94.3599091 | 94.9413317 | 2.11486276 | 0 |
| 122 | 90.7761917 | 91.6170883 | 91.0892258 | 94.6796417 | 93.9237061 | 92.0548401 | 94.2929993 | 93.2704392 | 95.2556229 | 91.9093781 | 93.3397675 | 90.9340591 | 91.788971 | 92.6870716 | 1.51866137 | 0 |
| 123 | 93.9917297 | 95.3355103 | 95.0731201 | 95.9082336 | 97.2987442 | 78.4758606 | 96.214798 | 91.372467 | 96.4045334 | 96.9833603 | 98.0271454 | 94.7224426 | 96.2402725 | 94.3114014 | 5.04518515 | 0 |
| 124 | 91.9934235 | 97.324707 | 98.5738754 | 94.5124893 | 98.6461639 | 98.1703644 | 92.5051422 | 94.9507446 | 97.0036926 | 94.0263977 | 95.3665619 | 95.3941116 | 96.4664078 | 95.7641602 | 2.17634977 | 0 |
| 125 | 96.2320023 | 97.0381546 | 93.1942139 | 96.400322 | 97.5218582 | 91.6876221 | 95.8588409 | 94.9715653 | 97.4042816 | 96.3787079 | 97.014885 | 94.7677918 | 96.8883591 | 95.7968157 | 1.73375292 | 0 |
| 126 | 94.2953186 | 95.0807266 | 91.2809753 | 96.2732468 | 96.8562622 | 93.7508316 | 96.6441574 | 94.0703735 | 96.9233093 | 91.4387436 | 95.2805176 | 94.9064789 | 96.9754715 | 94.9058779 | 1.93186887 | 0 |
| 127 | 94.3159943 | 96.3414078 | 95.7288284 | 94.1770935 | 96.2384186 | 93.9336014 | 90.1986237 | 92.6733475 | 95.2957459 | 94.8701935 | 96.7833023 | 95.3201828 | 94.93396 | 94.6777461 | 1.74772138 | 0 |
| 128 | 95.2313614 | 96.2208252 | 97.7276611 | 97.7477112 | 98.2456207 | 90.830658 | 95.5747299 | 92.6707611 | 98.0845947 | 96.5918961 | 98.3395081 | 96.1305313 | 96.3906708 | 96.1374254 | 2.23237157 | 0 |
| 129 | 91.7865677 | 95.1645889 | 97.1396561 | 95.7060852 | 97.6280365 | 95.3310852 | 93.9440308 | 93.4984665 | 96.408493 | 91.5501251 | 96.9042969 | 96.0098877 | 95.5382919 | 95.1238163 | 1.92436049 | 0 |
| 130 | 94.1039658 | 91.0901642 | 96.0177536 | 94.4220276 | 96.3115616 | 78.8516312 | 94.3029633 | 94.0405502 | 96.0060577 | 93.7213287 | 96.2363358 | 70.1721115 | 93.3129196 | 91.4299516 | 7.84810887 | 0 |
| 131 | 94.7205429 | 97.2465973 | 97.2464218 | 96.2952728 | 98.5004425 | 96.5610428 | 96.054863 | 93.5487976 | 96.471405 | 94.584549 | 97.5405197 | 95.4081955 | 96.5319061 | 96.2085043 | 1.35637608 | 0 |
| 132 | 89.4248047 | 92.0357971 | 92.9684677 | 94.2960205 | 95.5448608 | 69.4303055 | 90.3760147 | 83.3251495 | 94.7010498 | 92.5102158 | 95.3363876 | 91.0951996 | 90.4533997 | 90.1152056 | 7.00241816 | 0 |
| 133 | 93.4061813 | 78.9731522 | 94.9062424 | 91.352829 | 96.3231354 | 93.1641464 | 86.5089493 | 94.3524628 | 95.6034012 | 93.8134613 | 94.3601532 | 91.1967773 | 91.7196655 | 91.9754275 | 4.64427076 | 0 |
| 134 | 91.0321121 | 94.3079605 | 98.1666946 | 95.1770248 | 98.4867554 | 96.7687073 | 97.3939056 | 93.7309494 | 97.5755081 | 95.322731 | 97.6181793 | 95.7850952 | 98.1799164 | 96.1188877 | 2.17859063 | 0 |
| 135 | 93.3950043 | 95.5536041 | 96.8646393 | 93.3939438 | 96.0620422 | 95.515892 | 92.917923 | 94.3131867 | 94.5800171 | 92.8435822 | 97.2312241 | 79.5303879 | 93.864708 | 93.5435503 | 4.45314589 | 0 |
| 136 | 95.1804428 | 93.4657059 | 98.3247147 | 95.7393875 | 97.6456833 | 75.8268967 | 77.9290772 | 92.4627533 | 96.254982 | 94.1169434 | 97.009285 | 96.2941971 | 96.0369568 | 92.7913097 | 7.25817211 | 0 |
| 137 | 69.0859757 | 90.4301682 | 71.3594894 | 90.6636658 | 93.4226837 | 93.1757736 | 93.273819 | 93.1670227 | 87.9250488 | 91.0165634 | 90.1739426 | 72.114708 | 94.0582733 | 86.9128565 | 9.33570738 | 0 |
| 138 | 92.4208984 | 95.1311874 | 93.1233902 | 93.6492004 | 97.1724014 | 75.198204 | 86.8914948 | 91.9235382 | 94.5511246 | 91.1521454 | 95.4573059 | 94.5783081 | 93.5887528 | 91.9106117 | 5.60779299 | 0 |
| 139 | 95.2440033 | 95.5479279 | 97.1371689 | 91.3196106 | 94.8390274 | 94.4826202 | 91.356041 | 91.4277725 | 97.9252243 | 95.9150085 | 97.1287537 | 92.1994019 | 96.5226898 | 94.6957885 | 2.37504358 | 0 |
| 140 | 91.2005997 | 94.0129852 | 96.7651138 | 93.4580383 | 96.3194199 | 94.5147553 | 93.8952179 | 94.0205841 | 96.9237213 | 94.92202 | 97.7075577 | 92.0170288 | 95.0051193 | 94.6740124 | 1.90642242 | 0 |
| 141 | 95.6356583 | 96.9384079 | 97.2955627 | 96.8047409 | 98.9904709 | 91.3052673 | 95.5993195 | 92.0195541 | 97.8715897 | 95.5816422 | 98.114212 | 95.3268814 | 96.3441391 | 95.9867266 | 2.21818927 | 0 |
| 142 | 92.3983078 | 95.5207672 | 97.3954544 | 93.2137146 | 97.3871994 | 95.5409699 | 93.7137451 | 91.4870071 | 96.8329468 | 90.6113663 | 96.7887039 | 63.6735077 | 94.1091919 | 92.2056063 | 8.86192986 | 0 |
| 143 | 94.7035294 | 95.5751267 | 97.0336075 | 94.5355148 | 97.5877686 | 93.9973068 | 91.9309769 | 91.4133987 | 96.0519943 | 92.3262024 | 95.6590805 | 93.0390472 | 94.8227539 | 94.5135621 | 1.92018801 | 0 |
| 144 | 97.4618988 | 96.8417206 | 98.2914505 | 97.451767 | 97.8702927 | 94.1576233 | 97.6395416 | 95.8188171 | 98.063591 | 97.2895584 | 98.6182556 | 95.9511795 | 96.3682098 | 97.0633774 | 1.22582792 | 0 |
| 145 | 95.5605087 | 97.4469757 | 94.9435501 | 97.2137909 | 97.2496109 | 95.0670319 | 96.6979446 | 94.9010925 | 95.903862 | 95.435791 | 97.5149612 | 93.9427185 | 95.1952515 | 95.9286992 | 1.17045458 | 0 |
| 146 | 94.4359818 | 97.7347717 | 91.6577835 | 96.3455124 | 96.559639 | 93.1816177 | 96.7929001 | 95.9219055 | 96.9530563 | 95.3722076 | 97.4910736 | 95.8697586 | 97.5070419 | 95.8325577 | 1.79746281 | 0 |
| 147 | 95.4339294 | 96.9371033 | 98.243866 | 95.9487686 | 98.65345 | 96.607666 | 96.4864807 | 93.5682144 | 97.4201813 | 96.7198868 | 98.1091766 | 97.0660706 | 97.0720291 | 96.7897556 | 1.32154288 | 0 |
| 148 | 96.8230896 | 97.3060608 | 95.3931198 | 97.7717514 | 98.6892777 | 90.1036987 | 97.245697 | 94.976059 | 97.7519684 | 95.2509079 | 97.982399 | 96.533844 | 98.3847961 | 96.4778976 | 2.25560666 | 0 |
| 149 | 93.6109085 | 95.3694 | 94.8400192 | 94.2425537 | 97.1647949 | 92.559166 | 95.1350327 | 93.3812256 | 96.8886337 | 94.4064255 | 96.7586441 | 89.8557358 | 94.1680298 | 94.490813 | 1.97368566 | 0 |
| 150 | 95.4730988 | 96.7687683 | 98.0844193 | 95.8602676 | 98.4454346 | 94.1046219 | 94.4815674 | 91.1113739 | 95.9118805 | 95.8190613 | 97.1350021 | 95.1747513 | 96.7134247 | 95.775667 | 1.89020248 | 0 |
| 151 | 91.6898193 | 95.1864243 | 96.1135864 | 93.1264572 | 96.9250946 | 93.3850327 | 91.2518616 | 91.1782532 | 96.8853455 | 91.9927139 | 97.0735016 | 95.2804337 | 91.4524841 | 93.9646929 | 2.35120031 | 0 |
| 152 | 93.2525406 | 91.290657 | 95.764885 | 94.0725479 | 95.5775909 | 94.4814606 | 94.1241455 | 95.3014679 | 94.6063538 | 94.1666184 | 96.3722 | 74.0934067 | 94.4776764 | 92.8908885 | 5.78778073 | 0 |
| 153 | 79.6732407 | 94.9591141 | 96.5930557 | 94.2978439 | 97.8732605 | 94.2446899 | 95.4346161 | 93.5562744 | 94.6815186 | 94.6390457 | 96.1842804 | 93.8214722 | 95.2232971 | 93.9370546 | 4.44809291 | 0 |
| 154 | 93.2165833 | 90.7826004 | 97.3267822 | 92.2762222 | 95.307457 | 93.6895523 | 92.5348511 | 94.3712845 | 98.3465271 | 98.151001 | 97.2188416 | 91.7067795 | 96.7677002 | 94.7458602 | 2.60233394 | 0 |
| 155 | 93.9667664 | 96.448616 | 96.6094589 | 95.3214798 | 97.7693787 | 74.9197388 | 95.7715836 | 85.465683 | 94.9714737 | 92.4381867 | 96.4798966 | 95.1291428 | 96.6884155 | 93.2292169 | 6.31863734 | 0 |
| 156 | 92.8314743 | 92.5169678 | 94.7978668 | 93.3409271 | 95.256218 | 93.7001572 | 91.5099258 | 93.6233139 | 95.6706924 | 19.4497185 | 96.9695892 | 92.7010269 | 94.8393097 | 88.2467067 | 20.7245569 | 0 |
| 157 | 94.4676514 | 96.5693512 | 97.3874283 | 97.0787354 | 97.7712631 | 92.3182831 | 94.9916611 | 90.2952652 | 96.9563904 | 94.9542389 | 98.046669 | 95.3416672 | 96.1355438 | 95.5626268 | 2.2393475 | 0 |
| 158 | 92.5766449 | 93.268898 | 97.88237 | 94.2720032 | 97.8449478 | 93.870903 | 92.9832611 | 90.50177 | 94.790062 | 93.000946 | 96.3321686 | 92.0981979 | 93.9314194 | 94.1041225 | 2.16642611 | 0 |
| 159 | 93.3084412 | 92.3122864 | 96.0980988 | 96.5429382 | 97.9563217 | 66.3410416 | 90.933876 | 94.4196625 | 96.6099396 | 93.9180145 | 97.5004959 | 93.8609924 | 94.703682 | 92.6542916 | 8.16745189 | 0 |
| 160 | 94.7386169 | 90.3318634 | 97.7498932 | 93.1555176 | 96.7503433 | 92.6065292 | 92.9461365 | 95.3927918 | 96.1155319 | 96.5334168 | 95.627388 | 90.933403 | 96.1488953 | 94.5407944 | 2.3290583 | 0 |
| 161 | 90.1131668 | 95.0073013 | 93.8151321 | 95.4319153 | 96.8074951 | 96.2149658 | 93.3292847 | 92.844696 | 92.5620499 | 92.3655548 | 96.4157028 | 90.7623673 | 93.1963348 | 93.7589205 | 2.11594539 | 0 |
| 162 | 92.1349335 | 91.390625 | 93.7603455 | 90.0923767 | 94.9565125 | 90.3398743 | 92.1039352 | 94.0599899 | 93.9378662 | 94.3470764 | 95.7016678 | 67.764473 | 94.117836 | 91.1313471 | 7.23349798 | 0 |
| 163 | 96.4288788 | 96.7819977 | 97.664238 | 96.839653 | 98.0626297 | 93.2142944 | 95.3250961 | 96.2995529 | 97.223053 | 95.4051437 | 97.4919739 | 94.5327454 | 95.8790207 | 96.2421752 | 1.36353471 | 0 |
| 164 | 95.040802 | 95.2221527 | 95.449379 | 95.1829987 | 96.4482193 | 92.22715 | 96.9529877 | 94.4262695 | 97.2755737 | 95.04422 | 96.8339386 | 91.9579163 | 95.3120499 | 95.182589 | 1.62474556 | 0 |
| 165 | 97.5197296 | 96.3240738 | 98.4493103 | 96.9140778 | 98.0911026 | 94.447113 | 93.2668304 | 93.7537079 | 97.844696 | 98.1734695 | 98.3260269 | 95.9786301 | 96.6911774 | 96.5984573 | 1.77738021 | 0 |
| 166 | 94.430275 | 96.3696518 | 96.0519333 | 96.5600128 | 97.8108749 | 91.4028854 | 93.5066299 | 93.0317535 | 95.2403107 | 95.88517 | 97.5854034 | 93.8863831 | 94.3358383 | 95.084394 | 1.86635528 | 0 |
| 167 | 94.1326828 | 96.4535751 | 97.4096832 | 96.1879807 | 98.2922058 | 94.8237991 | 93.0661392 | 91.756958 | 97.2034226 | 93.3108215 | 96.9698639 | 95.7138367 | 95.6951218 | 95.4627762 | 1.93664676 | 0 |
| 168 | 94.0816116 | 96.2160111 | 93.797554 | 96.133606 | 96.105835 | 90.4635086 | 95.1810608 | 95.3678742 | 95.6203537 | 93.0856781 | 95.8083038 | 93.2729111 | 96.2046585 | 94.718382 | 1.71098524 | 0 |
| 169 | 93.1265564 | 93.9885025 | 98.186409 | 94.3453827 | 96.6050186 | 94.9105072 | 97.8084106 | 96.7666473 | 97.0233841 | 97.1659546 | 95.1960068 | 90.0195999 | 97.370163 | 95.5778879 | 2.30238033 | 0 |
| 170 | 92.3266068 | 95.4211426 | 96.1794281 | 94.0074387 | 98.1319656 | 85.6806259 | 90.9868774 | 92.1654282 | 95.788559 | 92.8513947 | 95.6801224 | 93.6262741 | 93.0283661 | 93.5287869 | 3.08001337 | 0 |
| 171 | 97.4575501 | 95.7034836 | 98.4937134 | 95.5840225 | 97.2783127 | 96.6246643 | 96.8710556 | 93.6648789 | 96.6894684 | 95.9854736 | 97.4573059 | 94.8115616 | 97.8010864 | 96.4940444 | 1.32042965 | 0 |
| 172 | 96.9395447 | 97.0676422 | 93.1776505 | 96.8187408 | 97.5591888 | 91.3898239 | 94.1897354 | 93.0719757 | 97.0302734 | 93.891777 | 95.7348938 | 95.1709137 | 95.570488 | 95.2009729 | 1.92228625 | 0 |
| 173 | 97.5610428 | 96.2701569 | 97.6848831 | 97.892067 | 98.6606979 | 93.623703 | 96.6762085 | 94.6912613 | 97.6989975 | 95.7950745 | 98.4900436 | 97.8514481 | 97.242775 | 96.9337199 | 1.48862828 | 0 |
| 174 | 93.3072968 | 90.0509415 | 97.7613831 | 93.2435608 | 95.6438904 | 90.6121674 | 95.014946 | 90.295517 | 96.8111801 | 97.3540649 | 97.8584518 | 93.3087769 | 93.7878952 | 94.2346209 | 2.7924825 | 0 |
| 175 | 95.4175263 | 95.791069 | 95.6163178 | 95.504036 | 98.4006348 | 92.2182465 | 93.8033142 | 93.1808777 | 96.7009049 | 94.0788651 | 96.509697 | 94.6508713 | 95.0605545 | 95.1486858 | 1.61597915 | 0 |
| 176 | 93.0400391 | 93.4820023 | 96.5650559 | 96.6813507 | 98.2898712 | 90.8347244 | 94.8216934 | 92.1856842 | 96.4900131 | 96.497612 | 98.0217133 | 94.5620194 | 96.9395828 | 95.2624124 | 2.3129959 | 0 |
| 177 | 95.4631958 | 96.4927597 | 98.0454483 | 95.9971237 | 97.925766 | 94.7254181 | 93.9154892 | 95.4230118 | 95.1301498 | 94.1843109 | 97.5547333 | 92.7100983 | 95.6604385 | 95.6329187 | 1.59121695 | 0 |
| 178 | 92.9929047 | 96.9997788 | 96.6005707 | 96.3364639 | 98.7278442 | 94.0178757 | 96.01194 | 95.698616 | 96.8695374 | 94.8197479 | 96.7584152 | 93.4031525 | 96.7778091 | 95.8472812 | 1.62651692 | 0 |
| 179 | 94.1610107 | 95.894577 | 97.0103073 | 95.327713 | 97.634964 | 95.5510712 | 96.6223373 | 96.4043732 | 95.4868164 | 94.0735474 | 97.054596 | 92.2554474 | 93.5095673 | 95.4604868 | 1.57633178 | 0 |
| 180 | 96.0253143 | 95.0403519 | 97.5426483 | 96.7247315 | 98.5758057 | 90.4280777 | 93.8496704 | 95.5869446 | 94.1644592 | 95.0482636 | 96.9596252 | 96.357872 | 95.3607178 | 95.5126525 | 2.01963771 | 0 |
| 181 | 93.8497772 | 93.8163071 | 96.2615356 | 94.8350525 | 98.0663376 | 92.3158417 | 94.2052536 | 93.8700943 | 94.7215729 | 92.7903137 | 95.1140137 | 95.0020752 | 94.9164047 | 94.5972754 | 1.46429893 | 0 |
| 182 | 95.2386704 | 96.4586029 | 98.3738251 | 96.3296509 | 98.4024582 | 95.3053665 | 96.6584396 | 93.9619446 | 97.535614 | 96.5711517 | 98.5012741 | 93.7195358 | 95.074646 | 96.3177831 | 1.60882626 | 0 |
| 183 | 97.2525482 | 96.0689087 | 96.9924851 | 95.9294357 | 96.9984207 | 95.0542526 | 96.5610809 | 93.9726181 | 96.9250565 | 94.8027039 | 96.1835861 | 96.1255493 | 96.3878784 | 96.0965019 | 0.96816417 | 0 |
| 184 | 94.8980866 | 90.2108688 | 96.7833481 | 92.3063202 | 93.4106445 | 81.307251 | 90.4588547 | 91.4771042 | 95.0248566 | 93.1230087 | 95.7865067 | 74.8566742 | 93.5371018 | 91.0138943 | 6.20423966 | 0 |
| 185 | 94.3912659 | 92.2789841 | 97.6326981 | 94.8816299 | 96.6609802 | 93.7840118 | 97.3022385 | 96.0758514 | 96.7690048 | 94.4156189 | 97.2345352 | 90.7836151 | 97.8296127 | 95.3876959 | 2.19297111 | 0 |
| 186 | 95.7065811 | 95.518425 | 97.0928497 | 96.8458633 | 98.3927383 | 93.1067658 | 94.9039154 | 91.3888702 | 97.9778519 | 96.0060196 | 97.5330353 | 95.9982224 | 94.5516129 | 95.7709808 | 1.96777894 | 0 |
| 187 | 97.242981 | 97.8046112 | 94.1799774 | 96.2526779 | 97.8666 | 93.4906769 | 96.6784439 | 92.6664734 | 97.6690521 | 95.9415817 | 97.6365433 | 95.1160049 | 97.6384659 | 96.1680069 | 1.78180697 | 0 |
| 188 | 97.3116608 | 96.3085022 | 96.56176 | 96.4587555 | 98.485611 | 96.2074814 | 94.619278 | 94.719101 | 97.5878372 | 96.3873367 | 98.0077133 | 95.5558929 | 96.318306 | 96.5022489 | 1.14860358 | 0 |
| 189 | 96.3233719 | 98.0734863 | 99.3482056 | 97.5783234 | 99.1406708 | 98.1399536 | 98.057663 | 96.3347855 | 97.7923889 | 95.5499191 | 98.0766449 | 95.5985489 | 97.3778153 | 97.4916752 | 1.21640739 | 0 |
| 190 | 95.6344452 | 94.9575119 | 96.7065964 | 95.6323776 | 97.3510666 | 94.9687424 | 93.7282944 | 82.0352631 | 95.8857422 | 93.3049316 | 97.1341553 | 90.2709961 | 93.0440369 | 93.8964738 | 4.05706691 | 0 |
| 191 | 95.6742249 | 94.9864044 | 97.0132752 | 94.7056809 | 97.8547592 | 87.4640579 | 94.8816528 | 93.7048645 | 96.5163498 | 95.684967 | 97.6007233 | 95.0427246 | 95.7088394 | 95.1414249 | 2.59811861 | 0 |
| 192 | 92.5176926 | 91.8901291 | 96.2963409 | 90.4561005 | 93.94104 | 96.3739471 | 96.1213913 | 93.7365036 | 97.269577 | 94.8016663 | 96.7137833 | 91.6407623 | 96.693779 | 94.4963625 | 2.29141539 | 0 |
| 193 | 96.7852097 | 96.8674164 | 97.8353424 | 97.0020294 | 99.0035934 | 95.0002213 | 95.6350708 | 93.6443329 | 97.4917755 | 96.8036194 | 98.4348068 | 96.8286667 | 95.0161133 | 96.6421691 | 1.48711778 | 0 |
| 194 | 96.9651184 | 91.9428406 | 97.4828262 | 93.4500732 | 94.6578751 | 81.4931107 | 95.5814896 | 95.19104 | 97.1227188 | 96.594017 | 96.3524551 | 92.025116 | 96.5752869 | 94.2641514 | 4.26097231 | 0 |
| 195 | 95.4562149 | 93.4676819 | 97.9359055 | 95.0417709 | 98.1117935 | 92.6306152 | 92.7221069 | 91.4917603 | 97.7637711 | 93.4780426 | 96.6337128 | 87.5645752 | 93.9483566 | 94.326639 | 2.98736374 | 0 |
| 196 | 91.625557 | 94.0287933 | 97.9333878 | 94.5002976 | 97.5132828 | 95.0226746 | 89.1008835 | 92.6095734 | 97.017273 | 96.2696228 | 96.4193497 | 93.1564179 | 95.2115707 | 94.6468218 | 2.54187078 | 0 |
| 197 | 94.3570633 | 94.9882355 | 98.224968 | 93.9621429 | 96.8477173 | 94.4427414 | 95.5759735 | 76.774231 | 95.7883911 | 94.1501312 | 96.1379547 | 94.0776367 | 94.7711182 | 93.8537157 | 5.28035993 | 0 |
| 198 | 93.7548523 | 95.0320892 | 96.8159866 | 96.9810104 | 98.2184753 | 78.400856 | 96.4065476 | 91.372345 | 94.2369843 | 95.0566788 | 97.5470429 | 95.7752457 | 95.4917908 | 94.237685 | 5.08398741 | 0 |
| 199 | 95.7977295 | 97.2261734 | 97.696579 | 97.557724 | 98.2777176 | 92.5531464 | 95.8153458 | 92.3020325 | 94.5917358 | 96.4557648 | 97.8726349 | 97.2965241 | 97.0368271 | 96.1907642 | 1.94952453 | 0 |
| 200 | 96.5287628 | 97.5478134 | 93.0754547 | 96.1792145 | 97.600174 | 92.7419968 | 94.3059082 | 92.9810715 | 96.5086517 | 94.1047287 | 97.3693466 | 92.3476028 | 95.8536682 | 95.1649534 | 1.96726944 | 0 |
| 201 | 90.1483688 | 96.3056641 | 98.3628998 | 95.1108704 | 98.556572 | 96.3321304 | 94.4183731 | 93.2973862 | 96.7893143 | 93.3621597 | 97.6263351 | 94.1103287 | 96.8336792 | 95.4810832 | 2.38314207 | 0 |
| Average | 93.2165062 | 95.1761747 | 96.1999158 | 95.2252444 | 97.3016687 | 91.8545734 | 93.7072825 | 92.8925097 | 96.3444797 | 94.4032014 | 97.0226833 | 93.0181436 | 95.3790171 | 94.7512463 | | |
| StDev | 7.27011575 | 2.33312568 | 3.12544188 | 2.18772355 | 1.37061749 | 5.45352917 | 4.26278887 | 3.52575048 | 1.37261128 | 5.68044992 | 1.13337276 | 5.45760806 | 2.06259366 | | 4.28361442 | |
| Count | 198 | 200 | 200 | 200 | 200 | 200 | 200 | 200 | 200 | 198 | 199 | 198 | 200 | | | |
| Not Recognised | 3 | 1 | 1 | 1 | 1 | 1 | 1 | 1 | 1 | 3 | 2 | 3 | 1 | 20 | | 20 |

*Note:* *The original data are in the format of real numbers in the range between 0 and 100 with 6 decimal digits.*
*The data shows the similairity between the mutant images and the original images.The higher the score is, the more similar of the image to the original. The cell is left empty if failed to recognise a face.*
*The data in the red cells are the overall average and standard deviation of similarity scores, respectively.*

### 3. Results of Testing Face++ with Datamorphisms

| Image ID | Bald | Bangs | Black_Hair | Blond_Hair | Brown_Hair | Bushy_Eyebr | Eyeglasses | Male | Mouth_Sligh | Mustache | No_Beard | Pale_Skin | Young | Average | StDev |
|---|---|---|---|---|---|---|---|---|---|---|---|---|---|---|---|
| 1 | 92.806 | 94.503 | 94.081 | 95.887 | 95.963 | 91.825 | 94.462 | 92.723 | 95.304 | 95.124 | 95.878 | 94.890 | 95.502 | 94.5345 | 1.3385 |
| 2 | 90.068 | 94.734 | 94.516 | 92.596 | 94.435 | 94.052 | 92.288 | 91.309 | 93.414 | 93.725 | 95.548 | 87.382 | 86.232 | 92.3307 | 2.8771 |
| 3 | 91.394 | 95.613 | 92.908 | 95.508 | 96.247 | 91.384 | 92.325 | 87.675 | 95.293 | 91.843 | 95.873 | 95.203 | 94.654 | 93.5323 | 2.5237 |
| 4 | 94.509 | 95.765 | 96.276 | 94.745 | 96.553 | 91.342 | 92.830 | 91.640 | 95.994 | 93.708 | 95.605 | 94.554 | 94.479 | 94.4615 | 1.6841 |
| 5 | 92.118 | 90.665 | 91.936 | 83.611 | 89.787 | 91.480 | 83.407 | 84.225 | 91.930 | 78.724 | 94.131 | 86.555 | 90.015 | 88.3526 | 4.5923 |
| 6 | 94.056 | 93.245 | 95.669 | 91.374 | 95.915 | 91.784 | 92.011 | 87.068 | 93.228 | 94.727 | 94.998 | 94.252 | 91.656 | 93.0756 | 2.3701 |
| 7 | 93.567 | 94.725 | 96.377 | 92.553 | 96.152 | 94.093 | 91.924 | 86.527 | 95.766 | 94.602 | 95.615 | 94.200 | 93.505 | 93.8158 | 2.5688 |
| 8 | 91.583 | 95.027 | 90.991 | 94.758 | 95.798 | 87.233 | 86.543 | 83.512 | 95.769 | 93.665 | 95.818 | 94.593 | 94.489 | 92.2907 | 4.0856 |
| 9 | 90.520 | 95.188 | 96.204 | 94.063 | 96.444 | 95.195 | 94.126 | 83.545 | 94.253 | 88.211 | 95.803 | 91.749 | 94.846 | 93.0882 | 3.7295 |
| 10 | 93.177 | 89.592 | 95.808 | 94.372 | 95.003 | 93.114 | 90.161 | 91.727 | 94.922 | 94.598 | 93.232 | 91.888 | 94.677 | 93.2516 | 1.9289 |
| 11 | 84.282 | 92.508 | 94.651 | 92.715 | 95.243 | 93.641 | 92.124 | 92.320 | 94.525 | 93.633 | | 90.949 | 92.042 | 92.3861 | 2.8453 |
| 12 | 92.509 | 91.763 | 94.734 | 92.981 | 93.329 | 92.392 | 88.779 | 91.589 | 92.224 | 94.039 | 91.870 | 91.895 | 92.986 | 92.3915 | 1.4269 |
| 13 | 94.168 | 95.092 | 95.418 | 94.372 | 96.257 | 95.236 | 92.388 | 88.455 | 93.159 | 94.138 | 95.897 | 94.353 | 92.949 | 93.9909 | 2.0185 |
| 14 | 89.806 | 95.396 | 95.775 | 94.049 | 96.565 | 94.531 | 92.179 | 92.121 | 92.846 | 93.102 | 95.588 | 94.491 | 91.245 | 93.6688 | 1.9888 |
| 15 | 88.291 | 91.834 | 94.499 | 94.019 | 95.295 | 94.519 | 87.461 | 92.470 | 94.096 | 91.937 | 95.751 | 92.851 | 92.982 | 92.7696 | 2.4931 |
| 16 | 87.305 | 92.982 | 93.833 | 93.608 | 95.424 | 90.627 | 87.297 | 88.455 | 95.479 | 94.434 | 95.975 | 92.762 | 92.649 | 92.3715 | 3.0315 |
| 17 | 94.504 | 94.972 | 89.443 | 93.697 | 95.042 | 92.687 | 94.588 | 90.694 | 95.070 | 95.024 | 96.147 | 95.580 | 95.527 | 94.0750 | 1.9937 |
| 18 | 85.770 | 93.781 | 94.785 | 95.081 | 96.184 | 93.617 | 93.265 | 89.784 | 96.193 | 91.880 | 94.617 | 90.785 | 94.434 | 93.0905 | 2.9191 |
| 19 | 91.710 | 93.984 | 95.358 | 92.086 | 96.086 | 94.298 | 91.158 | 88.292 | 94.319 | 93.695 | 95.653 | 92.063 | 94.145 | 93.2959 | 2.1623 |
| 20 | 93.250 | 94.501 | 92.022 | 94.940 | 93.144 | 93.500 | 90.716 | 90.646 | 93.507 | 91.502 | 94.610 | 95.646 | 94.092 | 93.2366 | 1.5975 |
| 21 | 92.559 | 93.841 | 94.871 | 94.251 | 95.876 | 93.573 | 89.562 | 89.741 | 94.296 | 93.935 | 95.976 | 89.062 | 91.905 | 93.0345 | 2.3293 |
| 22 | 93.504 | 95.167 | 96.022 | 94.770 | 96.279 | 95.537 | 93.573 | 87.689 | 95.276 | 90.586 | 95.717 | 94.089 | 91.749 | 93.8429 | 2.4974 |
| 23 | 91.933 | 94.310 | 95.817 | 95.232 | 96.816 | 93.151 | 90.308 | 85.102 | 93.814 | 92.147 | 94.731 | 92.981 | 91.209 | 92.8885 | 2.9919 |
| 24 | 88.676 | 91.566 | 92.864 | 90.754 | 91.075 | 90.196 | 82.977 | 91.278 | 93.571 | 92.323 | 94.483 | 91.555 | 89.256 | 90.8134 | 2.8540 |
| 25 | 92.763 | 92.098 | 94.686 | 93.060 | 94.555 | 92.256 | 88.913 | 92.640 | 93.562 | 93.339 | 94.303 | 91.182 | 91.771 | 92.7022 | 1.5629 |
| 26 | 89.587 | 95.409 | 96.312 | 94.621 | 96.487 | 93.678 | 93.101 | 90.527 | 94.881 | 93.265 | 95.435 | 94.067 | 91.080 | 93.7269 | 2.1849 |
| 27 | 83.621 | 95.766 | 95.388 | 95.890 | 96.497 | 96.078 | 88.721 | 94.650 | 95.626 | 92.582 | 95.846 | 94.619 | 94.924 | 93.8622 | 3.6995 |
| 28 | 94.455 | 93.627 | 95.865 | 95.210 | 96.125 | 92.306 | 95.167 | 93.238 | 95.701 | 92.602 | 95.983 | 94.006 | 94.331 | 94.5089 | 1.2955 |
| 29 | 93.101 | 91.554 | 93.339 | 92.775 | 94.667 | 90.713 | 93.676 | 92.184 | 94.291 | 91.054 | 94.511 | 92.702 | 92.520 | 92.8528 | 1.2640 |
| 30 | 91.811 | 94.208 | 94.073 | 94.647 | 96.408 | 90.027 | 93.842 | 91.761 | 95.355 | 93.314 | 95.347 | 93.635 | 94.302 | 93.7485 | 1.7176 |
| 31 | 93.063 | 93.488 | 89.023 | 94.088 | 92.284 | 89.264 | 92.079 | 92.324 | 94.433 | 94.551 | 95.403 | 94.142 | 93.164 | 92.8697 | 1.9227 |
| 32 | 92.031 | 92.685 | 93.956 | 92.128 | 94.971 | 92.710 | 93.934 | 93.154 | 94.445 | 93.295 | 94.632 | 90.481 | 94.001 | 93.2633 | 1.2537 |
| 33 | 94.490 | 94.832 | 96.741 | 94.006 | 96.282 | 93.998 | 91.912 | 88.843 | 95.426 | 94.669 | 96.424 | 93.080 | 94.238 | 94.2262 | 2.1056 |
| 34 | 93.600 | 94.042 | 95.822 | 92.966 | 95.722 | 91.819 | 95.084 | 92.476 | 95.207 | 90.372 | 95.722 | 95.395 | 93.848 | 94.0058 | 1.7157 |
| 35 | 96.077 | 91.744 | 96.166 | 94.576 | 96.362 | 89.612 | 93.464 | 94.052 | 96.153 | 96.614 | 96.444 | 94.471 | 93.999 | 94.5949 | 2.0939 |
| 36 | 78.287 | 86.981 | | | 92.331 | 88.083 | 88.690 | 89.360 | | | 91.449 | | | 87.8830 | 4.6237 |
| 37 | 94.795 | 94.643 | 96.364 | 94.798 | 96.357 | 95.598 | 92.359 | 88.952 | 95.339 | 91.822 | 95.485 | 94.872 | 95.009 | 94.3379 | 2.0935 |
| 38 | 91.964 | 92.905 | 95.072 | 90.937 | 94.300 | 94.928 | 80.552 | 81.944 | 94.736 | 90.273 | 96.005 | 93.103 | 92.073 | 91.4455 | 4.8405 |
| 39 | 93.356 | 93.008 | 95.652 | 92.481 | 94.708 | 91.058 | 88.926 | 92.368 | 94.400 | 95.121 | 94.787 | 91.162 | 93.998 | 93.1558 | 1.9327 |
| 40 | 91.360 | 94.638 | 95.502 | 94.311 | 95.606 | 92.214 | 94.689 | 94.866 | 94.604 | 89.811 | 94.265 | 93.399 | 93.059 | 93.7172 | 1.7057 |
| 41 | 91.238 | 91.206 | 95.263 | 93.753 | 95.546 | 94.821 | 92.785 | 93.094 | 94.416 | 93.420 | 94.353 | 93.906 | 89.866 | 93.3590 | 1.7040 |
| 42 | 89.008 | 95.227 | 95.714 | 93.575 | 96.061 | 93.618 | 92.769 | 88.664 | 94.419 | 92.247 | 95.251 | 93.726 | 93.799 | 93.3906 | 2.3099 |
| 43 | 83.716 | 86.666 | 93.984 | 91.750 | 94.600 | 92.503 | 71.616 | 85.157 | 93.609 | 93.352 | 95.188 | 91.895 | 86.093 | 89.2407 | 6.5887 |
| 44 | 93.386 | 95.792 | 95.625 | 95.037 | 96.136 | 93.378 | 93.771 | 92.267 | 95.361 | 91.091 | 94.566 | 94.835 | 94.638 | 94.2987 | 1.4688 |
| 45 | 93.844 | 93.963 | 95.491 | 91.747 | 95.001 | 90.726 | 89.596 | 81.923 | 93.871 | 91.275 | 94.480 | 91.754 | 92.505 | 92.0135 | 3.5113 |
| 46 | 94.683 | 96.110 | 90.671 | 95.016 | 95.326 | 89.867 | 92.738 | 93.325 | 96.189 | 95.513 | 96.279 | 93.775 | 94.997 | 94.1915 | 2.0568 |
| 47 | 94.645 | 95.134 | 93.407 | 94.973 | 95.612 | 92.756 | 93.277 | 91.080 | 95.777 | 93.817 | 96.017 | 93.533 | 94.374 | 94.1848 | 1.3955 |
| 48 | 94.220 | 92.568 | 95.569 | 93.109 | 94.285 | 94.335 | 91.924 | 91.517 | 94.635 | 92.465 | 93.951 | 86.284 | 91.946 | 92.8314 | 2.3236 |
| 49 | 94.740 | 95.742 | 96.731 | 94.996 | 96.480 | 95.865 | 93.914 | 91.503 | 95.077 | 94.582 | 95.848 | 94.594 | 95.042 | 95.0088 | 1.3237 |
| 50 | 93.479 | 95.289 | 95.652 | 95.121 | 95.504 | 94.954 | 95.154 | 93.301 | 94.993 | 93.111 | 95.741 | 88.262 | 95.142 | 94.2848 | 2.0185 |
| 51 | 82.813 | 92.839 | 94.442 | 94.360 | 96.173 | 94.219 | 94.269 | 91.784 | 93.106 | 93.743 | 94.881 | 91.200 | 94.692 | 92.9632 | 3.3236 |
| 52 | 90.994 | 86.336 | 94.867 | 93.692 | 95.508 | 91.783 | 94.254 | 93.410 | 95.517 | 93.951 | 94.807 | 88.037 | 89.044 | 92.4769 | 3.0158 |
| 53 | 94.288 | 94.136 | 87.771 | 93.287 | 94.021 | 91.902 | 92.309 | 88.428 | 95.034 | 93.472 | 96.020 | 93.541 | 92.791 | 92.8462 | 2.3719 |
| 54 | 94.402 | 94.055 | 90.129 | 95.185 | 94.789 | 89.226 | 91.583 | 88.903 | 96.063 | 94.152 | 94.893 | 94.163 | 93.849 | 93.1840 | 2.3850 |
| 55 | 91.882 | 95.222 | 95.823 | 94.731 | 96.398 | 90.572 | 87.686 | 85.237 | 94.375 | 91.531 | 95.288 | 93.991 | 93.676 | 92.8009 | 3.3330 |
| 56 | 90.325 | 87.438 | 93.698 | 91.272 | 93.656 | 87.876 | 92.700 | 90.168 | 94.189 | 92.769 | 94.554 | 90.871 | 90.640 | 91.5505 | 2.2918 |
| 57 | 90.881 | 94.664 | 95.705 | 93.096 | 94.882 | 93.234 | 90.153 | 87.082 | 92.312 | 93.016 | 95.825 | 92.177 | 94.036 | 92.8510 | 2.4363 |
| 58 | 95.440 | 95.328 | 92.362 | 94.048 | 95.132 | 91.089 | 91.123 | 90.865 | 95.724 | 94.353 | 95.332 | 91.531 | 93.991 | 93.5629 | 1.8900 |
| 59 | 89.137 | 94.542 | 93.666 | 94.423 | 95.745 | 93.288 | 92.390 | 88.816 | 93.421 | 86.948 | 93.555 | 89.423 | 93.614 | 92.2283 | 2.7050 |
| 60 | 94.023 | 92.997 | 95.657 | 93.609 | 95.845 | 94.168 | 89.422 | 88.910 | 95.199 | 90.699 | 95.447 | 92.846 | 92.958 | 93.2138 | 2.2916 |
| 61 | 69.332 | 89.224 | 93.836 | 87.519 | 92.922 | 90.731 | 81.554 | 81.858 | 94.010 | 89.566 | 94.111 | 88.773 | 88.836 | 87.8671 | 6.9055 |
| 62 | 91.882 | 91.188 | 93.673 | 94.101 | 94.201 | 92.862 | 91.639 | 86.458 | 93.509 | 90.946 | 93.375 | 93.792 | 91.243 | 92.2207 | 2.0966 |
| 63 | 92.767 | 93.966 | 95.613 | 93.636 | 95.341 | 91.701 | 92.073 | 92.573 | 94.234 | 93.644 | 95.612 | 92.555 | 95.146 | 93.7585 | 1.3697 |
| 64 | 93.229 | 94.650 | 95.171 | 95.425 | 96.201 | 89.584 | 88.045 | 81.776 | 93.584 | 88.131 | 95.575 | 92.885 | 91.991 | 92.0190 | 4.1333 |
| 65 | 95.065 | 94.660 | 95.101 | 94.927 | 96.571 | 94.362 | 91.269 | 88.477 | 95.148 | 92.228 | 95.415 | 94.187 | 94.540 | 93.9962 | 2.1434 |
| 66 | 93.119 | 91.864 | 95.044 | 93.406 | 95.741 | 93.365 | 93.367 | 90.734 | 94.505 | 91.584 | 95.123 | 91.313 | 93.839 | 93.3080 | 1.5721 |
| 67 | 94.926 | 92.807 | 95.954 | 94.914 | 96.165 | 90.180 | 95.443 | 94.403 | 94.714 | 95.449 | 95.849 | 85.440 | 91.944 | 93.7068 | 3.0396 |
| 68 | 95.188 | 94.849 | 87.928 | 93.478 | 95.216 | 87.436 | 93.877 | 87.880 | 95.208 | 94.773 | 95.274 | 92.777 | 92.173 | 92.7736 | 3.0314 |
| 69 | 91.531 | 89.055 | 94.941 | 90.393 | 94.857 | 88.252 | 81.045 | 85.340 | 94.277 | 91.905 | 94.774 | 88.042 | 91.506 | 90.4552 | 4.1279 |
| 70 | 88.822 | 95.283 | 93.515 | 95.609 | 95.352 | 93.879 | 94.958 | 92.391 | 95.591 | 94.229 | 96.111 | 94.333 | 95.619 | 94.2840 | 1.9441 |
| 71 | 86.774 | 92.556 | 94.093 | 89.679 | 93.426 | 90.827 | 88.731 | 92.864 | 94.956 | 93.779 | 93.286 | 90.211 | 93.461 | 91.8956 | 2.4321 |
| 72 | 92.514 | 93.898 | 95.162 | 93.949 | 95.942 | 91.293 | 91.472 | 87.188 | 94.823 | 95.912 | 96.251 | 94.496 | 93.303 | 93.5541 | 2.5002 |
| 73 | 93.180 | 94.986 | 96.025 | 92.931 | 96.055 | 95.226 | 92.107 | 86.409 | 94.925 | 87.867 | 93.820 | 93.122 | 90.765 | 92.8783 | 2.9889 |
| 74 | 93.986 | 95.171 | 95.372 | 93.890 | 95.857 | 93.981 | 92.836 | 89.986 | 94.673 | 89.412 | 95.240 | 93.124 | 92.607 | 93.5488 | 1.9850 |
| 75 | 91.934 | 96.139 | 96.074 | 95.125 | 96.597 | 95.489 | 95.375 | 93.120 | 96.373 | 94.684 | 96.052 | 95.309 | 96.349 | 95.2785 | 1.3641 |
| 76 | 94.094 | 95.543 | 94.219 | 95.273 | 96.672 | 92.822 | 94.566 | 89.604 | 93.201 | 94.321 | 95.199 | 93.783 | 95.075 | 94.1825 | 1.7122 |
| 77 | 95.304 | 94.573 | 94.755 | 95.151 | 95.789 | 93.332 | 95.736 | 95.380 | 96.037 | 94.816 | 96.024 | 95.120 | 94.326 | 95.1033 | 0.7603 |
| 78 | 94.974 | 93.670 | 95.602 | 95.008 | 96.141 | 93.883 | 92.580 | 89.643 | 94.405 | 92.558 | 95.845 | 93.481 | 94.284 | 94.0057 | 1.7377 |
| 79 | 93.140 | 94.333 | 95.995 | 95.674 | 96.364 | 94.925 | 93.814 | 90.343 | 95.367 | 95.152 | 95.682 | 94.289 | 93.547 | 94.5096 | 1.5922 |
| 80 | 91.280 | 89.680 | 95.247 | 90.901 | 92.663 | 90.584 | 83.982 | 87.098 | 93.798 | 85.488 | 94.561 | 91.418 | 90.964 | 90.5895 | 3.3685 |
| 81 | 93.998 | 95.377 | 90.994 | 94.000 | 95.201 | 92.815 | 93.700 | 89.599 | 94.918 | 93.460 | 94.474 | 93.120 | 93.492 | 93.4729 | 1.6334 |
| 82 | 93.882 | 95.233 | 96.052 | 93.101 | 95.604 | 94.952 | 92.559 | 86.578 | 95.277 | 89.165 | 95.349 | 94.794 | 94.469 | 93.6165 | 2.7833 |
| 83 | 94.013 | 92.745 | 95.644 | 92.866 | 95.376 | 91.156 | 90.459 | 93.995 | 95.954 | 91.797 | 95.269 | 88.499 | 93.426 | 93.1692 | 2.2357 |
| 84 | 93.793 | 96.008 | 96.139 | 92.718 | 96.494 | 96.273 | 95.523 | 94.883 | 94.464 | 95.231 | 96.040 | 94.444 | 95.513 | 95.1941 | 1.1039 |
| 85 | 94.427 | 95.564 | 91.983 | 95.882 | 95.743 | 92.339 | 92.845 | 92.968 | 95.003 | 94.203 | 95.640 | 91.657 | 94.882 | 94.0874 | 1.5368 |
| 86 | 93.781 | 93.558 | 93.831 | 93.356 | 95.239 | 92.576 | 91.598 | 92.953 | 95.827 | 93.837 | 94.043 | 93.244 | 94.098 | 93.6878 | 1.0722 |
| 87 | 94.587 | 95.450 | 95.962 | 95.496 | 96.572 | 93.133 | 95.318 | 90.325 | 95.932 | 94.145 | 96.189 | 94.875 | 95.050 | 94.8488 | 1.6379 |
| 88 | 92.544 | 91.339 | 91.286 | 90.727 | 93.269 | 91.340 | 89.941 | 87.843 | 93.196 | 90.471 | 94.446 | 92.650 | 92.026 | 91.6214 | 1.7015 |
| 89 | 94.814 | 93.546 | 95.748 | 94.195 | 94.855 | 94.547 | 94.740 | 93.570 | 95.206 | 90.155 | 94.499 | 92.202 | 92.382 | 93.8815 | 1.5243 |
| 90 | 94.429 | 94.349 | 94.229 | 95.711 | 95.590 | 90.932 | 94.486 | 90.611 | 95.230 | 85.088 | 93.304 | 94.350 | 93.615 | 93.2249 | 2.8987 |
| 91 | 91.740 | 93.855 | 95.660 | 91.288 | 95.323 | 94.698 | 90.896 | 92.370 | 92.330 | 94.852 | 95.140 | 93.362 | 91.408 | 93.3017 | 1.7168 |
| 92 | 84.365 | 90.708 | 94.032 | 85.271 | 92.663 | 89.788 | 86.762 | 86.364 | 91.992 | 93.846 | 93.804 | 87.631 | 92.140 | 89.9512 | 3.4774 |
| 93 | 94.206 | 94.635 | 85.810 | 95.112 | 91.718 | 93.581 | 94.432 | 93.637 | 95.184 | 93.360 | 94.401 | 95.748 | 94.650 | 93.5749 | 2.5423 |
| 94 | 88.387 | 94.789 | 94.381 | 93.714 | 96.279 | 94.232 | 91.823 | 85.165 | 93.561 | 92.002 | 95.295 | 94.842 | 92.775 | 92.8650 | 3.0513 |
| 95 | 86.622 | 90.876 | 94.119 | 92.780 | 95.097 | 93.188 | | 93.362 | 93.715 | 89.652 | 93.074 | 85.227 | 91.387 | 91.5916 | 3.0451 |
| 96 | 94.229 | 90.163 | 95.141 | 94.537 | 95.873 | 86.678 | 88.276 | 92.243 | 95.058 | 92.782 | 95.476 | 93.679 | 94.205 | 92.9492 | 2.8854 |
| 97 | 94.822 | 95.831 | 96.386 | 94.034 | 96.772 | 95.377 | 93.040 | 91.563 | 95.008 | 93.959 | 95.586 | 94.391 | 94.060 | 94.6792 | 1.4100 |
| 98 | 95.895 | 93.877 | 95.734 | 95.864 | 96.118 | 92.555 | 95.987 | 93.238 | 95.978 | 94.200 | 95.401 | 94.169 | 95.348 | 94.9511 | 1.1969 |
| 99 | 94.014 | 94.343 | 96.596 | 95.267 | 96.505 | 93.875 | 91.873 | 89.369 | 95.483 | 92.553 | 96.074 | 92.608 | 92.251 | 93.9085 | 2.1254 |
| 100 | 92.412 | 94.333 | 95.933 | 90.876 | 94.976 | 94.318 | 93.692 | 90.553 | 94.868 | 92.495 | 95.008 | 91.855 | 93.634 | 93.4579 | 1.6859 |
| 101 | 93.663 | 93.599 | 96.064 | 93.649 | 96.309 | 94.339 | 92.412 | 88.566 | 95.463 | 92.428 | 95.845 | 89.933 | 93.010 | 93.4831 | 2.3133 |
| 102 | 85.006 | 94.489 | 93.828 | 92.947 | 95.315 | 94.297 | 83.320 | 92.568 | 93.731 | 92.454 | 94.668 | 92.186 | 91.764 | 92.0441 | 3.6704 |
| 103 | 88.084 | 91.229 | 91.379 | 94.212 | 94.812 | 90.985 | 91.734 | 93.097 | 93.220 | 82.797 | 86.422 | 91.438 | 94.388 | 91.0613 | 3.4554 |
| 104 | 93.197 | 94.712 | 96.338 | 95.090 | 96.843 | 94.015 | 92.721 | 92.928 | 94.776 | 94.292 | 95.734 | 94.208 | 94.262 | 94.5474 | 1.2445 |
| 105 | 90.372 | 91.242 | 92.585 | 92.297 | 94.801 | 93.058 | 86.920 | 93.807 | 92.243 | 94.315 | 88.045 | 88.049 | 93.237 | 91.6132 | 2.5487 |
| 106 | 90.286 | 94.148 | 94.337 | 94.762 | 95.874 | 87.205 | 92.261 | 88.512 | 92.290 | 90.777 | 94.685 | 93.717 | 91.549 | 92.3387 | 2.6033 |
| 107 | 91.385 | 91.954 | 95.684 | 94.491 | 95.153 | 91.434 | 88.065 | 88.780 | 93.483 | 89.699 | 94.701 | 93.793 | 92.983 | 92.4312 | 2.4635 |
| 108 | 92.711 | 93.413 | 95.344 | 91.642 | 95.233 | 94.249 | 83.317 | 85.250 | 94.430 | 91.635 | 94.809 | 92.196 | 92.752 | 92.0755 | 3.7052 |
| 109 | 92.600 | 92.432 | 95.695 | 95.501 | 96.452 | 93.379 | 91.331 | 92.782 | 95.114 | 95.076 | 94.323 | 95.075 | 93.483 | 94.0956 | 1.5372 |
| 110 | 92.088 | 90.794 | 95.282 | 93.789 | 95.193 | 93.946 | 88.956 | 92.034 | 93.571 | 93.736 | 94.929 | 89.197 | 92.899 | 92.8011 | 2.1029 |
| 111 | 90.883 | 93.350 | 95.809 | 94.787 | 95.433 | 93.511 | 88.189 | 91.930 | 93.840 | 91.219 | 95.429 | 94.874 | 91.638 | 93.1455 | 2.2552 |
| 112 | 93.212 | 93.976 | 95.812 | 93.316 | 96.155 | 93.684 | 92.187 | 88.066 | 92.823 | 93.031 | 96.111 | 92.385 | 91.541 | 93.2538 | 2.1566 |
| 113 | 95.511 | 95.852 | 93.765 | 94.432 | 95.630 | 92.473 | 93.904 | 90.612 | 96.149 | 93.676 | 95.672 | 93.194 | 93.779 | 94.2038 | 1.5840 |
| 114 | 93.128 | 92.097 | 95.605 | 90.961 | 95.650 | 91.425 | 91.381 | 92.000 | 93.563 | 94.119 | 95.608 | 88.788 | 93.575 | 92.9154 | 2.0666 |
| 115 | 92.533 | 94.895 | 92.822 | 95.885 | 96.503 | 95.158 | 91.778 | 91.093 | 94.551 | 88.525 | 94.721 | 93.756 | 93.907 | 93.5482 | 2.1814 |
| 116 | 90.974 | 90.604 | 95.230 | 93.300 | 94.716 | 93.253 | 86.762 | 92.897 | 94.315 | 93.396 | 95.390 | 92.889 | 93.669 | 92.8765 | 2.3220 |
| 117 | 90.685 | 94.575 | 95.221 | 95.360 | 96.380 | 93.246 | 94.949 | 92.583 | 95.188 | 94.966 | 95.966 | 89.899 | 95.441 | 94.1892 | 2.0092 |
| 118 | 91.606 | 95.275 | 93.437 | 93.007 | 95.324 | 92.812 | 91.852 | 86.704 | 91.058 | 91.466 | 94.932 | 94.177 | 91.795 | 92.5727 | 2.3098 |
| 119 | 92.903 | 94.714 | 95.259 | 95.275 | 96.267 | 91.030 | 93.512 | 88.439 | 95.663 | 90.871 | 95.573 | 93.338 | 95.049 | 93.6841 | 2.3331 |
| 120 | 92.967 | 93.548 | 96.125 | 92.547 | 96.441 | 93.487 | 91.822 | 87.405 | 94.195 | 93.641 | 95.990 | 88.873 | 94.020 | 93.1585 | 2.6389 |
| 121 | 92.188 | 94.990 | 96.109 | 93.709 | 96.238 | 90.832 | 91.340 | 90.090 | 94.472 | 90.905 | 95.080 | 93.798 | 93.020 | 93.2901 | 2.0741 |
| 122 | | 86.974 | 86.978 | 92.861 | 84.709 | 84.882 | 86.706 | 87.490 | 93.172 | 85.073 | 82.165 | 87.971 | 90.274 | 87.4379 | 3.2885 |
| 123 | 91.733 | 93.800 | 94.784 | 93.362 | 95.898 | 91.580 | 94.124 | 90.673 | 94.792 | 95.163 | 96.230 | 93.545 | 93.688 | 93.7978 | 1.6701 |
| 124 | 90.421 | 95.405 | 95.985 | 93.510 | 96.448 | 95.998 | 91.705 | 92.004 | 95.150 | 89.188 | 94.309 | 93.033 | 93.636 | 93.5994 | 2.2685 |
| 125 | 94.862 | 95.276 | 89.170 | 94.713 | 95.054 | 91.660 | 92.924 | 92.638 | 95.914 | 93.786 | 95.026 | 92.809 | 95.015 | 93.7575 | 1.8765 |
| 126 | 91.696 | 93.699 | 88.171 | 94.768 | 94.191 | 91.972 | 90.107 | 88.673 | 94.501 | 89.454 | 93.956 | 94.119 | 94.847 | 92.3195 | 2.4595 |
| 127 | 92.763 | 94.185 | 94.887 | 92.413 | 95.232 | 93.491 | 86.373 | 91.385 | 94.500 | 93.933 | 95.035 | 94.544 | 92.058 | 93.1384 | 2.3699 |
| 128 | 90.375 | 94.769 | 96.083 | 94.537 | 96.244 | 89.177 | 94.229 | 89.405 | 94.818 | 93.523 | 95.339 | 92.826 | 92.292 | 93.3552 | 2.4049 |
| 129 | 90.548 | 93.808 | 95.054 | 94.480 | 95.907 | 94.808 | 88.435 | 87.740 | 95.642 | 91.038 | 95.189 | 94.802 | 94.060 | 93.1932 | 2.7810 |
| 130 | 93.983 | 90.028 | 95.399 | 93.301 | 95.104 | 91.740 | 92.567 | 93.354 | 94.356 | 92.463 | 94.394 | 84.278 | 92.011 | 92.5368 | 2.8911 |
| 131 | 92.469 | 94.981 | 95.362 | 94.070 | 95.962 | 94.840 | 94.176 | 88.579 | 94.535 | 93.433 | 95.341 | 93.642 | 94.082 | 93.9594 | 1.8632 |
| 132 | 88.413 | 91.548 | 91.766 | 93.776 | 93.973 | 89.753 | 86.776 | 82.794 | 92.826 | 87.152 | 93.168 | 88.096 | 86.807 | 89.7575 | 3.4114 |
| 133 | 90.369 | 87.052 | 89.987 | | 91.559 | 91.116 | 81.639 | 89.303 | 93.394 | 86.693 | 94.234 | 90.257 | 84.313 | 89.1597 | 3.6583 |
| 134 | 90.379 | 94.044 | 96.542 | 93.868 | 96.448 | 95.538 | 95.873 | 92.787 | 95.842 | 90.634 | 95.430 | 94.191 | 95.912 | 94.4222 | 2.0630 |
| 135 | 93.792 | 94.022 | 94.471 | 85.151 | 94.201 | 94.268 | 90.443 | 90.258 | 93.136 | 92.953 | 95.698 | 87.575 | 92.021 | 92.1530 | 3.0402 |
| 136 | 94.099 | 92.795 | 95.923 | 94.440 | 95.396 | 89.858 | 83.817 | 91.292 | 95.140 | 93.822 | 95.574 | 94.803 | 94.839 | 93.2152 | 3.3246 |
| 137 | 86.150 | 90.937 | 91.280 | 92.395 | 94.148 | 93.287 | 91.487 | 93.972 | 88.747 | 90.197 | 92.379 | 81.571 | 90.991 | 90.5801 | 3.4641 |
| 138 | 87.031 | 92.728 | 90.123 | 92.079 | 94.768 | 83.868 | 81.699 | 87.438 | 92.930 | 88.977 | 94.307 | 92.312 | 90.906 | 89.9358 | 3.9864 |
| 139 | 92.644 | 90.032 | 95.350 | 89.305 | 92.872 | 93.215 | 90.241 | 88.469 | 95.459 | 92.171 | 94.826 | 90.618 | 93.920 | 92.2402 | 2.3345 |
| 140 | 91.198 | 90.558 | 95.345 | 92.112 | 94.557 | 94.831 | 90.869 | 89.153 | 94.041 | 94.375 | 95.988 | 90.733 | 92.336 | 92.7766 | 2.1880 |
| 141 | 92.331 | 95.422 | 93.335 | 94.066 | 96.513 | 91.773 | 91.893 | 88.714 | 95.327 | 92.970 | 96.064 | 94.258 | 94.371 | 93.6182 | 2.1272 |
| 142 | 90.368 | 94.081 | 94.996 | 87.362 | 95.880 | 94.152 | 89.171 | 83.314 | 94.808 | 88.508 | 94.821 | 88.936 | 91.570 | 91.3821 | 3.7998 |
| 143 | 94.103 | 94.461 | 95.559 | 94.650 | 95.797 | 92.167 | 91.446 | 84.097 | 94.664 | 91.843 | 94.964 | 93.299 | 93.841 | 93.1455 | 3.0455 |
| 144 | 94.433 | 93.282 | 95.803 | 94.269 | 95.648 | 92.845 | 95.035 | 93.113 | 95.861 | 93.562 | 95.595 | 93.677 | 93.414 | 94.3490 | 1.1179 |
| 145 | 95.129 | 94.245 | 93.443 | 94.767 | 94.994 | 94.080 | 94.879 | 93.560 | 94.784 | 91.615 | 94.981 | 95.111 | 92.451 | 94.1568 | 1.1095 |
| 146 | 91.988 | 95.979 | 89.878 | 94.804 | 94.411 | 93.750 | 95.248 | 92.769 | 94.846 | 95.139 | 95.845 | 94.530 | 95.396 | 94.1987 | 1.7330 |
| 147 | 94.340 | 95.609 | 96.406 | 93.596 | 96.580 | 95.813 | 94.057 | 90.178 | 95.355 | 93.623 | 95.134 | 95.392 | 93.618 | 94.5924 | 1.6832 |
| 148 | 94.523 | 94.436 | 93.392 | 95.878 | 96.340 | 91.625 | 94.230 | 91.092 | 95.074 | 93.290 | 95.477 | 94.607 | 95.688 | 94.2809 | 1.5816 |
| 149 | 93.018 | 94.588 | 95.225 | 94.464 | 96.051 | 92.351 | 92.720 | 93.048 | 95.589 | 93.443 | 95.132 | 94.900 | 93.679 | 94.1698 | 1.1955 |
| 150 | 91.647 | 94.493 | 95.289 | 91.947 | 96.296 | 90.923 | 89.571 | 86.739 | 92.703 | 93.181 | 94.574 | 93.416 | 92.760 | 92.5799 | 2.5344 |
| 151 | 87.980 | 93.720 | 95.251 | 91.126 | 95.461 | 93.690 | 88.872 | 80.921 | 95.294 | 93.108 | 95.448 | 94.456 | 90.004 | 91.9485 | 4.2008 |
| 152 | 89.600 | 90.175 | 93.994 | 92.877 | 94.667 | 92.980 | 91.212 | 92.074 | 93.965 | 93.239 | 95.019 | 84.211 | 92.465 | 92.0368 | 2.8565 |
| 153 | 86.456 | 92.599 | 94.725 | 93.204 | 95.619 | 92.946 | 92.075 | 83.102 | 94.231 | 93.431 | 95.089 | 92.516 | 92.762 | 92.2119 | 3.5352 |
| 154 | 91.426 | 87.102 | 95.430 | 91.171 | 93.010 | 91.369 | 88.950 | 91.098 | 95.987 | 94.608 | 94.891 | 83.581 | 93.789 | 91.7240 | 3.5643 |
| 155 | 91.617 | 94.980 | 94.486 | 93.967 | 95.928 | 84.934 | 92.674 | 86.123 | 93.749 | 92.109 | 94.719 | 94.549 | 93.952 | 92.5990 | 3.3607 |
| 156 | 92.094 | 91.506 | 94.450 | 94.596 | 95.189 | 93.131 | 90.245 | 93.745 | 95.282 | 90.820 | 96.088 | 89.536 | 94.827 | 93.1930 | 2.1410 |
| 157 | 92.152 | 95.387 | 94.877 | 95.816 | 96.219 | 93.401 | 92.633 | 88.216 | 94.680 | 93.418 | 95.931 | 94.617 | 93.257 | 93.8926 | 2.1472 |
| 158 | 89.802 | 91.296 | 95.941 | 91.683 | 95.943 | 94.532 | 91.044 | 82.356 | 94.506 | 92.954 | 94.373 | 92.790 | 90.417 | 92.1259 | 3.5658 |
| 159 | 92.094 | 90.669 | 93.760 | 94.691 | 95.783 | 84.620 | 85.074 | 90.266 | 93.163 | 90.628 | 95.170 | 89.762 | 90.894 | 91.2749 | 3.4707 |
| 160 | 94.640 | 88.119 | 94.373 | 89.307 | 95.416 | 91.055 | 86.985 | 93.349 | 94.010 | 94.377 | 93.307 | 81.769 | 92.868 | 91.5058 | 3.9591 |
| 161 | 89.367 | 90.495 | 91.766 | 92.756 | 94.461 | 92.476 | 84.485 | 89.879 | 93.285 | 89.723 | 94.300 | 91.031 | 89.569 | 91.0456 | 2.6491 |
| 162 | 91.187 | 89.794 | 93.886 | 90.197 | 91.527 | 90.541 | 91.681 | 92.409 | 92.495 | 92.485 | 93.926 | 81.396 | 90.095 | 90.8938 | 3.1532 |
| 163 | 95.009 | 94.490 | 95.938 | 95.331 | 96.269 | 92.651 | 94.879 | 94.270 | 94.348 | 92.092 | 94.626 | 90.577 | 94.381 | 94.2201 | 1.5807 |
| 164 | 94.246 | 92.091 | 93.717 | 93.083 | 94.762 | 89.902 | 94.099 | 92.593 | 94.105 | 92.341 | 94.685 | 91.220 | 92.406 | 93.0192 | 1.4430 |
| 165 | 94.878 | 94.835 | 96.395 | 95.025 | 95.853 | 91.483 | 89.377 | 92.057 | 95.307 | 95.779 | 95.555 | 94.238 | 94.247 | 94.2330 | 2.0387 |
| 166 | 93.002 | 94.268 | 94.445 | 94.132 | 95.716 | 90.546 | 90.643 | 89.425 | 93.150 | 93.918 | 95.103 | 90.525 | 90.175 | 92.6960 | 2.1390 |
| 167 | 92.874 | 94.061 | 95.237 | 94.447 | 95.492 | 93.111 | 86.590 | 83.766 | 95.513 | 91.145 | 95.348 | 93.801 | 92.491 | 92.6058 | 3.5936 |
| 168 | 91.067 | 94.245 | 93.169 | 95.086 | 95.273 | 91.677 | 91.185 | 92.459 | 93.066 | 92.431 | 94.963 | 92.427 | 95.089 | 93.2413 | 1.5353 |
| 169 | | 92.267 | 95.813 | 93.833 | 95.300 | 95.226 | 95.024 | 94.707 | 94.738 | 95.388 | 93.188 | | 95.280 | 94.6149 | 1.0802 |
| 170 | 90.789 | 94.440 | 94.013 | 91.291 | 95.460 | 90.796 | 88.029 | 90.463 | 94.226 | 90.750 | 94.520 | 91.261 | 92.197 | 92.1719 | 2.1717 |
| 171 | 94.944 | 94.445 | 96.062 | 93.524 | 95.415 | 94.893 | 92.614 | 91.459 | 94.805 | 94.381 | 95.971 | 93.839 | 95.019 | 94.4132 | 1.2996 |
| 172 | 94.045 | 94.782 | 88.174 | 93.810 | 95.254 | 90.053 | 92.008 | 90.600 | 95.278 | 93.368 | 94.618 | 94.138 | 93.029 | 93.0121 | 2.1977 |
| 173 | 94.405 | 94.612 | 95.570 | 95.670 | 96.388 | 94.942 | 92.823 | 88.226 | 95.522 | 93.692 | 95.946 | 95.469 | 93.980 | 94.4035 | 2.1085 |
| 174 | 92.271 | 90.844 | 95.929 | 93.397 | 95.058 | 90.933 | 93.381 | 90.718 | 95.305 | 95.438 | 95.681 | 92.251 | 94.481 | 93.5144 | 1.9480 |
| 175 | 93.062 | 95.005 | 94.429 | 94.388 | 96.469 | 94.745 | 93.584 | 89.272 | 94.884 | 91.485 | 94.237 | 93.344 | 94.223 | 93.7790 | 1.7891 |
| 176 | 92.043 | 93.916 | 95.605 | 94.764 | 96.297 | 93.534 | 93.248 | 90.120 | 95.748 | 93.964 | 95.737 | 94.463 | 95.269 | 94.2083 | 1.7186 |
| 177 | 93.975 | 93.789 | 96.080 | 92.764 | 95.744 | 94.131 | 90.922 | 92.316 | 94.190 | 91.745 | 95.377 | 91.191 | 93.489 | 93.5164 | 1.6682 |
| 178 | 88.478 | 94.983 | 95.410 | 93.922 | 96.474 | 92.349 | 92.580 | 89.698 | 93.832 | 93.954 | 95.516 | 90.070 | 94.077 | 93.1802 | 2.4472 |
| 179 | 92.333 | 93.871 | 95.630 | 94.388 | 95.894 | 93.943 | 93.657 | 93.954 | 91.675 | 93.272 | 95.080 | 92.678 | 91.235 | 93.6623 | 1.4238 |
| 180 | 93.350 | 92.367 | 94.321 | 94.226 | 95.406 | 88.095 | 90.127 | 90.123 | 93.330 | 90.189 | 94.563 | 93.973 | 92.251 | 92.4862 | 2.2113 |
| 181 | 88.305 | 94.071 | 94.799 | 93.670 | 95.741 | 91.801 | 88.695 | 90.849 | 94.110 | 92.283 | 94.600 | 93.553 | 90.875 | 92.5655 | 2.3392 |
| 182 | 93.699 | 92.715 | 95.934 | 92.821 | 95.792 | 93.084 | 92.111 | 89.441 | 94.418 | 93.812 | 95.974 | 91.639 | 91.684 | 93.3172 | 1.9205 |
| 183 | 94.663 | 93.213 | 95.178 | 94.075 | 95.019 | 91.487 | 94.765 | 91.907 | 94.941 | 92.920 | 94.721 | 94.764 | 94.938 | 94.0455 | 1.2510 |
| 184 | 92.180 | 85.197 | 94.730 | 90.207 | 92.321 | 87.242 | 90.499 | 87.822 | 94.358 | 90.924 | 94.106 | 81.294 | 91.138 | 90.1552 | 3.8913 |
| 185 | 93.919 | 88.545 | 95.845 | 93.808 | 95.562 | 94.366 | 95.094 | 91.469 | 95.634 | 92.381 | 94.887 | 89.456 | 95.898 | 93.6049 | 2.4440 |
| 186 | 93.852 | 94.406 | 96.059 | 95.310 | 96.668 | 94.621 | 92.869 | 87.526 | 95.190 | 93.940 | 95.453 | 93.929 | 94.135 | 94.1506 | 2.2379 |
| 187 | 93.911 | 94.870 | 91.525 | 94.363 | 95.461 | 91.507 | 93.269 | 90.982 | 95.870 | 92.441 | 94.820 | 93.455 | 94.945 | 93.6476 | 1.6122 |
| 188 | 93.854 | 95.804 | 94.617 | 94.526 | 96.176 | 94.973 | 92.149 | 90.563 | 95.469 | 93.264 | 95.623 | 92.785 | 92.655 | 94.0352 | 1.6698 |
| 189 | 94.572 | 95.883 | 96.655 | 94.779 | 96.749 | 96.447 | 95.553 | 93.250 | 96.355 | 94.945 | 96.434 | 94.261 | 94.940 | 95.4479 | 1.0843 |
| 190 | 93.796 | 93.411 | 95.537 | 95.075 | 95.775 | 93.804 | 91.018 | 80.831 | 94.485 | 89.815 | 94.862 | 91.253 | 91.391 | 92.3887 | 3.9634 |
| 191 | 93.513 | 92.892 | 95.082 | 93.681 | 95.578 | 92.122 | 91.765 | 91.262 | 95.197 | 90.860 | 94.998 | 93.840 | 93.699 | 93.4222 | 1.5629 |
| 192 | 92.340 | 89.926 | 95.407 | 88.360 | 93.710 | 95.527 | 93.948 | 92.710 | 94.987 | 93.044 | 95.533 | 84.960 | 94.379 | 92.6793 | 3.1709 |
| 193 | 94.327 | 95.241 | 94.867 | 94.550 | 96.486 | 95.334 | 92.865 | 87.157 | 95.063 | 89.936 | 95.781 | 94.756 | 90.199 | 93.5817 | 2.7740 |
| 194 | 94.287 | 88.414 | 95.163 | 92.129 | 92.301 | 85.909 | 93.266 | 90.643 | 94.172 | 93.810 | 93.822 | 88.181 | 94.189 | 92.0220 | 2.8785 |
| 195 | 93.038 | 92.731 | 95.744 | 93.896 | 96.179 | 93.875 | 88.278 | 83.853 | 95.711 | 92.242 | 95.444 | 90.451 | 91.348 | 92.5223 | 3.4809 |
| 196 | 91.764 | 91.807 | 96.096 | 92.141 | 95.939 | 94.189 | 89.555 | 89.776 | 94.583 | 93.898 | 94.756 | 93.936 | 93.463 | 93.2233 | 2.0902 |
| 197 | 93.899 | 94.271 | 96.383 | 92.313 | 95.305 | 95.224 | 93.348 | 86.599 | 94.601 | 92.223 | 95.466 | 94.225 | 94.052 | 93.6853 | 2.4400 |
| 198 | 93.279 | 94.697 | 94.833 | 94.153 | 96.293 | 90.614 | 93.217 | 90.672 | 94.949 | 94.514 | 95.843 | 94.626 | 94.144 | 93.9872 | 1.7104 |
| 199 | 94.124 | 94.863 | 94.973 | 94.996 | 96.424 | 91.768 | 91.064 | 90.479 | 93.038 | 94.424 | 95.962 | 95.624 | 95.422 | 94.0893 | 1.9167 |
| 200 | 94.588 | 95.043 | 88.354 | 93.257 | 95.374 | 93.592 | 91.836 | 88.921 | 94.136 | 91.931 | 95.523 | 90.297 | 94.218 | 92.8515 | 2.4088 |
| 201 | 89.649 | 93.652 | 96.204 | 94.518 | 96.519 | 95.252 | 90.611 | 91.703 | 92.603 | 91.559 | 95.367 | 87.782 | 94.302 | 93.0555 | 2.6738 |
| Average | 92.0156 | 93.3047 | 94.4530 | 93.5080 | 95.3116 | 92.4935 | 91.2252 | 89.9723 | 94.5040 | 92.4441 | 94.9514 | 92.1766 | 93.0918 | 93.0349 | |
| StDev | 3.1692 | 2.2083 | 2.0514 | 1.9216 | 1.4102 | 2.3420 | 3.4766 | 3.1471 | 1.1154 | 2.3952 | 1.4625 | 2.9391 | 1.8642 | 2.7958 | |
| Count | 199 | 201 | 200 | 199 | 201 | 201 | 200 | 201 | 200 | 200 | 200 | 199 | 200 | | |
| Not Recognised | 2 | 0 | 1 | 2 | 0 | 0 | 1 | 0 | 1 | 1 | 1 | 2 | 1 | 12 | 12 |

*Note:* *The original data are in the format of real numbers in the range between 0 and 100 with 3 decimal digits.*
*The data shows the similairity between the mutant images and the original images.The higher the score is, the more similar of the image to the original. The cell is left empty if it fails to recognise a face*
*The data in the red cells are the overall average and standard deviation of similarity scores, respectively.*

**4. Results of Testing SeetaFace with Datamorphisms**

| Image ID | Bald | Bangs | Black_Hair | Blond_Hair | Brown_Hair | Bushy_Eye | Eyeglasses | Male | Mouth_Sligl | Mustache | No_Beard | Pale_Skin | Young | Average | StDev |
|---|---|---|---|---|---|---|---|---|---|---|---|---|---|---|---|
| 182638.png | 0.757941 | 0.874569 | 0.861861 | 0.914225 | 0.947356 | 0.541035 | 0.888468 | 0.806494 | 0.909724 | 0.870651 | 0.921411 | 0.859388 | 0.895153 | 0.84986738 | 0.105227452 |
| 182639.png | 0.761847 | 0.86689 | 0.855713 | 0.835764 | 0.852629 | 0.78793 | 0.691434 | 0.785498 | 0.864919 | 0.83428 | 0.903342 | 0.524355 | 0.664994 | 0.78689192 | 0.105383259 |
| 182640.png | 0.831675 | 0.709237 | 0.798906 | 0.893598 | 0.907156 | 0.498822 | 0.85721 | 0.748655 | 0.938686 | 0.811263 | 0.933073 | 0.917308 | 0.844626 | 0.82232423 | 0.119876952 |
| 182641.png | 0.837939 | 0.916556 | 0.942128 | 0.882037 | 0.948755 | 0.581622 | 0.872229 | 0.767812 | 0.878194 | 0.870375 | 0.910247 | 0.85952 | 0.887882 | 0.85809969 | 0.095144608 |
| 182642.png | 0.829869 | 0.66784 | 0.906597 | 0.726399 | 0.838199 | 0.49705 | 0.710076 | 0.719839 | 0.911563 | 0.524239 | 0.820693 | 0.690022 | 0.68195 | 0.73264123 | 0.128729165 |
| 182643.png | 0.831272 | 0.821634 | 0.869628 | 0.608501 | 0.88363 | 0.690097 | 0.697371 | 0.715206 | 0.852231 | 0.871785 | 0.881999 | 0.787462 | 0.772924 | 0.79105692 | 0.088608155 |
| 182644.png | 0.816542 | 0.831017 | 0.904069 | 0.622136 | 0.876786 | 0.641605 | 0.726932 | 0.62433 | 0.940284 | 0.824429 | 0.865481 | 0.7821 | 0.812645 | 0.78987354 | 0.105907568 |
| 182645.png | 0.7221 | 0.906048 | 0.869963 | 0.92417 | 0.942168 | 0.429261 | 0.560779 | 0.454949 | 0.925881 | 0.80344 | 0.8614 | 0.805519 | 0.824556 | 0.77155646 | 0.178259513 |
| 182646.png | 0.77528 | 0.889152 | 0.925356 | 0.773652 | 0.926285 | 0.881135 | 0.663544 | 0.605929 | 0.895599 | 0.618162 | 0.896522 | 0.738009 | 0.818572 | 0.80055362 | 0.115284657 |
| 182647.png | 0.825718 | 0.736639 | 0.889589 | 0.782875 | 0.806276 | 0.691786 | 0.633232 | 0.709678 | 0.830097 | 0.794711 | 0.657668 | 0.587487 | 0.890594 | 0.75664231 | 0.096106477 |
| 182648.png | 0.762059 | | 0.888481 | 0.785292 | 0.890639 | 0.775478 | 0.676925 | 0.779639 | 0.853876 | 0.736114 | | | 0.850854 | 0.7999357 | 0.069469136 |
| 182649.png | 0.855998 | 0.771998 | 0.926725 | 0.865278 | 0.822909 | 0.749623 | 0.640084 | 0.837039 | 0.843749 | 0.904419 | 0.865759 | 0.799343 | 0.785074 | 0.82061523 | 0.074211373 |
| 182650.png | 0.836421 | 0.884325 | 0.926352 | 0.826459 | 0.959771 | 0.789276 | 0.733114 | 0.725972 | 0.854906 | 0.762938 | 0.884767 | 0.83181 | 0.805201 | 0.83240862 | 0.070507832 |
| 182651.png | 0.795122 | 0.841949 | 0.910472 | 0.800615 | 0.961797 | 0.735137 | 0.743696 | 0.725474 | 0.854132 | 0.765773 | 0.862834 | 0.829993 | 0.632676 | 0.80459 | 0.086492093 |
| 182652.png | 0.534406 | 0.656192 | 0.847193 | 0.76442 | 0.826506 | 0.701712 | 0.618718 | 0.587958 | 0.857436 | 0.719379 | 0.837919 | 0.685089 | 0.757041 | 0.722613 | 0.104616317 |
| 182653.png | 0.662275 | | 0.710308 | 0.767983 | 0.906774 | | 0.662101 | 0.616145 | 0.860805 | 0.785361 | 0.877171 | 0.812829 | | 0.7661752 | 0.100469854 |
| 182654.png | 0.815192 | 0.815087 | 0.626695 | 0.834405 | 0.800476 | 0.508848 | 0.795147 | 0.70251 | 0.852183 | 0.864716 | 0.901296 | 0.794765 | 0.870292 | 0.78320092 | 0.109714047 |
| 182655.png | 0.639699 | 0.7592 | 0.852674 | 0.8004 | 0.93108 | 0.747811 | 0.84095 | 0.739334 | 0.943617 | 0.676211 | 0.822278 | 0.690325 | 0.747806 | 0.78395269 | 0.092758496 |
| 182656.png | 0.721376 | 0.850029 | 0.899221 | 0.70213 | 0.904287 | 0.804978 | 0.729388 | 0.71878 | 0.879033 | 0.795543 | 0.898267 | 0.683293 | 0.840567 | 0.80206862 | 0.082524915 |
| 182657.png | 0.796195 | 0.696127 | 0.711602 | 0.790054 | 0.869555 | 0.565484 | 0.77388 | 0.68502 | 0.806331 | 0.813372 | 0.887004 | 0.852157 | 0.732348 | 0.76762531 | 0.088466415 |
| 182658.png | 0.797026 | 0.801041 | 0.837423 | 0.74434 | 0.901262 | 0.730559 | 0.702759 | 0.60573 | 0.856461 | 0.731 | 0.84883 | 0.591245 | 0.745352 | 0.76100215 | 0.09305401 |
| 182659.png | 0.520512 | 0.797467 | 0.902699 | 0.811804 | 0.908884 | 0.812274 | 0.553146 | 0.635925 | 0.857341 | 0.79508 | 0.887032 | 0.837349 | 0.67778 | 0.76902254 | 0.130196921 |
| 182660.png | 0.784011 | 0.871617 | 0.91583 | 0.879424 | 0.958185 | 0.687258 | 0.818872 | 0.603416 | 0.817641 | 0.768774 | 0.875428 | 0.863568 | 0.729193 | 0.81332438 | 0.098135213 |
| 182661.png | 0.78433 | 0.791262 | 0.89337 | 0.754433 | 0.83155 | 0.632554 | 0.570978 | 0.768857 | 0.899577 | 0.805372 | 0.828328 | 0.717661 | 0.784709 | 0.77407546 | 0.092463558 |
| 182662.png | 0.878016 | 0.850389 | 0.898084 | 0.886723 | 0.877925 | 0.753925 | 0.709653 | 0.870214 | 0.900236 | 0.815578 | 0.871544 | 0.806126 | 0.925969 | 0.84956785 | 0.062201459 |
| 182663.png | 0.634154 | 0.829349 | 0.902447 | 0.847469 | 0.961975 | 0.700059 | 0.685299 | 0.616334 | 0.884884 | 0.840475 | 0.903595 | 0.821843 | 0.74388 | 0.79782792 | 0.110818642 |
| 182664.png | 0.646583 | 0.95049 | 0.758005 | 0.863144 | 0.959383 | 0.922034 | 0.684424 | 0.834199 | 0.907289 | 0.821346 | 0.899387 | 0.817487 | 0.811453 | 0.83655569 | 0.096465263 |
| 182665.png | 0.882174 | 0.881045 | 0.935651 | 0.944945 | 0.943174 | 0.616689 | 0.898621 | 0.835495 | 0.949125 | 0.830587 | 0.904053 | 0.82227 | 0.877458 | 0.87086823 | 0.088046101 |
| 182666.png | 0.821848 | 0.761891 | 0.780704 | 0.826101 | 0.841911 | 0.612367 | 0.831834 | 0.776917 | 0.903088 | 0.747409 | 0.785479 | 0.813569 | 0.713952 | 0.78592846 | 0.07077559 |
| 182667.png | 0.838934 | 0.727519 | 0.830207 | 0.819019 | 0.93477 | 0.486317 | 0.771953 | 0.6689 | 0.927271 | 0.783357 | 0.872876 | 0.815668 | 0.841772 | 0.79373562 | 0.117559701 |
| 182668.png | | 0.761402 | 0.483557 | 0.818713 | 0.693973 | 0.507035 | 0.635305 | | 0.866824 | 0.814526 | 0.883985 | 0.790702 | | 0.7256022 | 0.142453451 |
| 182669.png | 0.914275 | 0.80447 | 0.910017 | 0.767188 | 0.880063 | 0.697253 | 0.812584 | 0.862098 | 0.897377 | 0.796783 | 0.839742 | 0.689675 | 0.816448 | 0.82215177 | 0.073243748 |
| 182670.png | 0.866263 | 0.841669 | 0.957608 | 0.831701 | 0.938657 | 0.68644 | 0.789209 | 0.638851 | 0.911241 | 0.892218 | 0.948233 | 0.817471 | 0.862583 | 0.84478031 | 0.096256788 |
| 182671.png | 0.873738 | 0.739129 | 0.870013 | 0.750963 | 0.854233 | 0.662556 | 0.85493 | 0.773221 | 0.876755 | 0.726008 | 0.840065 | 0.83621 | 0.825879 | 0.80643846 | 0.068574077 |
| 182672.png | 0.887129 | 0.751681 | 0.906459 | 0.804873 | 0.95476 | 0.533156 | 0.828678 | 0.877546 | 0.932027 | 0.95284 | 0.923622 | 0.798725 | 0.753679 | 0.83885962 | 0.116141289 |
| 182673.png | | | | | | | | | | | | | | #DIV/0! | #DIV/0! |
| 182674.png | 0.888554 | 0.873919 | 0.929058 | 0.845196 | 0.963285 | 0.826067 | 0.696423 | 0.648694 | 0.911415 | 0.753547 | 0.888524 | 0.812213 | 0.810536 | 0.83441777 | 0.091394714 |
| 182675.png | 0.777011 | 0.708122 | 0.884209 | 0.66479 | 0.856765 | 0.803206 | 0.4285 | 0.558333 | 0.898907 | 0.673725 | 0.850505 | 0.813768 | 0.710423 | 0.74063569 | 0.136830017 |
| 182676.png | 0.842777 | 0.841864 | 0.931118 | 0.867106 | 0.916171 | 0.730482 | 0.606554 | 0.8416 | 0.883023 | 0.892698 | 0.875658 | 0.762377 | 0.868756 | 0.83539877 | 0.088421708 |
| 182677.png | 0.728428 | 0.908026 | 0.904991 | 0.871951 | 0.913068 | 0.735339 | 0.816481 | 0.8459 | 0.859024 | 0.639177 | 0.734477 | 0.80707 | 0.828975 | 0.814839 | 0.083668093 |
| 182678.png | | 0.735599 | 0.889655 | 0.72999 | 0.859433 | 0.78588 | 0.671323 | | 0.757609 | 0.771861 | 0.711749 | | | 0.76812211 | 0.069469036 |
| 182679.png | 0.701513 | 0.826142 | 0.847053 | 0.869561 | 0.927645 | 0.483232 | 0.739313 | 0.68069 | 0.878421 | 0.733418 | 0.838462 | 0.799268 | 0.813158 | 0.77983662 | 0.115086748 |
| 182680.png | | | | | | | | | | | | | | #DIV/0! | #DIV/0! |
| 182681.png | 0.858324 | 0.832549 | 0.895911 | 0.848606 | 0.944728 | 0.576314 | 0.745202 | 0.588544 | 0.879035 | 0.776175 | 0.847407 | 0.862417 | 0.802659 | 0.80445162 | 0.110840572 |
| 182682.png | 0.854736 | 0.726638 | 0.917673 | 0.730431 | 0.87915 | 0.535538 | 0.671735 | 0.357636 | 0.874536 | 0.746259 | 0.882553 | 0.698711 | 0.748446 | 0.74031092 | 0.157392093 |
| 182683.png | 0.900564 | 0.953962 | 0.775768 | 0.909739 | 0.901873 | 0.588589 | 0.734962 | 0.823128 | 0.959619 | 0.901408 | 0.937807 | 0.769052 | 0.88346 | 0.84922546 | 0.107027974 |
| 182684.png | 0.85939 | 0.872474 | 0.804593 | 0.880548 | 0.926348 | 0.574879 | 0.812407 | 0.721124 | 0.903165 | 0.80157 | 0.88142 | 0.801899 | 0.837077 | 0.82129954 | 0.091770848 |
| 182685.png | 0.835716 | 0.83902 | 0.937316 | 0.727577 | 0.830026 | 0.824135 | 0.807892 | 0.808596 | 0.866623 | 0.846995 | 0.875377 | 0.666016 | 0.644725 | 0.80846262 | 0.082653433 |
| 182686.png | 0.798037 | 0.853998 | 0.927532 | 0.725933 | 0.913393 | 0.814501 | 0.744905 | 0.481375 | 0.878432 | 0.815119 | 0.926756 | 0.758965 | 0.802917 | 0.80322023 | 0.11772156 |
| 182687.png | 0.822639 | 0.90122 | 0.909073 | 0.850381 | 0.906511 | 0.814802 | 0.864304 | 0.875716 | 0.941894 | 0.869121 | 0.934561 | 0.606678 | 0.87828 | 0.85962923 | 0.085108184 |
| 182688.png | | 0.815787 | 0.849306 | 0.738134 | 0.90397 | 0.831061 | 0.842591 | 0.800801 | 0.844916 | 0.661989 | 0.786548 | | 0.855397 | 0.81186364 | 0.065469808 |
| 182689.png | 0.818143 | | 0.900735 | | 0.906331 | 0.626638 | 0.862015 | | 0.932076 | 0.858899 | 0.868847 | 0.683267 | 0.729588 | 0.8186539 | 0.103516541 |
| 182690.png | 0.872504 | 0.895709 | 0.728204 | 0.861422 | 0.89311 | 0.64255 | 0.782254 | 0.794799 | 0.961222 | 0.814174 | 0.924625 | 0.853725 | 0.827194 | 0.83473015 | 0.08505716 |
| 182691.png | 0.876454 | 0.895403 | 0.718847 | 0.871608 | 0.863867 | 0.398038 | 0.75382 | 0.586581 | 0.917402 | 0.814391 | 0.835265 | 0.837481 | 0.834285 | 0.78488015 | 0.145726764 |
| 182692.png | 0.856194 | 0.885196 | 0.92344 | 0.92699 | 0.975203 | 0.547362 | 0.657653 | 0.635741 | 0.857392 | 0.723206 | 0.874478 | 0.784394 | 0.730269 | 0.79827062 | 0.130176472 |
| 182693.png | 0.818368 | | 0.857415 | 0.808793 | 0.853409 | 0.644345 | 0.784365 | 0.772922 | 0.893766 | 0.694848 | 0.789732 | 0.735398 | 0.790516 | 0.78698975 | 0.070036651 |
| 182694.png | 0.869684 | 0.868201 | 0.89226 | 0.824012 | 0.80603 | 0.825319 | 0.784754 | 0.643018 | 0.782652 | 0.75585 | 0.937687 | 0.711182 | 0.764522 | 0.80501315 | 0.078622209 |
| 182695.png | 0.893724 | 0.880631 | 0.778873 | 0.879879 | 0.886902 | 0.499176 | 0.865523 | 0.72392 | 0.958196 | 0.855225 | 0.873375 | 0.734813 | 0.806268 | 0.81819269 | 0.116792784 |
| 182696.png | 0.680284 | 0.802469 | 0.722968 | 0.831043 | 0.891956 | 0.788457 | 0.641814 | 0.725874 | 0.842114 | 0.749272 | 0.861621 | 0.820668 | 0.804503 | 0.78177254 | 0.073379504 |
| 182697.png | 0.749742 | 0.71983 | 0.891443 | 0.726003 | 0.85397 | 0.545491 | 0.576419 | 0.545136 | 0.825357 | 0.730214 | 0.853967 | 0.674675 | 0.642694 | 0.71807238 | 0.117983571 |
| 182698.png | 0.420726 | 0.562486 | 0.802266 | 0.663329 | 0.801856 | 0.481727 | 0.532942 | 0.53852 | 0.838143 | 0.71259 | 0.784199 | 0.796935 | 0.647906 | 0.66027885 | 0.141020009 |
| 182699.png | | | 0.876345 | | 0.858766 | | | 0.684898 | 0.804668 | | | 0.915331 | 0.664119 | 0.80068783 | 0.104225054 |
| 182700.png | 0.687309 | 0.780895 | 0.873663 | 0.713843 | 0.90641 | 0.588453 | 0.788671 | 0.685088 | 0.927394 | 0.823912 | 0.907505 | 0.727821 | 0.779529 | 0.78388408 | 0.102442128 |
| 182701.png | 0.874831 | 0.802641 | 0.914839 | 0.901778 | 0.922496 | 0.574363 | 0.752584 | 0.550489 | 0.90928 | 0.594327 | 0.856688 | 0.757033 | 0.761471 | 0.78252462 | 0.13428588 |
| 182702.png | 0.924342 | 0.950609 | 0.933617 | 0.932268 | 0.979353 | 0.760974 | 0.720276 | 0.652898 | 0.919163 | 0.727377 | 0.884672 | 0.908018 | 0.866761 | 0.85848677 | 0.105465963 |
| 182703.png | 0.872674 | 0.856261 | 0.911454 | 0.806957 | 0.901159 | 0.776947 | 0.830281 | 0.740576 | 0.911912 | 0.725926 | 0.870247 | 0.56798 | 0.812799 | 0.81424408 | 0.095931968 |
| 182704.png | 0.90505 | 0.822163 | 0.907283 | 0.932152 | 0.932075 | 0.555274 | 0.897712 | 0.872385 | 0.907745 | 0.897546 | 0.890714 | 0.601639 | 0.765699 | 0.83749515 | 0.123902749 |
| 182705.png | 0.861946 | 0.864478 | 0.823934 | 0.903326 | 0.939646 | 0.385451 | 0.767233 | 0.794486 | 0.941174 | 0.87437 | 0.893146 | 0.848139 | 0.858379 | 0.82736215 | 0.14195935 |
| 182706.png | 0.772853 | 0.670974 | 0.86948 | 0.694941 | 0.830609 | 0.568206 | 0.529067 | 0.663991 | 0.871856 | 0.741051 | 0.823759 | 0.696866 | 0.71549 | 0.72685715 | 0.106946628 |
| 182707.png | 0.687671 | 0.87825 | 0.673256 | 0.822646 | 0.850586 | 0.772445 | 0.794434 | 0.751744 | 0.875454 | 0.873292 | 0.934429 | 0.790689 | 0.880945 | 0.81429546 | 0.078588526 |
| 182708.png | 0.587792 | | 0.725717 | 0.854571 | 0.807848 | 0.765802 | 0.64869 | 0.787695 | 0.902049 | 0.797657 | 0.849518 | 0.707472 | 0.828467 | 0.77193983 | 0.090754545 |
| 182709.png | 0.866515 | 0.838235 | 0.884533 | 0.840733 | 0.895496 | 0.700708 | 0.667894 | 0.689984 | 0.933332 | 0.916469 | 0.923892 | 0.899773 | 0.791801 | 0.83456654 | 0.093276305 |
| 182710.png | 0.843678 | 0.858743 | 0.941209 | 0.842409 | 0.898773 | 0.851911 | 0.77484 | 0.677156 | 0.935298 | 0.678712 | 0.811073 | 0.69585 | 0.807221 | 0.81668254 | 0.089299056 |
| 182711.png | 0.865368 | 0.854377 | 0.902066 | 0.826069 | 0.950989 | 0.712873 | 0.793319 | 0.645527 | 0.851779 | 0.6771 | 0.86351 | 0.829188 | 0.705307 | 0.80595938 | 0.093004589 |
| 182712.png | 0.603535 | 0.915704 | 0.91721 | 0.855272 | 0.935278 | 0.831693 | 0.858942 | 0.700187 | 0.920497 | 0.791597 | 0.876204 | 0.751178 | 0.910618 | 0.83599346 | 0.099873129 |
| 182713.png | 0.825062 | 0.901819 | 0.824079 | 0.906864 | 0.962336 | 0.816871 | 0.817892 | 0.79853 | 0.886491 | 0.881727 | 0.898851 | 0.719493 | 0.903366 | 0.85718315 | 0.063602522 |
| 182714.png | 0.875999 | 0.896064 | 0.851907 | 0.920851 | 0.963142 | 0.747494 | 0.863896 | 0.888078 | 0.917376 | 0.842258 | 0.888136 | 0.876207 | 0.884054 | 0.87811246 | 0.050331592 |
| 182715.png | 0.917389 | 0.7864 | 0.93174 | 0.881662 | 0.933136 | 0.690278 | 0.750835 | 0.61343 | 0.883417 | 0.804186 | 0.890122 | 0.729484 | 0.771766 | 0.81414192 | 0.101343705 |
| 182716.png | 0.869168 | 0.863579 | 0.951461 | 0.916538 | 0.962515 | 0.816855 | 0.864405 | 0.807186 | 0.923259 | 0.862126 | 0.924765 | 0.849075 | 0.818108 | 0.87915692 | 0.051754421 |
| 182717.png | 0.730879 | 0.660876 | 0.82474 | 0.672726 | 0.806183 | 0.61901 | 0.526473 | 0.608446 | 0.879514 | 0.719783 | 0.85457 | 0.679242 | 0.735286 | 0.71674831 | 0.103944718 |
| 182718.png | 0.853915 | 0.849774 | 0.723676 | 0.779968 | 0.925976 | 0.540953 | 0.756905 | 0.625541 | 0.872737 | 0.814213 | 0.851708 | 0.813852 | 0.768893 | 0.78293162 | 0.105046803 |
| 182719.png | 0.851194 | 0.909755 | 0.948342 | 0.784028 | 0.924994 | 0.788744 | 0.714041 | 0.742473 | 0.914869 | 0.636998 | 0.871031 | 0.808992 | 0.813748 | 0.82378531 | 0.091788306 |
| 182720.png | 0.84306 | 0.843786 | | 0.855621 | 0.880999 | 0.688939 | 0.737036 | 0.865917 | 0.906824 | 0.789214 | 0.835414 | 0.663937 | 0.754831 | 0.80546483 | 0.078071549 |
| 182721.png | 0.751917 | 0.908705 | 0.921895 | 0.780479 | 0.936935 | 0.902413 | 0.909498 | 0.817396 | 0.939522 | 0.861938 | 0.942261 | 0.74603 | 0.859326 | 0.86756269 | 0.071607825 |
| 182722.png | 0.858627 | 0.899867 | 0.757421 | 0.909342 | 0.909856 | 0.522248 | 0.832472 | 0.831865 | 0.879855 | 0.86895 | 0.904973 | 0.767511 | 0.844923 | 0.82983923 | 0.104925545 |
| 182723.png | 0.833465 | | 0.863552 | 0.841632 | 0.922572 | 0.602362 | 0.75691 | 0.839042 | 0.911436 | 0.853658 | 0.85715 | 0.847809 | 0.805905 | 0.82795775 | 0.083045394 |
| 182724.png | 0.81722 | 0.847944 | 0.908736 | 0.870706 | 0.940974 | 0.718451 | 0.877597 | 0.764583 | 0.904966 | 0.83525 | 0.914189 | 0.803653 | 0.87446 | 0.85220992 | 0.063664732 |
| 182725.png | 0.856042 | 0.800081 | 0.808598 | 0.769607 | 0.872431 | 0.454504 | 0.618585 | 0.703935 | 0.812646 | 0.72196 | 0.846031 | 0.756042 | 0.755141 | 0.75196946 | 0.112995399 |
| 182726.png | 0.932982 | 0.875933 | 0.918357 | 0.84572 | 0.914041 | 0.714118 | 0.855571 | 0.853022 | 0.853871 | 0.815149 | 0.82637 | | 0.64311 | 0.83735367 | 0.083951899 |
| 182727.png | 0.893226 | 0.877018 | 0.90426 | 0.929361 | 0.931428 | 0.541547 | 0.815305 | 0.799642 | 0.898373 | 0.64657 | 0.773798 | 0.888548 | 0.831668 | 0.82544185 | 0.115782248 |
| 182728.png | 0.784312 | 0.878352 | 0.912753 | 0.715949 | 0.910456 | 0.811544 | 0.827811 | 0.824799 | 0.884987 | 0.934164 | 0.904947 | 0.810007 | 0.832432 | 0.84865485 | 0.06224882 |
| 182729.png | 0.695691 | 0.631191 | 0.859757 | 0.745771 | 0.891638 | 0.634688 | 0.721287 | 0.670005 | 0.807285 | 0.859378 | 0.775101 | 0.707176 | 0.842754 | 0.75705554 | 0.089151826 |
| 182730.png | 0.784869 | 0.797009 | | 0.82198 | 0.727153 | 0.617744 | 0.78303 | 0.828082 | 0.903063 | 0.832794 | 0.867888 | 0.886984 | 0.902139 | 0.81272792 | 0.081286435 |
| 182731.png | | 0.914463 | 0.770664 | 0.802116 | 0.913959 | 0.884066 | | 0.696313 | 0.890925 | | | 0.923652 | 0.817569 | 0.84596967 | 0.078754799 |
| 182732.png | 0.716196 | 0.76784 | 0.865526 | 0.749283 | 0.893955 | | | | 0.828933 | | 0.87001 | 0.557474 | | 0.78115213 | 0.110496431 |
| 182733.png | 0.868205 | 0.753711 | 0.904318 | 0.844727 | 0.919085 | 0.5273 | 0.583077 | 0.803386 | 0.891803 | 0.840737 | 0.900531 | 0.810213 | 0.809405 | 0.804346 | 0.120910407 |
| 182734.png | 0.887269 | 0.901505 | 0.954744 | 0.793957 | 0.956236 | 0.833205 | 0.729851 | 0.748056 | 0.915132 | 0.769081 | 0.923222 | 0.889633 | 0.805506 | 0.85441515 | 0.078527588 |
| 182735.png | 0.932264 | 0.823275 | 0.917572 | 0.899388 | 0.919033 | 0.587346 | 0.905153 | 0.801814 | 0.933589 | 0.839923 | 0.897177 | 0.814563 | 0.836302 | 0.85441531 | 0.093143027 |
| 182736.png | 0.697023 | 0.673816 | 0.838852 | 0.746501 | 0.928283 | 0.669987 | 0.746133 | 0.619688 | 0.851032 | 0.806392 | 0.847322 | 0.784633 | 0.686545 | 0.76124669 | 0.090434735 |
| 182737.png | 0.848526 | 0.901003 | 0.967224 | 0.761298 | 0.94141 | 0.8681 | 0.821073 | 0.817224 | 0.928634 | 0.840572 | 0.928133 | 0.830842 | 0.846254 | 0.86925331 | 0.059684223 |
| 182738.png | 0.805526 | 0.861196 | 0.906412 | 0.863213 | 0.92597 | 0.524889 | 0.695035 | 0.615601 | 0.872271 | 0.813368 | 0.883766 | 0.76326 | 0.759201 | 0.791516 | 0.119082339 |
| 182739.png | 0.515014 | 0.799114 | 0.713224 | 0.759406 | 0.904978 | 0.81177 | 0.660296 | 0.837615 | 0.752849 | 0.758241 | 0.766659 | | 0.742635 | 0.75181675 | 0.096764311 |
| 182740.png | | | | | | | | | | | | | | #DIV/0! | #DIV/0! |
| 182741.png | 0.865359 | 0.878833 | 0.969343 | 0.918757 | 0.977731 | 0.731266 | 0.814348 | 0.828475 | 0.940371 | 0.864738 | 0.930496 | 0.83869 | 0.863803 | 0.87863154 | 0.068744761 |
| 182742.png | 0.796181 | 0.752925 | 0.815883 | 0.815167 | 0.828531 | 0.808797 | 0.597198 | 0.840162 | | 0.882921 | 0.807171 | 0.717816 | | 0.78752291 | 0.076364298 |
| 182743.png | 0.772308 | 0.859544 | 0.890948 | 0.881808 | 0.914931 | 0.585873 | 0.819985 | 0.71591 | 0.902461 | 0.833867 | 0.884086 | 0.904069 | 0.829049 | 0.83037223 | 0.093186284 |
| 182744.png | 0.80989 | 0.553318 | 0.881888 | 0.806242 | 0.892533 | 0.573816 | 0.708292 | 0.7827 | 0.837805 | 0.792131 | 0.929128 | 0.739822 | 0.839133 | 0.78051523 | 0.113471019 |
| 182745.png | 0.843925 | 0.696446 | 0.882506 | 0.706212 | 0.879308 | 0.82419 | 0.384548 | 0.611691 | 0.907668 | 0.80692 | 0.876005 | 0.684522 | 0.79618 | 0.76154777 | 0.145698652 |
| 182746.png | 0.768039 | 0.830111 | 0.873907 | 0.86708 | 0.937176 | 0.681859 | 0.539419 | 0.765615 | 0.859879 | 0.844132 | 0.810262 | 0.749022 | 0.777433 | 0.79261031 | 0.100741439 |
| 182747.png | 0.702106 | 0.599662 | 0.844255 | 0.712198 | 0.87912 | 0.747534 | 0.673921 | 0.721827 | 0.80688 | 0.825324 | 0.861857 | 0.598836 | 0.705228 | 0.74451908 | 0.093253115 |
| 182748.png | 0.790832 | 0.701928 | 0.870887 | 0.881681 | 0.947554 | 0.713918 | 0.388853 | 0.708965 | 0.87751 | 0.803296 | 0.905775 | 0.859129 | 0.730077 | 0.78310808 | 0.144277955 |
| 182749.png | 0.898289 | 0.870331 | 0.951574 | 0.824047 | 0.957753 | 0.694314 | 0.786401 | 0.796784 | 0.852281 | 0.783945 | 0.899831 | 0.81523 | 0.835647 | 0.84357131 | 0.073138521 |
| 182750.png | 0.906364 | 0.917193 | 0.893616 | 0.915853 | 0.940903 | 0.659944 | 0.782322 | 0.77003 | 0.932595 | 0.83935 | 0.889182 | 0.827707 | 0.832023 | 0.85439092 | 0.08079347 |
| 182751.png | 0.881002 | 0.702618 | 0.915696 | 0.819327 | 0.928609 | 0.658164 | 0.795494 | 0.751855 | 0.870811 | 0.777353 | 0.878507 | 0.724264 | 0.801067 | 0.808059 | 0.084351863 |
| 182752.png | 0.674622 | 0.852351 | 0.690839 | 0.825142 | 0.941059 | | 0.789129 | 0.742311 | 0.876437 | | | | 0.855193 | 0.80523144 | 0.088737622 |
| 182753.png | 0.82074 | 0.751227 | 0.930883 | 0.798013 | 0.887947 | 0.674463 | 0.513513 | 0.774768 | 0.843971 | 0.821378 | 0.918631 | 0.871176 | 0.791706 | 0.79987815 | 0.110960703 |
| 182754.png | 0.727125 | 0.818265 | 0.741136 | 0.855512 | 0.934108 | 0.798856 | 0.833421 | 0.779266 | 0.894668 | 0.863483 | 0.868065 | 0.759608 | 0.862083 | 0.82581508 | 0.062029158 |
| 182755.png | 0.757923 | 0.887657 | 0.858194 | 0.829137 | 0.918442 | 0.774309 | 0.787895 | 0.720044 | 0.830826 | 0.791571 | 0.874342 | 0.78923 | 0.818006 | 0.81827508 | 0.055949583 |
| 182756.png | 0.838745 | 0.836289 | 0.894871 | 0.853631 | 0.959585 | 0.569493 | 0.865095 | 0.649709 | 0.951055 | 0.801239 | 0.900493 | 0.821495 | 0.872001 | 0.83182315 | 0.110120899 |
| 182757.png | 0.861527 | 0.810502 | 0.926714 | 0.83099 | 0.938773 | 0.659275 | 0.816882 | 0.761184 | 0.930662 | 0.831503 | 0.943898 | 0.698573 | 0.835425 | 0.83430062 | 0.089807626 |
| 182758.png | 0.885662 | 0.807546 | 0.94447 | 0.798453 | 0.933052 | 0.69567 | 0.675478 | 0.642818 | 0.901066 | 0.655429 | 0.830002 | 0.847605 | 0.685961 | 0.79255477 | 0.109405263 |
| 182759.png | | | | | | | | | | | | | | #DIV/0! | #DIV/0! |
| 182760.png | 0.83391 | 0.847663 | 0.8283 | 0.791799 | 0.930349 | 0.581174 | 0.807522 | 0.656371 | 0.915986 | 0.888464 | 0.896362 | 0.836985 | 0.774471 | 0.81456585 | 0.099865369 |
| 182761.png | 0.584706 | 0.857173 | 0.904989 | 0.670109 | 0.939196 | 0.874353 | 0.668034 | 0.661747 | 0.877599 | 0.721453 | 0.790021 | 0.773927 | 0.809144 | 0.77941931 | 0.110489562 |
| 182762.png | 0.909712 | 0.8697 | 0.690809 | 0.84567 | 0.888728 | 0.57003 | 0.833359 | 0.818265 | 0.895247 | 0.871871 | 0.840835 | 0.808408 | 0.868851 | 0.82396038 | 0.094145858 |
| 182763.png | 0.738661 | 0.804358 | | 0.807292 | 0.740974 | 0.663996 | | | 0.906661 | | 0.847961 | | 0.872401 | 0.797788 | 0.079906537 |
| 182764.png | 0.850178 | 0.807345 | 0.878223 | 0.766899 | 0.861293 | 0.733824 | 0.632576 | 0.684734 | 0.820425 | 0.78912 | 0.890179 | 0.803271 | 0.75734 | 0.79041592 | 0.075577614 |
| 182765.png | 0.760208 | 0.87489 | 0.925785 | 0.875549 | 0.953372 | 0.56942 | 0.783285 | 0.721503 | 0.908799 | 0.872194 | 0.931531 | 0.888644 | 0.858209 | 0.84026069 | 0.106605329 |
| 182766.png | 0.694223 | 0.763402 | 0.893451 | 0.808222 | 0.945058 | 0.75604 | 0.383675 | 0.702924 | 0.935683 | 0.717144 | 0.886388 | 0.881867 | 0.804621 | 0.78251523 | 0.147913419 |
| 182767.png | 0.832786 | | 0.851121 | 0.7689 | 0.885458 | 0.555615 | 0.745565 | 0.776007 | 0.908792 | 0.816888 | 0.865726 | 0.485929 | 0.76764 | 0.77170225 | 0.128537522 |
| 182768.png | 0.769725 | 0.873462 | 0.889 | 0.85528 | 0.931192 | 0.78398 | 0.820246 | 0.648365 | 0.873583 | 0.810239 | 0.885697 | 0.819056 | 0.819976 | 0.82921546 | 0.071116428 |
| 182769.png | 0.608238 | 0.728721 | 0.75793 | 0.72746 | 0.887831 | 0.544481 | 0.553696 | 0.558256 | 0.819025 | 0.6482 | 0.765021 | 0.612754 | 0.569267 | 0.67545231 | 0.112644428 |
| 182770.png | | | | | | | | | | | | | | #DIV/0! | #DIV/0! |
| 182771.png | 0.511737 | 0.653463 | 0.887173 | 0.613653 | 0.803099 | 0.901051 | 0.900909 | 0.821027 | 0.882831 | | | 0.79661 | 0.763156 | 0.77588264 | 0.130219232 |
| 182772.png | 0.742904 | 0.784491 | 0.885674 | 0.73253 | 0.898074 | 0.792343 | 0.598724 | 0.762195 | 0.800206 | 0.748532 | 0.847374 | 0.557247 | 0.694724 | 0.75730908 | 0.09925714 |
| 182773.png | 0.890614 | 0.740235 | 0.936094 | 0.868722 | 0.935893 | 0.632684 | 0.480923 | 0.77514 | 0.933721 | 0.869269 | 0.902565 | 0.856006 | 0.85477 | 0.82127969 | 0.134587493 |
| 182774.png | 0.528594 | 0.6715 | 0.781135 | 0.656327 | 0.856373 | 0.750241 | 0.701124 | 0.64657 | 0.591901 | 0.702197 | 0.658989 | 0.570311 | 0.691613 | 0.67745192 | 0.087639167 |
| 182775.png | 0.700688 | 0.793181 | 0.685736 | 0.773321 | 0.929611 | 0.398895 | 0.593057 | 0.529918 | 0.886471 | 0.681558 | 0.86314 | 0.764823 | 0.657465 | 0.71214338 | 0.148277098 |
| 182776.png | 0.842775 | 0.744946 | 0.88154 | 0.677249 | 0.779647 | 0.730807 | 0.674944 | 0.784558 | 0.926986 | 0.781695 | 0.812654 | 0.656303 | 0.825157 | 0.77840469 | 0.081392207 |
| 182777.png | 0.706593 | 0.685476 | 0.682723 | 0.81165 | 0.827349 | 0.583655 | 0.690044 | 0.723915 | 0.863798 | 0.808903 | 0.821901 | 0.713924 | 0.811947 | 0.748606 | 0.081140853 |
| 182778.png | 0.874431 | 0.856397 | 0.886753 | 0.900446 | 0.974378 | 0.56146 | 0.828182 | 0.71928 | 0.942805 | 0.6325 | 0.888537 | 0.737273 | 0.837365 | 0.81844669 | 0.121843661 |
| 182779.png | 0.802319 | 0.880066 | 0.846058 | 0.704858 | 0.91875 | 0.81764 | 0.673589 | 0.591217 | 0.880714 | 0.580055 | 0.8621 | 0.371425 | 0.794594 | 0.74795269 | 0.157799669 |
| 182780.png | 0.863101 | 0.865468 | 0.924442 | 0.817932 | 0.940707 | 0.644817 | 0.638422 | 0.517205 | 0.913085 | 0.801937 | 0.902891 | 0.88091 | 0.815461 | 0.80972138 | 0.130054885 |
| 182781.png | 0.954909 | 0.883154 | 0.916605 | 0.864564 | 0.917046 | 0.615609 | 0.849103 | 0.823478 | 0.936238 | 0.844085 | 0.890739 | 0.893013 | 0.768385 | 0.85822523 | 0.088343654 |
| 182782.png | 0.934623 | 0.873761 | 0.888036 | 0.930437 | 0.929183 | 0.662849 | 0.908332 | 0.8766 | 0.935441 | 0.836594 | | 0.910913 | 0.789986 | 0.87306292 | 0.079462566 |
| 182783.png | 0.763901 | 0.896364 | 0.540569 | 0.785477 | 0.745067 | 0.79872 | 0.895574 | | 0.915696 | | | 0.78149 | 0.898943 | 0.8021801 | 0.112412215 |
| 182784.png | 0.860166 | 0.895378 | 0.947744 | 0.833333 | 0.963573 | 0.814542 | 0.777622 | 0.761212 | 0.897449 | 0.861629 | 0.921312 | 0.867222 | 0.833251 | 0.86418715 | 0.061036026 |
| 182785.png | 0.876417 | 0.845935 | 0.910461 | 0.944684 | 0.965434 | 0.548442 | 0.856725 | 0.799139 | 0.946214 | 0.875933 | 0.909779 | 0.902347 | 0.927144 | 0.86989646 | 0.106956987 |
| 182786.png | 0.823277 | 0.82768 | 0.814895 | 0.793756 | 0.89388 | 0.576441 | 0.797193 | 0.69966 | 0.85576 | 0.835163 | 0.853344 | 0.69806 | 0.617911 | 0.77592462 | 0.097314428 |
| 182787.png | 0.793102 | 0.862853 | 0.907559 | 0.787959 | 0.920762 | 0.714939 | 0.74752 | 0.694905 | 0.900212 | 0.812709 | 0.910723 | 0.86092 | 0.889144 | 0.83102362 | 0.077963322 |
| 182788.png | 0.754866 | 0.775851 | 0.902656 | 0.773954 | 0.886248 | 0.622525 | 0.608793 | 0.752273 | 0.913129 | 0.70226 | 0.817145 | 0.837808 | 0.613027 | 0.766195 | 0.106375295 |
| 182789.png | 0.809406 | 0.710065 | 0.873031 | 0.850226 | 0.873141 | 0.825993 | 0.725953 | 0.8408 | 0.831004 | 0.742875 | 0.773298 | 0.643018 | 0.800716 | 0.79227123 | 0.069528547 |
| 182790.png | 0.618416 | 0.786697 | 0.918688 | 0.801393 | 0.932668 | 0.767624 | 0.792352 | 0.679275 | 0.877617 | 0.839759 | 0.868557 | 0.789731 | 0.717 | 0.79921362 | 0.091201322 |
| 182791.png | 0.80869 | 0.568298 | 0.90459 | 0.703037 | 0.853743 | 0.749907 | 0.646121 | 0.856592 | 0.940053 | 0.906029 | 0.761411 | 0.667633 | 0.841414 | 0.78519369 | 0.113693133 |
| 182792.png | 0.839277 | 0.854478 | 0.901309 | 0.916649 | 0.95925 | 0.543747 | 0.753099 | 0.595616 | 0.908966 | 0.692842 | 0.802069 | 0.884075 | 0.859337 | 0.80851646 | 0.127928048 |
| 182793.png | 0.813286 | 0.735758 | 0.846456 | 0.795923 | 0.859138 | 0.7284 | 0.56973 | 0.752708 | 0.888492 | 0.424837 | 0.860766 | 0.66789 | 0.826807 | 0.75155315 | 0.132379829 |
| 182794.png | 0.799847 | 0.856997 | 0.897532 | 0.877416 | 0.94756 | 0.680519 | 0.756935 | 0.695077 | 0.864407 | 0.710779 | 0.889134 | 0.861726 | 0.820814 | 0.81990331 | 0.08505611 |
| 182795.png | 0.756604 | 0.872262 | 0.938791 | 0.778407 | 0.903373 | 0.826331 | 0.741734 | 0.653538 | 0.884453 | 0.744909 | 0.870809 | 0.73303 | 0.723456 | 0.80213054 | 0.085890312 |
| 182796.png | 0.873153 | 0.687592 | 0.845716 | 0.855954 | 0.953058 | 0.401551 | 0.730997 | 0.666112 | 0.849491 | 0.794809 | 0.860854 | 0.825099 | 0.740115 | 0.77573085 | 0.138690989 |
| 182797.png | 0.875709 | | 0.927768 | 0.762548 | 0.927023 | 0.706355 | 0.661746 | 0.844357 | 0.882124 | 0.890861 | 0.742041 | 0.613195 | 0.83531 | 0.80575308 | 0.105980339 |
| 182798.png | | | | | | | | | | | | | | #DIV/0! | #DIV/0! |
| 182799.png | 0.810484 | 0.644715 | 0.788813 | 0.699511 | 0.800552 | 0.556437 | 0.662327 | 0.794141 | 0.783682 | 0.771905 | 0.758902 | 0.490162 | 0.753274 | 0.71653115 | 0.101289143 |
| 182800.png | 0.878594 | 0.833663 | 0.924778 | 0.852926 | 0.946288 | 0.568022 | 0.77827 | 0.7761 | 0.9111 | 0.827997 | 0.864008 | 0.782171 | 0.808285 | 0.82709246 | 0.095745697 |
| 182801.png | 0.880152 | 0.828249 | 0.809628 | 0.879211 | 0.860997 | 0.471658 | 0.875597 | 0.821501 | 0.933287 | 0.874801 | 0.887308 | 0.854468 | 0.741946 | 0.82452331 | 0.115757634 |
| 182802.png | 0.893293 | 0.878614 | 0.920061 | 0.840151 | 0.927712 | 0.673847 | 0.801459 | 0.784299 | 0.873251 | 0.913033 | 0.879164 | 0.92226 | 0.791524 | 0.85374369 | 0.073747054 |
| 182803.png | 0.811036 | 0.83205 | 0.932863 | 0.844638 | 0.949022 | 0.69677 | 0.684495 | 0.669031 | 0.892327 | 0.810046 | 0.858495 | 0.750108 | 0.748874 | 0.806135 | 0.091657443 |
| 182804.png | 0.816368 | 0.845029 | 0.920642 | 0.888137 | 0.930028 | 0.677932 | 0.697768 | 0.581139 | 0.947743 | 0.772882 | 0.930531 | 0.871912 | 0.751215 | 0.81779431 | 0.114678641 |
| 182805.png | 0.752488 | 0.8068 | 0.663065 | 0.851065 | 0.843917 | 0.648703 | 0.73559 | 0.78549 | 0.903808 | 0.724387 | 0.837014 | 0.607565 | 0.831192 | 0.76854492 | 0.089371081 |
| 182806.png | 0.783521 | | 0.877579 | 0.808733 | 0.822983 | 0.924574 | 0.868212 | 0.891556 | 0.92103 | 0.887634 | 0.82328 | | 0.816175 | 0.85684336 | 0.04799721 |
| 182807.png | 0.677787 | 0.734406 | 0.803402 | 0.71969 | 0.903425 | 0.506157 | 0.47271 | 0.638969 | 0.83421 | 0.716149 | 0.821314 | 0.67973 | 0.627707 | 0.70274277 | 0.124374697 |
| 182808.png | 0.868569 | 0.904099 | 0.943379 | 0.880607 | 0.918432 | 0.771054 | 0.816099 | 0.769613 | 0.886586 | 0.757112 | 0.87814 | 0.770769 | 0.847771 | 0.84709462 | 0.063499378 |
| 182809.png | 0.90993 | 0.893471 | 0.71568 | 0.859609 | 0.909109 | 0.50109 | 0.69395 | 0.743189 | 0.930161 | 0.833159 | 0.824722 | 0.826603 | 0.827471 | 0.80524185 | 0.11795774 |
| 182810.png | 0.852455 | 0.871558 | 0.938871 | 0.908438 | 0.964523 | 0.636025 | 0.762511 | 0.72607 | 0.841691 | 0.865739 | 0.927792 | 0.93729 | 0.83032 | 0.85102177 | 0.094894346 |
| 182811.png | 0.793554 | 0.473354 | 0.858672 | 0.692863 | 0.81051 | 0.665359 | 0.770885 | 0.688978 | 0.874868 | 0.880571 | 0.893948 | 0.657748 | 0.691839 | 0.75024223 | 0.120327272 |
| 182812.png | 0.773629 | 0.818774 | 0.814371 | 0.799187 | 0.930496 | 0.752406 | 0.704431 | 0.630986 | 0.901568 | 0.691926 | 0.776428 | 0.803499 | 0.749857 | 0.78058138 | 0.08084183 |
| 182813.png | 0.720472 | 0.81311 | 0.913224 | 0.824034 | 0.945152 | 0.608653 | 0.882799 | 0.762704 | 0.934134 | 0.874258 | 0.914496 | 0.852055 | 0.902622 | 0.84213177 | 0.096869208 |
| 182814.png | 0.872185 | 0.799869 | 0.910319 | 0.827459 | 0.900491 | 0.701484 | 0.786393 | 0.828203 | 0.896525 | 0.785729 | 0.87972 | 0.803266 | 0.816653 | 0.83140738 | 0.059343539 |
| 182815.png | 0.756583 | 0.903137 | 0.886001 | 0.856317 | 0.939097 | 0.776661 | 0.790029 | 0.726586 | 0.871107 | 0.849324 | 0.871373 | 0.635077 | 0.80676 | 0.82061938 | 0.083052778 |
| 182816.png | 0.779382 | 0.817192 | 0.870407 | 0.86468 | 0.920384 | 0.800909 | 0.869646 | 0.819228 | 0.816407 | 0.71527 | 0.779445 | 0.675677 | 0.81292 | 0.81088823 | 0.065528326 |
| 182817.png | 0.819444 | 0.808464 | 0.857825 | 0.799146 | 0.918287 | 0.493433 | 0.621095 | 0.775811 | 0.857914 | 0.71941 | 0.849496 | 0.847974 | 0.745333 | 0.77797169 | 0.113886578 |
| 182818.png | 0.872948 | 0.836451 | 0.921261 | 0.898495 | 0.916131 | 0.64843 | 0.743664 | 0.706163 | 0.882384 | 0.751987 | 0.829187 | 0.828822 | 0.67505 | 0.80853638 | 0.09351236 |
| 182819.png | 0.836135 | 0.822529 | 0.88961 | 0.756694 | 0.933825 | 0.852094 | 0.793967 | 0.754875 | 0.948014 | 0.816459 | 0.923665 | 0.784225 | 0.767676 | 0.83690523 | 0.067991453 |
| 182820.png | 0.837409 | 0.815362 | 0.882311 | 0.852279 | 0.909727 | 0.581571 | 0.874081 | 0.787641 | 0.904179 | 0.751707 | 0.777189 | 0.815596 | 0.808051 | 0.81516177 | 0.085421038 |
| 182821.png | 0.793856 | 0.635815 | 0.86885 | 0.614679 | 0.807131 | 0.622351 | 0.781862 | 0.721498 | 0.851368 | 0.731439 | 0.778818 | 0.475859 | 0.620739 | 0.71571269 | 0.114314452 |
| 182822.png | 0.706545 | 0.63989 | 0.874934 | 0.670598 | 0.855233 | 0.753442 | 0.912033 | 0.700707 | 0.869054 | 0.781036 | 0.883318 | 0.643102 | 0.900642 | 0.78388723 | 0.103198252 |
| 182823.png | 0.897445 | 0.860044 | 0.943 | 0.894293 | 0.947035 | 0.726744 | 0.774055 | 0.636414 | 0.91698 | 0.842729 | 0.903112 | 0.878514 | 0.77271 | 0.84562115 | 0.092580971 |
| 182824.png | 0.837576 | 0.88827 | 0.762732 | 0.862164 | 0.902541 | 0.743048 | 0.796343 | 0.782815 | 0.94124 | 0.743246 | 0.858244 | 0.762106 | 0.852929 | 0.82563492 | 0.065051461 |
| 182825.png | 0.828524 | 0.874801 | 0.809451 | 0.851038 | 0.904648 | 0.77863 | 0.666844 | 0.767971 | 0.929573 | 0.863134 | 0.908536 | 0.875495 | 0.797195 | 0.83506462 | 0.071615816 |
| 182826.png | 0.88238 | 0.884404 | 0.963564 | 0.829226 | 0.960938 | 0.931709 | 0.883056 | 0.734223 | 0.948785 | 0.847265 | 0.921544 | 0.845975 | 0.915935 | 0.88838492 | 0.063988898 |
| 182827.png | 0.863042 | 0.802858 | 0.862915 | 0.79628 | 0.929506 | 0.668614 | 0.661143 | 0.536964 | 0.86442 | 0.693315 | 0.814674 | 0.706093 | 0.621701 | 0.75550192 | 0.115856546 |
| 182828.png | 0.820524 | 0.82478 | 0.91745 | 0.803148 | 0.931973 | 0.707762 | 0.710291 | 0.715303 | 0.911084 | 0.814037 | 0.930081 | 0.828944 | 0.744921 | 0.82002292 | 0.083717276 |
| 182829.png | 0.772772 | 0.787114 | 0.906778 | 0.635411 | 0.824885 | 0.861856 | 0.774252 | 0.78182 | 0.888851 | 0.816775 | 0.838594 | 0.645657 | 0.772167 | 0.79284092 | 0.08084602 |
| 182830.png | 0.861994 | 0.894236 | 0.933151 | 0.890455 | 0.95983 | 0.805363 | 0.847307 | 0.70072 | 0.900891 | 0.662413 | 0.920225 | 0.830592 | 0.790993 | 0.84601308 | 0.088204506 |
| 182831.png | 0.916886 | 0.635079 | 0.895176 | | 0.800789 | 0.562761 | 0.823527 | 0.754633 | 0.847657 | 0.852081 | 0.833055 | | 0.817358 | 0.79445473 | 0.107169817 |
| 182832.png | 0.879033 | 0.686462 | 0.932082 | 0.818073 | 0.952459 | 0.761404 | 0.592922 | 0.54248 | 0.914222 | 0.716208 | 0.82596 | 0.535451 | 0.744931 | 0.76166823 | 0.14303259 |
| 182833.png | 0.777995 | 0.813499 | 0.906175 | 0.766989 | 0.922316 | 0.725017 | 0.589754 | 0.745475 | 0.854353 | 0.797519 | 0.876374 | 0.839323 | 0.751887 | 0.79743662 | 0.088203571 |
| 182834.png | 0.760359 | 0.843019 | 0.88916 | 0.783622 | 0.828626 | 0.713893 | 0.784245 | 0.571734 | 0.807729 | 0.695698 | 0.823781 | 0.753413 | 0.748783 | 0.76954323 | 0.079825829 |
| 182835.png | 0.738084 | 0.700294 | 0.890745 | 0.807178 | 0.936342 | 0.489006 | 0.814079 | 0.640928 | 0.899178 | 0.795237 | 0.892333 | 0.794677 | 0.785607 | 0.78336062 | 0.121754577 |
| 182836.png | 0.900381 | 0.934168 | 0.939279 | 0.897834 | 0.955313 | 0.57279 | 0.843187 | 0.709229 | 0.851329 | 0.842493 | 0.948285 | 0.849343 | 0.878289 | 0.85553231 | 0.107160826 |
| 182837.png | 0.833111 | 0.911745 | 0.82824 | 0.933649 | 0.94869 | 0.667676 | 0.723513 | 0.755654 | 0.928146 | 0.781436 | 0.864769 | 0.779039 | 0.802206 | 0.82752877 | 0.087193016 |
| 182838.png | 0.721211 | 0.85955 | 0.90159 | 0.735305 | 0.912501 | 0.897623 | 0.844125 | 0.800087 | 0.904589 | 0.82719 | 0.886533 | | 0.830024 | 0.84336067 | 0.064596239 |
| Average | 0.80131875 | 0.81183941 | 0.86549331 | 0.81386751 | 0.90197095 | 0.67931162 | 0.73816701 | 0.72305273 | 0.88580421 | 0.78989265 | 0.86489281 | 0.76933406 | 0.79244646 | 0.80323691 | |
| StDev | 0.09178972 | 0.08576633 | 0.07843309 | 0.07372486 | 0.05198184 | 0.12144575 | 0.10943726 | 0.09428389 | 0.04595572 | 0.07878397 | 0.05155757 | 0.10013829 | 0.07072669 | 0.07072669 | |
| Count | 190 | 185 | 192 | 192 | 195 | 191 | 191 | 189 | 194 | 188 | 188 | 185 | 190 | | |
| Not Recognised | 11 | 16 | 9 | 9 | 6 | 10 | 10 | 12 | 7 | 13 | 13 | 16 | 11 | 143 | |

*Note:* *The original data are in the format of real numbers in the range between 0 and 1 with 6 decimal digits.*
*The data shows the similairity between the mutant images and the original images.The higher the score is, the more similar of the image to the original. The cell is left empty if it fails to recognise a face.*
*The data in the red cells are the overall average and standard deviation of similarity scores, respectively.*

# Appendix B. Results of Testing on Real Images

## 1. Results of Testing Tencent Face Recognition on Real Images

| ID | 1 | 2 | 3 | 4 | 5 | 6 | 7 | 8 | 9 | 10 | 11 | 12 | Average | StDev | Not Recognised |
|---|---|---|---|---|---|---|---|---|---|---|---|---|---|---|---|
| 1 | 94 | 100 | 98 | 100 | 100 | 86 | 98 | 99 | 96 | 100 | 100 | 100 | 97.5833 | 4.1222 | 0 |
| 2 | 100 | 100 | 100 | 100 | 99 | 98 | 100 | 98 | 80 | 100 | 100 | 78 | 96.0833 | 8.0279 | 0 |
| 3 | 100 | 100 | 100 | 100 | 100 | 100 | 100 | 100 | 100 | 100 | 100 | 100 | 100.0000 | 0.0000 | 0 |
| 4 | 97 | 99 | 100 | 82 | 100 | 94 | 100 | 100 | 94 | 100 | 93 | 98 | 96.4167 | 5.2649 | 0 |
| 5 | 100 | 90 | 100 | 100 | 99 | 100 | 100 | 100 | 100 | 100 | 100 | 100 | 99.0833 | 2.8749 | 0 |
| 6 | 100 | 100 | 100 | 100 | 100 | 100 | 93 | 100 | 98 | 100 | 96 | 100 | 98.9167 | 2.2344 | 0 |
| 7 | 100 | 100 | 100 | 100 | 100 | 100 | 100 | 97 | 100 | 100 | 98 | 100 | 99.5833 | 0.9962 | 0 |
| 8 | 99 | 100 | 100 | 100 | 98 | 100 | 100 | 100 | 97 | 94 | 100 | 100 | 99.0000 | 1.8586 | 0 |
| 9 | 99 | 100 | 98 | 96 | 100 | 78 | 94 | 98 | 100 | 100 | 100 | 100 | 96.9167 | 6.2589 | 0 |
| 10 | 94 | 98 | 99 | 100 | 100 | 80 | 98 | 100 | 100 | 97 | 100 | 100 | 97.1667 | 5.7022 | 0 |
| 11 | 100 | 99 | 97 | 100 | 100 | 100 | 97 | 100 | 97 | 100 | 100 | 100 | 99.1667 | 1.3371 | 0 |
| 12 | 100 | 100 | 100 | 100 | 100 | 100 | 100 | 100 | 100 | 100 | 98 | 100 | 99.8333 | 0.5774 | 0 |
| 13 | 100 | 97 |  | 96 | 100 | 96 | 100 | 94 | 91 | 96 | 94 | 100 | 96.7273 | 3.0361 | 1 |
| 14 | 98 | 98 | 100 | 100 | 97 | 100 | 97 | 98 | 99 | 100 | 90 | 100 | 98.0833 | 2.8110 | 0 |
| 15 | 79 | 96 | 94 | 90 | 97 | 95 | 99 | 100 | 96 | 94 | 98 | 85 | 93.5833 | 6.1416 | 0 |
| 16 | 100 | 100 | 100 | 100 | 100 | 100 | 100 | 99 | 100 | 100 | 100 | 100 | 99.9167 | 0.2887 | 0 |
| 17 | 100 | 100 | 94 | 89 | 100 | 100 | 100 | 100 | 99 | 100 | 100 | 96 | 98.1667 | 3.4859 | 0 |
| 18 | 100 | 100 | 100 | 99 | 94 | 100 | 100 | 100 | 100 | 100 | 99 | 100 | 99.3333 | 1.7233 | 0 |
| 19 | 94 | 91 | 96 | 94 | 90 | 100 | 100 | 77 | 100 | 100 | 97 | 100 | 94.9167 | 6.7212 | 0 |
| 20 | 92 | 100 | 100 | 100 | 98 | 100 | 100 | 100 | 100 | 100 | 100 | 100 | 99.1667 | 2.3290 | 0 |
| 21 | 97 | 94 | 92 | 94 | 94 | 94 | 91 | 94 | 76 | 100 | 87 | 81 | 91.1667 | 6.7667 | 0 |
| 22 | 100 | 94 | 100 | 99 | 100 | 99 | 97 | 98 | 93 | 100 | 99 | 100 | 98.2500 | 2.4168 | 0 |
| 23 | 94 | 91 | 100 | 100 | 100 | 82 | 91 | 100 | 94 | 100 | 100 | 100 | 96.0000 | 5.7683 | 0 |
| 24 | 100 | 100 | 100 | 100 | 100 | 99 | 93 | 100 | 99 | 100 | 96 | 95 | 98.5000 | 2.4309 | 0 |
| 25 | 97 | 93 | 98 | 98 | 99 | 100 | 91 | 93 | 97 | 98 | 97 | 94 | 96.2500 | 2.8002 | 0 |
| 26 | 100 | 96 | 100 | 100 | 97 | 100 | 100 | 100 | 100 | 100 | 100 | 98 | 99.2500 | 1.4222 | 0 |
| 27 | 70 | 100 | 94 | 100 | 100 | 98 | 69 | 98 | 94 | 100 | 76 | 100 | 91.5833 | 12.3101 | 0 |
| 28 | 100 | 97 | 100 | 100 | 100 | 100 | 100 | 100 | 100 | 100 | 100 | 100 | 99.7500 | 0.8660 | 0 |
| 29 | 88 |  | 100 | 96 | 99 | 76 | 97 | 88 | 99 | 79 | 99 | 91 | 92.0000 | 8.4261 | 1 |
| 30 | 95 | 100 | 100 | 100 | 100 | 100 | 94 | 100 | 100 | 94 | 96 | 100 | 98.2500 | 2.6328 | 0 |
| 31 | 100 | 80 | 97 | 100 | 100 | 97 | 92 | 97 | 99 | 94 | 94 | 93 | 95.2500 | 5.5780 | 0 |
| 32 | 100 | 95 | 93 | 100 | 94 | 100 | 100 | 100 | 94 | 98 | 93 | 100 | 97.2500 | 3.1370 | 0 |
| 33 | 100 | 100 | 100 | 79 | 96 | 98 | 100 | 65 | 100 | 94 | 85 | 100 | 93.0833 | 11.1719 | 0 |
| 34 | 100 | 100 | 93 | 100 | 100 | 100 | 100 | 100 | 100 | 100 | 100 | 100 | 99.4167 | 2.0207 | 0 |
| 35 | 96 | 100 | 100 | 100 | 100 | 100 | 96 | 94 | 93 | 99 | 100 | 100 | 98.1667 | 2.6572 | 0 |
| 36 | 100 | 90 | 88 | 89 | 92 | 91 | 94 | 94 | 90 | 94 | 81 | 99 | 91.8333 | 5.0422 | 0 |
| 37 | 100 | 93 | 100 | 100 | 100 | 100 | 94 | 94 | 100 | 98 | 97 | 89 | 97.0833 | 3.7285 | 0 |
| 38 | 100 | 100 | 100 | 100 | 94 | 100 | 100 | 91 | 99 | 100 | 90 | 100 | 97.8333 | 3.8337 | 0 |
| 39 | 100 | 94 | 96 | 100 | 80 | 94 | 100 | 99 | 94 | 100 | 94 | 94 | 95.4167 | 5.5996 | 0 |
| 40 | 100 | 100 | 100 | 89 | 97 | 100 | 94 | 99 | 94 | 31 | 94 |  | 90.7273 | 20.1350 | 1 |
| 41 | 99 | 98 | 88 | 100 | 100 | 87 | 72 | 100 | 97 | 97 | 97 | 95 | 94.1667 | 8.2333 | 0 |
| 42 | 100 | 94 | 97 | 100 | 97 | 94 | 98 | 79 | 100 | 100 | 99 | 93 | 95.9167 | 5.9154 | 0 |
| 43 | 90 | 97 | 94 | 100 | 100 | 100 | 100 | 96 | 99 | 97 | 100 | 100 | 97.7500 | 3.1659 | 0 |
| 44 | 100 | 100 | 94 | 100 | 100 | 100 | 98 | 100 | 100 | 99 | 100 | 99 | 99.1667 | 1.7495 | 0 |
| 45 | 100 | 99 | 94 | 100 | 95 | 94 | 100 | 100 | 100 | 97 | 100 | 98 | 98.0833 | 2.4664 | 0 |
| 46 | 100 | 100 | 99 | 100 | 100 | 100 | 89 | 94 | 98 | 99 | 100 | 99 | 98.1667 | 3.3530 | 0 |
| 47 | 98 | 100 | 100 | 100 | 94 | 99 | 98 | 100 | 100 | 97 | 97 | 100 | 98.5833 | 1.8809 | 0 |
| 48 | 100 | 97 | 100 | 97 | 92 | 100 | 94 | 100 | 100 | 100 | 91 | 100 | 97.5833 | 3.4234 | 0 |
| 49 | 94 | 96 | 71 | 100 | 94 | 84 | 94 | 91 | 100 | 90 | 96 | 94 | 92.0000 | 7.8971 | 0 |
| 50 | 96 | 97 | 100 | 100 | 100 | 97 | 100 | 99 | 98 | 90 | 96 | 100 | 97.7500 | 2.9271 | 0 |
| 51 | 100 | 100 | 100 | 100 | 100 | 100 | 100 | 100 | 99 | 100 | 100 | 100 | 99.9167 | 0.2887 | 0 |
| 52 | 100 | 94 | 100 | 94 | 95 | 100 | 99 | 100 | 100 | 100 | 99 | 100 | 98.4167 | 2.5030 | 0 |
| 53 | 97 | 100 | 100 | 100 | 96 | 93 | 87 | 100 | 98 | 100 | 96 | 100 | 97.2500 | 3.9572 | 0 |
| 54 | 95 | 89 | 75 | 94 | 99 | 100 | 93 | 100 | 97 | 91 | 99 | 89 | 93.4167 | 7.0641 | 0 |
| 55 | 100 | 100 | 100 | 100 | 100 | 100 | 100 | 98 | 81 | 100 | 100 | 100 | 98.2500 | 5.4627 | 0 |
| 56 | 100 | 94 | 97 | 100 | 100 | 100 | 98 | 100 | 98 | 100 | 100 | 100 | 98.9167 | 1.8809 | 0 |
| 57 | 100 | 100 | 100 | 94 | 100 | 100 | 100 | 100 | 100 | 97 | 100 | 100 | 99.2500 | 1.8647 | 0 |
| 58 | 100 | 85 | 94 | 94 | 92 | 93 | 90 | 96 | 98 | 100 | 83 | 96 | 93.4167 | 5.3506 | 0 |
| 59 | 97 | 89 | 98 | 100 | 100 | 94 | 94 | 100 | 100 | 84 | 94 | 92 | 95.1667 | 5.0602 | 0 |
| 60 | 100 | 96 | 97 | 100 | 94 | 94 | 100 | 100 | 100 | 83 |  | 100 | 96.7273 | 5.1786 | 1 |
| 61 | 100 | 92 | 96 | 100 | 83 | 97 | 94 | 100 | 100 | 100 | 99 | 94 | 96.2500 | 5.0655 | 0 |
| 62 | 97 | 95 | 96 | 98 | 100 | 81 | 100 | 89 | 100 | 95 | 93 | 96 | 95.0000 | 5.4439 | 0 |
| 63 | 99 | 97 | 100 | 100 | 100 | 100 | 94 | 96 | 95 | 94 | 88 | 97 | 96.6667 | 3.6013 | 0 |
| 64 | 100 | 100 | 100 | 96 | 100 | 100 | 100 | 98 | 100 | 98 | 100 | 100 | 99.3333 | 1.3027 | 0 |
| 65 | 100 | 100 | 100 | 100 | 100 | 100 | 100 | 100 | 100 | 100 | 100 | 100 | 100.0000 | 0.0000 | 0 |
| 66 | 97 | 100 | 95 | 96 | 100 | 88 | 100 | 100 | 100 | 96 | 100 | 100 | 97.6667 | 3.6265 | 0 |
| 67 | 100 | 100 | 96 | 100 | 99 | 98 | 99 | 100 | 100 | 100 | 100 | 94 | 98.8333 | 1.9462 | 0 |
| 68 | 98 | 100 | 100 | 99 | 96 | 94 | 97 | 97 | 96 | 97 | 74 | 94 | 95.1667 | 6.9522 | 0 |
| 69 | 100 | 95 | 97 | 93 | 100 | 95 | 99 | 100 | 100 | 98 | 95 | 98 | 97.5000 | 2.4680 | 0 |
| 70 | 100 | 97 | 99 | 91 | 100 | 100 | 100 | 94 | 92 | 100 | 100 | 100 | 97.7500 | 3.4411 | 0 |
| 71 | 83 | 100 | 95 | 92 | 100 | 94 | 96 | 86 | 100 | 99 | 91 | 100 | 94.6667 | 5.7735 | 0 |
| 72 | 94 | 100 | 100 | 100 | 100 | 100 | 98 | 100 | 100 | 97 | 100 | 99 | 99.0000 | 1.8586 | 0 |
| 73 | 100 | 100 | 100 | 100 | 99 | 100 | 100 | 100 | 100 | 100 | 100 | 100 | 99.9167 | 0.2887 | 0 |
| 74 | 100 | 97 | 100 | 100 | 94 | 94 | 94 | 100 | 96 | 94 | 100 | 100 | 97.4167 | 2.8431 | 0 |
| 75 | 100 | 95 | 100 | 81 | 98 | 96 | 98 | 100 | 100 | 70 | 98 | 98 | 94.5000 | 9.3176 | 0 |
| 76 | 96 | 98 | 100 | 100 | 100 | 100 | 94 | 94 | 100 | 100 | 97 | 98 | 98.0833 | 2.3533 | 0 |
| 77 | 95 | 99 | 100 | 90 | 90 | 85 | 99 | 98 | 91 | 93 | 100 | 100 | 95.0000 | 5.0990 | 0 |
| 78 | 90 | 81 | 90 | 90 | 75 | 66 | 75 | 84 | 100 | 79 | 81 | 81 | 82.6667 | 8.9477 | 0 |
| 79 | 98 | 100 | 98 | 99 | 96 | 86 | 88 | 100 | 100 | 96 | 98 | 94 | 96.0833 | 4.6409 | 0 |
| 80 | 100 | 99 | 96 | 100 | 99 | 100 | 100 | 100 | 100 | 94 | 94 | 98 | 98.3333 | 2.3484 | 0 |
| 81 | 97 | 100 | 97 | 100 | 95 | 100 | 100 | 97 | 100 | 100 | 100 | 100 | 98.8333 | 1.8007 | 0 |
| 82 | 100 | 100 | 100 | 100 | 100 | 100 | 100 | 99 | 98 | 100 | 100 | 100 | 99.7500 | 0.6216 | 0 |
| 83 | 99 | 94 | 100 | 100 | 100 | 100 | 97 | 90 | 100 | 98 | 98 | 100 | 98.0000 | 3.1042 | 0 |
| 84 | 96 | 100 | 94 | 99 | 97 | 98 | 100 | 100 | 100 | 100 | 99 | 98 | 98.4167 | 1.9287 | 0 |
| 85 | 85 | 76 | 76 | 78 | 100 | 94 | 94 | 86 | 97 | 97 | 85 | 94 | 88.5000 | 8.6392 | 0 |
| 86 | 63 | 100 | 100 | 100 | 100 | 100 | 100 | 78 | 100 | 100 | 100 | 100 | 95.0833 | 11.9199 | 0 |
| 87 | 100 | 94 | 100 | 100 | 100 | 100 | 100 | 100 | 87 | 97 | 100 | 100 | 98.1667 | 3.9734 | 0 |
| 88 | 93 | 93 | 100 | 100 | 100 | 100 | 95 | 95 | 100 | 100 | 100 | 100 | 98.0000 | 3.0151 | 0 |
| 89 | 99 | 98 | 94 | 100 | 99 | 100 | 99 | 100 | 96 | 100 | 96 | 94 | 97.9167 | 2.3143 | 0 |
| 90 | 100 | 100 | 100 | 100 | 100 | 100 | 99 | 96 | 100 | 100 | 100 | 97 | 99.3333 | 1.3707 | 0 |
| 91 | 100 | | 100 | 100 | 92 | 100 | 98 | 100 | 99 | 100 | 100 | 100 | 99.0000 | 2.4083 | 1 |
| 92 | 69 | 89 | 74 | 82 | 81 | 94 | 80 | 86 | 94 | 84 | 81 | 89 | 83.5833 | 7.4767 | 0 |
| 93 | 100 | 100 | 99 | 90 | 94 | 88 | 98 | 100 | 100 | 94 | 98 | 100 | 96.7500 | 4.2453 | 0 |
| 94 | 86 | 97 | 98 | 78 | 96 | 100 | 96 | 99 | 95 | 93 | 89 | 91 | 93.1667 | 6.3365 | 0 |
| 95 | 98 | 97 | 92 | 92 | 80 | 93 | 94 | 86 | 94 | 95 | 97 | 94 | 92.6667 | 5.0692 | 0 |
| 96 | 100 | 98 | 96 | 94 | 96 | 100 | 100 | 99 | 100 | 100 | 94 | 97 | 97.8333 | 2.3677 | 0 |
| 97 | 100 | 100 | 98 | 98 | 100 | 100 | 100 | 100 | 97 | 100 | 100 | 100 | 99.4167 | 1.0836 | 0 |
| 98 | 100 | 100 | 100 | 100 | 100 | 100 | 100 | 100 | 100 | 100 | 100 | 99 | 99.9167 | 0.2887 | 0 |
| 99 | 100 | 94 | 100 | 100 | 94 | 100 | 100 | 100 | 100 | 100 | 100 | 94 | 98.5000 | 2.7136 | 0 |
| 100 | 99 | 100 | 100 | 100 | 100 | 94 | 100 | 94 | 100 | 100 | 94 | 100 | 98.4167 | 2.6785 | 0 |
| 101 | 95 | 100 | 95 | 94 | 86 | 100 | 100 | 100 | 94 | 98 | 97 | 100 | 96.5833 | 4.1661 | 0 |
| 102 | 94 | 100 | 93 | 100 | 100 | 97 | 94 | 100 | 100 | 94 | 91 | 100 | 96.9167 | 3.4761 | 0 |
| 103 | 100 | 99 | 100 | 100 | 98 | 100 | 96 | 99 | 100 | 100 | 100 | 100 | 99.3333 | 1.2309 | 0 |
| 104 | 97 | 98 | 94 | 99 | 100 | 100 | 94 | 90 | 100 | 100 | 95 | 96 | 96.9167 | 3.2039 | 0 |
| 105 | 98 | 94 | 82 | 83 | 96 | 94 | 92 | 94 | 81 | 94 | 90 | 89 | 90.5833 | 5.7122 | 0 |
| 106 | 96 | 93 | 93 | 94 | 89 | 94 | 93 | 87 | 74 | 99 | 90 | 90 | 91.0000 | 6.2523 | 0 |
| 107 | 100 | 97 | 100 | 100 | 100 | 100 | 100 | 100 | 100 | 100 | 96 | 99 | 99.3333 | 1.3707 | 0 |
| 108 | 94 | 94 | 64 | 94 | 97 | 95 | 78 | 77 | 100 | 84 | 96 | 58 | 85.9167 | 13.8397 | 0 |
| 109 | 100 | 96 | 100 | 94 | 98 | 94 | 95 | 92 | 100 | 91 | 99 | 100 | 96.5833 | 3.3428 | 0 |
| 110 | 85 | 99 | 100 | 94 | 94 | 100 | 96 | 96 | 97 | 95 | 96 | 100 | 96.0000 | 4.1341 | 0 |
| 111 | 100 | 100 | 91 | 100 | 94 | 100 | 100 | 100 | 100 | 100 | 100 | 94 | 98.2500 | 3.2509 | 0 |
| 112 | 98 | 92 | 90 | 87 | 80 | 90 | 94 | 94 | 90 | 75 | 94 | 100 | 90.3333 | 7.0625 | 0 |
| 113 | 82 | 99 | 100 | 95 | 97 | 98 | 98 | 100 | 100 | 100 | 98 | 78 | 95.4167 | 7.4034 | 0 |
| 114 | 98 | 96 | 100 | 94 | 88 | 100 | 100 | 97 | 94 | 94 | 87 | 100 | 95.6667 | 4.5193 | 0 |
| 115 | 96 | 97 | 100 | 100 | 96 | 100 | 100 | 96 | 100 | 100 | 96 | 100 | 98.4167 | 1.9752 | 0 |
| 116 | 90 | 100 | 100 | 100 | 100 | 98 | 100 | 99 | 100 | 100 | 99 | 94 | 98.3333 | 3.1431 | 0 |
| 117 | 97 | 88 | 99 | 94 | 94 | 94 | 94 | 94 | 90 | 96 | 97 | 99 | 94.6667 | 3.2845 | 0 |
| 118 | 99 | 100 | 100 | 99 | 92 | 100 | 95 | 100 | 93 | 98 | 100 | 85 | 96.7500 | 4.6928 | 0 |
| 119 | 100 | 100 | 100 | 94 | 100 | 100 | 100 | 100 | 97 | 100 | 73 | 100 | 97.0000 | 7.7811 | 0 |
| 120 | 100 | 100 | 96 | 98 | 100 | 100 | 100 | 100 | 100 | 92 | 100 | 97 | 98.5833 | 2.5030 | 0 |
| 121 | 100 | 100 | 100 | 100 | 100 | 85 | 100 | 100 | 100 | 100 | 100 | 100 | 98.7500 | 4.3301 | 0 |
| 122 | 93 | 100 | 100 | 93 | 96 | 100 | 88 | 100 | 91 | 99 | 100 | 100 | 96.6667 | 4.3345 | 0 |
| 123 | 97 | 93 | 98 | 100 | 100 | 97 | 94 | 91 | 99 | 87 | 100 | 95 | 95.9167 | 4.0778 | 0 |
| 124 | 100 | 100 | 98 | | 99 | 100 | 94 | 100 | 100 | 98 | 97 | 96 | 98.3636 | 2.0136 | 1 |
| 125 | 96 | 91 | 87 | 94 | 94 | 90 | 100 | 93 | 78 | 73 | 74 | 94 | 88.6667 | 8.8967 | 0 |
| 126 | 89 | 100 | 100 | 100 | 100 | 97 | 99 | 97 | 79 | 94 | 94 | 94 | 95.2500 | 6.1663 | 0 |
| 127 | 98 | 81 | 90 | 100 | 91 | 96 | 82 | 97 | 80 | 100 | 100 | 94 | 92.4167 | 7.6332 | 0 |
| 128 | 100 | 100 | 98 | 97 | 100 | 100 | 99 | 100 | 100 | 95 | 100 | | 99.0000 | 1.6733 | 1 |
| 129 | 100 | 100 | 100 | 100 | 100 | 100 | 100 | 100 | 100 | 98 | 100 | 100 | 99.8333 | 0.5774 | 0 |
| 130 | 99 | 100 | 95 | 96 | 96 | 100 | 100 | 94 | 96 | 100 | 97 | 98 | 97.5833 | 2.1933 | 0 |
| 131 | 100 | 100 | 96 | 100 | 100 | 100 | 97 | 97 | 100 | 92 | 100 | 100 | 98.5000 | 2.5406 | 0 |
| 132 | 100 | 100 | 100 | 100 | 100 | 96 | 98 | 97 | 100 | 100 | 99 | 94 | 98.6667 | 2.0151 | 0 |
| 133 | 100 | 100 | 94 | 83 | 94 | 96 | 91 | 94 | 100 | 94 | 99 | 97 | 95.1667 | 4.8586 | 0 |
| 134 | 99 | 93 | 99 | 97 | | 89 | 100 | 99 | 99 | 100 | 100 | 100 | 97.7273 | 3.5522 | 1 |
| 135 | 100 | 98 | 94 | 98 | 92 | 100 | | 85 | 98 | 94 | 100 | 94 | 95.7273 | 4.5627 | 1 |
| 136 | 100 | 100 | 100 | 100 | | 100 | 94 | 100 | 100 | 100 | 95 | 100 | 99.0000 | 2.2361 | 1 |
| 137 | 100 | 100 | 95 | 95 | 96 | 97 | 94 | 100 | 98 | 94 | 100 | 94 | 96.9167 | 2.5746 | 0 |
| 138 | 96 | 94 | 100 | 94 | 96 | 100 | 97 | 100 | 86 | 99 | 98 | 94 | 96.1667 | 3.9734 | 0 |
| 139 | 90 | 82 | 83 | 73 | 90 | 89 | 94 | 88 | 94 | 84 | 88 | 74 | 85.7500 | 6.8639 | 0 |
| 140 | 98 | 100 | 93 | 96 | 97 | 100 | 82 | 100 | 100 | 94 | 94 | 85 | 94.9167 | 5.9461 | 0 |
| 141 | | 88 | 93 | | 91 | 90 | 96 | 99 | 90 | 94 | 94 | 87 | 92.2000 | 3.7059 | 2 |
| 142 | 100 | 83 | 78 | 93 | 93 | | 100 | 94 | 94 | 90 | 100 | 94 | 92.6364 | 6.9465 | 1 |
| 143 | 99 | 90 | 97 | 93 | 97 | 98 | 98 | 90 | 91 | 94 | 98 | 80 | 93.7500 | 5.4627 | 0 |
| 144 | 98 | 89 | 94 | 99 | 96 | 94 | 99 | 100 | 100 | 100 | 99 | 100 | 97.3333 | 3.4466 | 0 |
| 145 | 98 | 81 | 98 | 99 | 99 | 76 | 99 | 88 | 98 | 94 | 97 | 94 | 93.4167 | 7.7279 | 0 |
| 146 | 94 | 94 | 93 | 100 | 97 | 100 | 100 | 100 | 100 | 100 | 99 | 94 | 97.5833 | 2.9683 | 0 |
| 147 | 100 | 100 | 99 | 100 | 100 | 100 | 100 | 100 | 96 | 98 | 100 | 97 | 99.1667 | 1.4035 | 0 |
| 148 | 92 | 97 | 100 | 96 | 100 | 96 | 92 | 96 | 91 | 99 | 95 | 100 | 96.1667 | 3.2427 | 0 |
| 149 | 100 | 100 | 100 | 100 | 100 | 97 | 96 | 100 | 100 | 100 | 100 | 100 | 99.4167 | 1.3790 | 0 |
| 150 | 94 | 82 | 97 | 93 | 90 | 88 | 99 | 100 | 93 | 96 | 96 | 92 | 93.3333 | 4.9970 | 0 |
| 151 | 100 | 92 | 100 | 100 | 98 | 100 | 100 | 98 | 100 | 100 | 100 | 100 | 99.0000 | 2.3355 | 0 |
| 152 | 98 | 100 | 100 | 100 | 100 | 100 | 100 | 99 | 97 | 99 | 100 | 100 | 99.4167 | 0.9962 | 0 |
| 153 | 99 | 100 | 100 | 92 | 94 | 96 | 100 | 96 | 99 | 100 | 100 | 74 | 95.8333 | 7.3957 | 0 |
| 154 | 98 | 96 | 97 | 90 | 89 | 94 | 96 | 100 | 100 | 100 | 77 | 96 | 94.4167 | 6.5569 | 0 |
| 155 | 100 | 90 | 100 | 94 | 100 | 100 | 94 | 98 | 100 | 100 | 100 | 97 | 97.7500 | 3.3609 | 0 |
| 156 | 100 | 100 | 99 | 100 | 99 | 100 | 97 | 100 | 100 | 100 | 100 | 100 | 99.5833 | 0.9003 | 0 |
| 157 | 98 | 98 | 97 | 97 | 96 | 99 | 95 | 100 | 94 | 100 | 98 | 100 | 97.6667 | 1.9695 | 0 |
| 158 | 85 | 100 | 100 | 100 | 100 | 100 | 100 | 100 | 100 | 100 | 100 | 100 | 98.7500 | 4.3301 | 0 |
| 159 | 86 | 72 | 87 | 99 | 88 | 100 | 97 | 96 | 97 | 100 | 93 | 98 | 92.7500 | 8.2586 | 0 |
| 160 | 99 | 100 | 100 | 100 | 100 | 100 | 100 | 94 | 100 | 100 | 100 | 100 | 99.4167 | 1.7299 | 0 |
| 161 | 98 | 100 | 100 | 97 | 100 | 100 | 96 | 16 | 100 | 100 | 100 | 100 | 92.2500 | 24.0534 | 0 |
| 162 | 100 | 100 | 100 | 100 | 100 | 100 | 99 | 100 | 100 | 100 | 77 | 100 | 98.0000 | 6.6195 | 0 |
| 163 | 91 | 100 | 99 | 99 | 99 | 99 | 100 | 98 | 94 | 99 | 99 | 93 | 97.5000 | 3.0302 | 0 |
| 164 | 99 | 92 | 83 | 93 | 83 | 100 | 94 | 94 | 99 | 95 | 98 | 93 | 93.5833 | 5.6320 | 0 |
| 165 | 100 | | 81 | 94 | 100 | 94 | 100 | 74 | 100 | 89 | 75 | 89 | 90.5455 | 9.9636 | 1 |
| 166 | 94 | 94 | 97 | 98 | 96 | 92 | 76 | 100 | 94 | 95 | 98 | 94 | 94.0000 | 6.1051 | 0 |
| 167 | 100 | 100 | 99 | 100 | 100 | 100 | 99 | 98 | 100 | 95 | 100 | 100 | 99.2500 | 1.4848 | 0 |
| 168 | 94 | 100 | 93 | 97 | 78 | 100 | 100 | 79 | 74 | 98 | 58 | 80 | 87.5833 | 13.5275 | 0 |
| 169 | 100 | 98 | 100 | 96 | 97 | 100 | 96 | 100 | 100 | 100 | 100 | 100 | 98.9167 | 1.6765 | 0 |
| 170 | 100 | 100 | 93 | 100 | 99 | 94 | 94 | 91 | 83 | 90 | 95 | 100 | 94.9167 | 5.2822 | 0 |
| 171 | 97 | 96 | 97 | 100 | 100 | 100 | 96 | 88 | 100 | 100 | 96 | 98 | 97.3333 | 3.4201 | 0 |
| 172 | 94 | 100 | 97 | 97 | 95 | 100 | 100 | 96 | 100 | 100 | 100 | 99 | 98.1667 | 2.2496 | 0 |
| 173 | 99 | 98 | 100 | 100 | 100 | 100 | 100 | 99 | 100 | 100 | 99 | 98 | 99.4167 | 0.7930 | 0 |
| 174 | 94 | 95 | 98 | 93 | 99 | 99 | 100 | 92 | 96 | 100 | 82 | 94 | 95.1667 | 5.0061 | 0 |
| 175 | 100 | 100 | 86 | 94 | 100 | 100 | 100 | 99 | 99 | 100 | 100 | 95 | 97.7500 | 4.2453 | 0 |
| 176 | 100 | 100 | 100 | 98 | 100 | 100 | 100 | 100 | 100 | 97 | 100 | 100 | 99.5833 | 0.9962 | 0 |
| 177 | 100 | 96 | 100 | 99 | 99 | 95 | 100 | 100 | 97 | 100 | 100 | 100 | 98.8333 | 1.8007 | 0 |
| 178 | 99 | 98 | 95 | 100 | 95 | 100 | 100 | 96 | 94 | 100 | 94 | 67 | 94.8333 | 9.1038 | 0 |
| 179 | 72 | 100 | 94 | 100 | 100 | 94 | 99 | 100 | 100 | 100 | 91 | 100 | 95.8333 | 8.1445 | 0 |
| 180 | 100 | 100 | 94 | 97 | 95 | | 100 | 83 | 100 | 91 | 92 | 94 | 95.0909 | 5.2432 | 1 |
| 181 | 91 | 99 | 94 | 97 | 98 | 97 | 94 | 96 | 100 | 100 | 100 | 100 | 97.1667 | 2.9491 | 0 |
| 182 | 100 | 82 | 94 | 100 | 96 | 85 | 89 | 98 | 96 | 99 | 94 | 100 | 94.4167 | 6.0672 | 0 |
| 183 | 100 | 100 | 98 | 100 | 99 | 100 | 36 | 100 | 100 | 94 | 100 | 75 | 91.8333 | 18.9777 | 0 |
| 184 | 100 | 100 | 100 | 100 | 100 | 100 | 98 | 97 | 100 | 97 | 100 | 98 | 99.1667 | 1.2673 | 0 |
| 185 | 100 | 100 | 100 | 100 | 100 | 96 | 100 | 100 | 100 | 100 | 99 | 100 | 99.5833 | 1.1645 | 0 |
| 186 | 99 | 100 | 100 | 100 | 99 | 100 | 100 | 100 | 100 | 100 | 100 | 100 | 99.8333 | 0.3892 | 0 |
| 187 | 86 | 94 | 84 | 100 | 100 | 87 | 100 | 99 | 94 | 100 | 94 | 98 | 94.6667 | 5.9747 | 0 |
| 188 | 100 | 100 | 100 | 100 | 100 | 100 | 100 | 100 | 100 | 96 | 100 | 95 | 99.2500 | 1.7645 | 0 |
| 189 | 100 | 100 | 94 | 100 | 100 | 100 | 100 | 100 | 98 | 100 | 100 | 100 | 99.3333 | 1.7753 | 0 |
| 190 | 100 | 98 | 100 | 98 | 75 | 100 | 99 | 100 | 100 | 98 | 100 | 100 | 97.3333 | 7.0882 | 0 |
| 191 | 84 | 100 | 100 | 80 | 98 | | 94 | 100 | | 97 | 100 | 97 | 95.0000 | 7.1802 | 2 |
| 192 | 77 | 81 | 100 | 86 | 77 | 97 | 83 | 84 | 97 | 93 | 96 | 89 | 88.3333 | 8.1501 | 0 |
| 193 | 100 | 97 | 100 | 100 | 94 | 100 | 96 | 100 | 100 | 96 | 100 | 94 | 98.0833 | 2.5030 | 0 |
| 194 | 100 | 95 | 100 | 100 | 100 | 100 | 98 | 100 | 100 | 100 | 100 | 100 | 99.4167 | 1.5050 | 0 |
| 195 | 92 | 93 | 80 | 96 | 94 | 83 | 90 | 89 | 91 | 94 | 96 | 78 | 89.6667 | 6.1101 | 0 |
| 196 | 83 | 94 | 94 | 88 | 94 | 94 | 94 | 86 | 83 | 93 | 81 | 94 | 89.8333 | 5.2541 | 0 |
| 197 | 100 | 100 | 94 | 100 | 100 | 100 | 100 | 94 | 84 | 100 | 100 | 100 | 97.6667 | 4.8866 | 0 |
| 198 | 100 | 96 | 100 | 96 | 98 | 100 | 100 | 100 | 100 | 100 | 100 | 99 | 99.0833 | 1.5643 | 0 |
| 199 | 94 | 97 | 100 | 100 | 100 | 100 | 100 | 100 | 100 | 100 | 100 | 90 | 98.4167 | 3.2322 | 0 |
| 200 | 97 | 98 | 94 | 100 | 97 | 96 | 100 | 99 | 95 | 98 | 98 | 95 | 97.2500 | 1.9598 | 0 |
| 201 | 98 | 92 | 92 | 100 | 79 | 72 | 100 | 71 | 90 | 84 | 86 | 99 | 88.5833 | 10.3876 | 0 |
| 202 | 100 | 80 | 100 | 81 | 100 | 75 | 100 | 72 | 100 | 77 | 74 | 90 | 87.4167 | 11.9655 | 0 |
| 203 | 96 | 100 | 94 | 96 | 100 | 96 | 86 | 98 | 100 | 98 | 100 | 94 | 96.5000 | 4.0113 | 0 |
| 204 | 100 | 100 | 94 | 99 | 75 | 100 | 94 | 100 | 100 | 94 | 100 | 100 | 96.3333 | 7.2153 | 0 |
| 205 | 95 | 100 | 100 | 100 | 100 | 86 | 97 | 100 | 100 | 97 | 93 | 100 | 97.3333 | 4.2923 | 0 |
| 206 | 100 | 100 | 100 | 100 | 100 | 100 | 100 | 100 | 100 | 100 | 100 | 100 | 100.0000 | 0.0000 | 0 |
| 207 | 94 | 94 | 100 | 100 | 94 | 100 | 98 | 98 | 100 | 99 | 100 | 100 | 98.0833 | 2.5746 | 0 |
| 208 | 98 | 95 | 94 | 100 | 82 | 100 | 100 | 100 | 99 | 100 | 100 | 100 | 97.3333 | 5.2628 | 0 |
| | | | | | | | | | | | | Overall | 96.3848 | 6.2191 | 17 |

## 2. Results of Testing Baidu Face Recognition on Real Images

| ID | 1 | 2 | 3 | 4 | 5 | 6 | 7 | 8 | 9 | 10 | 11 | 12 | Average | StDev | Not Recognised |
|---|---|---|---|---|---|---|---|---|---|---|---|---|---|---|---|
| 1 | 91.13899994 | 92.20657349 | 81.75003052 | 93.19892883 | 92.23316193 | 82.29689026 | 92.44201660 | 90.24936676 | 90.45787048 | 92.86672974 | 93.73857880 | 93.24377441 | 90.48524348 | 4.100036 | 0 |
| 2 | 92.81726837 | 95.20304871 | 95.02879333 | 94.96638489 | 91.53098297 | 85.43309021 | 93.16026306 | 90.76249695 | 87.20800018 | 94.6255188 | 90.5656662 | 40.03432083 | 87.61131954 | 15.30402 | 0 |
| 3 | 93.63815308 | 92.45935822 | 95.31562042 | 84.79029846 | 93.16909027 | 93.96163177 | 93.62574005 | 91.53869629 | 93.76976776 | 93.15830231 | 95.54188538 | 94.75302124 | 92.97679710 | 2.813709 | 0 |
| 4 | 72.20942688 | 79.45475006 | 90.47218323 | 87.83020782 | 90.32827759 | 72.94982147 | 91.53079224 | 94.01856995 | 90.55123138 | 91.96697998 | 79.40697479 | 86.25363922 | 85.58107122 | 7.616032 | 0 |
| 5 | 91.48065186 | 75.23606873 | 91.1407547 | 95.3356781 | 90.67670441 | 95.04718018 | 93.10852051 | 94.37010956 | 90.14400482 | 94.63623047 | 93.14458466 | 91.31062317 | 91.30259260 | 5.371558 | 0 |
| 6 | 90.58377838 | 92.07763672 | 92.27616882 | 90.66506958 | 91.42993927 | 92.4080658 | 90.1541214 | 92.4410553 | 86.16692352 | 91.7914505 | 71.81710815 | 92.73401642 | 89.54544449 | 5.859114 | 0 |
| 7 | 94.56093597 | 94.45449066 | 90.6867218 | 95.7253952 | 91.59957123 | 94.12969208 | 95.19355774 | 92.47018433 | 87.78852844 | 95.14299011 | 90.90582275 | 90.93009949 | 92.79899915 | 2.442018 | 0 |
| 8 | 79.04377747 | 93.90643311 | 91.61978149 | 93.87457275 | 91.08500671 | 94.01587677 | 92.73442841 | 89.3994751 | 80.37497711 | 75.98330688 | 80.29970551 | 94.38482666 | 88.06018066 | 6.976282 | 0 |
| 9 | 78.44857788 | 92.80300903 | 91.32108307 | 87.05062866 | 90.40294647 | 70.6399231 | 76.30343628 | 91.4154129 | 93.6473465 | 92.43470001 | 91.90374756 | 93.03578949 | 87.45055008 | 7.806671 | 0 |
| 10 | 78.13105011 | 76.53978729 | 91.17879486 | 91.03935242 | 90.18136597 | 61.57236099 | 81.32639313 | 88.35040283 | 90.21851349 | 90.63728333 | 96.50809479 | 92.48571777 | 85.68075975 | 9.703576 | 0 |
| 11 | 91.46522522 | 91.45001221 | 83.91022491 | 93.94487 | 92.65849304 | 95.89089966 | 84.68341064 | 94.23703766 | 92.085289 | 90.61629486 | 95.88682556 | 95.14056396 | 91.83076223 | 3.936701 | 0 |
| 12 | 80.08392334 | 94.31486511 | 95.81632996 | 85.45957184 | 93.6726532 | 93.09955597 | 93.94435883 | 94.27680206 | 95.15303802 | 90.52774048 | 90.65518188 | 92.16339874 | 91.59728495 | 4.56929 | 0 |
| 13 | 98.70271301 | 90.47911835 | 76.27961731 | 76.6364975 | 90.1395874 | 66.94986725 | 89.32935333 | 73.41426849 | 72.45907593 | 75.63781738 | 77.31587982 | 90.67633057 | 81.50167720 | 9.808651 | 0 |
| 14 | 90.61492157 | 90.38392639 | 91.34771729 | 92.87245178 | 82.83576965 | 92.96913147 | 90.50476074 | 74.82739258 | 76.03274536 | 91.94976044 | 64.63387299 | 90.00102997 | 85.74779002 | 9.177781 | 0 |
| 15 | 68.08177948 | 87.45146942 | 83.62006378 | 72.74493408 | 90.03251648 | 78.52381897 | 91.82714081 | 93.88178253 | 90.45614624 | 90.85411072 | 91.84896851 | 65.93959808 | 83.77186076 | 9.969213 | 0 |
| 16 | 92.48267365 | 93.15507507 | 93.51398468 | 94.43414307 | 92.02745819 | 90.34900665 | 95.01608276 | 82.47200012 | 91.58596802 | 92.16107941 | 92.13006592 | 86.4216156 | 91.31242943 | 3.542599 | 0 |
| 17 | 92.19421387 | 91.99486542 | 63.80133057 | 64.97790527 | 91.51752472 | 89.11940765 | 90.36925507 | 92.47707367 | 88.16589355 | 92.17217255 | 76.17449951 | 76.79092407 | 84.14625549 | 10.84239 | 0 |
| 18 | 91.56578827 | 93.63811493 | 89.11323547 | 92.44651794 | 70.91323853 | 92.45866394 | 91.01399231 | 90.99525452 | 92.77687836 | 92.19022369 | 90.70484924 | 94.53157806 | 90.19569461 | 6.237046 | 0 |
| 19 | 75.25994873 | 70.44718933 | 82.88812256 | 77.35879517 | 77.54871368 | 70.25727844 | 93.6428833 | 58.05994415 | 93.92308044 | 93.14814758 | 92.18292999 | 93.63198853 | 81.52908516 | 11.94833 | 0 |
| 20 | 76.88140106 | 93.65145874 | 90.50083923 | 91.6186142 | 90.9673996 | 91.81568146 | 89.5546875 | 85.49853516 | 94.49447632 | 90.9500885 | 92.54841614 | 93.04358673 | 90.12709872 | 4.764095 | 0 |
| 21 | 90.22648621 | 88.49141693 | 84.72833252 | 91.1607666 | 90.33460999 | 76.58563232 | 89.28075409 | 90.24185181 | 90.15908813 | 99.21356964 | 64.43460083 | 62.15340424 | 84.75087611 | 11.26296 | 0 |
| 22 | 95.47677612 | 70.7727356 | 96.32958984 | 92.54813385 | 93.29399109 | 90.03689575 | 91.7480011 | 91.82561493 | 84.89165497 | 91.0429306 | 91.42458344 | 92.53285217 | 90.16031329 | 6.733896 | 0 |
| 23 | 80.04039001 | 71.83474731 | 91.66339874 | 94.00511169 | 90.29909515 | 76.4302063 | 72.24642944 | 93.09759521 | 77.66886902 | 92.32785034 | 94.88637543 | 90.62457275 | 85.42705345 | 8.983781 | 0 |
| 24 | 95.26113129 | 90.95696259 | 90.57489777 | 90.40900421 | 90.17835999 | 67.38109589 | 54.14163208 | 95.00627136 | 69.5633316 | 92.93047333 | 70.82970428 | 69.06321716 | 81.35800680 | 14.10194 | 0 |
| 25 | 90.31498718 | 71.21186066 | 92.24220276 | 91.62867737 | 90.67053986 | 92.96694946 | 79.5925827 | 77.0052948 | 92.08900452 | 90.09459686 | 90.13679504 | 79.76610565 | 86.47663307 | 7.430633 | 0 |
| 26 | 92.62988281 | 78.19561768 | 89.92152405 | 92.96047974 | 85.55204773 | 91.03885651 | 91.68564606 | 93.09785461 | 91.88578796 | 95.99376678 | 92.89884949 | 90.23510742 | 90.50795174 | 4.604107 | 0 |
| 27 | 21.09319305 | 77.5927124 | 90.26381683 | 92.01334381 | 92.64994812 | 78.93662262 | 8.814346313 | 90.42854309 | 71.80289459 | 91.04647827 | 61.31411362 | 94.00839233 | 72.49703375 | 28.79687 | 0 |
| 28 | 92.58769226 | 78.13761902 | 90.73581696 | 99.2535553 | 98.57417297 | 93.23950195 | 86.48878479 | 92.81977844 | 91.21879578 | 74.54720306 | 94.02094269 | 88.09692383 | 89.97673225 | 7.372739 | 0 |
| 29 | 79.98940277 | 71.98374939 | 95.25115967 | 87.57870483 | 85.72814941 | 90.34570312 | 91.27024841 | 60.63890457 | 90.05529785 | 66.43496704 | 90.59555054 | 75.00833893 | 82.07334804 | 11.14665 | 0 |
| 30 | 78.66640472 | 90.81607819 | 87.37384796 | 94.56677246 | 95.62345886 | 96.49462891 | 82.65346527 | 92.77085114 | 95.15257263 | 90.26781464 | 90.19956207 | 94.39337921 | 90.74823634 | 5.490199 | 0 |
| 31 | 75.52829742 | 61.90235901 | 62.48466492 | 92.64581299 | 93.71495819 | 59.38487244 | 62.11199951 | 71.35514069 | 79.01092529 | 58.19340134 | 58.94996262 | 64.24731445 | 69.96080907 | 12.72205 | 0 |
| 32 | 93.19649506 | 85.75315094 | 68.56978607 | 95.08135223 | 67.76183319 | 90.35971832 | 90.52993774 | 83.66162109 | 76.09274292 | 89.16812134 | 66.79420471 | 93.30596161 | 83.35624377 | 10.71243 | 0 |
| 33 | 90.47586823 | 91.81677246 | 92.17004395 | 60.44752884 | 91.1085434 | 91.32807922 | 92.40560913 | | 92.66756439 | 90.49465942 | 71.3261795 | 91.61642456 | 86.89611574 | 10.69147 | 1 |
| 34 | 96.00920105 | 90.44780731 | 73.34806061 | 95.83316803 | 93.91851807 | 94.6616745 | 92.01012421 | 93.80960083 | 94.42453766 | 91.22639465 | 92.78603363 | 81.20121765 | 90.80636152 | 6.753599 | 0 |
| 35 | 78.73547363 | 91.00615692 | 90.21269226 | 91.85171509 | 93.10147095 | 91.93112183 | 62.99486542 | 84.11475372 | 75.06195068 | 90.53240204 | 91.77416229 | 91.65723419 | 86.08116659 | 9.311154 | 0 |
| 36 | 90.81489563 | 51.28044891 | 64.77722168 | 79.04251099 | 51.42263031 | 71.70809937 | 73.51424408 | 91.53872681 | 63.71672821 | 61.5450058 | 70.60326385 | 90.84498596 | 71.73406347 | 14.22798 | 0 |
| 37 | 91.53606415 | 67.86155701 | 90.53399658 | 90.66407013 | 91.57498932 | 76.69728088 | 62.74118042 | 70.24333954 | 91.41931152 | 74.33836365 | 59.80150604 | 66.94477844 | 77.86303647 | 12.53769 | 0 |
| 38 | 91.49758911 | 91.42371368 | 92.13222504 | 91.63397217 | 80.81217194 | 93.32994843 | 91.580513 | 81.88270569 | 92.48509216 | 92.25543976 | 68.69563293 | 93.84700775 | 88.46466764 | 7.550018 | 0 |
| 39 | 92.16876984 | 75.3615799 | 75.08148956 | 91.68531799 | 72.98965454 | 70.3769455 | 91.86431885 | 90.95909882 | 80.28845978 | 91.61663055 | 90.62612915 | 84.97412109 | 83.99937630 | 8.600687 | 0 |
| 40 | 92.98152161 | 92.25360107 | 97.19294739 | 75.9855423 | 64.42077637 | 86.67726898 | 90.14215851 | 91.75001526 | 85.38253021 | 2.219932556 | 90.03036499 | 81.60415649 | 79.22006798 | 25.81966 | 0 |
| 41 | 90.19129944 | 77.56938171 | 54.23184204 | 90.70894623 | 90.07777405 | 68.64528656 | 52.3010025 | 94.08773804 | 83.33605957 | 91.24472046 | 71.08477783 | 83.62783813 | 78.92555555 | 14.46313 | 0 |
| 42 | 93.73964691 | 91.52861786 | 86.8799057 | 93.34933472 | 90.12126923 | 76.58133698 | 90.04823303 | 76.76387787 | 93.30238342 | 94.8141098 | 93.05166626 | 90.35210419 | 89.21104050 | 6.242055 | 0 |
| 43 | 90.78205109 | 80.26676178 | 90.43428802 | 87.88082123 | 92.37871552 | 94.56521606 | 91.91027832 | 90.40973663 | 91.04174805 | 90.74505615 | 96.37759399 | 96.31263733 | 91.09207535 | 4.244735 | 0 |
| 44 | 92.03813934 | 92.61645508 | 92.14435577 | 92.81058502 | 93.66166687 | 93.61849213 | 77.57873535 | 90.38807678 | 93.6051178 | 90.25858307 | 94.53831482 | 94.11959839 | 91.44817670 | 4.571669 | 0 |
| 45 | 92.57567596 | 80.80990601 | 91.07579041 | 91.86438751 | 75.53632355 | 75.10942078 | 91.21203613 | 92.646698 | 92.37579346 | 90.67553711 | 92.09179688 | 88.79877472 | 87.89767838 | 6.701484 | 0 |
| 46 | 93.17129517 | 91.51241302 | 90.80935669 | 94.39311981 | 94.40540314 | 92.43292999 | 81.56694031 | 86.21755219 | 93.67858124 | 93.42855835 | 93.23214722 | 91.86746216 | 91.39297994 | 3.801742 | 0 |
| 47 | 90.09017944 | 90.30887604 | 79.9546051 | 91.3458252 | 71.74826813 | 90.55433655 | 91.55043793 | 93.82463074 | 91.25476074 | 75.71099854 | 73.2466507 | 90.39552307 | 85.83209102 | 8.154268 | 0 |
| 48 | 94.36129761 | 91.22502899 | 93.1858139 | 80.73033142 | 71.71977234 | 92.29409027 | 74.46237183 | 93.91642761 | 91.94200897 | 93.17324066 | 77.29344177 | 90.51000977 | 87.06781960 | 8.44861 | 0 |
| 49 | 90.95961761 | 91.23497009 | 72.36174774 | 92.9906311 | 83.39485931 | 82.62114716 | 75.0619278 | 88.86786652 | 95.22241211 | 71.47123718 | 91.19747162 | 91.75466919 | 85.59487979 | 8.459751 | 0 |
| 50 | 90.87644958 | 93.91077423 | 90.43612671 | 93.90148163 | 94.25444794 | 92.91184998 | 94.58152771 | 92.25437164 | 92.30832672 | 91.49315643 | 92.42712402 | 94.77774811 | 92.84444873 | 1.455738 | 0 |
| 51 | 96.48636627 | 95.5866394 | 92.49056244 | 93.66002655 | 92.47823334 | 96.58280945 | 93.22129059 | 93.08872986 | 87.59075928 | 92.58942413 | 95.61528778 | 93.94841003 | 93.61154493 | 2.43807 | 0 |
| 52 | 92.57041168 | 71.31087494 | 92.90229034 | 65.78855896 | 76.26084137 | 79.08187103 | 79.68932343 | 82.23358154 | 92.3272171 | 90.2872467 | 63.95791245 | 92.24101257 | 81.55426184 | 10.69346 | 0 |
| 53 | 91.26165771 | 93.15252686 | 97.14388275 | 93.444664 | 90.77118683 | 79.49752808 | 79.62433624 | 95.11009979 | 93.33205414 | 93.58309937 | 90.03684235 | 94.94042969 | 90.99152565 | 5.686092 | 0 |
| 54 | 90.55990601 | 71.44787598 | 67.0631485 | 75.4059906 | 92.57984924 | 91.44813538 | 73.87346649 | 92.95427704 | 90.17823029 | 78.74599457 | 91.92396545 | 61.05097961 | 81.43598493 | 11.47903 | 0 |
| 55 | 93.88773346 | 94.44064331 | 94.39685059 | 95.1547699 | 92.25452423 | 92.66769409 | 92.2706604 | 90.56058502 | 69.65062714 | 93.10417938 | 98.86726379 | 91.40188599 | 91.55478478 | 7.221972 | 0 |
| 56 | 92.95436096 | 89.86755371 | 91.20291901 | 92.33178711 | 86.83785248 | 93.5909729 | 91.27058411 | 91.89810181 | 90.15272522 | 94.06385803 | 90.96219635 | 93.61608124 | 91.56241608 | 2.022327 | 0 |
| 57 | 92.55442047 | 92.53214264 | 92.32698059 | 74.47357941 | 94.58944702 | 97.17832947 | 94.64881897 | 91.98672485 | 96.29603577 | 91.19033813 | 94.76544952 | 92.97142029 | 92.12614059 | 5.845912 | 0 |
| 58 | 89.90530396 | 80.28354645 | 81.69233704 | 90.63363647 | 78.55112457 | 75.90796661 | 75.21307373 | 90.17404938 | 92.78848267 | 83.82562256 | 74.34780884 | 86.18081665 | 83.29198074 | 6.589857 | 0 |
| 59 | 65.8806839 | 76.48790741 | 75.22919464 | 83.72672272 | 91.15529633 | 65.66061401 | 75.78760529 | 100 | 92.59440613 | 52.05716324 | 66.95809174 | 73.6268692 | 76.59704622 | 13.50862 | 0 |
| 60 | 91.27627563 | 72.75670624 | 82.59416199 | 91.90270233 | 51.90981293 | 83.35053253 | 88.74822998 | 91.28998566 | 92.6084671 | 64.79587555 | 69.31340027 | 90.12723541 | 80.88944880 | 13.24494 | 0 |
| 61 | 95.37789154 | 74.56234741 | 91.58177185 | 93.21347046 | 91.54372406 | 88.16007233 | 91.96168518 | 92.72747803 | 94.47651672 | 93.8915329 | 92.72414398 | 89.17371368 | 90.78286235 | 5.502557 | 0 |
| 62 | 84.44602203 | 72.86437225 | 72.40504456 | 69.05072784 | 78.53936768 | 60.42453003 | 70.74200439 | 69.52318573 | 79.96740723 | 78.10396576 | 64.67797852 | 66.14124298 | 72.24048742 | 6.986779 | 0 |
| 63 | 93.49077606 | 92.7335968 | 92.61971283 | 91.9865036 | 95.00485229 | 92.99172974 | 70.10507965 | 92.36899567 | 82.57984924 | 77.00734711 | 75.43708038 | 90.02243042 | 87.19566282 | 8.567705 | 0 |
| 64 | 93.04086304 | 91.77228546 | 92.66345978 | 90.41427612 | 92.24895477 | 92.11827087 | 92.84907532 | 91.72211456 | 94.16622925 | 91.24043274 | 83.39389801 | 92.5626297 | 91.51604080 | 2.727045 | 0 |
| 65 | 93.70383453 | 91.03276062 | 93.36389923 | 94.84812164 | 85.91873169 | 93.83403778 | 93.42124939 | 88.0575943 | 93.48114777 | 92.68911743 | 92.08443451 | 92.42827606 | 92.07193375 | 2.602722 | 0 |
| 66 | 77.85135651 | 92.38587952 | 83.22960663 | 83.26685333 | 90.88777161 | 69.22947693 | 92.39168549 | 92.60709381 | 90.73396301 | 85.02354431 | 89.39866638 | 91.52489471 | 86.54423269 | 7.222805 | 0 |
| 67 | 84.66773987 | 92.34989929 | 78.55558014 | 80.56903839 | 72.98712158 | 72.79904175 | 74.75328827 | 89.38478088 | 77.17306519 | 68.78444672 | 90.19126129 | 68.57830811 | 79.23279762 | 8.281349 | 0 |
| 68 | 76.20707703 | 76.44856262 | 77.79268646 | 70.14237976 | 73.85729218 | 69.7568512 | 82.05164337 | 74.1913147 | 69.39924622 | 74.7197113 | 48.83739471 | 73.0929718 | 72.20809428 | 8.202859 | 0 |
| 69 | 90.19691467 | 77.50572968 | 79.80355072 | 65.85733032 | 85.85173035 | 77.50054932 | 84.71512604 | 73.08101654 | 93.25642395 | 63.05294037 | 76.76390839 | 62.30806351 | 77.49110699 | 10.14318 | 0 |
| 70 | 96.09123993 | 79.83537292 | 91.75284576 | 91.89858246 | 94.33786011 | 92.45185089 | 91.23249817 | 89.86477661 | 80.61711884 | 93.2375946 | 93.9052887 | 91.80799866 | 90.58608564 | 5.102446 | 0 |
| 71 | 56.9157486 | 92.14630127 | 70.95136261 | 69.85509491 | 92.30934143 | 49.40030289 | 61.63667297 | 18.50128555 | 94.64020538 | 91.11056519 | 60.91035461 | 91.13710022 | 70.79286130 | 23.12289 | 0 |
| 72 | 72.81713867 | 92.66835785 | 90.67237854 | 92.08350372 | 93.02868652 | 80.65153503 | 77.83679199 | 90.2900238 | 79.10838318 | 85.32900238 | 90.28974915 | 91.06108093 | 86.31971931 | 6.947074 | 0 |
| 73 | 92.77325439 | 92.46874237 | 93.18659973 | 94.24253082 | 92.96386719 | 93.54473114 | 93.82447052 | 91.74665833 | 91.49440002 | 92.08576965 | 93.02685547 | 91.16902924 | 92.71057574 | 0.94924 | 0 |
| 74 | 91.69698334 | 91.16654205 | 92.27367401 | 93.3923111 | 85.38687134 | 77.35031891 | 90.48548126 | 83.94109344 | 90.00885773 | 79.14586639 | 94.67383575 | 92.98491669 | 88.54222933 | 5.738427 | 0 |
| 75 | 91.92569733 | 75.43379974 | 90.31774902 | 69.36482239 | 78.26548767 | 78.38782501 | 90.01097107 | 90.89951324 | 91.16126251 | 62.78225327 | 91.98955536 | 78.63653564 | 82.43128935 | 10.01345 | 0 |
| 76 | 93.08241272 | 73.99866486 | 94.17906189 | 92.87263489 | 91.30293274 | 92.28883362 | 84.567276 | 91.40129089 | 94.19580078 | 93.67271423 | 87.74885559 | 92.01055145 | 90.11008581 | 5.797595 | 0 |
| 77 | 78.42805481 | 68.23078156 | 93.45928192 | pic not has face | 70.38267517 | 77.01475525 | 85.4575119 | 80.93855286 | 66.96542358 | 70.81687164 | 88.47503662 | 90.95760345 | 79.19332261 | 9.450794 | 1 |
| 78 | 73.26277924 | 70.59150696 | 79.1091156 | 75.44836426 | 64.53292847 | 56.4335556 | 54.27122879 | 69.64031219 | 96.36404419 | 65.54769897 | 63.72635269 | 80.56963348 | 70.79146004 | 11.45569 | 0 |
| 79 | 90.14264679 | 90.95870209 | 90.9256897 | 89.4397049 | 86.06814575 | 67.04470062 | 65.86461639 | 93.72970581 | 92.43926239 | 92.31739044 | 74.59063721 | 85.94760895 | 84.95573425 | 10.00363 | 0 |
| 80 | 90.14744568 | 88.75056458 | 83.23794556 | 91.28455353 | 91.77117157 | 100 | 91.74846649 | 91.37568665 | 91.90055084 | 52.57830811 | 76.71828461 | 90.03540039 | 86.62903150 | 12.06454 | 0 |
| 81 | 60.86476135 | 69.83053589 | 64.14024353 | 79.05500031 | 63.70748138 | 79.80182648 | 90.98853302 | 91.30591583 | 89.87935638 | 86.53692627 | 81.05150604 | 91.57665253 | 79.06156158 | 11.68117 | 0 |
| 82 | 95.17974854 | 92.85874939 | 91.34516144 | 93.99323273 | 91.44525146 | 91.59660339 | 95.91093445 | 92.75785828 | 90.65667725 | 94.13087463 | 85.70558929 | 92.70664978 | 92.35727755 | 2.637537 | 0 |
| 83 | 91.34546661 | 74.12103271 | 85.09242249 | 92.37715912 | 78.76114655 | 92.11018372 | 91.01258087 | 58.14719391 | 93.15506744 | 74.87680817 | 89.30384827 | 90.21019745 | 84.20942561 | 10.73187 | 0 |
| 84 | 90.7013092 | 86.52986908 | 78.83419037 | 91.24912262 | 91.17253876 | 80.38995361 | 92.3563385 | 94.18048096 | 95.8026123 | 90.82368469 | 91.17633057 | 91.01301575 | 89.51912053 | 5.13021 | 0 |
| 85 | 55.2754097 | 39.93399048 | 66.32513428 | 55.79650116 | 90.22766876 | 64.27989197 | 71.8144455 | 45.03155899 | 68.89710999 | 66.31493378 | 64.27288055 | 76.95767975 | 63.76060041 | 13.59632 | 0 |
| 86 | 41.02762604 | 91.48685455 | 82.19952393 | 91.99256897 | 93.24819946 | 94.07810974 | 92.54924011 | 56.99526596 | 91.96789551 | 90.98581696 | 90.23484802 | 92.57718658 | 84.11192799 | 17.01289 | 0 |
| 87 | 90.98056793 | 91.95469666 | 93.82397461 | 95.58740997 | 92.62238312 | 94.53123474 | 91.30493164 | 92.48121643 | 83.66134644 | 79.34107208 | 93.88670349 | 93.87324524 | 91.17073186 | 4.7991 | 0 |
| 88 | 79.0329361 | 79.0329361 | 93.63404083 | 93.63404083 | 92.75675201 | 92.75675201 | 71.01743317 | 71.01743317 | 94.30992126 | 94.30992126 | 93.23277283 | 93.23277283 | 87.33064270 | 9.416559 | 0 |
| 89 | 92.86303711 | 92.65820312 | 82.14259338 | 93.68614197 | 92.15242004 | 94.12085724 | 85.01078033 | 92.13241577 | 90.35997009 | 92.47416687 | 91.37624359 | 90.54177856 | 90.79321734 | 3.599643 | 0 |
| 90 | 93.75042725 | 78.66557312 | 95.4211731 | 93.4493866 | 93.1047821 | 95.73173523 | 92.55613708 | 91.1192627 | 92.53462219 | 91.32865143 | 93.44678497 | 91.56060028 | 91.88909467 | 4.40849 | 0 |
| 91 | 88.85773468 | 69.90031433 | 90.49817657 | 78.16612244 | 59.20693588 | 92.95256805 | 84.02680969 | 91.07263947 | 81.47956085 | 67.95555878 | 91.95237732 | 94.60865784 | 82.55645466 | 11.51316 | 0 |
| 92 | 38.71838379 | 54.82052612 | 34.38047791 | 47.73306274 | 52.88723755 | 51.40404892 | 54.8521347 | 52.73948669 | 61.353405 | 46.8151474 | 57.66731262 | 61.8283844 | 51.26663399 | 8.30917 | 0 |
| 93 | 92.58336639 | 94.07152557 | 92.80206299 | 76.86883545 | 90.76576233 | 80.92073059 | 92.09552765 | 94.71081543 | 94.82453156 | 91.17160797 | 91.48395538 | 92.97186279 | 90.43921534 | 5.612491 | 0 |
| 94 | 71.00434875 | 64.89015961 | 91.80780029 | 52.7899704 | 79.60535431 | 91.74720764 | 90.9306488 | 91.53018188 | 74.91678619 | 83.73339081 | 74.73038483 | 69.24416351 | 78.07753309 | 12.49909 | 0 |
| 95 | 74.01161957 | 71.48765564 | 69.93703461 | 73.76198578 | 57.83447647 | 66.35694122 | 77.61089325 | 67.58681488 | 70.52330017 | 72.09716034 | 77.64976501 | 65.94673157 | 70.40036488 | 5.520046 | 0 |
| 96 | 90.93857574 | 90.89173889 | 75.79458618 | 78.52605438 | 78.57198334 | 90.67583466 | 87.13252258 | 75.74081421 | 91.26567078 | 93.39442444 | 75.55128479 | 66.48112488 | 82.91371791 | 8.791352 | 0 |
| 97 | 92.62542725 | 91.62783051 | 90.50595093 | 88.5797348 | 91.04930115 | 90.15349579 | 90.82136536 | 90.14667511 | 91.91983032 | 91.40891266 | 91.92073059 | 91.65909576 | 91.03486252 | 1.082837 | 0 |
| 98 | 96.62099457 | 93.5484314 | 93.32350922 | 95.28806305 | 95.89811707 | 96.84469604 | 96.22546387 | 97.41938782 | 92.91980743 | 95.16632843 | 93.40248108 | 92.37780762 | 94.91959063 | 1.729378 | 0 |
| 99 | 94.99536133 | 88.44452667 | 94.77546692 | 95.30641174 | 90.60515594 | 94.26930237 | 95.76629639 | 94.89653015 | 95.08396149 | 90.58184814 | 94.10050964 | 84.74103546 | 92.79720052 | 3.449897 | 0 |
| 100 | 90.22657776 | 92.86607361 | 92.77589417 | 91.66124725 | 92.76292419 | 73.69167328 | 93.70024109 | 68.82228851 | 93.19813538 | 93.89047241 | 90.14990234 | 91.09619141 | 88.73680178 | 8.32277 | 0 |
| 101 | 60.84676361 | 90.96109009 | 71.24279022 | 72.59329987 | 59.22344971 | 91.72647095 | 91.84751892 | 91.81021881 | 71.98945618 | 76.17948151 | 75.85669708 | 94.89894867 | 79.09801547 | 12.70543 | 0 |
| 102 | 78.74700165 | 92.50204468 | 90.43186951 | 93.21163177 | 92.25443268 | 92.20687103 | 88.65795135 | 93.44463348 | 92.05918884 | 81.41246796 | 77.05711365 | 93.82650757 | 88.81764285 | 6.107979 | 0 |
| 103 | 78.42930603 | 90.46524048 | 86.9136734 | 91.76707458 | 72.94530487 | 94.61209869 | 69.03701019 | 81.641922 | 92.75711823 | 75.65750885 | 75.71398163 | 91.072258 | 83.41770808 | 8.877007 | 0 |
| 104 | 79.42029572 | 92.01766205 | 62.59767151 | 91.08003235 | 94.09882355 | 90.19275665 | 71.81607056 | 66.0052948 | 92.95619965 | 91.10973358 | 90.79598236 | 86.44137573 | 84.04432488 | 11.23679 | 0 |
| 105 | 92.8923111 | 90.45961761 | 74.94863892 | 74.376091 | 90.0565567 | 79.42073822 | 91.9625473 | 89.52790833 | 73.92350769 | 91.29217529 | 77.82923126 | 75.52977753 | 83.51825841 | 8.026704 | 0 |
| 106 | 74.42759705 | 70.75283813 | 67.53173828 | 73.58776855 | 68.91642761 | 72.90477753 | 67.2461319 | 63.92389679 | 45.6576004 | 83.22888184 | 62.7667923 | 77.34413147 | 69.02404849 | 9.320705 | 0 |
| 107 | 92.30949402 | 72.84716797 | 91.28887939 | 90.42179871 | 92.4969635 | 90.16604614 | 92.47432709 | 92.77320099 | 92.99639893 | 92.98842621 | 74.28027344 | 83.05562592 | 88.17488353 | 7.353036 | 0 |
| 108 | 81.12245178 | 76.5996933 | 40.98032379 | 76.0388031 | 93.37274933 | 83.16758728 | 65.1399765 | 71.89355469 | 92.31965637 | 87.49199677 | 66.43951416 | 31.31715775 | 72.15695540 | 19.18829 | 0 |
| 109 | 87.91713715 | 66.31819153 | 91.62644958 | 91.11332703 | 63.71058273 | 74.75700378 | 76.87438202 | 70.73054504 | 79.23040771 | 58.00179291 | 62.9307251 | 69.36386108 | 74.38120047 | 11.31061 | 0 |
| 110 | 74.68869019 | 90.3595047 | 95.45928192 | 90.80705261 | 85.12161255 | 91.85719299 | 86.59301758 | 77.35742188 | 92.32761383 | 91.42720795 | 91.56393433 | 91.87742615 | 88.28666306 | 6.338266 | 0 |
| 111 | 90.20005035 | 91.25969696 | 54.56466675 | 92.03517914 | 64.57595062 | 90.9681778 | 88.58420563 | 90.90211487 | 90.30788422 | 92.8008728 | 90.50321198 | 73.73936462 | 84.20344798 | 12.72425 | 0 |
| 112 | 91.55438232 | 67.0876236 | 81.97421265 | 67.6140976 | 69.41828156 | 82.40742493 | 75.53529358 | 90.61135864 | 72.65598297 | 67.91120911 | 90.01774597 | 93.74751282 | 79.21126048 | 10.40018 | 0 |
| 113 | 61.16001511 | 81.74226379 | 91.92977905 | 91.75881195 | 72.4549942 | 92.64192963 | 87.22109985 | 91.25497437 | 91.01328278 | 93.54850006 | 77.81417084 | 63.81154633 | 83.02928066 | 11.66079 | 0 |
| 114 | 71.90158844 | 70.73038483 | 90.21482849 | 83.28530884 | 71.20048523 | 75.31326294 | 76.07601166 | 69.17276001 | 76.61094666 | 56.33848572 | 62.9572258 | 91.52695465 | 74.61068694 | 10.20166 | 0 |
| 115 | 90.28141785 | 68.19804382 | 90.56729126 | 88.85817719 | 76.39228058 | 86.87122345 | 90.88939667 | 61.20427704 | 80.29649353 | 90.85490417 | 88.48254395 | 85.29798126 | 83.18283590 | 9.833107 | 0 |
| 116 | 79.53054047 | 92.69199371 | 90.99800873 | 92.70027161 | 91.80710602 | 91.3815918 | 95.06798553 | 90.61180115 | 91.78723907 | 91.49266815 | 92.63936615 | 78.48474121 | 89.93277613 | 5.233742 | 0 |
| 117 | 78.84977722 | 86.14324951 | 90.62020874 | 75.28889465 | 90.0832901 | 70.18941498 | 79.93156433 | 68.28878021 | 73.31448364 | 77.49992371 | 75.08327484 | 91.8943634 | 79.76560211 | 8.118782 | 0 |
| 118 | 92.31182098 | 90.12355804 | 93.79438782 | 87.730896 | 83.49808502 | 93.51411438 | 74.43669891 | 93.15045929 | 71.42227173 | 91.6605072 | 93.42102814 | 68.2848587 | 86.11239052 | 9.445776 | 0 |
| 119 | 90.82080841 | 94.86637115 | 93.68097687 | 76.84237671 | 90.51638794 | 94.74719238 | 91.53102112 | 92.48826599 | 80.10601044 | 92.37784576 | 50.2224884 | 95.10797119 | 86.94230970 | 12.94063 | 0 |
| 120 | 91.16506195 | 91.34729767 | 63.94282532 | 75.23547363 | 72.85822296 | 79.59251404 | 92.98847961 | 90.95223236 | 93.41345215 | 70.06083679 | 92.38059998 | 81.62362671 | 82.96338526 | 10.45779 | 0 |
| 121 | 91.34613037 | 79.75399017 | 81.68216705 | 90.87773132 | 92.01961517 | 71.87277985 | 93.1880722 | 94.56822968 | 91.03947449 | 92.90409851 | 90.95384979 | 90.24808502 | 88.37118530 | 6.869605 | 0 |
| 122 | 77.59100342 | 91.85566711 | 91.36390686 | 79.03340149 | 90.86995697 | 93.0266037 | 77.92888641 | 92.97805023 | 73.45667267 | 84.57956696 | 95.13339233 | 93.80105591 | 86.80151367 | 7.776127 | 0 |
| 123 | 76.14285278 | 60.2264595 | 90.1368103 | 91.20794678 | 91.09915924 | 92.13226318 | 80.18270874 | 71.70406342 | 91.17910767 | 54.00524139 | 82.43363953 | 90.87399292 | 80.94368712 | 13.09983 | 0 |
| 124 | 90.71604919 | 80.51058197 | 91.52262878 | 67.42729187 | 75.98653412 | 85.50564575 | 90.13996124 | 92.45684052 | 92.98592377 | 81.22423553 | 78.85826874 | 75.43663025 | 83.56421598 | 8.258149 | 0 |
| 125 | 79.14809418 | 61.47752762 | 56.63985062 | 56.63656235 | 54.93479538 | 63.33121872 | 97.78868866 | 40.72254944 | 48.41725922 | 60.91141891 | 61.70664978 | 73.7124939 | 62.95225907 | 14.92143 | 0 |
| 126 | 78.19139099 | 90.78387451 | 91.67318726 | 92.24700928 | 94.25954437 | 91.42822266 | 91.79854584 | 86.13938904 | 72.10910797 | 90.83255005 | 78.09221649 | 90.99189758 | 87.37891134 | 7.181694 | 0 |
| 127 | 93.39350128 | 89.0463562 | 79.25411987 | 95.97327423 | 84.15267181 | 92.55096436 | 91.67805481 | 91.29782867 | 90.7177887 | 95.18501282 | 94.86972809 | 90.28266144 | 90.70016352 | 4.806143 | 0 |
| 128 | 91.04705048 | 90.57145691 | 72.1444397 | 64.08535767 | 88.1109314 | 92.68327332 | 90.55726624 | 91.58660889 | 91.68391418 | 92.23817444 | 91.62123871 | 72.2694931 | 85.71660042 | 10.04573 | 0 |
| 129 | 92.5524826 | 95.1884613 | 80.1646347 | 90.78990936 | 93.1078949 | 90.13026428 | 93.61352539 | 90.26300049 | 82.17980957 | 68.80334473 | 93.87299347 | 95.48381042 | 88.84584427 | 7.93804 | 0 |
| 130 | 90.5632019 | 92.41786194 | 78.23413849 | 83.37824249 | 76.51243591 | 91.71656799 | 86.6523819 | 73.4362793 | 92.73905945 | 95.02549744 | 91.39431763 | 85.51636505 | 86.46552912 | 7.159013 | 0 |
| 131 | 93.99916077 | 94.80766296 | 80.10844421 | 92.04715729 | 80.83743286 | 92.752388 | 81.55271912 | 91.72919464 | 92.11965179 | 62.26901245 | 83.4283905 | 92.47202301 | 86.51026980 | 9.456477 | 0 |
| 132 | 92.42273712 | 91.24158478 | 93.23188019 | 93.26422882 | 94.87219238 | 77.08338928 | 90.66288757 | 90.96479797 | 91.71199036 | 95.65075684 | 90.38576508 | 74.11551666 | 89.63397725 | 6.783527 | 0 |
| 133 | 79.08649445 | 67.82990265 | 63.74990845 | 56.90262604 | 78.70965576 | 61.48138809 | 65.08140564 | 62.98988342 | 91.36251831 | 55.74825287 | 90.75032043 | 69.31785583 | 70.25085100 | 12.0958 | 0 |
| 134 | 68.64472198 | 66.30476379 | 92.42777252 | 90.53217316 | 93.15357971 | 71.36889648 | 92.97397614 | 86.74616241 | 87.97131348 | 83.21169281 | 93.77568817 | 91.96201324 | 84.92272949 | 10.26531 | 0 |
| 135 | 93.93184662 | 93.7323761 | 69.36442566 | 76.95002747 | 82.48747253 | 92.34171295 | 90.76072693 | 90.29528809 | 91.1578064 | 72.58295441 | 93.58522034 | 91.29423523 | 86.54034106 | 8.86104 | 0 |
| 136 | 90.63315582 | 92.46065521 | 90.459198 | 91.68927765 | 73.91981506 | 88.77437592 | 80.00227356 | 91.28014374 | 84.41890717 | 91.79223633 | 91.01651001 | 91.88300323 | 88.19412931 | 5.807469 | 0 |
| 137 | 77.3527832 | 73.37957764 | 73.8119812 | 88.19787598 | 76.03016663 | 59.83588791 | 63.03430176 | 92.75857544 | 79.35870361 | 90.08687592 | 91.58420563 | 65.36508179 | 77.56633473 | 11.31676 | 0 |
| 138 | 81.33589172 | 81.26706696 | 91.17641449 | 90.22537231 | 90.71129608 | 92.11484528 | 77.77562714 | 92.07286072 | 67.06684875 | 91.0585022 | 90.56521606 | 86.71951294 | 86.00745455 | 7.736114 | 0 |
| 139 | 79.8182373 | 81.40397644 | 73.41854858 | 43.34049606 | 68.32370758 | 74.09525299 | 91.47057343 | 70.79003906 | 74.92382812 | 88.26119232 | 78.84625244 | 57.49648666 | 73.51571592 | 13.05248 | 0 |
| 140 | 91.18717194 | 78.91210938 | 48.78417969 | 61.2954483 | 82.75289154 | 91.20883179 | 37.28106689 | 91.33660126 | 91.57450867 | 51.88730621 | 71.21661377 | 43.88885498 | 70.11046537 | 20.60249 | 0 |
| 141 | 76.12297821 | 57.47561264 | 66.13661194 | 72.28222656 | 55.65233994 | 68.75341797 | 90.42710114 | 92.65611267 | 66.39450073 | 71.56906128 | 61.86211395 | 63.40478897 | 70.22807217 | 11.58769 | 0 |
| 142 | 92.08668518 | 73.27174377 | 75.71865082 | 78.55480957 | 88.43484497 | 89.09685516 | 90.93730164 | 90.59698486 | 91.93818665 | 71.77812958 | 92.11474609 | 91.31882477 | 85.48731359 | 8.097536 | 0 |
| 143 | 91.96934509 | 72.68772125 | 84.94029999 | 56.45042038 | 90.78689575 | 82.37239075 | 91.5519104 | 70.81956482 | 74.68346405 | 77.46333313 | 76.272995 | 39.69049454 | 75.80740293 | 15.30879 | 0 |
| 144 | 90.73147583 | 63.51114655 | 90.2100296 | 90.72432709 | 91.10223389 | 79.23717499 | 90.48098755 | 91.68264771 | 91.45948029 | 93.73382568 | 90.77940369 | 92.31832123 | 87.99758784 | 8.511046 | 0 |
| 145 | 95.17651367 | 71.30027771 | 95.22912598 | 95.25941467 | 94.34407043 | 79.66043091 | 94.48635864 | 77.03101349 | 93.22598267 | 91.43824005 | 93.94638824 | 91.93807983 | 89.41965802 | 8.384753 | 0 |
| 146 | 90.36077881 | 81.14003754 | 81.93341064 | 93.51611328 | 92.90168762 | 94.2796936 | 95.05749512 | 95.76364899 | 92.49782562 | 93.55130768 | 93.92598724 | 90.87232971 | 91.31669299 | 4.822453 | 0 |
| 147 | 94.36276245 | 96.13618469 | 92.87160492 | 92.21310425 | 93.49226379 | 92.99240112 | 91.2092514 | 94.74884796 | 92.40459442 | 82.51808167 | 92.19256592 | 91.29871368 | 92.20336469 | 3.370892 | 0 |
| 148 | 76.99146271 | 90.60358429 | 92.69708252 | 68.58545685 | 91.65283966 | 81.0602951 | 87.62696075 | 91.12462616 | 70.16816711 | 91.61384583 | 82.94790649 | 91.29314423 | 84.69711431 | 8.711814 | 0 |
| 149 | 93.27004242 | 92.75907135 | 92.77709198 | 90.78349304 | 92.19168091 | 75.21055603 | 74.14482117 | 92.50845337 | 90.46316528 | 91.67034149 | 93.46295166 | 86.78783417 | 88.83579191 | 6.854938 | 0 |
| 150 | 63.52578354 | 60.62754822 | 70.59189606 | 64.94725037 | 61.09558868 | 40.43305969 | 75.12435913 | 88.76734161 | 64.75301361 | 60.68122101 | 73.15254211 | 59.30578232 | 65.25044886 | 11.50328 | 0 |
| 151 | 94.2250061 | 76.7696991 | 91.31758881 | 92.24198151 | 86.56034088 | 91.63756561 | 91.17862701 | 90.66243744 | 92.05528259 | 93.05300903 | 90.34803009 | 84.64838409 | 89.55816269 | 4.823788 | 0 |
| 152 | 91.68051147 | 93.79412842 | 94.42835236 | 96.17629242 | 95.64451599 | 91.74519348 | 93.29537964 | 91.25131226 | 90.55588531 | 90.57881165 | 91.99576569 | 96.87033081 | 93.16803996 | 2.209459 | 0 |
| 153 | 91.28072357 | 90.17985535 | 96.05524445 | 85.54380798 | 85.2193222 | 71.6200943 | 96.2590332 | 90.22863007 | 92.32452393 | 93.17990112 | 95.08963013 | 48.66077423 | 86.30346171 | 13.63311 | 0 |
| 154 | 90.77508545 | 74.96572876 | 90.75023651 | 85.60519409 | 69.09178925 | 73.90166473 | 78.1131134 | 90.05786133 | 91.5490036 | 91.17582703 | 69.30373383 | 74.83692169 | 81.67717997 | 9.120021 | 0 |
| 155 | 93.05962372 | 67.75787354 | 91.40283966 | 78.46641541 | 90.8319397 | 92.6457901 | 77.84262848 | 91.84454346 | 93.66433716 | 92.88683319 | 91.65463257 | 85.61174774 | 87.30576706 | 8.282569 | 0 |
| 156 | 95.25730896 | 95.25240326 | 72.12083435 | 90.55297089 | 77.11780548 | 95.17176056 | 76.72190857 | 87.14357758 | 92.77649689 | 93.97761536 | 94.87680054 | 91.38937378 | 88.52990469 | 8.397359 | 0 |
| 157 | 92.32241821 | 87.29851532 | 83.04275513 | 67.80034637 | 77.94908142 | 69.01095581 | 90.23027039 | 91.54272461 | 64.89697266 | 92.64939117 | 90.14631653 | 90.58750916 | 83.12310473 | 10.49747 | 0 |
| 158 | 65.82151794 | 94.07353973 | 92.96514893 | 95.43548584 | 96.55517578 | 93.67391205 | 94.3536377 | 94.983284 | 89.87185669 | 97.01290894 | 97.20715332 | 97.63039398 | 92.46533458 | 8.66662 | 0 |
| 159 | 70.14866638 | 54.04009628 | 75.4950943 | 90.39573669 | 67.83263397 | 91.46446228 | 90.07757568 | 71.2979126 | 80.37306976 | 95.60955811 | 59.5256424 | 73.29241943 | 76.62940566 | 13.23821 | 0 |
| 160 | 85.67914581 | 94.78929138 | 91.86434174 | 88.99382019 | 93.59057617 | 90.21617889 | 94.87400055 | 70.17073059 | 92.58358002 | 90.37986755 | 90.72638702 | 85.06542206 | 89.07777850 | 6.724994 | 0 |
| 161 | 73.11358643 | 79.93835449 | 87.03002167 | 77.08592224 | 91.55700684 | 78.60404205 | 72.59270477 | | 90.24940491 | 80.15275574 | 91.37191772 | 92.70098877 | 83.12697324 | 7.641154 | 1 |
| 162 | 90.73101044 | 81.9539032 | 91.40354919 | 92.17435455 | 96.34143829 | 91.49435425 | 90.31957245 | 91.79852295 | 91.58259583 | 91.46474457 | 54.20212555 | 90.6018219 | 87.83899943 | 11.07091 | 0 |
| 163 | 76.63031769 | 95.89757538 | 91.73105621 | 90.2248764 | 90.77144623 | 92.37619019 | 94.02892303 | 92.80229187 | 90.50983429 | 92.57546997 | 93.83972931 | 88.15502167 | 90.79522769 | 4.902274 | 0 |
| 164 | 91.04216766 | 69.95093536 | 65.23937988 | 77.27312469 | 69.87181854 | 90.66147614 | 69.00076294 | 76.13511658 | 91.1765213 | 65.35103607 | 76.20484924 | 74.51629639 | 76.36862373 | 9.658396 | 0 |
| 165 | 91.13726044 | 59.7576828 | 51.26145172 | 60.40395355 | 90.04363251 | 79.32700348 | 86.54225922 | 37.09171295 | 94.20767975 | 70.10150909 | 43.04165268 | 50.69981384 | 67.80130100 | 20.15284 | 0 |
| 166 | 62.52458954 | 78.67895508 | 76.1552124 | 91.49183655 | 85.56053925 | 76.65098572 | 52.38471222 | 94.45557404 | 75.64495087 | 62.57110214 | 91.25910187 | 63.1557045 | 75.87777202 | 13.47388 | 0 |
| 167 | 90.43473053 | 90.78868866 | 91.18167114 | 90.50480652 | 90.40261841 | 93.98925018 | 94.10762024 | 90.10964203 | 91.45246887 | 85.05410767 | 93.792099 | 92.73486328 | 91.21271388 | 2.453307 | 0 |
| 168 | 72.75621796 | 93.64150238 | 67.50913239 | 73.17499542 | 54.80192184 | 90.68373108 | 92.26756287 | 56.98497009 | 61.98896027 | 91.22514343 | 48.61102676 | 65.06949615 | 72.39288839 | 16.05277 | 0 |
| 169 | 90.79830933 | 91.02519226 | 91.28240967 | 57.90578079 | 81.712677 | 85.79925537 | 68.78744507 | 91.52352142 | 83.70652771 | 93.20511627 | 78.31874084 | 90.73509216 | 83.73333899 | 10.81922 | 0 |
| 170 | 93.78837585 | 93.25659943 | 63.64699173 | 81.0032959 | 91.21357727 | 72.26163483 | 80.79984283 | 75.08655548 | 67.86834717 | 71.33812714 | 85.2348938 | 92.25131989 | 80.64579678 | 10.61471 | 0 |
| 171 | 90.0019455 | 90.1332016 | 90.28683472 | 93.74391174 | 92.54884338 | 92.2168045 | 83.20926666 | 74.90362549 | 92.77967834 | 89.28900909 | 71.05176544 | 90.67095947 | 87.56965383 | 7.363184 | 0 |
| 172 | 90.47836304 | 92.96092987 | 89.5035553 | 90.47228241 | 86.29782867 | 90.7151947 | 90.59407043 | 79.51470184 | 90.57486725 | 90.89009094 | 93.94826508 | 90.71105194 | 89.72176679 | 3.696012 | 0 |
| 173 | 93.20383453 | 90.64937592 | 93.2924881 | 91.06100464 | 95.35333252 | 92.64788055 | 93.56539917 | 91.65348053 | 92.96678925 | 92.4617691 | 91.99469757 | 81.28020477 | 91.67752139 | 3.501621 | 0 |
| 174 | 91.24334717 | 90.83500671 | 76.78113556 | 92.44355774 | 91.73561096 | 91.30471802 | 90.67564392 | 73.09976196 | 91.77513123 | 93.75050354 | 70.29695892 | 89.19940948 | 86.92839877 | 8.34807 | 0 |
| 175 | 92.84223938 | 93.01473999 | 88.24784851 | 91.08901978 | 94.74085236 | 95.95207977 | 94.01286316 | 93.03101349 | 93.36241913 | 93.52107239 | 93.39444733 | 90.76319885 | 92.83098285 | 2.005697 | 0 |
| 176 | 91.75711823 | 92.43859863 | 93.2224884 | 90.3820343 | 91.69367981 | 93.39713287 | 90.24547577 | 90.58161163 | 82.66336823 | 78.76783752 | 94.29756165 | 93.77648926 | 90.26861636 | 4.731182 | 0 |
| 177 | 93.27180481 | 90.52959442 | 92.69556427 | 91.01992798 | 88.65965271 | 90.37028503 | 90.91430664 | 77.49882507 | 91.74330902 | 92.42744446 | 94.24958801 | 90.7172699 | 90.34146436 | 4.311152 | 0 |
| 178 | 91.42423248 | 91.59488678 | 67.42691803 | 92.99164581 | 90.91542053 | 92.08986664 | 93.88484955 | 91.23034668 | 78.98073578 | 93.57688141 | 90.07182312 | 42.51337051 | 84.72508144 | 15.37467 | 0 |
| 179 | 63.06943893 | 93.6726532 | 87.81951141 | 91.51829529 | 92.73273468 | 90.34049225 | 93.76592255 | 92.33627319 | 93.92891693 | 94.27142334 | 64.60482788 | 90.44163513 | 87.37517707 | 11.1565 | 0 |
| 180 | 89.90105438 | 92.70065308 | 62.88571167 | 90.01757812 | 77.79956055 | 63.55423355 | 90.11406708 | 72.75589752 | 91.95301056 | 71.86637115 | 72.84144592 | 84.1115799 | 80.04176362 | 11.11847 | 0 |
| 181 | 67.76013947 | 70.48856354 | 74.91324615 | 73.51989746 | 89.42760468 | 73.51624298 | 77.92267609 | 90.11653137 | 75.54730225 | 90.98894501 | 91.45730591 | 91.97198486 | 80.63586998 | 9.32278 | 0 |
| 182 | 69.01503754 | 50.49621582 | 61.29045105 | 92.54999542 | 75.45968628 | 56.31494904 | 59.86749649 | 70.34643555 | 67.52814484 | 70.39491272 | 62.00566864 | 90.81152344 | 68.84004307 | 12.69485 | 0 |
| 183 | 95.57605743 | 99.50543213 | 91.99427032 | 95.07514191 | 91.79299927 | 93.21807098 | 30.46087837 | 93.33428192 | 93.18935394 | 80.72299957 | 93.00008392 | 51.08000565 | 84.07913128 | 21.14903 | 0 |
| 184 | 90.00712585 | 53.62437439 | 84.89454651 | 88.15531158 | 94.65200806 | 93.83476257 | 64.20162964 | 70.29705811 | 93.30435181 | 74.71738434 | 91.25915527 | 74.19203186 | 81.09497833 | 13.44052 | 0 |
| 185 | 90.01197815 | 93.84011841 | 93.1400528 | 92.01313782 | 92.69001007 | 90.96070099 | 92.65827179 | 74.20576477 | 94.17192841 | 90.05211639 | 90.40719604 | 92.83489227 | 90.58218066 | 5.350324 | 0 |
| 186 | 80.98882294 | 92.49758911 | 92.77004242 | 92.45917511 | 88.45423126 | 93.17955017 | 94.95089722 | 82.70277405 | 92.54404449 | 93.39251709 | 90.73339081 | 92.25707245 | 90.57750893 | 4.380028 | 0 |
| 187 | 68.896698 | 72.54081726 | 75.81065369 | 91.57958221 | 91.83358002 | 70.95108795 | 93.26823425 | 90.49986267 | 72.69963837 | 92.50932312 | 71.76354218 | 90.62217712 | 81.91459974 | 10.38018 | 0 |
| 188 | 95.20894623 | 94.64646912 | 95.35945129 | 93.54623413 | 94.49734497 | 94.72010803 | 92.15393829 | 94.45360565 | 86.4861908 | 78.17718506 | 94.02554321 | 91.06575012 | 92.02839724 | 5.019016 | 0 |
| 189 | 82.2000351 | 79.52097321 | 67.01928711 | 90.45346832 | 88.59707642 | 84.9126358 | 85.15546417 | 95.53453064 | 75.38194275 | 91.04618073 | 91.88813782 | 92.25582886 | 85.33046341 | 8.216941 | 0 |
| 190 | 93.64673615 | 89.59598541 | 92.48449707 | 90.20322418 | 67.9548111 | 95.82064819 | 90.40126038 | 88.7593689 | 94.03927612 | 91.61820984 | 94.10485077 | 91.36711884 | 89.99966558 | 7.254463 | 0 |
| 191 | 68.7567215 | 92.57752991 | 90.95036316 | 66.79261017 | 91.91512299 | 90.637146 | 90.44287109 | 92.47449493 | 92.24062347 | 90.94017792 | 92.50270081 | 68.69152069 | 85.74349022 | 10.68816 | 0 |
| 192 | 53.63869476 | 44.67890167 | 66.82517242 | 53.27009583 | 60.03029251 | 69.22815704 | 60.04159546 | 65.07427216 | 68.29523468 | 64.1662674 | 58.18265533 | 63.34817886 | 60.56495984 | 7.228971 | 0 |
| 193 | 92.64118958 | 92.45787048 | 94.02452087 | 94.46689606 | 91.81411743 | 93.82389832 | 90.71237946 | 95.10473633 | 93.81090546 | 92.53115845 | 94.80475616 | 91.33080292 | 93.12693596 | 1.417415 | 0 |
| 194 | 90.10527039 | 70.33589172 | 91.45926666 | 93.07957458 | 94.05786896 | 90.45579529 | 90.45024872 | 92.16731262 | 91.26291656 | 86.40240479 | 92.51403046 | 92.59983063 | 89.57420095 | 6.362653 | 0 |
| 195 | 73.63304901 | 82.99177551 | 60.73117828 | 83.23488617 | 75.3836441 | 63.56268692 | 72.1427536 | 69.11024475 | 63.48394012 | 73.34049988 | 77.59626007 | 63.31758499 | 71.54404195 | 7.664744 | 0 |
| 196 | 76.1417923 | 77.84631348 | 75.82518768 | 77.79409027 | 77.24337769 | 82.04828644 | 75.87126923 | 76.76396179 | 65.55329132 | 82.46533966 | 67.86833954 | 86.32607269 | 76.81227684 | 5.730102 | 0 |
| 197 | 92.34545135 | 92.60436249 | 64.75856018 | 92.63710022 | 92.84590912 | 90.04393768 | 93.37864685 | 87.78282166 | 74.16426849 | 83.59027863 | 86.65362549 | 91.69884491 | 86.87531726 | 8.891141 | 0 |
| 198 | 94.88121796 | 85.91305542 | 95.10985565 | 85.88301086 | 91.51934052 | 95.98854828 | 94.95883179 | 94.26094818 | 94.47174835 | 95.73170471 | 91.90357971 | 92.33899689 | 92.74673653 | 3.5151 | 0 |
| 199 | 75.98686218 | 80.19326019 | 93.0703125 | 96.90345001 | 94.89302063 | 94.64968872 | 94.99729156 | 95.91226959 | 94.80918121 | 96.15399933 | 95.22499847 | 91.00969696 | 91.98366928 | 6.723192 | 0 |
| 200 | 90.83827209 | 94.13799286 | 72.76012421 | 93.39524078 | 93.65697479 | 91.14246368 | 93.01178741 | 90.2233963 | 79.64883423 | 90.61516571 | 81.39682007 | 87.96705627 | 88.23284403 | 6.732315 | 0 |
| 201 | 81.87548828 | 72.55408478 | 74.76532745 | 91.12232208 | 57.11278534 | 45.1871109 | 91.00787354 | 51.93016434 | 71.95816803 | 77.61634827 | 67.25695801 | 79.84117889 | 71.85231749 | 14.43186 | 0 |
| 202 | 93.24660492 | 73.77728271 | 93.42840576 | 77.96829987 | 91.37887573 | 60.43974304 | 93.70790863 | 75.01023865 | 90.91101074 | 72.76364899 | 57.05251312 | 85.62975311 | 80.44285711 | 12.97156 | 0 |
| 203 | 70.92973328 | 92.51715088 | 62.830513 | 73.96786499 | 93.51596832 | 73.04527283 | 67.55775452 | 88.0100174 | 91.38464355 | 78.49401093 | 90.97528839 | 55.38738251 | 78.21796672 | 12.93394 | 0 |
| 204 | 93.24520111 | 94.93048859 | 75.51580811 | 90.58663177 | 81.3802948 | 93.06251526 | 91.21869659 | 92.05951691 | 94.29122162 | 83.80484009 | 93.47382355 | 94.45762634 | 89.83555540 | 6.199634 | 0 |
| 205 | 77.37841797 | 92.44213867 | 93.55820465 | 94.56983948 | 92.58685303 | 73.28787231 | 77.19313049 | 94.59385681 | 94.81210327 | 91.2326889 | 79.12820435 | 90.99614716 | 87.64828809 | 8.242995 | 0 |
| 206 | 93.59928894 | 93.76081848 | 92.11310577 | 93.38204193 | 94.60978699 | 94.2171936 | 92.52793121 | 91.34207916 | 94.81497192 | 90.98029327 | 91.60609436 | 91.35483551 | 92.85903676 | 1.371777 | 0 |
| 207 | 73.17631531 | 76.30054474 | 97.891716 | 95.71002197 | 77.81060028 | 91.04879761 | 84.91254425 | 90.48722839 | 92.93045044 | 86.9807663 | 91.96130371 | 93.56860352 | 87.73157438 | 8.056234 | 0 |
| 208 | 87.27664948 | 62.39595795 | 90.04115295 | 93.32073975 | 81.10861206 | 93.74013519 | 92.67444611 | 93.49027252 | 90.96984863 | 93.54983521 | 93.98394775 | 92.94442749 | 88.79133542 | 9.112239 | 0 |
| | | | | | | | | | | | | Overall | 84.50225834 | 11.85391 | 3 |

### 3. Results of Testing Face++ on Real Images

| ID | 1 | 2 | 3 | 4 | 5 | 6 | 7 | 8 | 9 | 10 | 11 | 12 | Average | StDev | Count |
|---|---|---|---|---|---|---|---|---|---|---|---|---|---|---|---|
| 1 | 86.353 | 91.209 | 86.97 | 91.066 | 88.709 | 84.505 | 88.395 | 84.096 | 88.999 | 90.706 | 92.804 | 90.161 | 88.6644 | 2.7363 | 0 |
| 2 | 93.261 | 92.973 | 93.269 | 93.122 | 90.457 | 90.804 | 91.802 | 89.851 | 84.275 | 93.072 | 91.792 | 66.274 | 89.2460 | 7.6607 | 0 |
| 3 | 92.383 | 88.381 | 90.639 | 90.383 | 90.039 | 90.402 | 92.083 | 88.786 | 90.243 | 91.555 | 94.29 | 92.857 | 91.0034 | 1.7001 | 0 |
| 4 | 86.263 | 87.298 | | 84.093 | 88.371 | 85.246 | 90.876 | 93.138 | 90.096 | 90.003 | 84.545 | 88.371 | 88.0273 | 2.8573 | 1 |
| 5 | 91.456 | 82.273 | 91.379 | 93.929 | 89.343 | 94.007 | 93.81 | 92.465 | 90.096 | 93.654 | 91.26 | 91.592 | 91.2720 | 3.2237 | 0 |
| 6 | 85.304 | 91.597 | 89.896 | 89.415 | 88.637 | 87.579 | 87.038 | 90.201 | 86.545 | 89.334 | 84.145 | 87.733 | 88.1187 | 2.1493 | 0 |
| 7 | 93.047 | 92.319 | 91.267 | 94.702 | 90.81 | 92.903 | 94.769 | 89.589 | 90.577 | 93.672 | 90.392 | 87.401 | 91.7873 | 2.1901 | 0 |
| 8 | 89.453 | 92.823 | 92.103 | 93.407 | 88.523 | 93.035 | 92.546 | 89.449 | 84.49 | 84.796 | 86.696 | 93.024 | 90.0288 | 3.3025 | 0 |
| 9 | 85.741 | 92.955 | 89.021 | 88.945 | 90.869 | 83.043 | 86.511 | 90.084 | 91.348 | 92.429 | 92.156 | 92.335 | 89.6198 | 3.1095 | 0 |
| 10 | 84.7 | 87.347 | 89.486 | 90.484 | 89.783 | 80.905 | 89.306 | 89.558 | 89.603 | 88.308 | 94.209 | 90.522 | 88.6843 | 3.2933 | 0 |
| 11 | 89.3 | 85.89 | 84.012 | 89.915 | 91.754 | 94.444 | 87.766 | 89.985 | 85.825 | 88.684 | 92.328 | 92.595 | 89.3748 | 3.1289 | 0 |
| 12 | 81.391 | 91.708 | 91.895 | 85.875 | 90.416 | 91.551 | 91.263 | 91.956 | 93.804 | 88.651 | 86.649 | 88.554 | 89.4761 | 3.4635 | 0 |
| 13 | 96.642 | 81.254 | 78.881 | 81.128 | 86.139 | 77.912 | 90.612 | 77.058 | 83.718 | 84.843 | 83.967 | 83.507 | 83.8051 | 5.5263 | 0 |
| 14 | 89.858 | 90.1 | 89.196 | 92.116 | 83.153 | 91.528 | 88.849 | 81.55 | 88.215 | 91.413 | 80.202 | 89.846 | 88.0022 | 4.0498 | 0 |
| 15 | 75.384 | 79.613 | 86.513 | 83.003 | 85.328 | 85.343 | 88.994 | 93.163 | 86.47 | 89.032 | 87.323 | 79.375 | 84.9618 | 4.9176 | 0 |
| 16 | 89.162 | 91.654 | 93.128 | 94.099 | 90.187 | 87.06 | 92.468 | 88.042 | 92.235 | 87.839 | 91.85 | 87.806 | 90.4608 | 2.4117 | 0 |
| 17 | 91.015 | 89.126 | 78.26 | 84.345 | 86.894 | 88.823 | 87.996 | 91.421 | 88.03 | 88.93 | 90.357 | 85.067 | 87.5220 | 3.6240 | 0 |
| 18 | 90.546 | 92.614 | 89.24 | 90.432 | 82.493 | 87.504 | 88.981 | 88.679 | 91.534 | 90.249 | 88.684 | 91.104 | 89.3383 | 2.5827 | 0 |
| 19 | 86.478 | 76.45 | 86.79 | 86.233 | 85.41 | 86.526 | 91.386 | 83.53 | 90.582 | 89.822 | 89.419 | 93.252 | 87.1565 | 4.3867 | 0 |
| 20 | | 91.18 | 86.508 | 90.134 | 84.846 | 91.472 | 88.448 | 80.363 | 93.433 | 90.216 | 90.144 | 91.807 | 88.9592 | 3.7526 | 1 |
| 21 | 81.451 | 80.702 | 74.047 | 84.106 | 82.93 | 77.289 | 73.731 | 82.306 | 77.002 | 96.97 | 76.572 | 66.793 | 79.4916 | 7.3577 | 0 |
| 22 | 92.585 | 86.871 | 94.413 | 87.653 | 89.757 | 88.965 | 89.766 | 90.483 | 87.153 | 90.35 | 87.705 | 90.411 | 89.6760 | 2.2417 | 0 |
| 23 | 81.331 | 79.82 | 90.789 | 91.588 | 87.526 | 84.259 | 84.19 | 91.446 | 81.794 | 89.232 | 93.302 | 86.59 | 86.8223 | 4.5341 | 0 |
| 24 | 94.562 | 89.899 | 89.791 | 90.121 | 86.001 | 84.988 | 73.402 | 94.145 | 88.164 | 91.536 | 87.651 | 84.774 | 87.9195 | 5.5701 | 0 |
| 25 | 81.845 | 80.99 | 86.948 | 86.603 | 85.916 | 91.294 | 85.842 | 82.132 | 89.826 | 84.214 | 89.493 | 84.907 | 85.8342 | 3.2689 | 0 |
| 26 | 91.179 | 85.984 | 88.708 | 90.598 | 87.974 | 91.461 | 91.154 | 91.095 | 91.4 | 91.317 | 91.843 | 87.241 | 89.9962 | 1.9776 | 0 |
| 27 | 40.904 | 88.266 | 81.566 | 87.444 | 87.665 | 84.316 | 36.303 | 87.474 | 83.507 | 87.221 | 70.334 | 89.516 | 77.0430 | 18.6861 | 0 |
| 28 | 89.235 | 79.017 | 88.024 | 97.055 | 96.591 | 90.263 | 81.571 | 90.228 | 86.946 | 80.355 | 91.832 | 83.031 | 87.8457 | 5.9397 | 0 |
| 29 | 83.94 | 82.182 | 93.648 | 86.372 | 82.039 | | 82.708 | 85.629 | 89.326 | 78.983 | 89.136 | 77.429 | 84.6720 | 4.7782 | 1 |
| 30 | 89.964 | 87.992 | 88.462 | 91.947 | 93.5 | 94.646 | 84.46 | 91.291 | 91.948 | 84.46 | 86.728 | 93.951 | 89.9458 | 3.5310 | 0 |
| 31 | 87.661 | 79.053 | 81.928 | 92.683 | 91.959 | 84.924 | 84.353 | 83.758 | 88.336 | 79.402 | 85.355 | 82.606 | 85.1682 | 4.3619 | 0 |
| 32 | 92.457 | 85.066 | 82.084 | 95.57 | 78.414 | 89.747 | 88.487 | 89.086 | 85.627 | 86.52 | 87.101 | 93.68 | 87.8199 | 4.8396 | 0 |
| 33 | 86.891 | 88.089 | 89.452 | 80.489 | 85.099 | 87.879 | 89.371 | 39.283 | 90.225 | 88.356 | 81.732 | 90.462 | 83.1107 | 14.1645 | 0 |
| 34 | 94.574 | 89.487 | 85.28 | 94.417 | 93.346 | 94.166 | 90.351 | 92.432 | 94.865 | 89.351 | 90.927 | 90.418 | 91.6345 | 2.8703 | 0 |
| 35 | 83.532 | 89.903 | 88.303 | 91.847 | 89.588 | 85.44 | 69.571 | 86.966 | 86.361 | 85.508 | 90.715 | 87.935 | 86.3058 | 5.8003 | 0 |
| 36 | 89.589 | 78.596 | 82.664 | 82.102 | 75.904 | 86.691 | 86.206 | 82.7 | 84.689 | 86.164 | 78.021 | 84.923 | 83.1874 | 4.0340 | 0 |
| 37 | 90.163 | 78.956 | 88.867 | 86.02 | 87.28 | 88.136 | 79.233 | 79.79 | 84.847 | 81.157 | 81.604 | 82.098 | 84.0126 | 4.0188 | 0 |
| 38 | 91.671 | 89.801 | 93.414 | 90.813 | 85.776 | 90.814 | 89.029 | 82.824 | 89.439 | 93.937 | 79.003 | 92.405 | 89.0772 | 4.4559 | 0 |
| 39 | 91.844 | 87.559 | 87.043 | 92.036 | 83.355 | 85.264 | 92.643 | 90.76 | 84.617 | 92.604 | 84.587 | 90.394 | 88.5588 | 3.5251 | 0 |
| 40 | 90.231 | 92.571 | 96.364 | 83.503 | 87.434 | 87.355 | 90.264 | 90.503 | 88.413 | 26.703 | 86.984 | 82.805 | 83.5942 | 18.2925 | 0 |
| 41 | 85.837 | 81.88 | 81.753 | 90.498 | 87.284 | 82.721 | 69.052 | 92.04 | 84.456 | 89.647 | 84.317 | 84.697 | 84.5152 | 5.9172 | 0 |
| 42 | 93.358 | 87.332 | 86.373 | | 84.872 | 82.074 | 88.097 | 78.925 | 91.328 | 92.929 | 88.872 | 82.408 | 86.9607 | 4.6324 | 1 |
| 43 | 88.497 | 87.734 | 88.158 | 91.609 | 90.453 | 93.284 | 88.963 | 86.772 | 90.122 | 91.555 | 95.225 | 94.301 | 90.5561 | 2.7032 | 0 |
| 44 | 90.598 | 90.606 | 88.159 | 90.505 | 91.06 | 90.874 | 86.398 | 91.428 | 90.839 | 88.374 | 92.41 | 89.316 | 90.0473 | 1.6751 | 0 |
| 45 | 90.898 | 85.393 | 89.191 | 88.39 | 86.218 | 87.018 | 88.733 | 88.713 | 90.703 | 85.588 | 87.942 | 88.226 | 88.0844 | 1.7849 | 0 |
| 46 | 92.715 | 90.684 | 90.365 | 92.804 | 91.638 | 92.507 | 80.41 | 85.976 | 92.216 | 92.607 | 92.486 | 89.143 | 90.2959 | 3.6993 | 0 |
| 47 | 91.325 | 86.681 | 88.862 | 92.082 | 83.304 | 87.512 | 90.36 | 92.997 | 91.277 | 89.226 | 84.855 | 90.575 | 89.0880 | 2.9786 | 0 |
| 48 | 94.039 | 87.168 | 90.855 | 84.683 | 84.495 | 92.507 | 78.367 | 92.374 | 91.119 | 92.6 | 80.018 | 88.594 | 88.0683 | 5.1851 | 0 |
| 49 | 84.168 | 81.182 | 77.858 | 87.412 | 85.786 | 75.011 | 80.518 | 82.383 | 90.772 | 79.475 | 85.277 | 80.843 | 82.5571 | 4.3591 | 0 |
| 50 | 80.511 | 90.033 | 89.169 | 89.967 | 92.14 | 86.314 | 92.605 | 87.417 | 88.747 | 88.537 | 90.637 | 92.544 | 89.0518 | 3.3324 | 0 |
| 51 | 94.02 | 94.734 | 91.221 | 91.46 | 89.98 | 94.93 | 93.513 | 91.651 | 87.629 | 90.994 | 92.841 | 91.202 | 92.0146 | 2.1103 | 0 |
| 52 | 92.152 | 83.699 | 90.345 | 77.18 | 84.478 | 84.709 | 83.963 | 82.323 | 87.089 | 89.309 | 81.885 | 87.191 | 85.3603 | 4.1236 | 0 |
| 53 | 89.401 | 90.698 | 95.061 | 90.845 | 87.208 | 80.079 | 82.066 | 90.771 | 87.859 | 91.097 | 87.337 | 91.232 | 88.6378 | 4.1458 | 0 |
| 54 | 87.771 | 87.073 | 81.438 | 84.584 | 89.805 | 89.67 | 84.14 | 90.802 | 86.645 | 84.68 | 89.551 | 78.207 | 86.1972 | 3.7731 | 0 |
| 55 | 92.959 | 93.016 | 93.124 | 92.679 | 91.268 | 89.593 | 88.117 | 88.662 | 71.059 | 90.212 | 97.04 | 87.558 | 89.6073 | 6.4271 | 0 |
| 56 | 91.291 | 86.702 | 89.873 | 90.966 | 88.052 | 91.141 | 90.92 | 89.055 | 88.338 | 91.432 | 89.534 | 91.886 | 89.9325 | 1.6198 | 0 |
| 57 | 91.533 | 91.18 | 88.92 | 86.982 | 92.288 | 94.947 | 92.408 | 91.885 | 92.632 | 85.45 | 93.966 | 93.054 | 91.2704 | 2.8025 | 0 |
| 58 | 90.109 | 85.112 | 87.023 | 86.738 | 84.582 | 85.288 | 84.703 | 87.825 | 88.715 | 86.747 | 81.795 | 88.392 | 86.4191 | 2.2567 | 0 |
| 59 | 82.026 | 82.604 | 83.74 | 84.166 | 87.024 | 82.873 | 85.454 | 97.389 | 88.161 | 70.408 | 79.443 | 76.975 | 83.3553 | 6.5012 | 0 |
| 60 | 88.624 | 87.496 | 85.301 | 92.46 | 79.937 | 82.253 | 87.65 | 89.742 | 91.9 | 85.854 | 83.155 | 87.483 | 86.8213 | 3.7647 | 0 |
| 61 | 93.773 | 85.762 | 86.04 | 87.043 | 89.759 | 89.813 | 84.765 | 89.953 | 88.631 | 91.578 | 91.314 | 83.969 | 88.5333 | 3.0267 | 0 |
| 62 | 86.408 | 86.181 | 85.336 | 84.778 | 87.189 | 61.929 | 84.2 | 78.142 | 82.259 | 83.572 | 79.72 | 82.545 | 81.8549 | 6.8293 | 0 |
| 63 | 90.693 | 92.089 | 90.117 | 91.936 | 93.378 | 91.92 | 84.769 | 89.073 | 88.081 | 85.59 | 75.994 | 84.286 | 88.1605 | 4.8922 | 0 |
| 64 | 87.621 | 89.864 | 88.383 | 85.471 | 88.242 | 87.708 | 86.08 | 89.429 | 90.799 | 90.039 | 87.926 | 89.065 | 88.3856 | 1.5777 | 0 |
| 65 | 90.37 | 89.513 | 88.918 | 91.865 | 87.277 | 92.806 | 91.49 | 89.478 | 92.863 | 90.008 | 91.846 | 91.721 | 90.6796 | 1.7010 | 0 |
| 66 | 88.444 | 91.505 | 84.45 | 84.902 | 89.223 | 78.593 | 89.338 | 90.792 | 89.511 | 88.831 | 89.194 | 87.464 | 87.6873 | 3.5390 | 0 |
| 67 | 88.366 | 90.767 | 88.698 | 91.066 | 88.546 | 82.371 | 89.617 | 89.236 | 89.207 | 84.77 | 91.436 | 73.794 | 87.3228 | 4.9891 | 0 |
| 68 | 85.143 | 84.763 | 83.472 | 83.269 | 85.226 | 82.676 | 85.416 | 88.044 | 82.303 | 87.514 | 78.172 | 84.199 | 84.1831 | 2.5787 | 0 |
| 69 | 87.042 | 85.802 | 87.046 | 82.752 | 87.675 | 77.545 | 84.207 | 85.829 | 91.31 | 82.732 | 87.54 | 83.901 | 85.2818 | 3.4347 | 0 |
| 70 | 93.843 | 87.517 | 88.367 | | 90.947 | 91.51 | 91.248 | 80.928 | 85.642 | 91.504 | 91.858 | 90.47 | 89.4395 | 3.6347 | 1 |
| 71 | 77.588 | 91.154 | 78.714 | 72.028 | 91.112 | 70.525 | 70.524 | 33.001 | 93.22 | 85.227 | 71.849 | 88.934 | 76.9897 | 16.3679 | 0 |
| 72 | 79.99 | 88.456 | 87.087 | 88.829 | 92.275 | 86.591 | 81.727 | 87.927 | 84.404 | 85.488 | 89.945 | 87.655 | 86.6978 | 3.4152 | 0 |
| 73 | 91.322 | 86.45 | 90.411 | 93.191 | 88 | 89.455 | 89.948 | 87.534 | 90.04 | 91.251 | 91.455 | 88.547 | 89.8003 | 1.9164 | 0 |
| 74 | 90.787 | 87.865 | 90.483 | 92.385 | 86.628 | 78.789 | 82.598 | 88.993 | 86.973 | 86.463 | 93.633 | 91.711 | 88.1090 | 4.2517 | 0 |
| 75 | 90.464 | 83.076 | 88.835 | 77.963 | 83.162 | 88.272 | 86.231 | 88.237 | 90.31 | 66.572 | 90.671 | 83.764 | 84.7964 | 6.9074 | 0 |
| 76 | 88.583 | 84.277 | 93.074 | 92.001 | 87.961 | 90.188 | 85.291 | 86.883 | 91.31 | 90.676 | 83.364 | 88.226 | 88.4862 | 3.1065 | 0 |
| 77 | 86.807 | 84.623 | 89.796 | 85.286 | 87.748 | 84.925 | 88.218 | 88.514 | 78.747 | 85.569 | 86.548 | 88.852 | 86.3028 | 2.9084 | 0 |
| 78 | 81.029 | 82.704 | 85.197 | 81.304 | 79.32 | 68.419 | 79.817 | 83.254 | 95.103 | 79.61 | 80.771 | 83.046 | 81.6312 | 5.9428 | 0 |
| 79 | 87.341 | 89.111 | 84.324 | 83.586 | 82.515 | 79.641 | 70.797 | 91.509 | 89.99 | 89.818 | 85.522 | 87.834 | 85.1657 | 5.7201 | 0 |
| 80 | 89.357 | 88.978 | 86.834 | 88.155 | 87.496 | 97.389 | 91.29 | 91.725 | 89.503 | 79.887 | 84.251 | 83.593 | 88.2048 | 4.4552 | 0 |

| | | | | | | | | | | | | | | |
|---|---|---|---|---|---|---|---|---|---|---|---|---|---|---|
| 81 | 84.54 | 88.092 | 87.943 | 91.006 | 88.699 | 84.229 | 89.103 | 88.614 | 90.025 | 89.436 | 91.266 | 90.935 | 88.6573 | 2.2879 | 0 |
| 82 | 90.216 | 89.147 | 87.171 | 89.662 | 87.941 | 89.154 | 91.641 | 90.534 | 83.665 | 90.6 | 83.31 | 90.011 | 88.5877 | 2.6652 | 0 |
| 83 | 90.549 | 89.396 | 92.371 | 91.655 | 89.167 | 92.11 | 88.479 | 81.659 | 92.502 | 86.452 | 91.884 | 91.16 | 89.7820 | 3.1512 | 0 |
| 84 | 90.176 | 92.523 | 90.253 | 89.561 | 91.331 | 90.902 | 91.563 | 93.463 | 93.977 | 89.157 | 90.162 | 88.298 | 90.9472 | 1.7151 | 0 |
| 85 | 73.783 | 55.382 | 70.542 | 74.932 | 84.712 | 80.091 | 83.773 | 75.405 | 79.474 | 79.466 | 74.911 | 86.594 | 76.5888 | 8.2445 | 0 |
| 86 | 71.837 | 91.934 | 86.965 | 90.596 | 90.949 | 91.707 | 89.217 | 77.505 | 92.586 | 86.377 | 88.522 | 90.296 | 87.3743 | 6.3468 | 0 |
| 87 | 88.881 | 88.412 | 92.049 | 93.587 | 88.643 | 91.911 | 90.872 | 91.025 | 79.173 | 86.655 | 92.056 | 92.357 | 89.6351 | 3.8699 | 0 |
| 88 | 80.618 | 80.609 | 88.768 | 88.768 | 90.14 | 90.14 | 81.902 | 81.897 | 90.554 | 90.571 | 90.997 | 91.006 | 87.1642 | 4.4381 | 0 |
| 89 | 88.765 | 88.5 | 83.22 | 88.956 | 90.473 | 90.75 | 84.042 | 89.421 | 80.652 | 89.242 | 33.083 | 88.564 | 82.9723 | 16.0237 | 0 |
| 90 | 91.5 | 85.433 | 94.049 | 92.37 | 92.309 | 92.702 | 88.492 | 88.789 | 91.067 | 88.388 | 92.56 | 82.84 | 90.0416 | 3.3443 | 0 |
| 91 | 90.504 | 87.804 | 90.594 | 89.278 | 78.317 | 92.393 | 90.6 | 88.755 | 83.99 | 88.342 | 91.504 | 94.089 | 88.8475 | 4.1778 | 0 |
| 92 | 67.194 | 68.892 | 67.007 | 66.611 | 69.631 | 66.738 | 60.233 | 70.033 | 70.093 | 71.199 | 68.441 | 70.165 | 68.0198 | 2.9008 | 0 |
| 93 | 90.788 | 90.639 | 89.855 | 85.686 | 86.222 | 80.245 | 89.375 | 90.97 | 92.566 | 81.217 | 87.592 | 89.851 | 87.9172 | 3.9070 | 0 |
| 94 | 80.011 | 82.765 | 88.963 | 68.052 | 86.633 | 90.985 | 88.391 | 90.126 | 84.355 | 86.009 | 83.367 | 82.663 | 84.3600 | 6.1329 | 0 |
| 95 | 91.69 | 90.485 | 86.731 | 90.092 | 86.15 | 88.995 | 88.582 | 87.511 | 89.64 | 85.336 | 90.437 | 87.673 | 88.6102 | 1.9571 | 0 |
| 96 | 91.093 | 87.948 | 87.97 | 86.587 | 86.72 | 91.421 | 88.377 | 82.051 | 89.8 | 90.662 | 87.123 | 82.069 | 87.6518 | 3.0934 | 0 |
| 97 | 90.606 | 90.681 | 88.805 | 87.307 | 87.498 | 89.179 | 90.296 | 86.458 | 91.121 | 86.784 | 90.964 | 88.026 | 88.9771 | 1.7307 | 0 |
| 98 | 94 | 91.679 | 92.075 | 93.691 | 94.632 | 94.887 | 95.136 | 95.167 | 91.057 | 94.55 | 91.089 | 89.802 | 93.1471 | 1.8916 | 0 |
| 99 | 91.776 | 82.421 | 91.302 | 92.79 | 85.609 | 89.863 | 93.84 | 91.335 | 92.681 | 88.387 | 91.738 | 83.874 | 89.6347 | 3.7494 | 0 |
| 100 | 89.541 | 88.772 | 84.213 | 83.84 | 91.193 | 79.568 | 90.192 | 81.882 | 91.4 | 88.031 | 87.693 | 88.583 | 87.0757 | 3.8125 | 0 |
| 101 | 77.062 | 89.259 | 79.043 | 82.929 | 72.762 | 90.428 | 92.823 | 91.701 | 86.186 | 86.798 | 83.287 | 93.135 | 85.4511 | 6.5890 | 0 |
| 102 | 86.089 | 91.467 | 88.805 | 93.336 | 92.064 | 91.063 | 84.771 | 94 | 92.294 | 84.438 | 85.396 | 93.689 | 89.7843 | 3.6827 | 0 |
| 103 | 88.524 | 88.847 | 89.791 | 91.255 | 88.205 | 93.334 | 85.55 | 88.258 | 92.609 | 86.203 | 87.369 | 90.449 | 89.1995 | 2.3924 | 0 |
| 104 | 89.716 | 91.278 | 79.809 | 89.919 | 92.476 | 89.655 | 81.239 | 79.77 | 89.268 | 89.719 | 88.796 | 80.022 | 86.8056 | 4.9764 | 0 |
| 105 | 88.331 | 83.325 | 78.142 | 78.934 | 85.337 | 80.204 | 81.11 | 78.805 | 81.487 | 77.924 | 81.592 | 81.261 | 81.3710 | 3.0796 | 0 |
| 106 | 87.163 | 84.12 | 75.731 | 79.655 | 79.794 | 80.495 | 82.707 | 81.325 | 72.186 | 87.535 | 84.043 | 85.16 | 81.6595 | 4.5102 | 0 |
| 107 | 88.642 | 83.334 | 87.757 | 87.066 | 87.972 | 87.134 | 85.785 | 88.546 | 90.085 | 90.028 | 83.63 | 81.008 | 86.7489 | 2.8080 | 0 |
| 108 | 87.712 | 82.492 | 69.512 | 84.369 | 91.456 | 83.767 | 77.851 | 84.314 | 91.851 | 81.701 | 85.756 | 67.258 | 82.3366 | 7.6083 | 0 |
| 109 | 90.647 | 83.321 | 90.263 | 87.546 | 85.036 | 78.788 | 85.531 | 85.761 | 89.589 | 83.351 | 84.42 | 85.387 | 85.8033 | 3.3691 | 0 |
| 110 | 84.319 | 89.203 | 93.094 | 85.494 | 88.442 | 89.324 | 87.179 | 79.09 | 89.556 | 89.445 | 88.496 | 90.63 | 87.8560 | 3.5759 | 0 |
| 111 | 91.087 | 90.482 | 80.922 | 89.448 | 76.108 | 88.022 | 87.408 | 88.81 | 88.507 | 91.341 | 86.689 | 82.761 | 86.7988 | 4.6081 | 0 |
| 112 | 91.495 | 77.402 | 87.404 | 83.161 | 81.865 | 86.71 | 86.888 | 91.689 | 85.428 | 77.811 | 90.679 | 88.678 | 85.7675 | 4.8684 | 0 |
| 113 | 80.498 | 84.171 | 90.334 | 88.42 | 82.078 | 86.607 | 80.168 | 85.593 | 88.835 | 85.121 | 81.72 | 82.449 | 84.6662 | 3.3922 | 0 |
| 114 | 82.742 | 85.225 | 87.014 | 85.864 | 81.065 | 87.711 | 86.061 | 81.341 | 86.947 | 79.571 | 72.731 | 87.773 | 83.6704 | 4.4474 | 0 |
| 115 | 86.763 | 84.964 | 86.842 | 89.987 | 85.497 | 87.008 | 88.041 | 81.032 | 87.859 | 90.225 | 84.076 | 89.518 | 86.8177 | 2.6584 | 0 |
| 116 | 86.281 | 89.211 | 88.478 | 90.925 | 87.909 | 87.398 | 92.73 | 85.636 | 91.563 | 88.338 | 89.076 | 81.942 | 88.2906 | 2.8792 | 0 |
| 117 | 87.848 | 89.719 | 89.84 | 82.972 | 89.208 | 85.251 | 88.295 | 86.57 | 88.461 | 88.762 | 89.744 | 90.936 | 88.1338 | 2.2367 | 0 |
| 118 | 86.333 | 85.197 | 92.885 | 82.041 | 85.072 | 85.834 | 80.561 | 88.884 | 79.928 | 85.734 | 88.102 | 75.681 | 84.6877 | 4.5706 | 0 |
| 119 | 91.692 | 93.265 | 91.524 | 82.663 | 87.259 | 92.274 | 88.065 | 93.286 | 87.925 | 91.661 | 86.283 | 93.611 | 89.9590 | 3.4484 | 0 |
| 120 | 88.041 | 89.354 | 84.589 | 83.811 | 83.737 | 87.446 | 91.731 | 88.287 | 90.099 | 82.047 | 90.039 | 82.916 | 86.8414 | 3.2647 | 0 |
| 121 | 87.702 | 86.76 | 78.526 | 86.677 | 87.107 | 81.157 | 89.514 | 92.946 | 89.151 | 90.314 | 87.417 | 82.595 | 86.6555 | 4.0682 | 0 |
| 122 | 85.156 | 90.985 | 89.187 | 85.158 | 85.644 | 93.37 | 85.893 | 91.579 | 81.373 | 87.94 | 94.077 | 92.072 | 88.5362 | 3.9555 | 0 |
| 123 | 85.283 | 79.772 | 84.998 | 88.966 | 89.763 | 88.116 | 81.41 | 82.939 | 87.808 | 66.331 | 88.54 | 83.729 | 83.9713 | 6.3998 | 0 |
| 124 | 84.481 | 85.465 | 89.603 | 78.502 | 88.251 | 83.747 | 84.767 | 86.964 | 91.416 | 87.151 | 88.117 | 83.199 | 85.9719 | 3.4027 | 0 |
| 125 | 79.183 | 82.62 | 70.349 | 76.307 | 75.709 | 80.434 | 96.223 | 74.509 | 74.104 | 68.222 | 79.83 | 79.15 | 78.0533 | 7.1032 | 0 |
| 126 | 81.854 | 87.746 | 87.848 | 89.086 | 90.812 | 86.341 | 90.004 | 87.098 | 81.217 | 82.243 | 83.002 | 83.867 | 85.9265 | 3.3595 | 0 |
| 127 | 90.273 | 80.364 | 85.377 | 94.042 | 84.158 | 88.779 | 87.082 | 89.373 | 81.552 | 92.051 | 92.784 | 87.726 | 87.7968 | 4.3213 | 0 |
| 128 | 89.06 | 86.76 | 83.047 | 77.637 | 90.711 | 89.338 | 88.529 | 89.456 | 90.427 | 84.846 | 90.027 | 80.238 | 86.6730 | 4.3190 | 0 |
| 129 | 90.09 | 93.125 | 89.463 | 88.114 | 90.196 | 90.59 | 90.399 | 86.836 | 89.88 | 81.199 | 92.772 | 91.722 | 89.5322 | 3.1575 | 0 |
| 130 | 86.51 | 91.737 | 87.755 | 88.236 | 86.509 | 89.771 | 92.13 | 84.195 | 90.435 | 94.045 | 90.2 | 88.972 | 89.2079 | 2.7625 | 0 |
| 131 | 93.256 | 92.385 | 82.157 | 92.249 | 91.667 | 88.715 | 89.135 | 85.407 | 90.925 | 70.171 | 87.736 | 89.122 | 87.7438 | 6.3817 | 0 |
| 132 | 88.079 | 90.508 | 91.731 | 91.629 | 91.037 | 79.634 | 83.893 | 87.928 | 86.817 | 91.084 | 86.371 | 82.65 | 87.6134 | 3.9297 | 0 |
| 133 | 82.878 | 86.37 | 83.046 | 72.047 | 83.639 | 83.323 | 79.988 | 84.327 | 90.97 | 83.718 | 90.552 | 84.572 | 83.7858 | 4.8507 | 0 |
| 134 | 81.913 | 77.124 | 90.966 | 88.938 | 90.529 | 81.577 | 91.474 | 86.205 | 89.593 | 90.521 | 91.072 | 90.699 | 87.5509 | 4.7783 | 0 |
| 135 | 90.747 | 89.491 | 71.544 | 88.919 | 85.551 | 90.401 | 85.454 | 86.058 | 85.546 | 78.994 | 88.965 | 79.725 | 85.1163 | 5.7222 | 0 |
| 136 | 90.065 | 91.874 | 89.176 | 90.506 | 87.037 | 91.084 | 87.126 | 88.285 | 85.763 | 91.684 | 86.758 | 89.716 | 89.0895 | 2.0679 | 0 |
| 137 | 87.084 | 85.551 | 86.395 | 88.638 | 86.012 | 78.971 | 78.535 | 55.602 | 85.478 | 77.316 | 91.159 | 72.931 | 81.1393 | 9.6637 | 0 |
| 138 | 90.478 | 87.246 | 89.581 | 85.05 | 89.453 | 91.685 | 88.829 | 93.225 | 80.584 | 90.482 | 87.034 | 87.175 | 88.4018 | 3.3361 | 0 |
| 139 | 79.473 | 77.146 | 78.348 | 58.945 | 78.108 | 81.702 | 88.916 | 85.977 | 82.73 | 82.723 | 83.237 | 68.113 | 78.7848 | 8.1189 | 0 |
| 140 | 87.516 | 83.981 | 73.675 | 83.076 | 87.289 | 91.525 | 73.57 | 87.15 | 91.11 | 77.889 | 77.733 | 66.297 | 81.7343 | 7.8909 | 0 |
| 141 | 87.4 | 80.203 | 83.441 | 79.29 | 67.074 | 82.6 | 85.567 | 85 | 74.647 | 82.804 | 75.457 | 78.674 | 80.1798 | 5.6849 | 0 |
| 142 | 89.423 | 82.622 | 74.919 | 83.425 | 85.839 | 77.585 | 88.972 | 82.608 | 90.2 | 82.404 | 90.581 | 89.71 | 84.8573 | 5.1644 | 0 |
| 143 | 86.773 | 81.921 | 86.587 | 82.374 | 82.268 | 83.32 | 88.488 | 78.425 | 82.364 | 85.277 | 84.286 | 52.238 | 81.1934 | 9.5094 | 0 |
| 144 | 90.037 | 73.923 | 87.477 | 90.021 | 89.378 | 85.892 | 90.826 | 90.626 | 90.805 | 91.994 | 91.006 | 92.094 | 88.6733 | 4.9733 | 0 |
| 145 | 88.423 | 80.868 | 87.568 | 84.279 | 88.204 | 76.897 | 87.809 | 79.822 | 87.72 | 83.592 | 86.473 | 84.971 | 84.7188 | 3.7795 | 0 |
| 146 | 84.692 | 86.417 | 87.449 | 92.093 | 89.569 | 91.348 | 92.43 | 92.584 | 88.972 | 90.575 | 90.354 | 89.374 | 89.6548 | 2.4672 | 0 |
| 147 | 91.588 | 93.231 | 89.581 | 90.802 | 91.241 | 90.535 | 88.868 | 92.658 | 91.963 | 86.687 | 89.838 | 85.819 | 90.2343 | 2.2488 | 0 |
| 148 | 84.597 | 84.085 | 91.444 | 81.085 | 87.114 | 81.817 | 85.28 | 86.554 | 71.116 | 88.061 | 87.961 | 91.469 | 85.0486 | 5.4578 | 0 |
| 149 | 88.691 | 92.356 | 91.043 | 89.54 | 91.094 | 80.067 | 88.325 | 92.604 | 86.257 | 91.176 | 91.533 | 88.183 | 89.2391 | 3.4655 | 0 |
| 150 | 74.975 | 82.435 | 77.086 | 78.703 | 77.414 | 78.036 | 85.738 | 87.729 | 85.731 | 85.284 | 84.772 | 78.728 | 81.3859 | 4.3345 | 0 |
| 151 | 92.393 | 86.313 | 87.74 | 89.742 | 88.22 | 90.528 | 89.48 | 85.133 | 90.27 | 90.822 | 85.843 | 86.073 | 88.5464 | 2.3354 | 0 |
| 152 | 87.585 | 91.33 | 91.314 | 94.782 | 92.662 | 88.594 | 91.545 | 88.978 | 89.175 | 88.112 | 90.201 | 94.666 | 90.7453 | 2.4136 | 0 |
| 153 | 90.478 | 90.232 | 93.285 | 85.997 | 86.701 | 87.438 | 94.137 | 88.994 | 90.54 | 91.904 | 93.407 | 77.228 | 89.1951 | 4.6078 | 0 |
| 154 | 87.955 | 86.134 | 88.019 | 85.598 | 75.516 | 87.204 | 86.758 | 85.958 | 89.498 | 87.752 | 77.335 | 86.711 | 85.3698 | 4.3290 | 0 |
| 155 | 90.527 | 84.996 | 92.551 | 90.141 | 89.782 | 90.496 | 86.324 | 88.361 | 91.614 | 91.615 | 89.921 | 85.352 | 89.3067 | 2.5117 | 0 |
| 156 | 92.22 | 93.949 | 86.304 | 91.282 | 88.582 | 92.338 | 82.466 | 89.918 | 91.277 | 84.731 | 92.911 | 91.299 | 89.7731 | 3.5549 | 0 |
| 157 | 91.075 | 80.832 | 88.005 | 84.295 | 86.747 | 88.547 | 88.675 | 89.056 | 77.607 | 87.978 | 89.948 | 88.418 | 86.7653 | 3.9507 | 0 |
| 158 | 81.477 | 92.497 | 88.762 | 93.3 | 94.817 | 91.626 | 92.91 | 93.707 | 85.677 | 94.982 | 94.842 | 95.099 | 91.6413 | 4.2638 | 0 |
| 159 | 82.159 | 80.538 | 82.339 | 91.714 | 82.93 | 92.063 | 91.853 | 84.781 | 88.976 | 93.993 | 83.908 | 84.612 | 86.6555 | 4.7334 | 0 |
| 160 | 88.535 | 93.042 | 91.038 | 91.188 | 91.577 | 92.932 | 93.981 | 87.019 | 92.347 | 86.576 | 90.668 | 93.453 | 91.0297 | 2.4603 | 0 |
| 161 | 86.591 | 87.461 | 90.835 | 86.807 | 91.707 | 85.035 | 78.389 | 20.199 | 85.956 | 90.131 | 91.972 | 92.056 | 82.2616 | 19.9304 | 0 |
| 162 | 91.152 | 89.055 | 90.051 | 91.387 | 94.403 | 92.691 | 89.775 | 88.876 | 91.994 | 91.723 | 72.995 | 89.68 | 89.4818 | 5.4363 | 0 |
| 163 | 82.908 | 93.87 | 90.072 | 82.89 | 88.959 | 87.515 | 91.082 | 87.646 | 85.119 | 86.807 | 89.787 | 73.086 | 86.6451 | 5.3503 | 0 |
| 164 | 84.904 | 83.636 | 78.69 | 75.618 | 82.392 | 90.18 | 74.254 | 81.718 | 85.254 | 82.303 | 84.522 | 79.109 | 81.8817 | 4.4228 | 0 |
| 165 | 90.269 | 83.697 | 79.405 | 82.541 | 86.172 | 86.94 | 87.612 | 53.581 | 91.984 | 82.107 | 63.168 | 63.226 | 79.2252 | 12.3348 | 0 |
| 166 | 76.107 | 84.188 | 84.063 | 88.707 | 88.04 | 82.719 | 74.795 | 90.477 | 81.59 | 87.325 | 88.774 | 81.159 | 83.9953 | 5.0169 | 0 |
| 167 | 88.069 | 90.686 | 88.909 | 88.463 | 91.321 | 90.193 | 88.917 | 91.526 | 89.102 | 90.133 | 89.35 | 89.721 | 89.6992 | 1.0985 | 0 |
| 168 | 87.679 | 88.043 | 87.861 | 86.913 | 76.726 | 86.548 | 88.759 | 74.578 | 82.1 | 91.361 | 71.201 | 76.579 | 83.1957 | 6.6947 | 0 |
| 169 | 92.463 | 86.174 | 89.916 | 81.522 | 87.004 | 89.787 | 86.084 | 90.829 | 89.744 | 92.209 | 83.892 | 90.091 | 88.3096 | 3.3791 | 0 |
| 170 | 91.894 | 92.684 | 82.832 | 87.697 | 89.656 | 83.792 | 80.974 | 81.565 | 78.654 | 81.478 | 85.57 | 89.631 | 85.5356 | 4.6778 | 0 |
| 171 | 88.016 | 85.206 | 82.668 | 91.682 | 88.468 | 87.042 | 85.436 | 77.089 | 85.314 | 86.533 | 75.85 | 82.392 | 84.6413 | 4.5689 | 0 |
| 172 | 90.665 | 93.005 | 84.074 | 89.921 | 85.06 | 87.613 | 91.607 | 84.84 | 90.873 | 88.878 | 92.529 | 87.226 | 88.8576 | 3.0801 | 0 |
| 173 | 89.358 | 88.036 | 90.217 | 88.141 | 91.687 | 90.886 | 91.326 | 89.243 | 92.476 | 91.334 | 89.878 | 86.928 | 89.9592 | 1.6808 | 0 |
| 174 | 87.775 | 86.126 | 83.916 | 87.097 | 89.482 | 87.343 | 87.231 | 83.587 | 85.229 | 89.081 | 81.518 | 88.353 | 86.3948 | 2.4087 | 0 |
| 175 | 91.349 | 91.608 | 81.674 | 83.487 | 91.853 | 90.911 | 93.066 | 87.259 | 89.809 | 91.545 | 88.498 | 88.455 | 89.1262 | 3.5062 | 0 |
| 176 | 91.629 | 88.506 | 91.505 | 89.302 | 91.684 | 91.422 | 88.639 | 88.057 | 87.094 | 86.781 | 92.345 | 91.711 | 89.8896 | 2.0273 | 0 |
| 177 | 90.106 | 83.625 | 90.669 | 90.483 | 86.99 | 85.601 | 91.048 | 85.065 | 87.667 | 90.168 | 90.337 | 87.943 | 88.3085 | 2.5348 | 0 |
| 178 | 90.294 | 88.189 | 87.709 | 88.879 | 89.871 | 89.157 | 89.93 | 86.796 | 84.992 | 88.614 | 84.455 | 50.011 | 84.9081 | 11.1462 | 0 |
| 179 | 73.559 | 93.311 | 85.749 | 91 | 89.946 | 83.952 | 87.459 | 91.434 | 90.532 | 92.403 | 70.606 | 88.364 | 86.5263 | 7.2957 | 0 |
| 180 | 87.757 | 89.386 | 83.204 | 87.953 | 81.263 | 80.194 | 89.617 | 83.071 | 89.467 | 80.311 | 81.856 | 88.17 | 85.1874 | 3.8416 | 0 |
| 181 | 83.537 | 86.502 | 85.505 | 89.396 | 88.873 | 87.04 | 88.387 | 87.222 | 89.492 | 90.245 | 91.486 | 92.586 | 88.3559 | 2.5532 | 0 |
| 182 | 86.127 | 74.345 | 83.692 | 92.78 | 83.221 | 71.329 | 78.228 | 85.838 | 72.925 | 87.372 | 83.796 | 89.829 | 82.4568 | 6.8241 | 0 |
| 183 | 91.772 | 97.232 | 88.316 | 91.076 | 87.124 | 88.314 | 27.039 | 88.652 | 88.471 | 84.777 | 88.127 | 51.383 | 81.0236 | 20.4360 | 0 |
| 184 | 89.066 | 83.24 | 86.592 | 91.33 | 94.902 | 93.885 | 80.728 | 83.017 | 94.303 | 84.705 | 92.087 | 85.567 | 88.2852 | 4.9459 | 0 |
| 185 | 88.162 | 89.858 | 90.244 | 84.027 | 88.524 | 86.096 | 89.578 | 87.082 | 92.025 | 84.431 | 86.81 | 87.487 | 87.8603 | 2.3725 | 0 |
| 186 | 88.612 | 92.872 | 89.054 | 89.971 | 89.097 | 92.005 | 91.975 | 84.996 | 89.631 | 90.322 | 85.945 | 91.038 | 89.6265 | 2.3536 | 0 |
| 187 | 84.112 | 88.39 | 88.243 | 93.424 | 90.239 | 88.683 | 91.67 | 88.479 | 91.711 | 88.555 | 85.929 | 90.93 | 89.1971 | 2.5881 | 0 |
| 188 | 91.588 | 88.233 | 93.865 | 91.943 | 91.117 | 92.901 | 88.06 | 91.348 | 89.045 | 87.247 | 92.5 | 86.077 | 90.3270 | 2.4945 | 0 |
| 189 | 89.591 | 84.978 | 80.498 | 91.673 | 90.252 | 90.798 | 90.249 | 92.899 | 86.553 | 91.833 | 91.821 | 94.147 | 89.6077 | 3.8323 | 0 |
| 190 | 92.77 | 84.417 | 87.539 | 83.507 | 74.028 | 93.482 | 82.195 | 83.837 | 90.572 | 83.53 | 90.955 | 88.587 | 86.2849 | 5.4993 | 0 |
| 191 | 83.592 | 90.927 | 90.032 | 80.079 | 90.572 | 86.241 | 85.011 | 90.544 | 90.762 | 87.287 | 92.22 | 86.992 | 87.8549 | 3.6518 | 0 |
| 192 | 70.341 | 72.487 | 88.951 | 77.582 | 69.88 | 87.284 | 77.362 | 82.891 | 85.735 | 82.029 | 86.506 | 78.834 | 79.9902 | 6.6551 | 0 |
| 193 | 89.432 | 90.383 | 90.507 | 92.649 | 86.704 | 91.036 | 86.289 | 92.487 | 90.996 | 88.558 | 90.411 | 86.377 | 89.6524 | 2.2270 | 0 |
| 194 | 87.525 | 81.259 | 90.133 | 91.256 | 91.599 | 85.945 | 81.125 | 88.685 | 90.045 | 88.956 | 92.197 | 90.884 | 88.3008 | 3.7619 | 0 |
| 195 | 82.052 | 84.315 | 70.496 | 82.969 | 81.198 | 79.048 | 80.577 | 80.89 | 75.219 | 75.424 | 86.679 | 76.895 | 79.6468 | 4.4923 | 0 |
| 196 | 82.88 | 84.151 | 85.132 | 83.009 | 84.364 | 85.038 | 84.648 | 82.83 | 82.791 | 85.793 | 76.02 | 85.15 | 83.4838 | 2.5765 | 0 |
| 197 | 88.823 | 91.417 | 72.48 | 90.813 | 90.546 | 89.342 | 89.725 | 83.452 | 81.094 | 82.554 | 82.739 | 89.148 | 86.0111 | 5.6565 | 0 |
| 198 | 88.445 | 85.106 | 91.736 | 84.626 | 87.357 | 92.986 | 90.482 | 90.464 | 89.76 | 93.714 | 87.707 | 88.002 | 89.1988 | 2.8605 | 0 |
| 199 | 80.682 | 88.922 | 91.344 | 94.608 | 93.144 | 93.672 | 92.622 | 92.912 | 93.05 | 92.75 | 92.649 | 88.245 | 91.2167 | 3.8013 | 0 |
| 200 | 86.054 | 92.727 | 85.939 | 91.545 | 91.537 | 85.852 | 88.281 | 89.628 | 84.502 | 88.202 | 87.619 | 88.262 | 88.3457 | 2.5912 | 0 |
| 201 | 82.759 | 86.206 | 83.575 | 86.813 | 77.663 | 78.963 | 89.034 | 74.767 | 76.966 | 82.301 | 81.69 | 87.294 | 82.3359 | 4.5321 | 0 |
| 202 | 92.684 | 76.379 | 91.819 | 78.428 | 90.585 | 67.721 | 90.471 | 70.81 | 89.409 | 77.434 | 73.657 | 76.521 | 81.3265 | 9.0423 | 0 |
| 203 | 81.863 | 91.869 | 76.4 | 85.642 | 93.341 | 88.449 | 76.859 | 87.258 | 89.89 | 84.177 | 88.86 | 80.721 | 85.4441 | 5.5569 | 0 |
| 204 | 91.123 | 92.323 | 82.666 | 85.58 | 81.165 | 88.563 | 86.747 | 88.89 | 92.911 | 83.027 | 89.794 | 89.097 | 87.6572 | 3.8613 | 0 |
| 205 | 88.771 | 91.28 | 93.726 | 92.442 | 91.755 | 84.303 | 81.428 | 93.006 | 92.783 | 87.592 | 77.386 | 85.943 | 88.3679 | 5.2066 | 0 |
| 206 | 93.484 | 92.287 | 90.922 | 92.807 | 94.091 | 93.61 | 91.321 | 91.249 | 93.617 | 90.971 | 92.137 | 91.615 | 92.3426 | 1.1489 | 0 |
| 207 | 86.154 | 82.795 | 96.199 | 93.43 | 87.632 | 90.516 | 89.765 | 88.321 | 86.427 | 85.806 | 91.57 | 91.045 | 89.1383 | 3.7007 | 0 |
| 208 | 87.35 | 85.47 | 79.036 | 91.444 | 83.566 | 92.107 | 91.428 | 91.189 | 89.194 | 91.125 | 92.382 | 91.604 | 88.8246 | 4.1856 | 0 |
| | | | | | | | | | | | | Overall | 86.8066 | 6.2876 | 5 |

### 4. Results of Testing SeetaFace on Real Images

| ID | Similarity Scores | | | | | | | | | | | | Average | StDev | Not Recognise |
|---|---|---|---|---|---|---|---|---|---|---|---|---|---|---|---|
| Aaron Eckhart | 0.532406 | 0.648035 | 0.669106 | 0.618295 | 0.700188 | 0.491662 | 0.547169 | 0.600206 | 0.609301 | 0.750952 | 0.719643 | 0.746034 | 0.6360831 | 0.0849824 | 0 |
| Abhishek Bachan | 0.789336 | 0.786134 | 0.808042 | 0.828407 | 0.827694 | 0.772499 | 0.761688 | 0.748213 | 0.706508 | 0.774594 | 0.790811 | 0.387369 | 0.7484413 | 0.1185848 | 0 |
| Adam Sandler | 0.701344 | 0.708784 | 0.740015 | 0.585824 | 0.694639 | 0.663177 | 0.684622 | 0.593249 | 0.696064 | 0.693087 | 0.771317 | 0.728852 | 0.6884145 | 0.0540239 | 0 |
| Adriana Lima | 0.578497 | 0.588112 | 0.575688 | 0.639735 | 0.619498 | 0.634913 | 0.628891 | 0.833481 | 0.637165 | 0.691925 | 0.563412 | 0.64169 | 0.6360839 | 0.0719876 | 0 |
| Alberto Gonzales | 0.616219 | 0.593046 | 0.647809 | 0.783136 | 0.635273 | 0.795002 | 0.72999 | 0.732845 | 0.692926 | 0.841334 | 0.755747 | 0.707496 | 0.7109019 | 0.0769002 | 0 |
| Alec Baldwin | 0.619433 | 0.719659 | 0.701677 | 0.684722 | 0.631634 | 0.722224 | 0.506835 | 0.682581 | 0.568556 | 0.709643 | 0.558805 | 0.72123 | 0.6522499 | 0.0740357 | 0 |
| Alex Rodriguez | 0.695637 | 0.764443 | 0.703712 | 0.777441 | 0.733979 | 0.728451 | 0.80332 | 0.729007 | 0.688233 | 0.803609 | 0.667814 | 0.608624 | 0.7253558 | 0.057372 | 0 |
| Ali Landry | 0.612121 | 0.647731 | 0.686598 | 0.739082 | 0.672826 | 0.767792 | 0.771226 | 0.657689 | 0.602738 | 0.571091 | 0.62416 | 0.741231 | 0.6745238 | 0.0674874 | 0 |
| Alicia Keys | 0.647743 | 0.687416 | 0.680896 | 0.638385 | 0.700775 | 0.529495 | 0.651189 | 0.700379 | 0.753469 | 0.705755 | 0.759368 | 0.710386 | 0.680438 | 0.0606316 | 0 |
| Alyssa Milano | 0.572298 | 0.597023 | 0.697666 | 0.697535 | 0.729361 | 0.443269 | 0.565479 | 0.7086 | 0.747936 | 0.718596 | 0.84686 | 0.677009 | 0.6668027 | 0.1059081 | 0 |
| Amanda Seyfried | 0.733666 | 0.669455 | 0.601344 | 0.80751 | 0.8197 | 0.871751 | 0.703117 | 0.801348 | 0.714318 | 0.737737 | 0.847169 | 0.758516 | 0.7554693 | 0.0781928 | 0 |
| Anderson Cooper | 0.523503 | 0.7275 | 0.827731 | 0.580292 | 0.759875 | 0.734607 | 0.712131 | 0.860764 | 0.862348 | 0.704486 | 0.638673 | 0.797678 | 0.7274657 | 0.1061535 | 0 |
| Angela Merkel | 0.901359 | 0.660717 | 0.440101 | 0.556594 | 0.685894 | 0.541804 | 0.645903 | 0.670848 | 0.568711 | 0.663665 | 0.482947 | 0.727765 | 0.628859 | 0.1224411 | 0 |
| Angelina Jolie | 0.671483 | 0.683978 | 0.718472 | 0.665715 | 0.548922 | 0.719704 | 0.668152 | 0.625714 | 0.67006 | 0.731268 | 0.617241 | 0.677824 | 0.6665444 | 0.0505566 | 0 |
| Anna Kournikova | 0.362814 | 0.578441 | 0.466773 | 0.56404 | 0.647869 | 0.556972 | 0.625052 | 0.616978 | 0.557644 | 0.620329 | 0.715088 | 0.351177 | 0.5552648 | 0.1104428 | 0 |
| Anna Paquin | 0.524824 | 0.621655 | 0.689142 | 0.764286 | 0.648559 | 0.560216 | 0.731323 | 0.554265 | 0.660587 | 0.648596 | 0.606563 | 0.658977 | 0.6390828 | 0.071069 | 0 |
| Antonio Banderas | 0.680054 | 0.75573 | 0.515128 | 0.658857 | 0.742065 | 0.693827 | 0.73157 | 0.701773 | 0.688301 | 0.718161 | 0.694352 | 0.660231 | 0.6866708 | 0.0618867 | 0 |
| Ashley Judd | 0.681573 | 0.670064 | 0.604572 | 0.699328 | 0.554796 | 0.579366 | 0.616254 | 0.654575 | 0.816495 | 0.635092 | 0.662195 | 0.706202 | 0.6567093 | 0.0686355 | 0 |
| Ashton Kutcher | 0.684031 | 0.508004 | 0.696622 | 0.575058 | 0.583177 | 0.479114 | 0.803684 | 0.588837 | 0.697452 | 0.768372 | 0.773482 | 0.805767 | 0.6636333 | 0.1142892 | 0 |
| Audrey Tautou | 0.697565 | 0.766468 | 0.749542 | 0.675741 | 0.573084 | 0.832925 | 0.61046 | 0.371056 | 0.739556 | 0.788473 | 0.742169 | 0.754419 | 0.6917882 | 0.124604 | 0 |
| Avril Lavigne | 0.543421 | 0.551439 | 0.48664 | 0.657299 | 0.635147 | 0.475304 | 0.562945 | 0.535489 | 0.545479 | 0.982067 | 0.409728 | 0.408379 | 0.5661114 | 0.1513877 | 0 |
| Barack Obama | 0.792371 | 0.52182 | 0.873315 | 0.698349 | 0.78072 | 0.718611 | 0.676622 | 0.746677 | 0.645017 | 0.630511 | 0.672912 | 0.709935 | 0.7055717 | 0.0895547 | 0 |
| Ben Affleck | 0.572798 | 0.416414 | 0.628926 | 0.72156 | 0.679035 | 0.454605 | 0.527196 | 0.680204 | 0.616341 | 0.624093 | 0.8553 | 0.580534 | 0.6130838 | 0.1181397 | 0 |
| Ben Stiller | 0.795642 | 0.601117 | 0.609283 | 0.654244 | 0.696729 | 0.576764 | 0.492079 | 0.847722 | 0.631986 | 0.715812 | 0.47753 | 0.590058 | 0.6407472 | 0.110299 | 0 |
| Beyonce Knowles | 0.714592 | 0.54537 | 0.705473 | 0.707342 | 0.715655 | 0.757266 | 0.599421 | 0.626381 | 0.806879 | 0.635673 | 0.605528 | 0.731472 | 0.6792543 | 0.076026 | 0 |
| Bill Clinton | 0.626488 | 0.728181 | 0.675093 | 0.710162 | 0.625542 | 0.70181 | 0.759131 | 0.698518 | 0.728727 | 0.774898 | 0.612233 | 0.681509 | 0.6935243 | 0.0521749 | 0 |
| Billy Crystal | 0.254939 | 0.61455 | 0.572859 | 0.694339 | 0.64412 | 0.57977 | 0.274555 | 0.595535 | 0.582137 | 0.539856 | 0.491679 | 0.685826 | 0.5441804 | 0.1423349 | 0 |
| Bob Dole | 0.63301 | 0.522013 | 0.648288 | 0.951458 | 0.897072 | 0.751288 | 0.617532 | 0.739605 | 0.658951 | 0.646842 | 0.770101 | 0.55412 | 0.69919 | 0.1288437 | 0 |
| Brad Pitt | 0.518399 | 0.605062 | 0.708685 | 0.547424 | 0.429634 | | 0.570066 | 0.520923 | 0.647506 | 0.473479 | 0.580262 | 0.424221 | 0.5477874 | 0.0877492 | 1 |
| Brendan Fraser | 0.617913 | 0.690618 | 0.690015 | 0.785119 | 0.772533 | 0.861638 | 0.522424 | 0.739379 | 0.763184 | 0.453593 | 0.549491 | 0.765968 | 0.6843229 | 0.1233646 | 0 |
| Bruce Willis | 0.615373 | 0.505309 | 0.512262 | 0.707109 | 0.836429 | 0.517069 | 0.515179 | 0.655077 | 0.610122 | 0.561874 | 0.579779 | 0.580232 | 0.5996512 | 0.0971089 | 0 |
| Cameron Diaz | 0.802229 | 0.61127 | 0.487008 | 0.8578 | 0.631108 | 0.747556 | 0.709584 | 0.651558 | 0.572733 | 0.767274 | 0.564138 | 0.759742 | 0.6801667 | 0.111219 | 0 |
| Carla Gugino | 0.602884 | 0.718747 | 0.657913 | 0.560064 | 0.559315 | 0.597865 | 0.752748 | 0.25784 | 0.707346 | 0.602273 | 0.487736 | 0.646599 | 0.5959442 | 0.1304638 | 0 |
| Carson Daly | 0.860282 | 0.689184 | 0.638616 | 0.835791 | 0.760462 | 0.80007 | 0.720861 | 0.702846 | 0.739952 | 0.691262 | 0.657897 | 0.619505 | 0.726394 | 0.0760934 | 0 |
| Cate Blanchett | 0.553339 | 0.641276 | 0.537207 | 0.696961 | 0.673441 | 0.666216 | 0.392532 | 0.564086 | 0.429043 | 0.634829 | 0.617567 | 0.671226 | 0.5898103 | 0.0981239 | 0 |
| Celine Dion | 0.657025 | 0.389854 | 0.602419 | 0.639146 | 0.362881 | 0.679378 | 0.519276 | 0.619975 | 0.432277 | 0.422645 | 0.454353 | 0.542855 | 0.5268403 | 0.1120996 | 0 |
| Charlize Theron | 0.735129 | 0.522889 | 0.701621 | 0.737028 | 0.6852 | 0.627455 | 0.537095 | 0.620798 | 0.671223 | 0.732126 | 0.556718 | 0.506258 | 0.6361283 | 0.0869899 | 0 |
| Chris Martin | 0.649889 | 0.661442 | 0.714121 | 0.705968 | 0.628495 | 0.60916 | 0.61894 | 0.507999 | 0.692909 | 0.756291 | 0.501719 | 0.726376 | 0.6477758 | 0.080642 | 0 |
| Christina Ricci | 0.625495 | 0.585198 | 0.559525 | 0.668836 | 0.416431 | 0.465637 | 0.668415 | 0.61502 | 0.506417 | 0.725567 | 0.606511 | 0.690388 | 0.5944533 | 0.0935516 | 0 |
| Christopher Walken | 0.682904 | 0.699315 | 0.837142 | 0.547218 | 0.427553 | 0.529755 | 0.554598 | 0.683103 | 0.617099 | 0.171013 | 0.562506 | 0.56342 | 0.5729688 | 0.1645093 | 0 |
| Cindy Crawford | 0.676796 | 0.597844 | 0.414521 | 0.715669 | 0.657221 | 0.540047 | 0.390362 | 0.779554 | 0.55695 | 0.673981 | 0.452259 | 0.509957 | 0.5804301 | 0.124099 | 0 |
| Claudia Schiffer | 0.764926 | 0.691466 | 0.513145 | 0.744327 | 0.570509 | 0.375198 | 0.572412 | 0.523568 | 0.70769 | 0.743352 | 0.653465 | 0.606637 | 0.6222246 | 0.1171912 | 0 |
| Clive Owen | 0.663018 | 0.575161 | 0.616441 | 0.716504 | 0.76856 | 0.83727 | 0.705272 | 0.613421 | 0.706697 | 0.738749 | 0.914563 | 0.875198 | 0.7275712 | 0.1063571 | 0 |
| Colin Farrell | 0.721575 | 0.675169 | 0.68424 | 0.582003 | 0.718163 | 0.7428 | 0.624201 | 0.703111 | 0.65967 | 0.642744 | 0.794939 | 0.723958 | 0.6893811 | 0.0574533 | 0 |
| Colin Powell | 0.785686 | 0.744823 | 0.677501 | 0.774022 | 0.664384 | 0.665375 | 0.718008 | 0.750903 | 0.690877 | 0.726367 | 0.785585 | 0.750404 | 0.7278279 | 0.044793 | 0 |
| Cristiano Ronaldo | 0.650657 | 0.589294 | 0.521878 | 0.726609 | 0.7383 | 0.732815 | 0.413405 | 0.535215 | 0.625581 | 0.555704 | 0.728844 | 0.475209 | 0.6077926 | 0.1106413 | 0 |
| Daisy Fuentes | 0.589641 | 0.54338 | 0.629284 | 0.769033 | 0.537709 | 0.585729 | 0.630386 | 0.807822 | 0.651348 | 0.534803 | 0.589698 | 0.707923 | 0.6313963 | 0.0893166 | 0 |
| Daniel Craig | 0.750171 | 0.803031 | 0.670864 | 0.402857 | 0.571584 | 0.741715 | 0.587419 | 0.731086 | 0.636299 | 0.680613 | 0.485933 | 0.630858 | 0.6410358 | 0.1157838 | 0 |
| Daniel Radcliffe | 0.680498 | 0.69381 | 0.440517 | 0.742237 | 0.726986 | 0.477436 | 0.614316 | 0.561073 | 0.787866 | 0.532533 | 0.634301 | 0.63509 | 0.6272219 | 0.1075238 | 0 |
| Danny Devito | 0.582332 | 0.706748 | 0.668402 | 0.677777 | 0.705531 | 0.676779 | 0.728686 | 0.601171 | 0.53953 | 0.578511 | 0.615341 | 0.825455 | 0.6588553 | 0.0795699 | 0 |
| Dave Chappelle | 0.865275 | 0.811389 | 0.747865 | 0.785831 | 0.727084 | 0.845247 | 0.759599 | 0.765763 | 0.635065 | 0.790656 | 0.883369 | 0.830437 | 0.7872983 | 0.0678966 | 0 |
| David Beckham | 0.785887 | 0.57549 | 0.733186 | 0.509582 | 0.547062 | 0.622332 | 0.655675 | 0.699892 | 0.769502 | 0.694215 | 0.514172 | 0.709187 | 0.6513485 | 0.0966265 | 0 |
| David Duchovny | 0.43114 | 0.676536 | 0.799122 | 0.783234 | 0.569056 | 0.399136 | 0.418404 | 0.686412 | 0.647631 | 0.703307 | 0.53595 | 0.688282 | 0.6115175 | 0.139019 | 0 |
| Denise Richards | 0.684129 | 0.664808 | 0.424451 | 0.649306 | 0.691 | 0.788769 | 0.72623 | 0.664644 | 0.612319 | 0.717035 | 0.775039 | 0.499434 | 0.658097 | 0.1055944 | 0 |
| Denzel Washington | 0.778096 | 0.808123 | 0.7309 | 0.788471 | 0.762714 | 0.710912 | 0.734672 | 0.672672 | 0.386695 | 0.648475 | 0.958515 | 0.666916 | 0.7205968 | 0.1335601 | 0 |
| Diane Sawyer | 0.610238 | 0.578775 | 0.712985 | 0.670408 | 0.613271 | 0.809362 | 0.717134 | 0.625329 | 0.638344 | 0.795136 | 0.736285 | 0.780147 | 0.6906178 | 0.0790193 | 0 |
| Donald Faison | 0.750525 | 0.722252 | 0.736211 | 0.509854 | 0.787815 | 0.875202 | 0.780678 | 0.656883 | 0.770649 | 0.683358 | 0.805169 | 0.675435 | 0.7295026 | 0.092332 | 0 |
| Donald Trump | 0.557012 | 0.55298 | 0.609918 | 0.582651 | 0.642412 | 0.497788 | 0.621636 | 0.625776 | 0.755665 | 0.620199 | 0.55376 | 0.65002 | 0.6058181 | 0.0651644 | 0 |
| Drew Barrymore | 0.386922 | 0.47014 | 0.519091 | 0.604726 | 0.555879 | 0.566196 | 0.626571 | 1 | 0.616471 | 0.409994 | 0.528073 | 0.451628 | 0.5613076 | 0.1590386 | 0 |
| Dustin Hoffman | 0.570769 | 0.559782 | 0.559744 | 0.736706 | 0.611086 | 0.674642 | 0.592884 | 0.622234 | 0.665919 | 0.508949 | 0.525208 | 0.675701 | 0.6086353 | 0.0688117 | 0 |
| Dwayne Johnson | 0.811626 | 0.502434 | 0.656583 | 0.650612 | 0.492688 | 0.424565 | 0.394136 | 0.556114 | 0.655621 | 0.710479 | 0.635694 | 0.418591 | 0.5757619 | 0.1311141 | 0 |
| Edie Falco | 0.675729 | 0.592091 | 0.470376 | 0.54995 | 0.6406 | 0.543359 | 0.623266 | 0.548063 | 0.607828 | 0.572556 | 0.545989 | 0.452235 | 0.5685035 | 0.06527 | 0 |
| Edward Norton | 0.746753 | 0.636874 | 0.708431 | 0.664418 | 0.794933 | 0.701141 | 0.467599 | 0.73233 | 0.585346 | 0.648838 | 0.627552 | 0.658278 | 0.6643744 | 0.0847988 | 0 |
| Ehud Olmert | 0.715074 | 0.788498 | 0.737076 | 0.684836 | 0.689572 | 0.786895 | 0.737232 | 0.704901 | 0.786921 | 0.719701 | 0.717341 | 0.741042 | 0.7340908 | 0.036576 | 0 |
| Eliot Spitzer | 0.773265 | 0.579096 | 0.695389 | 0.720895 | 0.675069 | 0.721944 | 0.66504 | 0.562902 | 0.671572 | 0.613973 | 0.729916 | 0.70206 | 0.6759268 | 0.0630428 | 0 |
| Eliza Dushku | 0.527811 | 0.726349 | 0.693587 | 0.607343 | 0.663123 | 0.539121 | 0.707094 | 0.743809 | 0.771902 | 0.650423 | 0.652598 | 0.759543 | 0.6702253 | 0.0803192 | 0 |
| Eva Mendes | 0.623287 | 0.64761 | 0.512696 | 0.665913 | 0.693831 | 0.498739 | 0.674126 | 0.57686 | 0.59294 | 0.484362 | 0.717306 | 0.481291 | 0.5974134 | 0.0858423 | 0 |
| Faith Hill | 0.544695 | 0.589276 | 0.586719 | 0.550446 | 0.573627 | 0.612073 | 0.746976 | 0.624367 | 0.56846 | 0.633782 | 0.433266 | 0.559366 | 0.5852544 | 0.0725013 | 0 |
| Famke Janssen | 0.765481 | 0.528412 | 0.599576 | 0.660135 | 0.701803 | 0.645977 | 0.72949 | 0.625261 | 0.760415 | 0.646053 | 0.594635 | 0.653808 | 0.6592538 | 0.0704882 | 0 |
| Gael Garcia Bernal | 0.875951 | 0.564546 | 0.596823 | 0.61736 | 0.672744 | 0.696988 | 0.720517 | 0.46779 | 0.648009 | 0.705927 | 0.669416 | 0.689133 | 0.6604337 | 0.0985723 | 0 |
| Gene Hackman | 0.464287 | 0.702045 | 0.616017 | 0.517838 | 0.737541 | 0.541937 | 0.585985 | 0.439783 | 0.746346 | 0.748182 | 0.513795 | 0.709598 | 0.6102795 | 0.1150269 | 0 |
| George Clooney | 0.436958 | 0.542979 | 0.615549 | 0.68385 | 0.737034 | 0.674902 | 0.483943 | 0.755416 | 0.675172 | 0.506284 | 0.583449 | 0.711892 | 0.6172857 | 0.1056411 | 0 |
| George W Bush | 0.763496 | 0.76497 | 0.70087 | 0.860304 | 0.695752 | 0.753928 | 0.782937 | 0.703 | 0.696262 | 0.72753 | 0.703959 | 0.70322 | 0.738019 | 0.0497424 | 0 |
| Gillian Anderson | 0.683061 | 0.636285 | 0.736585 | 0.751096 | 0.566862 | 0.571995 | 0.464554 | 0.65979 | 0.6432 | 0.505972 | 0.803902 | 0.716264 | 0.6449638 | 0.102267 | 0 |
| Gisele Bundchen | 0.71884 | 0.73508 | 0.741054 | 0.564005 | 0.712285 | 0.683072 | 0.72466 | 0.701555 | 0.746682 | 0.445164 | 0.760802 | 0.725769 | 0.6882473 | 0.0917693 | 0 |
| Gordon Brown | 0.623533 | 0.539698 | 0.755416 | 0.729686 | 0.69368 | 0.777135 | 0.462895 | 0.758004 | 0.806568 | 0.779783 | 0.454349 | 0.751613 | 0.6776967 | 0.1264301 | 0 |
| Gwyneth Paltrow | 0.594934 | 0.573376 | 0.580449 | | 0.46417 | 0.491055 | 0.55993 | 0.572292 | 0.479256 | 0.571862 | 0.577613 | 0.660889 | 0.5568933 | 0.0574006 | 1 |
| Halle Berry | 0.637229 | 0.490001 | 0.601546 | 0.596657 | 0.527848 | 0.442845 | 0.530858 | 0.555292 | 0.811659 | 0.539527 | 0.557895 | 0.526139 | 0.5681247 | 0.0922815 | 0 |
| Harrison Ford | 0.637999 | 0.642134 | 0.578723 | 0.639128 | 0.649942 | 0.512274 | 0.433407 | 0.555357 | 0.505302 | 0.660958 | 0.399654 | 0.464901 | 0.5566483 | 0.0922746 | 0 |
| Hilary Swank | 0.635818 | 0.605148 | 0.66872 | 0.657879 | 0.611395 | 1 | 0.766936 | 0.705436 | 0.678139 | 0.529507 | 0.530684 | 0.662888 | 0.6710458 | 0.1234948 | 0 |
| Holly Hunter | 0.439178 | 0.632781 | 0.62269 | 0.610816 | 0.484717 | 0.739255 | 0.639462 | 0.634869 | 0.552461 | 0.64739 | 0.622199 | 0.572164 | 0.5998318 | 0.0790848 | 0 |
| Hugh Grant | 0.740996 | 0.785429 | 0.506229 | 0.702984 | 0.673371 | 0.612603 | 0.820039 | 0.691733 | 0.659682 | 0.78431 | 0.658854 | 0.745081 | 0.6984426 | 0.086229 | 0 |
| Hugh Jackman | 0.617198 | 0.743526 | 0.742368 | 0.487568 | 0.607832 | 0.670893 | 0.650491 | 0.46973 | 0.167695 | 0.51866 | 0.65164 | 0.61934 | 0.5789118 | 0.1565596 | 0 |
| Hugh Laurie | 0.653108 | 0.647776 | 0.626791 | 0.599766 | 0.641016 | 0.645312 | 0.647191 | 0.772171 | 0.829625 | 0.694381 | 0.680178 | 0.709947 | 0.6789385 | 0.065229 | 0 |
| Jack Nicholson | 0.382475 | 0.418357 | 0.449911 | 0.482266 | 0.534403 | 0.453452 | 0.559054 | 0.389872 | 0.521872 | 0.515182 | 0.475338 | 0.532818 | 0.47625 | 0.058692 | 0 |
| James Franco | 0.425142 | 0.776853 | 0.631563 | 0.752841 | 0.763422 | 0.736134 | 0.794875 | 0.546834 | 0.746763 | 0.677084 | 0.620097 | 0.6949 | 0.6805423 | 0.1092526 | 0 |
| James Gandolfini | 0.581967 | 0.525608 | 0.642251 | 0.547812 | 0.591486 | 0.709183 | 0.591925 | 0.600094 | 0.412003 | 0.516167 | 0.682936 | 0.738705 | 0.5950114 | 0.0907626 | 0 |
| James Spader | 0.543228 | 0.674082 | 0.681757 | 0.524026 | 0.692032 | 0.686071 | 0.864355 | 0.729644 | 0.47821 | 0.756507 | 0.882356 | 0.899287 | 0.7009629 | 0.1381828 | 0 |
| Jared Leto | 0.675436 | 0.638038 | 0.499433 | 0.688266 | 0.625438 | 0.714344 | 0.67498 | 0.667349 | 0.650181 | 0.661406 | 0.705081 | 0.638433 | 0.6531988 | 0.0554062 | 0 |
| Jason Bateman | 0.704708 | 0.564005 | 0.788318 | 0.660347 | 0.671934 | 0.746563 | 0.654643 | 0.582325 | 0.709324 | 0.543718 | 0.603372 | 0.54536 | 0.6478848 | 0.0808586 | 0 |
| Jason Statham | 0.644712 | 0.608763 | 0.704784 | 0.748411 | 0.522602 | 0.740093 | 0.551134 | 0.676903 | 0.655327 | 0.559004 | 0.685075 | 0.749798 | 0.6538838 | 0.0789576 | 0 |
| Javier Bardem | 0.437026 | 0.57726 | 0.3056 | 0.590041 | 0.556634 | 0.618219 | 0.500026 | 0.532856 | 0.596177 | 0.603905 | 0.524141 | 0.520706 | 0.5302159 | 0.0874892 | 0 |
| Jay Leno | 0.700698 | 0.782666 | 0.657462 | 0.547856 | 0.592898 | 0.510676 | 0.643552 | 0.810438 | 0.782217 | 0.562548 | 0.685072 | 0.664178 | 0.6616884 | 0.0972505 | 0 |
| Jeff Bridges | 0.444181 | 0.439582 | 0.668385 | 0.390556 | 0.586275 | 0.674588 | 0.574998 | 0.638639 | 0.505924 | 0.624382 | 0.571528 | 0.49565 | 0.551224 | 0.0948788 | 0 |
| Jennifer Aniston | 0.649713 | 0.616515 | 0.558842 | 0.574306 | 0.470084 | 0.566515 | 0.522132 | 0.596076 | 0.570884 | 0.538092 | 0.531087 | 0.522725 | 0.5597476 | 0.0478009 | 0 |
| Jennifer Lopez | 0.557728 | 0.637695 | 0.461859 | 0.539652 | 0.615032 | 0.561533 | 0.651795 | 0.511514 | 0.638589 | 0.666704 | 0.469865 | 0.563493 | 0.5729549 | 0.0697919 | 0 |
| Jennifer Love Hewitt | 0.677908 | 0.727579 | 0.758366 | 0.686799 | 0.562429 | 0.712787 | 0.671626 | 0.619033 | 0.616171 | 0.6497 | 0.70666 | 0.645912 | 0.6695808 | 0.0545251 | 0 |
| Jeri Ryan | 0.909695 | 0.684204 | 0.781969 | 0.84631 | 0.840124 | 0.892763 | 0.890319 | 0.889726 | 0.800684 | 0.828978 | 0.775984 | 0.738134 | 0.8232408 | 0.0694768 | 0 |
| Jerry Seinfeld | 0.801466 | 0.561622 | 0.780811 | 0.802049 | 0.588223 | 0.815166 | 0.71656 | 0.769483 | 0.784056 | 0.675703 | 0.687991 | 0.542974 | 0.7105087 | 0.0992607 | 0 |
| Jessica Alba | 0.585871 | 0.729589 | 0.667623 | 0.625936 | 0.665769 | 0.562694 | 0.758251 | 0.563987 | 0.743949 | 0.776294 | 0.499499 | 0.611447 | 0.6492424 | 0.0890509 | 0 |
| Jessica Simpson | 0.36198 | 0.61986 | 0.467239 | 0.479299 | 0.295518 | 0.68024 | 0.59449 | 0.661491 | 0.633427 | 0.528548 | 0.568426 | 0.710758 | 0.5501063 | 0.1285641 | 0 |
| Jimmy Carter | 0.622635 | | 0.780779 | 0.842069 | 0.821844 | 0.824774 | 0.620095 | 0.877391 | 0.866949 | 0.60771 | 0.631513 | 0.852582 | 0.7589401 | 0.1127426 | 1 |
| Joaquin Phoenix | 0.711661 | 0.606802 | 0.671823 | 0.832725 | 0.702408 | 0.764465 | 0.61926 | 0.76383 | 0.808864 | 0.747979 | 0.681368 | 0.757921 | 0.7224255 | 0.0699396 | 0 |
| Jodie Foster | 0.721219 | 0.7475 | 0.499229 | 0.698264 | 0.701499 | 0.712433 | 0.45937 | 0.451717 | 0.720577 | 0.56707 | 0.57984 | 0.650545 | 0.6257719 | 0.1093713 | 0 |
| John Lennon | 0.666874 | 0.596898 | 0.548881 | 0.450866 | 0.538858 | 0.569806 | 0.55031 | 0.522755 | 0.468027 | 0.557761 | 0.560732 | 0.597721 | 0.5524574 | 0.0572446 | 0 |
| John Malkovich | 0.586192 | 0.61687 | 0.558532 | 0.531528 | 0.576747 | 0.582281 | 0.584934 | 0.494718 | 0.437592 | 0.536914 | 0.583082 | 0.600707 | 0.5575081 | 0.0504754 | 0 |
| John Travolta | 0.630103 | 0.482539 | 0.6911 | 0.668693 | 0.67492 | 0.65338 | 0.680462 | 0.707312 | 0.694335 | 0.685447 | 0.469056 | 0.455653 | 0.6244167 | 0.0959177 | 0 |
| Johnny Depp | 0.634816 | 0.53218 | 0.390254 | 0.711562 | 0.678458 | 0.563093 | 0.418553 | 0.542178 | 0.676575 | 0.614127 | 0.60891 | 0.411798 | 0.5652087 | 0.1099888 | 0 |
| Julia Roberts | 0.572281 | 0.404361 | 0.649738 | 0.706292 | 0.529735 | 0.550658 | 0.59989 | 0.606391 | 0.66812 | 0.423181 | 0.563499 | 0.658214 | 0.5776967 | 0.0928756 | 0 |
| Julia Stiles | 0.677278 | 0.678722 | 0.691298 | 0.764382 | 0.637386 | 0.720238 | 0.721131 | 0.747607 | 0.662448 | 0.623402 | 0.677136 | 0.674131 | 0.6895966 | 0.0419994 | 0 |
| Karl Rove | 0.597768 | 0.62859 | 0.543189 | 0.760007 | 0.615296 | 0.716248 | 0.691478 | 0.712964 | 0.738843 | 0.726306 | 0.735328 | | 0.6787288 | 0.0706346 | 1 |
| Kate Moss | 0.660572 | 0.526459 | 0.541396 | 0.499032 | 0.489238 | 0.661969 | 0.577321 | 0.682678 | 0.523767 | 0.464039 | 0.668417 | 0.71319 | 0.5840065 | 0.0878453 | 0 |
| Kate Winslet | 0.505241 | 0.55559 | 0.735442 | 0.661241 | 0.581885 | 0.643226 | 0.623863 | 0.504765 | 0.679617 | 0.612079 | 0.652219 | 0.530747 | 0.6071596 | 0.0727853 | 0 |
| Katherine Heigl | 0.602316 | 0.655591 | 0.655599 | 0.594925 | 0.481655 | 0.637821 | 0.69381 | 0.672954 | 0.648038 | 0.449792 | 0.451768 | 0.609816 | 0.5961738 | 0.0866208 | 0 |
| Katie Couric | 0.679285 | 0.553945 | 0.497752 | 0.653103 | 0.541635 | 0.662916 | 0.660551 | 0.524801 | 0.531995 | 0.620146 | 0.469917 | 0.678803 | 0.5895708 | 0.0770034 | 0 |
| Keanu Reeves | 0.486541 | 0.681395 | 0.661658 | 0.621432 | 0.617793 | 0.583967 | 0.75722 | 0.684632 | 0.675716 | 0.648864 | 0.65008 | 0.459555 | 0.6274044 | 0.0839669 | 0 |
| Keira Knightley | 0.519671 | 0.64514 | 0.679708 | 0.566536 | 0.631165 | 0.52347 | 0.511808 | 0.498213 | 0.610621 | 0.531595 | 0.588407 | 0.664028 | 0.5808635 | 0.0643581 | 0 |
| Kevin Bacon | 0.665938 | 0.5976 | 0.756516 | 0.482806 | 0.450294 | 0.776114 | 0.64195 | 0.732084 | 0.490814 | 0.482652 | 0.764911 | 0.604426 | 0.6205088 | 0.1217596 | 0 |
| Kiefer Sutherland | 0.613799 | 0.745324 | 0.724487 | 0.518287 | 0.725859 | 0.723034 | 0.654906 | 0.639231 | 0.53428 | 0.722298 | | 0.759575 | 0.6691891 | 0.0843189 | 1 |
| Kim Basinger | 0.739887 | 0.583337 | 0.563315 | 0.590472 | 0.445308 | 0.636634 | 0.789695 | 0.768839 | 0.702093 | 0.555187 | 0.661641 | 0.596349 | 0.6360631 | 0.1007619 | 0 |
| Lance Armstrong | 0.660213 | 0.598256 | 0.652167 | 0.73827 | 0.695504 | 0.421466 | 0.746474 | 0.772683 | 0.632942 | 0.791945 | 0.690691 | 0.666334 | 0.6722454 | 0.0980067 | 0 |
| Leelee Sobieski | 0.694742 | 0.816803 | 0.823565 | 0.752573 | 0.757404 | 0.833014 | 0.653269 | 0.827142 | 0.655663 | 0.761983 | 0.835914 | 0.749951 | 0.7635019 | 0.0670313 | 0 |
| Leonardo DiCaprio | 0.659183 | 0.441894 | 0.624912 | 0.635658 | 0.74799 | 0.565059 | 0.517259 | 0.5325 | 0.588871 | 0.517152 | 0.64891 | 0.671132 | 0.5958767 | 0.0848271 | 0 |
| Liam Neeson | 0.646812 | 0.633662 | 0.708498 | 0.568096 | 0.681663 | 0.577107 | 0.520587 | 0.658691 | 0.725704 | 0.640073 | 0.568151 | 0.525101 | 0.6211788 | 0.0685614 | 0 |
| Lindsay Lohan | 0.594572 | | 0.378979 | 0.520624 | 0.563367 | 0.501828 | 0.892247 | 0.461303 | 0.479485 | 0.459196 | 0.424853 | 0.44244 | 0.5198995 | 0.1378693 | 1 |
| Liv Tyler | 0.531367 | 0.608133 | 0.62195 | 0.70461 | 0.771853 | 0.67265 | 0.666001 | 0.558504 | 0.571059 | 0.59001 | 0.595152 | 0.64234 | 0.6278024 | 0.0678026 | 0 |
| Lucy Liu | 0.722804 | 0.567013 | 0.564491 | 0.81461 | 0.53036 | 0.745393 | 0.477985 | 0.716601 | 0.58707 | 0.828377 | 0.824923 | 0.636525 | 0.6680127 | 0.1231033 | 0 |
| Mariah Carey | 0.731432 | 0.68775 | 0.60316 | 0.596809 | 0.68676 | 0.664414 | 0.734509 | 0.768768 | 0.650232 | 0.737163 | 0.774223 | 0.503146 | 0.6781972 | 0.0803063 | 0 |
| Mark Ruffalo | 0.652454 | 0.835087 | 0.75887 | 0.700598 | 0.771809 | 0.698751 | 0.81186 | 0.673484 | 0.592717 | 0.370553 | 0.68519 | 0.764512 | 0.6929904 | 0.1229137 | 0 |
| Martha Stewart | 0.69516 | 0.713421 | 0.583627 | 0.689492 | 0.468522 | 0.562364 | 0.650901 | 0.581827 | 0.646723 | 0.742313 | 0.656834 | 0.59699 | 0.6323478 | 0.0768347 | 0 |
| Matt Damon | 0.605826 | 0.833073 | 0.462596 | 0.772838 | 0.706607 | 0.750045 | 0.687756 | 0.650985 | 0.753418 | 0.355191 | 0.649834 | 0.60338 | 0.6526291 | 0.1348449 | 0 |
| Matthew Broderick | 0.531902 | 0.621154 | 0.680469 | 0.73809 | 0.723527 | 0.463848 | 0.54339 | 0.703499 | 0.71635 | 0.742982 | 0.651208 | 0.555872 | 0.6393576 | 0.0945345 | 0 |
| Meg Ryan | 0.646257 | 0.637385 | 0.530135 | 0.467126 | 0.602208 | 0.547937 | 0.460945 | 0.584665 | 0.682628 | 0.55816 | 0.737913 | 0.644556 | 0.5916596 | 0.0837254 | 0 |
| Mel Gibson | 0.608311 | 0.573198 | 0.742234 | 0.618667 | 0.671992 | 0.550067 | 0.633677 | 0.535463 | 0.548817 | 0.560985 | 0.834634 | 0.760349 | 0.6365328 | 0.0968751 | 0 |
| Meryl Streep | 0.708824 | 0.627683 | 0.589835 | 0.680043 | 0.629484 | 0.747272 | 0.763052 | 0.63298 | 0.723417 | 0.508614 | 0.713112 | | 0.6658469 | 0.0761805 | 1 |
| Michael Bloomberg | 0.655565 | 0.777927 | 0.666434 | 0.680515 | 0.523698 | 0.79765 | 0.715739 | 0.679246 | 0.597035 | 0.765573 | 0.704366 | 0.693391 | 0.6880949 | 0.0763445 | 0 |
| Michael Douglas | 0.684409 | 0.636338 | 0.544088 | 0.565746 | 0.524801 | 0.478324 | 0.360044 | 0.657498 | 0.721768 | 0.580109 | 0.688755 | 0.497398 | 0.5782732 | 0.1053361 | 0 |
| Michelle Trachtenberg | 0.757804 | 0.787942 | 0.732372 | 0.731548 | 0.715863 | 0.801148 | 0.782728 | 0.807934 | 0.463949 | 0.767924 | 0.580234 | 0.688069 | 0.7181263 | 0.1013418 | 0 |
| Michelle Wie | 0.446794 | 0.515119 | 0.325017 | 0.361628 | 0.350148 | 0.519392 | 0.619141 | 0.509696 | 0.503355 | 0.40307 | 0.477912 | 0.424744 | 0.454668 | 0.0857288 | 0 |
| Mickey Rourke | 0.644291 | 0.60836 | 0.354038 | 0.480332 | 0.581321 | 0.601638 | 0.24453 | 0.574935 | 0.548016 | 0.442449 | 0.496084 | 0.39321 | 0.4974337 | 0.1202403 | 0 |
| Mikhail Gorbachev | 0.540927 | 0.465522 | 0.498655 | 0.635279 | 0.488315 | 0.495847 | 0.549906 | 0.639086 | 0.488349 | 0.484675 | 0.440145 | 0.433002 | 0.513309 | 0.0670628 | 0 |
| Miley Cyrus | 0.602754 | 0.443128 | 0.508179 | 0.388745 | 0.714108 | 0.492657 | 0.628173 | 0.635725 | 0.55195 | 0.601641 | 0.658931 | 0.66722 | 0.5744343 | 0.0985112 | 0 |
| Milla Jovovich | 0.629811 | 0.489954 | 0.605204 | 0.476031 | 0.63441 | 0.642002 | 0.64788 | 0.592547 | 0.624062 | 0.609039 | 0.481224 | 0.336179 | 0.5640286 | 0.0965463 | 0 |
| Minnie Driver | 0.694938 | 0.459659 | 0.722608 | 0.667222 | 0.707468 | 0.611732 | 0.565859 | 0.642468 | 0.741452 | 0.61685 | 0.622787 | 0.742723 | 0.6496472 | 0.0823077 | 0 |
| Monica Bellucci | 0.639222 | 0.471589 | 0.646792 | 0.672868 | 0.575079 | 0.513396 | 0.641672 | 0.610062 | 0.693278 | 0.568772 | 0.59187 | 0.523869 | 0.5957058 | 0.0679591 | 0 |
| Morgan Freeman | 0.469412 | 0.531904 | 0.647417 | 0.794249 | 0.667888 | 0.715725 | 0.770823 | 0.881459 | 0.752542 | 0.723561 | 0.809771 | 0.710833 | 0.7062987 | 0.1157204 | 0 |
| Nathan Lane | 0.659356 | 0.839734 | 0.64894 | 0.669568 | 0.617909 | 0.723771 | 0.710318 | 0.739498 | 0.706954 | 0.615244 | 0.649098 | 0.601126 | 0.681793 | 0.066948 | 0 |
| Nicolas Cage | 0.601569 | 0.503868 | 0.732466 | 0.641114 | 0.562262 | 0.626182 | 0.590554 | 0.585809 | 0.602734 | 0.7065 | 0.6236 | 0.728887 | 0.6254621 | 0.0685038 | 0 |
| Nicolas Sarkozy | 0.703919 | 0.860906 | 0.751019 | 0.776949 | 0.730017 | 0.490729 | 0.411803 | 0.71873 | 0.658456 | 0.725237 | 0.704405 | 0.692672 | 0.6854035 | 0.1215106 | 0 |
| Nicole Kidman | 0.411826 | 0.429621 | 0.576964 | 0.479542 | 0.48602 | 0.415264 | 0.711011 | 0.775254 | 0.541299 | 0.555405 | 0.57784 | 0.5216 | 0.5401372 | 0.1122922 | 0 |
| Nicole Richie | 0.660464 | 0.645651 | 0.566879 | 0.543565 | 0.574691 | 0.749325 | 0.686716 | 0.611434 | 0.703743 | 0.604523 | 0.675735 | 0.522395 | 0.6287601 | 0.0695612 | 0 |
| Noah Wyle | 0.732046 | 0.74031 | 0.85565 | 0.875143 | 0.860835 | 0.680746 | 0.8118 | 0.665267 | 0.600905 | 0.722125 | 0.688978 | 0.861508 | 0.7579428 | 0.0923065 | 0 |
| Oprah Winfrey | 0.651844 | 0.685554 | 0.854823 | 0.680982 | 0.570116 | 0.447707 | 0.829579 | 0.691764 | 0.735201 | 0.713628 | 0.766655 | 0.408819 | 0.6697227 | 0.136146 | 0 |
| Orlando Bloom | 0.546114 | 0.564778 | 0.503325 | 0.539812 | 0.414353 | 0.626024 | 0.543566 | 0.531597 | 0.651011 | 0.669293 | 0.427273 | 0.474523 | 0.5409724 | 0.0805043 | 0 |
| Owen Wilson | 0.716024 | 0.511014 | 0.745049 | 0.583309 | 0.661174 | 0.618502 | 0.557394 | 0.636894 | 0.695024 | 0.684244 | 0.679307 | 0.506732 | 0.6328889 | 0.0786788 | 0 |
| Paul Rudd | 0.782132 | 0.763224 | 0.460243 | 0.726308 | 0.640288 | 0.781597 | 0.57115 | 0.648113 | 0.689388 | 0.756854 | 0.724125 | 0.609409 | 0.6794026 | 0.0979236 | 0 |
| Philip Seymour Hoffman | 0.708695 | 0.656642 | 0.434776 | 0.506059 | 0.535893 | 0.538717 | 0.568312 | 0.646075 | 0.459388 | 0.652764 | 0.718318 | 0.622062 | 0.5873084 | 0.0940251 | 0 |
| Quincy Jones | 0.468901 | 0.647922 | 0.704995 | 0.732646 | 0.878202 | 0.697065 | 0.675737 | 0.772579 | 0.502067 | 0.858733 | 0.834307 | 0.88407 | 0.7214353 | 0.1367182 | 0 |
| Rachael Ray | 0.51705 | 0.456792 | 0.558182 | 0.689281 | 0.420971 | 0.667748 | 0.573692 | 0.51455 | 0.525848 | 0.79075 | 0.539966 | 0.515148 | 0.5641648 | 0.1040202 | 0 |
| Ralph Nader | 0.625233 | 0.818052 | 0.813525 | 0.745688 | 0.768566 | 0.766585 | 0.808889 | 0.752584 | 0.690681 | 0.660355 | 0.707523 | 0.797911 | 0.7462993 | 0.063128 | 0 |
| Ray Romano | 0.498654 | 0.59201 | 0.712582 | 0.674866 | 0.70721 | 0.608818 | 0.540354 | 0.230836 | 0.621949 | 0.63297 | 0.706891 | 0.711695 | 0.6032363 | 0.1366384 | 0 |
| Reese Witherspoon | 0.606051 | 0.681884 | 0.627192 | 0.729837 | 0.862183 | 0.709151 | 0.659511 | 0.737055 | 0.680544 | 0.659173 | 0.30549 | 0.694255 | 0.6626938 | 0.1299513 | 0 |
| Renee Zellweger | 0.52695 | 0.75867 | 0.567275 | 0.551084 | 0.71311 | 0.58871 | 0.758911 | 0.531418 | 0.67397 | 0.615661 | 0.737957 | 0.548931 | 0.6310539 | 0.0917698 | 0 |
| Ricky Martin | 0.539931 | 0.357511 | 0.519734 | 0.422149 | 0.438687 | 0.68303 | 0.556719 | 0.65405 | 0.54037 | 0.553896 | 0.499896 | 0.490838 | 0.5214009 | 0.0912927 | 0 |
| Robert De Niro | 0.72599 | 0.338548 | 0.425227 | 0.577459 | 0.602061 | 0.654448 | 0.648246 | 0.375158 | 0.770716 | 0.625378 | 0.353964 | 0.450148 | 0.5456119 | 0.1502771 | 0 |
| Robert Downey Jr | 0.575257 | 0.566798 | 0.628948 | 0.637706 | 0.592312 | 0.576926 | 0.47093 | 0.693349 | 0.598371 | 0.622922 | 0.634464 | 0.471088 | 0.5890893 | 0.0652579 | 0 |
| Robert Gates | 0.742221 | 0.692127 | 0.644617 | 0.716576 | 0.715358 | 0.68959 | 0.698263 | 0.643401 | 0.701081 | 0.601584 | 0.750562 | 0.659096 | 0.687873 | 0.0435776 | 0 |
| Rod Stewart | 0.624783 | 0.621442 | 0.520496 | 0.450047 | 0.395357 | 0.687431 | 0.567481 | 0.357747 | 0.339853 | 0.666939 | 0.350534 | 0.400032 | 0.4985118 | 0.1314006 | 0 |
| Rosario Dawson | 0.651913 | 0.7056 | 0.681831 | 0.575285 | 0.502571 | 0.725262 | 0.596073 | 0.78806 | 0.529217 | 0.833835 | 0.637186 | 0.62895 | 0.6546486 | 0.0989907 | 0 |
| Rosie Perez | 0.76075 | 0.730225 | 0.534708 | 0.505924 | 0.665343 | 0.420696 | 0.512416 | 0.540504 | 0.510285 | 0.589584 | 0.551234 | 0.706975 | 0.5857203 | 0.1057763 | 0 |
| Russell Crowe | 0.725968 | 0.769002 | 0.605269 | 0.826105 | 0.719752 | 0.817185 | 0.604273 | 0.525303 | 0.681496 | 0.810915 | 0.628542 | 0.641967 | 0.6963148 | 0.0977347 | 0 |
| Ryan Seacrest | 0.675864 | 0.732181 | 0.595895 | 0.654167 | 0.614749 | 0.707708 | 0.718968 | 0.58549 | 0.651672 | 0.789437 | 0.786058 | 0.627079 | 0.6782723 | 0.0692486 | 0 |
| Salma Hayek | 0.66814 | 0.552543 | 0.599343 | 0.630715 | 0.810265 | 0.681466 | 0.623734 | 0.723766 | 0.655003 | 0.623317 | 0.631501 | 0.581013 | 0.6484005 | 0.0683434 | 0 |
| Sania Mirza | 0.517776 | 0.574162 | 0.515646 | 0.517686 | 0.528713 | 0.560271 | 0.558861 | 0.477795 | 0.598509 | 0.574831 | 0.330128 | 0.609908 | 0.5303572 | 0.0738362 | 0 |
| Sarah Chalke | 0.735902 | 0.686772 | 0.64648 | 0.664532 | 0.756491 | 0.832279 | 0.751829 | 0.772773 | 0.738196 | 0.76056 | 0.670939 | 0.657733 | 0.7228738 | 0.0569979 | 0 |
| Sarah Palin | 0.714038 | 0.704252 | 0.68635 | 0.635649 | 0.737135 | 0.775819 | 0.633663 | 0.606977 | 0.656911 | 0.556177 | 0.788061 | 0.698489 | 0.6827934 | 0.0684038 | 0 |
| Scarlett Johansson | 0.729593 | 0.568248 | 0.641179 | 0.738195 | 0.557849 | 0.605102 | 0.592908 | 0.582203 | 0.6255 | 0.600257 | 0.677218 | 0.718756 | 0.6364173 | 0.064362 | 0 |
| Seth Rogen | 0.631706 | 0.602577 | 0.561364 | 0.563453 | 0.442597 | 0.647831 | 0.658235 | 0.509628 | 0.580104 | 0.606582 | 0.440175 | 0.213187 | 0.5381199 | 0.1250891 | 0 |
| Shahrukh Khan | 0.473072 | 0.802281 | 0.548701 | 0.690646 | 0.732385 | 0.63884 | 0.708495 | 0.636544 | 0.676851 | 0.75117 | 0.440166 | 0.618449 | 0.6431333 | 0.1096845 | 0 |
| Shakira | 0.637999 | 0.725947 | 0.416806 | 0.675399 | 0.671533 | 0.46929 | 0.617004 | 0.462519 | 0.749859 | 0.53101 | 0.540926 | 0.591613 | 0.5908254 | 0.1077502 | 0 |
| Shania Twain | 0.687675 | 0.702884 | 0.517282 | 0.713025 | 0.679167 | 0.562846 | 0.67009 | 0.669536 | 0.809347 | 0.65643 | 0.646377 | 0.733768 | 0.6707023 | 0.0752898 | 0 |
| Sharon Stone | 0.690514 | 0.444317 | 0.58437 | 0.813481 | 0.651041 | 0.536812 | 0.595076 | 0.63044 | 0.585423 | 0.727313 | 0.643629 | 0.751412 | 0.637819 | 0.0999473 | 0 |
| Shinzo Abe | 0.81754 | 0.979746 | 0.732508 | 0.812708 | 0.743254 | 0.793287 | 0.386883 | 0.780702 | 0.77008 | 0.711296 | 0.742922 | 0.476537 | 0.7289553 | 0.1560063 | 0 |
| Sigourney Weaver | 0.640804 | 0.452155 | 0.601154 | 0.627624 | 0.769449 | 0.660973 | 0.624109 | 0.556413 | 0.750314 | 0.471951 | 0.629129 | 0.480241 | 0.6053597 | 0.1013707 | 0 |
| Silvio Berlusconi | 0.546127 | 0.750818 | 0.708573 | 0.682386 | 0.732785 | 0.693668 | 0.66069 | 0.551196 | 0.726385 | 0.627601 | 0.666738 | 0.68974 | 0.6697256 | 0.0657799 | 0 |
| Simon Cowell | 0.550579 | 0.599559 | 0.632662 | 0.672458 | 0.632004 | 0.66533 | 0.729972 | 0.552758 | 0.677956 | 0.724143 | 0.551147 | 0.609742 | 0.6331925 | 0.0631921 | 0 |
| Stephen Colbert | 0.722838 | 0.551537 | 0.588623 | 0.663332 | 0.508918 | 0.639161 | 0.478722 | 0.694806 | 0.61219 | 0.671792 | 0.650901 | 0.596639 | 0.6149549 | 0.0740549 | 0 |
| Stephen Fry | 0.787591 | 0.836129 | 0.84556 | 0.724378 | 0.783141 | 0.782845 | 0.605429 | 0.753006 | 0.556808 | 0.678802 | 0.743311 | 0.641918 | 0.7282432 | 0.0905262 | 0 |
| Steve Carell | 0.644549 | 0.584612 | 0.566412 | 0.705378 | 0.698854 | 0.715445 | 0.622925 | 0.777264 | 0.665028 | 0.650915 | 0.741454 | 0.670793 | 0.6703024 | 0.0618168 | 0 |
| Steve Martin | 0.760097 | 0.629282 | 0.631227 | 0.663197 | 0.482497 | 0.828902 | 0.585559 | 0.592737 | 0.766697 | 0.580532 | 0.693667 | 0.630113 | 0.6537089 | 0.0958945 | 0 |
| Steven Spielberg | 0.519842 | 0.625692 | 0.606551 | 0.430115 | 0.650256 | 0.643456 | 0.61866 | 0.624977 | 0.704639 | 0.617677 | 0.652878 | 0.620544 | 0.6096073 | 0.0705483 | 0 |
| Susan Sarandon | 0.474915 | 0.480424 | 0.586694 | 0.503622 | 0.468999 | 0.556259 | 0.496676 | 0.55057 | 0.504615 | 0.561431 | 0.525612 | 0.512243 | 0.518505 | 0.0378125 | 0 |
| Tiger Woods | 0.738921 | 0.740937 | 0.67543 | 0.715656 | 0.628377 | 0.707402 | 0.582632 | 0.795894 | 0.771433 | 0.6913 | 0.811799 | 0.686673 | 0.7122045 | 0.0662383 | 0 |
| Tina Fey | 0.646739 | 0.45899 | 0.536409 | 0.52361 | 0.542046 | 0.461035 | 0.506898 | 0.571403 | 0.527792 | 0.567842 | 0.665034 | 0.531439 | 0.5449364 | 0.0624482 | 0 |
| Tom Cruise | 0.46837 | 0.492885 | 0.472296 | 0.482457 | 0.482787 | 0.401494 | 0.601154 | 0.479275 | 0.432814 | 0.557114 | 0.514882 | 0.446218 | 0.4859788 | 0.0533586 | 0 |
| Tom Hanks | 0.449361 | 0.513338 | 0.529962 | 0.488652 | 0.446293 | 0.573874 | 0.540535 | 0.421726 | 0.405742 | 0.565004 | 0.562641 | 0.46322 | 0.4966957 | 0.0588806 | 0 |
| Tony Blair | 0.61401 | 0.772579 | 0.398032 | 0.728808 | 0.818511 | 0.596895 | 0.605131 | 0.604153 | | 0.573404 | 0.761065 | 0.634771 | 0.6461235 | 0.1181608 | 1 |
| Tracy Morgan | 0.700349 | 0.659876 | 0.761034 | 0.650591 | 0.664479 | 0.892561 | 0.792505 | 0.717087 | 0.679639 | 0.880195 | 0.756379 | 0.64438 | 0.7332563 | 0.0857334 | 0 |
| Ty Pennington | 0.494786 | 0.571532 | 0.737795 | 0.839584 | 0.798748 | 0.795002 | 0.775392 | 0.789025 | 0.729404 | 0.797468 | 0.781796 | 0.673122 | 0.7319712 | 0.1033206 | 0 |
| Tyra Banks | 0.525826 | 0.714466 | 0.497446 | 0.619808 | 0.667465 | 0.531543 | 0.63962 | 0.608895 | 0.587997 | 0.674907 | 0.557008 | 0.567342 | 0.5993603 | 0.0667871 | 0 |
| Uma Thurman | 0.69697 | 0.676593 | 0.575628 | 0.72174 | 0.576884 | 0.31989 | 0.754368 | 0.318104 | 0.37941 | 0.612202 | 0.547135 | 0.794271 | 0.5810996 | 0.1646023 | 0 |
| Victoria Beckham | 0.681382 | 0.344536 | 0.704427 | 0.449284 | 0.642157 | 0.32819 | 0.62531 | 0.359086 | 0.539318 | 0.446348 | 0.293005 | 0.36786 | 0.4817419 | 0.1499848 | 0 |
| Viggo Mortensen | 0.550554 | 0.652592 | 0.398061 | 0.541809 | 0.739163 | 0.436335 | 0.492802 | 0.641278 | 0.673959 | 0.629996 | 0.65859 | 0.350997 | 0.5638447 | 0.1224161 | 0 |
| Will Smith | 0.716066 | 0.754584 | 0.535005 | 0.762129 | 0.502164 | 0.688145 | 0.708417 | 0.723645 | 0.638558 | 0.50864 | 0.616773 | 0.787438 | 0.661797 | 0.100762 | 0 |
| William Macy | 0.471155 | 0.786344 | 0.873863 | 0.829706 | 0.792938 | 0.524911 | 0.642664 | 0.870347 | 0.835373 | 0.802724 | 0.619297 | 0.721046 | 0.730864 | 0.1358071 | 0 |
| Wilmer Valderrama | 0.814186 | 0.683494 | 0.609299 | 0.765876 | 0.803971 | 0.686288 | 0.739048 | 0.694501 | 0.73226 | 0.556542 | 0.705452 | 0.626839 | 0.7014797 | 0.0771109 | 0 |
| Zac Efron | 0.488339 | 0.573861 | 0.875273 | 0.775339 | 0.43939 | 0.597712 | 0.627175 | 0.559487 | 0.649103 | 0.533154 | 0.680965 | 0.700859 | 0.6250548 | 0.122075 | 0 |
| Zach Braff | 0.584213 | 0.470521 | 0.330893 | 0.661083 | 0.373798 | 0.635854 | 0.707014 | 0.598238 | 0.629367 | 0.723289 | 0.654587 | 0.675004 | 0.5869884 | 0.1278166 | 0 |
| | | | | | | | | | | | | Overall | **0.6357443** | **0.1135416** | 8 |

# Appendix C. Summary and Comparisons of Test Results

1. Summary of The Results of Testing on Mutant Test Cases

Average Scores

| | Bald | Bangs | Black Hair | Blond Hair | Brown Hair | Bushy Eyebr | Eyeglasses | Male | Mouth Open | Mustache | Beard | Pale Skin | Young | Overall |
|---|---|---|---|---|---|---|---|---|---|---|---|---|---|---|
| Tencent | 99.20 | 99.63 | 99.91 | 99.87 | 99.96 | 99.66 | 99.07 | 99.59 | 99.95 | 99.89 | 99.89 | 99.52 | 99.96 | 99.70 |
| Baidu | 93.22 | 95.18 | 96.20 | 95.23 | 97.30 | 91.85 | 93.71 | 92.89 | 96.34 | 94.40 | 97.02 | 93.02 | 95.38 | 94.75 |
| Face++ | 92.02 | 93.30 | 94.45 | 93.51 | 95.31 | 92.49 | 91.23 | 89.97 | 94.50 | 92.44 | 94.95 | 92.18 | 93.09 | 93.03 |
| SeetaFace | 80.13 | 81.18 | 86.55 | 81.39 | 90.20 | 67.93 | 73.82 | 72.31 | 88.58 | 78.99 | 86.49 | 76.93 | 79.24 | 80.32 |

Standard Deviation of Scores

| | Bald | Bangs | Black Hair | Blond Hair | Brown Hair | Bushy Eyebr | Eyeglasses | Male | Mouth Open | Mustache | Beard | Pale Skin | Young | Overall |
|---|---|---|---|---|---|---|---|---|---|---|---|---|---|---|
| Tencent | 2.92 | 1.39 | 0.49 | 0.69 | 0.57 | 1.21 | 2.69 | 1.28 | 0.51 | 0.53 | 1.01 | 2.34 | 0.33 | 1.51 |
| Baidu | 7.27 | 2.33 | 3.13 | 2.19 | 1.37 | 5.45 | 4.26 | 3.53 | 1.37 | 5.68 | 1.13 | 5.46 | 2.06 | 4.28 |
| Face++ | 3.17 | 2.21 | 2.05 | 1.92 | 1.41 | 2.34 | 3.48 | 3.15 | 1.12 | 2.40 | 1.46 | 2.94 | 1.86 | 2.80 |
| SeetaFace | 9.18 | 8.58 | 7.84 | 7.37 | 5.20 | 12.14 | 10.94 | 9.43 | 4.60 | 7.88 | 5.16 | 10.01 | 7.07 | 7.07 |

Number of Not Recognised

| | Bald | Bangs | Black Hair | Blond Hair | Brown Hair | Bushy Eyebr | Eyeglasses | Male | Mouth Open | Mustache | Beard | Pale Skin | Young | Overall |
|---|---|---|---|---|---|---|---|---|---|---|---|---|---|---|
| Tencent | 9 | 5 | 6 | 8 | 3 | 4 | 7 | 4 | 4 | 10 | 5 | 14 | 6 | 85 |
| Baidu | 3 | 1 | 1 | 1 | 1 | 1 | 1 | 1 | 1 | 3 | 2 | 3 | 1 | 20 |
| Face++ | 2 | 0 | 1 | 2 | 0 | 0 | 1 | 0 | 1 | 1 | 1 | 2 | 1 | 12 |
| SeetaFace | 11 | 16 | 9 | 9 | 6 | 10 | 10 | 12 | 7 | 13 | 13 | 16 | 11 | 143 |

2. Comparison of Test Results between Using and Without Datamorphisms

Average scores

| | With DM | Without DM |
|---|---|---|
| Tencent | 99.70 | 96.38 |
| Baidu | 94.75 | 84.50 |
| Face++ | 93.03 | 86.81 |
| SeetaFace | 80.32 | 63.57 |
| Correlation | 0.99 | |

StDev

| | With DM | Without DM |
|---|---|---|
| Tencent | 1.51 | 6.22 |
| Baidu | 4.28 | 11.85 |
| Face++ | 2.80 | 6.29 |
| SeetaFace | 7.07 | 11.35 |
| Correlation | 0.82 | |

Rate of No Recognision

| | With DM | Without DM | NoRec without DM |
|---|---|---|---|
| Tencent | 3.25 | 0.68 | 17 |
| Baidu | 0.77 | 0.12 | 3 |
| Face++ | 0.46 | 0.20 | 5 |
| SeetaFace | 5.47 | 0.32 | 8 |
| Correlation | 0.49 | | |